\documentclass[12pt]{article}
\usepackage[a4paper,margin=1in]{geometry}
\usepackage{setspace}
\usepackage[round]{natbib} 
\usepackage{amsmath,amsthm,epsfig,amsfonts, color, epstopdf, bm}
\usepackage{amsbsy}
\allowdisplaybreaks[4]
\usepackage{amssymb}
\usepackage{rotating,comment}
\usepackage{array}
\usepackage[font=footnotesize,labelfont=bf]{caption}
\usepackage[english]{babel}
\usepackage{graphicx} 
\usepackage{longtable}
\usepackage{booktabs} 
\usepackage{adjustbox}
\usepackage{subcaption}
\usepackage{xcolor}
\usepackage{url}
\usepackage{afterpage}
\usepackage{diagbox}

\usepackage{lipsum}
\usepackage{multirow}
\usepackage{authblk}
\usepackage[flushleft]{threeparttable}
\usepackage{hyperref}
\usepackage{pdflscape}
\usepackage{floatrow}

\def\sym#1{\ifmmode^{#1}\else\(^{#1}\)\fi}

\newfloatcommand{capbtabbox}{table}[][\FBwidth]

\def\marginnote#1{\setbox0=\vtop{\hsize4pc
\small\raggedright\noindent\baselineskip9pt \rightskip=0.5pc plus
1.5pc #1}\leavevmode \vadjust{\dimen0=\dp0
\kern-\ht0\hbox{\kern-4.00pc\box0}\kern-\dimen0}}
\def\lboxit#1{\vbox{\hrule\hbox{\vrule\kern6pt
\vbox{\kern6pt#1\kern6pt}\kern6pt\vrule}\hrule}}

\begin{document}
\title{Small Area Causal Inference}
\author{Yilin Li, Jiangshan Zhang and Jiming Jiang}
 \affil{University of California, Davis, USA\\
{\footnotesize Address of correspondence: Jiming Jiang, Department of Statistics and Data Science, University of California, Davis, One Shields Avenue, Davis, CA 95616, USA. Email: {jimjiang@ucdavis.edu}}
}
\date{}
\maketitle

\begin{abstract}

We develop a framework for small area causal inference (SACI). Unlike the traditional doubly robust (DR) estimation approach, the new framework is based on the empirical best prediction (EBP) method. EBP takes into consideration not only the bias but also the variance factor, which is often a more dominant component in the mean squared prediction error (MSPE). Under a linear mixed model for SACI, we establish consistency and asymptotic normality of the maximum likelihood estimator of the model parameters. More relevant to our main objective, we establish asymptotic optimality of the EBP in terms of minimizing the MSPE of the area-specific causal-effect (ASCE) predictor. Moreover, in view of the traditional interest in causal inference, we establish asymptotic normality of the EBP of the ASCE , which is used to construct prediction intervals for the ASCEs. We carry out extensive simulation studies to investigate finite-sample performance of the proposed EBP and its comparison with the traditional DR and other existing methods. Our simulation results show that the proposed EBP method has significant advantage over the comparing methods  in terms of MSPE and coverage probability and length of the resulting prediction interval. A real-data example from agricultural-animal genetics is discussed.

\vspace{2mm}

\noindent{\bf Keywords:} asymptotic optimality, empirical best prediction, linear mixed model, maximum likelihood, MSPE, prediction interval, random effects, small area
\end{abstract}

\section{Introduction}
\hspace{4mm}
Causal inference has received much attention in the current literature of statistical data science. Early ground-breaking work can be traced back to Neyman (1923) and Rubin (1974). The current literature has largely followed a framework for causal inference set up by the latter reference. A central piece of this framework is the so-called potential outcomes. In situations of causal inference, the observed outcomes are either under ``treatment'' or under ``control''. Here, the terms treatment and control are understood in the broader sense; in particular, they do not have to be associated with medical studies. Any factor of interest, be it scientific, medical, economical, social, or political, can be regarded as the treatment, as long as any population unit can be classified under one, and only one, of the two categories under the treatment at the same time. Extensions of the concept may include multiple (more than two) categories. See Section 4 for more details. Let $Y(1)$ and $Y(0)$ denote the potential outcomes of an individual under the treatment and control, respectively. The mean of the difference, $Y(1)-Y(0)$, over the population, that is,
\begin{equation}
d={\rm E}\{Y(1)-Y(0)\},\label{eq:intr_1}
\end{equation}
is called causal effect, which is a characteristic of primary interest in causal inference. Another important part of the observed data is the treatment assignment. This is denoted by $W$ so that $W=1$ if the sample unit belongs to the treatment group, and $W=0$ if the sampled unit belongs to the control group. The observed data are there $Y=Y(W)$ and $W$ for the sampled units. Note that the causal effect, $d$, involves several parts of unknowns. For the sampled units, one observes either $Y(1)$, if $W=1$, or $Y(0)$, if $W=0$, but not both. For the non-sampled units, neither $Y(1)$ nor $Y(0)$ are observed. The contribution of these unknowns to the mean in (\ref{eq:intr_1}) is what causal inference is about.

In traditional causal inference, the characteristic of interest is one population mean difference, that is, (\ref{eq:intr_1}), where the mean is over the entire population. Oftentimes, there is heterogeneity in the population, which is desirable to take into consideration. For example, in animal breeding, sex and line effects on the weight of lambs at weaning (e.g., Posbergh {\it et al.} 2023) are of interest for causal inference. However, such effects depend heavily on genetic factors such as sire (male parent), or combination of sire and dam (female parent). Each sire, or sire/dam combination, defines naturally a subpopulation, which we call small area, borrowing a term from the subject field small area estimation (SAE; e.g., Rao and Molina 2015, Jiang and Rao 2025). The effects of certain factors such as lines and sex on the weight at weaning of lambs, who are offspring of the sire, for example, are of interest of causal inference. See Section 6 of the supplementary material for more details. We call such an object of interest small area causal inference (SACI).

Within the field of causal inference, a method known as doubly robust (DR) estimation (e.g., Robins and Rotnitzky 1995, Bang and Robins 2005) is highly influential. The method combines propensity scores and outcome regression to estimate treatment, which is naturally suggested by two identities that express the causal effect in terms of expectations involving either the conditional mean function or the inverse propensity score. Here, the term DR refers to the fact that the estimator is unbiased provided that either the outcome model or the propensity model is correctly specified. Some recent work on SACI has followed this line of developments (e.g., Ito and Sugasawa 2026).

While unbiasedness is a desirable property, estimation accuracy depends not only on the bias but also on the variance. In fact, oftentimes, variance is the dominant factor in the mean squared error (MSE) compared to the bias. To see this, note that, in the classical i.i.d. situation, a consistent estimator, $\hat{\theta}$, of a parameter $\theta$ typically satisfies $\hat{\theta}-\theta=O_{\rm P}(n^{-1/2})$, or $(\hat{\theta}-\theta)^{2}=O_{\rm P}(n^{-1})$, where $n$ is the sample size. It follows that, under regularity conditions, one has ${\rm MSE}(\hat{\theta})={\rm E}(\hat{\theta}-\theta)^{2}=O(n^{-1})$. On the other hand, it is known that $\hat{\theta}$ satisfies ${\rm bias}(\hat{\theta})={\rm E}(\hat{\theta}-\theta)=O(n^{-1})$; thus, again under regularity conditions, we have
\begin{equation}
O(n^{-1})={\rm MSE}(\hat{\theta})=\{{\rm bias}(\hat{\theta})\}^{2}+{\rm var}(\hat{\theta})=O(n^{-2})+{\rm var}(\hat{\theta}),\label{eq:intr_2}
\end{equation}
which implies ${\rm var}(\hat{\theta})=O(n^{-1})$. In summary, we have ${\rm var}(\hat{\theta})=O(n^{-1})$ and $\{{\rm bias}(\hat{\theta})\}^{2}=O(n^{-2})$; therefore, variance dominates the MSE. A similar pattern also holds for prediction, where the MSE is replaced by the mean squared prediction error (MSPE), that the MSPE is dominated by the variance, not the bias. Of course, the above facts hold under the assumption that the underlying model is correctly specified. However, as shown by Jiang, Nguyen and Rao (2011), even when the underlying model is moderately misspecified, a realistic situation, variance may still be a more dominant factor than bias.

The point to make is that one should not focus only on the bias when considering the overall performance in causal inference; instead, it is more natural to focus on the MSE, or MSPE, as an overall performance measure. Under the MSPE measure, the best predictor (BP) of an unobserved random variable, $\xi$, is its conditional expectation given the observed data, $Y$, that is, $\tilde{\xi}={\rm E}(\xi|Y)$. Typically, the conditional expectation involves some unknown parameters, say, $\psi$, that is, $\tilde{\xi}={\rm E}_{\psi}(\xi|Y)$. It is then customary to replace $\psi$ by $\hat{\psi}$, a consistent estimator, leading to the empirical BP, or EBP, $\hat{\xi}={\rm E}_{\hat{\psi}}(\xi|Y)$. The BP approach is particularly attractive under the joint normality distribution of $\xi$ and $Y$, in which case the conditional expectation has a simple (and intuitive) expression (e.g., Jiang and Nguyen 2021, sec. B.1); see below.

In fact, the BP idea has been incorporated in Ranjbar, Salvati and Pacini (2023) in the context of SACI. However, instead of originating from the BP, the latter authors' starting point was a DR estimator (DRE) following the traditional line. The expression of the DRE involves unobserved (non-sampled) quantities. The authors then replaced the unobserved quantities with their EBP. We refer to the latter method as partial EBP. This is expected to produce an estimator that is suboptimal, compared with an approach along the EBP line from the very beginning, which is our approach. More recently, Reluga {\it et al.} (2026) proposed a CSAE-AIPW approach to area-level causal inference. The method estimates the area-specific causal effects by combining outcome imputation with augmented inverse-propensity weighting. First, a mixed-effects model is used to predict unobserved factual outcomes for nonsampled population units. Separate M-quantile regressions (Chambers and Tzavidis 2006) for treatment and control units provide outcome predictions, while XGBoost (Chen and Guestrin 2016) method is used to estimate treatment propensity scores. These components are combined through an augmented inverse-propensity weighted estimator to obtain causal effect estimates for each area.

The BP approach is developed in Section 2 for estimating small-area quantities that are of interest for causal inference. In Section 3, we obtain explicit expressions of the BP in the case of linear mixed models (LMM) with normally distributed random effects and errors. In Section 4, we develop area-specific causal inference under the LMM based on the EBP with maximum likelihood estimators of the model parameters, and establish the associated asymptotic theory. Another advantage of the EBP is that it is robust to misspecification of the propensity model, and this is demonstrated in the next section. In Section 5, we carry out model-based simulation studies that compare the proposed method with a naive DR approach and the partial EBP in terms of predictive and inferential performance. In section 6, we carry out design-based simulation studies that compare the proposed EBP method with the CSAE-AIPW in terms of predictive performance. A real-data example on animal breeding, technical derivations and proofs, additional simulation results and discussions are deferred to supplementary material.
\section{Best prediction for causal inference}
\hspace{4mm}
Let $Y_{ij}(1)$ be the potential outcome of the $j$ subject in group $i$ who would receive the treatment, and $Y_{ij}(0)$ be the potential outcome of the same subject who would receive the control, $i=1,\dots,m$, $j=1,\dots,N_{i}$. Let ${\cal S}\subset{\cal P}=\{(i,j): i=1,\dots,m, j=1,\dots,N_{i}\}$ be the sampled indexes, and $W=(W_{ij})_{(i,j)\in{\cal S}}$ denote the collection of treatment assignments, that is,  $W_{ij}=1$ for receiving the treatment, and $W_{ij}=0$ for receiving the control. Let $Y(r)=[Y_{ij}(r)]_{(i,j)\in{\cal P}}$, $r=0,1$. We make the following assumption.

{\it A1.} Conditional on ${\cal S}$, $Y_{0,1}=[Y(0), Y(1)]$ and $W$ are independent.

We defer details of the derivation below to Section 1 of the supplement. Our characteristic of interest are the area-specific causal effects,
\begin{eqnarray}
d_{i}=\frac{1}{N_{i}}\sum_{j=1}^{N_{i}}\{Y_{ij}(1)-Y_{ij}(0)\}=\frac{1}{N_{i}}\left[\sum_{j=1}^{n_{i}}\{Y_{ij}(1)-Y_{ij}(0)\}+\sum_{j=n_{i}+1}^{N_{i}}\{Y_{ij}(1)-Y_{ij}(0)\}\right],\label{eq:bp_5}
\end{eqnarray}
$1\leq i\leq m$. The BP of $d_{i}$ is given by
\begin{eqnarray}
{\rm E}(d_{i}|Y_{\cal S},W)&=&\frac{1}{N_{i}}\left(\sum_{1\leq j\leq n_{i},W_{ij}=1}Y_{ij}-
\sum_{1\leq j\leq n_{i},W_{ij}=0}Y_{ij}\right)\nonumber\\
&&+\frac{1}{N_{i}}\left[\sum_{1\leq j\leq n_{i},W_{ij}=0}{\rm E}(Y_{ij}^{c}|Y_{\cal S},W)
-\sum_{1\leq j\leq n_{i},W_{ij}=1}{\rm E}(Y_{ij}^{c}|Y_{\cal S},W)\right.\nonumber\\
&&\left.+\sum_{j=n_{i}+1}^{N_{i}}{\rm E}\{Y_{ij}(1)-Y_{ij}(0)|Y_{\cal S},W\}\right],
\label{eq:bp_8}
\end{eqnarray}
and the conditional expectations involved in (\ref{eq:bp_8}) have the following expressions:
\begin{eqnarray}
{\rm E}(Y_{ij}^{c}|Y_{\cal S},W)&=&\left.\int y_{ij}^{c}f_{Y_{0,1;{\cal P}\setminus{\cal S}},
Y_{\cal S}^{c}|Y_{\cal S},W}(y_{0,1;{\cal P}\setminus{\cal S}},y^{c}|y,w)d y_{0,1;{\cal
P}\setminus{\cal S}}d y^{c}\right|_{y=Y_{\cal S},w=W}\nonumber\\
&=&\left.\int y_{ij}^{c}f_{Y_{0,1;{\cal P}\setminus{\cal S}},Y_{\cal S}^{c}(w)|Y_{\cal S}(w)}(y_{0,
1;{\cal P}\setminus{\cal S}},y^{c}|y)d y_{0,1;{\cal P}\setminus{\cal S}}d y^{c}\right|_{y=
Y_{\cal S}(w),w=W}\nonumber\\
&=&{\rm E}\{Y_{ij}^{c}(w_{ij})|Y_{\cal S}(w)=y\}|_{y=Y_{\cal S}(w),w=W}\nonumber\\
&=&{\rm E}\{Y_{ij}^{c}(w_{ij})|Y_{\cal S}(w)\}|_{w=W},\;\;\;1\leq j\leq n_{i};\label{eq:bp_9}\\
{\rm E}\{Y_{ij}(r)|Y_{\cal S},W)&=&\left.\int y_{ij}(r)f_{Y_{0,1;{\cal P}\setminus{\cal S}},
Y_{\cal S}^{c}|Y_{\cal S},W}(y_{0,1;{\cal P}\setminus{\cal S}},y^{c}|y,w)d y_{0,1;{\cal
P}\setminus{\cal S}}d y^{c}\right|_{y=Y_{\cal S},w=W}\nonumber\\
&=&\left.\int y_{ij}(r)f_{Y_{0,1;{\cal P}\setminus{\cal S}},Y_{\cal S}^{c}(w)|Y_{\cal S}(w)}(y_{0,
1;{\cal P}\setminus{\cal S}},y^{c}|y)d y_{0,1;{\cal P}\setminus{\cal S}}d y^{c}\right|_{y=
Y_{\cal S}(w),w=W}\nonumber\\
&=&{\rm E}\{Y_{ij}(r)|Y_{\cal S}(w)=y\}|_{y=Y_{\cal S}(w),w=W}\nonumber\\
&=&{\rm E}\{Y_{ij}(r)|Y_{\cal S}(w)\}|_{w=W},\;\;\;n_{i}+1\leq j\leq N_{i}, r=0,1.\label{eq:bp_10}
\end{eqnarray}

It remains to find further expressions for the conditional expectations in (\ref{eq:bp_9}) and (\ref{eq:bp_10}). In next section, we consider this under a setting of LMM.
\section{Linear mixed models}
\hspace{4mm}
Suppose that the hypothetical responses, $Y_{ij}(r), r=0,1$, satisfy
\begin{equation}
Y_{ij}(r)=x_{ij}'\beta_{0}+\alpha_{i0}+r(x_{ij}'\beta_{1}+\alpha_{i1})+\epsilon_{ij}(r)
=x_{ij}'(\beta_{0}+r\beta_{1})+\alpha_{i0}+r\alpha_{i1}+\epsilon_{ij}(r),\label{eq:lmm_1}
\end{equation}
$i=1,\dots,m, j=1,\dots,N_{i}, r=0, 1$, where $x_{ij}=(x_{ijk})_{1\leq k\leq p}$ is a $p$-dimensional vector of covariates not necessarily observed, and $\beta_{r}=
(\beta_{rk})_{1\leq k\leq p}$, $r=0, 1$ are vectors of unknown parameters (the fixed effects). Furthermore, $\alpha_{i}=(\alpha_{i0},\alpha_{i1})'$ is a bivariate group-specific random effect, and $\epsilon_{ij}(r), r=0, 1$ are additional errors. We make the following assumptions: (i) $\alpha_{i},\epsilon_{i}, W_{i}, i=1,\dots,m$ are independent, where $\epsilon_{i}=(\epsilon_{i0}',\epsilon_{i1}')'$ and $\epsilon_{ir}=[\epsilon_{ij}(r)]_{1\leq j\leq N_{i}}, r=0,1$;
\begin{equation}
{\rm (ii)}\;\;\alpha_{i}\sim N\left[\left(\begin{array}{c}0\\
0\end{array}\right),\left(\begin{array}{cc}\sigma_{0}^{2}&\rho\sigma_{0}\sigma_{1}\\
\rho\sigma_{0}\sigma_{1}&\sigma_{1}^{2}\end{array}\right)\right];\label{eq:lmm_2}
\end{equation}
and (iii) $\epsilon_{ij}(r), 1\leq j\leq N_{i}$, $r=0,1$ are independent $N(0,\tau^{2})$. Here, $\sigma_{r}^{2}, r=1, 2$, $\rho$, and $\tau^{2}$ are unknown variance components with $\rho={\rm cor}(\alpha_{0i},\alpha_{1i})$.

Under the assumed LMM, it is easy to see that, given $w$, $n_{i}+1\leq j\leq N_{i}$, and $r=0,1$, $Y_{ij}(r), Y_{\cal S}(w)$ are jointly multivariate normal. Furthermore, we have
$Y_{\cal S}(w)=[Y_{i,{\cal S}}(w)]_{1\leq i\leq m}$ with $Y_{i,{\cal S}}(w_{i})=[Y_{ij}(w_{ij})]_{1\leq j\leq n_{i}}$ and $w_{i}=(w_{ij})_{1\leq j\leq n_{i}}$. Thus, we have the following expressions:
\begin{equation}
Y_{i,{\cal S}}(w_{i})=X_{i,{\cal S}}\beta_{0}+T_{i,{\cal S}}(w_{i})\beta_{1}
+1_{n_{i}}\alpha_{i0}+w_{i}\alpha_{i1}+\epsilon_{i,{\cal S}}(w_{i}),\label{eq:lmm_3}
\end{equation}
where $X_{i,{\cal S}}=(x_{ij}')_{1\leq j\leq n_{i}}$, $T_{i,{\cal S}}(w_{i})=(w_{ij}x_{ij}')_{1\leq j\leq n_{i}}$, $1_{k}$ denotes the $k\times 1$ vector of $1$s, and $\epsilon_{i,{\cal
S}}(w_{i})=[\epsilon_{ij}(w_{ij})]_{1\leq j\leq n_{i}}$; and
\begin{equation}
Y_{\cal S}(w)=X_{\cal S}\beta_{0}+T_{\cal S}(w)\beta_{1}+Z_{\cal S}\alpha_{0}+W_{\cal
S}\alpha_{1}+\epsilon_{\cal S}(w),\label{eq:lmm_4}
\end{equation}
where $X_{\cal S}=(X_{i,{\cal S}})_{1\leq i\leq m}$, $T_{\cal S}(w)=[T_{i,{\cal S}}(w_{i})]_{1\leq i\leq m}$, $Z_{\cal S}={\rm diag}(1_{n_{i}},1\leq i\leq m)$, $W_{\cal S}={\rm diag}(w_{i}, 1\leq i\leq m)$, $\alpha_{r}=(\alpha_{ir})_{1\leq i\leq m}$, $r=0,1$, and
$\epsilon_{\cal S}(w)=[\epsilon_{i,{\cal S}}(w_{i})]_{1\leq i\leq m}$. It follows that ${\rm E}\{Y_{ij}(r)\}=x_{ij}'(\beta_{0}+r\beta_{1})$, ${\rm E}\{Y_{\cal S}(w)\}=X_{\cal S}\beta_{0}+T_{\cal S}(w)\beta_{1}$; ${\rm var}\{Y_{ij}(r)\}=\tau^{2}+\sigma_{[r]}^{2}$ with $\sigma_{[r]}^{2}=\sigma_{0}^{2}+2r\rho\sigma_{0}\sigma_{1}+r\sigma_{1}^{2}$, ${\rm
Var}\{Y_{\cal S}(w)\}={\rm diag}(\Sigma_{i}(w_{i}), 1\leq i\leq m)$, where
\begin{equation}
\Sigma_{i}(w_{i})=\tau^{2}I_{n_{i}}+\sigma_{0}^{2}J_{n_{i}}
+\rho\sigma_{0}\sigma_{1}(1_{n_{i}}w_{i}'+w_{i}1_{n_{i}}')+\sigma_{1}^{2}w_{i}w_{i}',
\label{eq:lmm_5}
\end{equation}
with $J_{n_{i}}=1_{n_{i}}1_{n_{i}}'$, and ${\rm Cov}\{Y_{ij}(r),Y_{\cal S}(w)\}=[0_{n_{1}}',\cdots,0_{n_{i-1}}',c_{ij,r}(w_{i})',0_{n_{i+1}}',\cdots,0_{n_{m}}']$, where $0_{k}$ denotes
the $k\times 1$ vector of $0$s, $c_{ij,r}(w_{i})=(\sigma_{0}^{2}+r\rho\sigma_{0}\sigma_{1})1_{n_{i}}+(\rho\sigma_{0}\sigma_{1}+r\sigma_{1}^{2})w_{i}$. It follows, by the normal theory (e.g., Jiang and Nguyen 2021, sec. B.1), that
\begin{eqnarray}
&&{\rm E}\{Y_{ij}(r)|Y_{\cal S}(w)\}\nonumber\\
&=&x_{ij}'(\beta_{0}+r\beta_{1})+[0_{n_{1}}',\cdots,0_{n_{i-1}}',
c_{ij,r}(w_{i})',0_{n_{i+1}}',\cdots,0_{n_{m}}']\nonumber\\
&&\times{\rm diag}(\Sigma_{i}^{-1}(w_{i}),1\leq i\leq m)\{Y_{\cal S}(w)-X_{\cal S}\beta_{0}
-T_{\cal S}(w)\beta_{1}\}\nonumber\\
&=&x_{ij}'(\beta_{0}+r\beta_{1})+[0_{n_{1}}',\cdots,0_{n_{i-1}}',
c_{ij,r}(w_{i})'\Sigma_{i}^{-1}(w_{i}),0_{n_{i+1}}',\cdots,0_{n_{m}}']\nonumber\\
&&\times[Y_{i,{\cal S}}(w_{i})-X_{i,{\cal S}}\beta_{0}-T_{i,{\cal S}}(w_{i})\beta_{1}]_{1\leq
i\leq m}\nonumber\\
&=&x_{ij}'(\beta_{0}+r\beta_{1})+c_{ij,r}(w_{i})'\Sigma_{i}^{-1}(w_{i})\{Y_{i,{\cal S}}(w_{i})
-X_{i,{\cal S}}\beta_{0}-T_{i,{\cal S}}(w_{i})\beta_{1}\}.\label{eq:lmm_6}
\end{eqnarray}

To obtain a more explicit expression of the the right side of (\ref{eq:lmm_6}), assume, without loss of generality, that $w_{i}=(1_{n_{1i}}', 0_{n_{0i}}')'$, where $n_{0i}+n_{1i}=n_{i}$.
Then, it is easy to derive the following expression for the covariance matrix in (\ref{eq:lmm_5}):
$$\Sigma_{i}(w_{i})=\left[\begin{array}{cc}\tau^{2}I_{n_{1i}}+(\sigma_{0}^{2}+2\rho\sigma_{0}
\sigma_{1}+\sigma_{1}^{2})J_{n_{1i}}&(\sigma_{0}^{2}+\rho\sigma_{0}\sigma_{1})1_{n_{1i}}
1_{n_{0i}}'\\
(\sigma_{0}^{2}+\rho\sigma_{0}\sigma_{1})1_{n_{0i}}1_{n_{1i}}'&\tau^{2}I_{n_{0i}}
+\sigma_{0}^{2}J_{n_{0i}}\end{array}\right]=\left(\begin{array}{cc}A_{11}&A_{12}\\
A_{21}&A_{22}\end{array}\right),$$
with the $A$s defined in obvious ways. By a well-known formula of block-matrix inversion (e.g., Lu and Shiou 2002), we have
\begin{equation}
\Sigma_{i}^{-1}(w_{i})=\left(\begin{array}{cc}A_{11}^{-1}+A_{11}^{-1}A_{12}
A_{*}^{-1}A_{21}A_{11}^{-1}&-A_{11}^{-1}A_{12}A_{*}^{-1}\\
-A_{*}^{-1}A_{21}A_{11}^{-1}&A_{*}^{-1}\end{array}\right)=\left(\begin{array}{cc}B_{11}&B_{12}\\
B_{21}&B_{22}\end{array}\right),\label{eq:lmm_7}
\end{equation}
where $A_{*}=A_{22}-A_{21}A_{11}^{-1}A_{12}$, and the $B$s are defined in obvious ways. Using a formula of matrix inversion (e.g., Jiang and Nguyen 2021, sec. A.3), it can be shown that $A_{*}=\tau^{2}I_{n_{0i}}+\gamma_{i}J_{n_{0i}}$, where $\gamma_{i}=\sigma_{0}^{2}\{\tau^{2}+n_{1i}(1-\rho^{2})\sigma_{1}^{2}\}/(\tau^{2}+n_{1i}\nu)$ with $\nu=\sigma_{0}^{2}+2\rho\sigma_{0}\sigma_{1}+\sigma_{1}^{2}$.
Thus, again, using the latter matrix-inversion formula, it can be shown that $B_{11}=\tau^{-2}\{I_{n_{1i}}+(\tau^{2}\phi_{i}-\nu)(\tau^{2}+n_{1i}\nu)^{-1}J_{n_{1i}}\}$, $B_{22}=A_{*}^{-1}=\tau^{-2}\{I_{n_{0i}}-\gamma_{i}(\tau^{2}
+n_{0i}\gamma_{i})^{-1}J_{n_{0i}}\}$, and
$B_{12}=B_{21}'=-(\sigma_{0}^{2}+\rho\sigma_{0}\sigma_{1})(\tau^{2}
+n_{0i}\gamma_{i})^{-1}(\tau^{2}+n_{1i}\nu)^{-1}1_{n_{1i}}1_{n_{0i}}'$, where $\phi_{i}=n_{0i}(\sigma_{0}^{2}+\rho\sigma_{0}\sigma_{1})^{2}/(\tau^{2}
+n_{0i}\gamma_{i})(\tau^{2}+n_{1i}\nu)$. Thus, the following expression can be obtained: $c_{ij,r}(w_{i})'\Sigma_{i}^{-1}(w_{i})=[a_{i,r,1}(w_{i})1_{n_{1i}}', a_{i,r,0}(w_{i})1_{n_{0i}}']$,
where
\begin{eqnarray}
a_{i,r,1}(w_{i})&=&\frac{\sigma_{0}^{2}+r\rho\sigma_{0}\sigma_{1}}{\tau^{2}+n_{1i}\nu}\left\{1
+n_{1i}\phi_{i}-\frac{n_{0i}(\sigma_{0}^{2}+\rho\sigma_{0}\sigma_{1})}{\tau^{2}
+n_{0i}\gamma_{i}}\right\}\nonumber\\
&&+(\rho\sigma_{0}\sigma_{1}+r\sigma_{1}^{2})\frac{1+n_{1i}\phi_{i}}{\tau^{2}+n_{1i}\nu},
\label{eq:lmm_8}\\
a_{i,r,0}(w_{i})&=&\frac{\sigma_{0}^{2}+r\rho\sigma_{0}\sigma_{1}}{\tau^{2}
+n_{0i}\gamma_{i}}\left\{1-\frac{n_{1i}(\sigma_{0}^{2}+\rho\sigma_{0}\sigma_{1})}{\tau^{2}
+n_{1i}\nu}\right\}\nonumber\\
&&-(\rho\sigma_{0}\sigma_{1}+r\sigma_{1}^{2})\frac{n_{1i}(\sigma_{0}^{2}+\rho\sigma_{0}
\sigma_{1})}{(\tau^{2}+n_{0i}\gamma_{i})(\tau^{2}+n_{1i}\nu)},\;\;r=0,1.\label{eq:lmm_9}
\end{eqnarray}
Note that $a_{i,r,s}(w_{i}), s=1,0$ depend on $w_{i}$ because $n_{si}, s=1,0$ depend on $w_{i}$, that is, $n_{si}=n_{si}(w_{i}), s=1,0$ in full notation. Thus, by (\ref{eq:lmm_6}), we obtain the following expression:
\begin{eqnarray}
{\rm E}\{Y_{ij}(r)|Y_{\cal S}(w)\}&=&x_{ij}'(\beta_{0}+r\beta_{1})
+a_{i,r,1}(w_{i})\sum_{k=1}^{n_{1i}(w_{i})}\{Y_{ik}(w_{ik})-x_{ik}'(\beta_{0}+w_{ik}\beta_{1})\}
\nonumber\\
&&+a_{i,r,0}(w_{i})\sum_{k=n_{1i}(w_{i})+1}^{n_{i}}\{Y_{ik}(w_{ik})-x_{ik}'(\beta_{0}
+w_{ik}\beta_{1})\}.\label{eq:lmm_10}
\end{eqnarray}

Similarly, given $w$, $Y_{ij}^{c}(w_{ij})=Y_{ij}(1-w_{ij})$ and $Y_{\cal S}(w)$ are jointly multivariate normal with ${\rm E}\{Y_{ij}^{c}(w_{ij})\}=x_{ij}'\{\beta_{0}+(1-w_{ij})\beta_{1}\}$
and ${\rm var}\{Y_{ij}^{c}(w_{ij})\}=\tau^{2}+\sigma_{[1-w_{ij}]}^{2}$; also, we have ${\rm Cov}\{Y_{ij}^{c}(w_{ij}),
Y_{\cal S}(w)\}=[0_{n_{1}}',\cdots,0_{n_{i-1}}',c_{ij,1-w_{ij}}(w_{ij})',0_{n_{i+1}}',\cdots,0_{n_{m}}']$, where $c_{ij,r}(w_{ij}), r=0,1$ was defined earlier. Then, following the above derivation, with $r=1-w_{ij}$, we have
\begin{eqnarray}
&&{\rm E}\{Y_{ij}^{c}(w_{ij})|Y_{\cal S}(w)\}=x_{ij}'\{\beta_{0}+(1-w_{ij})\beta_{1}\}\nonumber\\
&&+a_{i,1-w_{ij},1}(w_{i})\sum_{k=1}^{n_{1i}(w_{i})}\{Y_{ik}(w_{ik})-x_{ik}'(\beta_{0}
+w_{ik}\beta_{1})\}\nonumber\\
&&+a_{i,1-w_{ij},0}(w_{i})\sum_{k=n_{1i}(w_{i})+1}^{n_{i}}\{Y_{ik}(w_{ik})-x_{ik}'(\beta_{0}
+w_{ik}\beta_{1})\}.\label{eq:lmm_11}
\end{eqnarray}

Combining (\ref{eq:lmm_11}), (\ref{eq:lmm_10}) with (\ref{eq:bp_9}), (\ref{eq:bp_10}), we have
\begin{eqnarray}
{\rm E}(Y_{ij}^{c}|Y_{\cal S},W)&=&x_{ij}'\{\beta_{0}+(1-W_{ij})\beta_{1}\}\nonumber\\
&&+a_{i,1-W_{ij},1}(W_{i})\sum_{k=1}^{n_{1i}(W_{i})}\{Y_{ik}-x_{ik}'(\beta_{0}+W_{ik}\beta_{1})\}
\nonumber\\
&&+a_{i,1-W_{ij},0}(W_{i})\sum_{k=n_{1i}(W_{i})+1}^{n_{i}}\{Y_{ik}-x_{ik}'(\beta_{0}
+W_{ik}\beta_{1})\}\label{eq:lmm_12}
\end{eqnarray}
for $1\leq j\leq n_{i}$, and
\begin{eqnarray}
{\rm E}\{Y_{ij}(r)|Y_{\cal S},W\}&=&x_{ij}'(\beta_{0}+r\beta_{1})\nonumber\\
&&+a_{i,r,1}(W_{i})\sum_{k=1}^{n_{1i}(W_{i})}\{Y_{ik}-x_{ik}'(\beta_{0}+W_{ik}\beta_{1})\}
\nonumber\\
&&+a_{i,r,0}(W_{i})\sum_{k=n_{1i}(W_{i})+1}^{n_{i}}\{Y_{ik}-x_{ik}'(\beta_{0}+W_{ik}\beta_{1})\}
\label{eq:lmm_13}
\end{eqnarray}
for $n_{i}+1\leq j\leq N_{i}, r=0, 1$. The BP of $d_{i}$, $\tilde{d}_{i}={\rm E}(d_{i}|Y_{\cal S},W)$, is
given by (\ref{eq:bp_8}) with ${\rm E}(Y_{ij}^{c}|Y_{\cal S},W)$ and ${\rm E}\{Y_{ij}(1)-Y_{ij}(0)|Y_{\cal S},W)=
{\rm E}\{Y_{ij}(1)|Y_{\cal S},W\}-{\rm E}\{Y_{ij}(0)|Y_{\cal S},W\}$ given by (\ref{eq:lmm_12}) and (\ref{eq:lmm_13}), respectively.
\section{Estimation, prediction and inference}
\subsection{Maximum likelihood estimation}
\hspace{4mm}
Let $\psi$ be a vector of unknown parameters such that $f_{Y_{\cal S}(w)}(y)=f(y,w,\psi)$ for some function $f$ and any $y, w$. Also, suppose that $f_{W}(w)$ does not depend on $\psi$. It can be shown that the likelihood function is $L(\psi)=f_{Y_{\cal S},W}(Y_{\cal S},W)=f(Y_{\cal S}|W,\psi)f_{W}(W)$; hence, the log-likelihood function is $l(\psi)=\log f(Y_{\cal S}|W,\psi)+\log f_{W}(W)$. It follows that the maximum likelihood estimator (MLE) of $\psi$ is simply the maximizer of $\log f(Y_{\cal S},W,\psi)$.

In the case of LMM, this means to fit the LMM, (\ref{eq:lmm_4}), with $w=W$ fixed for estimating the parameters $\psi=(\beta_{0}',\beta_{1}',\sigma_{0}^{2},\sigma_{1}^{2},\rho,\tau^{2})'$. Denote the MLE by $\hat{\psi}=(\hat{\beta}_{0}',\hat{\beta}_{1}',\hat{\sigma}_{0}^{2},\hat{\sigma}_{1}^{2},\hat{\rho},
\hat{\tau}^{2})'$. 

Note that, although for any fixed $W=w$, the MLE of $\psi$ is the same as that obtained by fitting the LMM, (\ref{eq:lmm_4}), and asymptotic behavior of the MLE under a LMM is well established (e.g., Jiang 2001, Jiang, Wand and Bhaskaran 2022), when $W$ is also considered random, the asymptotic properties of the MLE does not follow straightforwardly from the known results. To establish the asymptotic properties of the MLE, we note the following result from Jiang (2001). Although the context of the latter result was generalized linear mixed models (GLMM; e.g., Jiang and Nguyen 2021, ch. 3), the result is more general that it is not restricted to GLMM. Suppose $\xi_{i}=(\xi_{ij})_{1\leq j\leq n_{i}}, 1\leq i\leq m$ are independent clusters of observations. Here, $\xi_{ij}$ may be vector-valued and, by independent clusters, it means that there are correlations among $\xi_{ij}$'s with the same $i$, but $\xi_{ij}$'s with different $i$ are independent; also $n_{i}$ is referred to as cluster size. Consider the estimating function, $l(\psi)=\sum_{i=1}^{m}l_{i}(\psi,\xi_{i},n_{i})$, where $l_{i}=l_{i}(\cdot,\cdot,\cdot)$, $1\leq i\leq m$ are known functions. Let
$\psi=(\psi_{k})_{1\leq k\leq r}$, $\psi_{*}$ the true parameter vector, and $\Psi$ the parameter space for $\psi$.

{\bf Theorem 1 (Jiang 2001).} Suppose that (i) $\psi_{*}$ lies in the interior of $\Psi$; (ii) each $l_{i}$ is three-times continuously differentiable with respect to $\psi$ in some neighborhood of $\psi_{*}$; (iii) ${\rm E}(\partial l_{i}/\partial\psi|_{\psi_{*}})=0$, $1\leq i\leq m$; (iv) the following quantities are bounded:
$${\rm E}\left(\left.\frac{\partial l_{i}}{\partial\psi_{u}}\right|_{\psi_{*}}\right)^{2},\;{\rm E}\left(\left.\frac{\partial^{2}l_{i}}{\partial\psi_{u}\partial\psi_{v}}\right|_{\psi_{*}}\right)^{2},\;{\rm E}\left(\sup_{|\psi-\psi_{*}|\leq\delta_{0}}\left|\frac{\partial^{3}l_{i}}{\partial\psi_{u}\partial\psi_{v}\partial\psi_{w}}\right|\right),$$
$1\leq u,v,w\leq r, 1\leq i\leq m$, where $\delta_{0}$ is some positive number; and (v)
\begin{equation}
\frac{1}{m}\sum_{i=1}^{m}{\rm E}\left(\left.\frac{\partial^{2}l_{i}}{\partial\psi\partial\psi'}\right|_{\psi_{*}}\right)\longrightarrow-B,\label{eq:thm1_1}
\end{equation}
where $B$ is positive definite. Then, with probability tending to one, there is a solution, $\hat{\psi}\in\Psi$, to the equation $\partial l/\partial\psi=0$ such that $\hat{\psi}\stackrel{\rm P}{\longrightarrow}\psi_{*}$. Furthermore, we have
\begin{equation}
\sqrt{m}(\hat{\psi}-\psi_{*})\stackrel{\rm d}{\longrightarrow}N(0,B^{-1})\label{eq:thm1_2}
\end{equation}
provided that the following additional conditions hold: (vi) there is $d>2$ such that ${\rm E}\{|(\partial l_{i}/\partial\psi_{u})|_{\psi_{*}}|^{d}\}$, $1\leq u\leq r, 1\leq i\leq m$ are bounded; and (vii) for $1\leq i\leq m$, we have
\begin{equation}
 {\rm E}\left(\left.\frac{\partial^{2}l_{i}}{\partial\psi\partial\psi'}\right|_{\psi_{*}}\right)=-{\rm Var}\left(\left.\frac{\partial l_{i}}{\partial\psi}\right|_{\psi_{*}}\right).\label{eq:thm1_3}
\end{equation}

Equation (\ref{eq:thm1_3}) is the so-called information identity. We now apply Theorem 1 to the LMM case. Let $\Sigma_{i,*}$ denote $\Sigma_{i}(W_{i})$ with $\psi=\psi_{*}$, the true $\psi$. Define ${\cal V}=\{\sigma_{0}^{2},\sigma_{1}^{2},\rho,\tau^{2}\}$,
\begin{eqnarray*}
B_{i,1}&=&\left[\begin{array}{cc}{\rm E}(X_{i,{\cal S}}'\Sigma_{i,*}^{-1}X_{i,{\cal S}})&{\rm E}(X_{i,{\cal S}}'\Sigma_{i,*}^{-1}T_{i,{\cal S}})\\
{\rm E}(T_{i,{\cal S}}'\Sigma_{i,*}^{-1}X_{i,{\cal S}})&{\rm E}(T_{i,{\cal S}}'\Sigma_{i,*}^{-1}T_{i,{\cal S}})\end{array}\right],\\\
B_{i,2}&=&\left[{\rm tr}\left(\Sigma_{i,*}^{-1}\frac{\partial\Sigma_{i,*}}{\partial\gamma}\Sigma_{i,*}^{-1}\frac{\partial\Sigma_{i,*}}{\partial\nu}\right)\right]_{\gamma,\nu\in{\cal V}},
\end{eqnarray*}
where $\partial\Sigma_{i,*}/\partial\gamma$ denotes $\partial\Sigma_{i}/\partial\gamma|_{\psi=\psi_{*}}$, etc. Let $\Psi$ denote the parameter space for $\psi$, which carries the restriction that $\sigma_{r}^{2}\geq 0$, $\tau^{2}\geq0$ and $|\rho|\leq 1$ (there is no restriction on the $\beta$'s).

{\bf Theorem 2.} Under the LMM setting, suppose that (I) the true variance components satisfy $\sigma_{r}^{2}>0$, $r=0,1$, $\tau^{2}>0$ and $|\rho|<1$; (II) $n_{i}$, $|x_{ijk}|$ are bounded; (III)
\begin{equation}
\frac{1}{m}\sum_{i=1}^{m}B_{i,r}\longrightarrow B_{r}>0\;({\rm positive}\;{\rm definite}),\;\;r=1,2.\label{eq:thm2_3}
\end{equation}
Then, the results of Theorem 1 hold with $B^{-1}={\rm diag}(B_{1}^{-1}/2,2B_{2}^{-1})$.

Note that, because we are considering small-area inference, the boundedness of $n_{i}$ in condition (II) is reasonable. Theorem 2 can be established by verifying the conditions of Theorem 1. The proof is deferred to Section 2.1 of the supplement.

Condition (\ref{eq:thm1_1}) of Theorem 1, or condition (\ref{eq:thm2_3}) of Theorem 2, can be weakened using a ``subsequence argument''. Denote the left side of (\ref{eq:thm1_1}) by $-B_{m}$.

{\bf Corollary 1.} Suppose that the conditions of Theorem 1 holds with (\ref{eq:thm1_1}) replaced by
\begin{equation}
\liminf\lambda_{\min}(B_{m})>0\;\;\;{\rm and}\;\;\;\limsup\lambda_{\max}(B_{m})<\infty.\label{eq:col1_1}
\end{equation}
Then, the conclusions of Theorem 1 hold with (\ref{eq:thm1_2}) replaced by
\begin{equation}
B_{m}^{1/2}\sqrt{m}(\hat{\psi}-\psi_{*})\stackrel{\rm d}{\longrightarrow}N(0,I_{2p+4}),\label{eq:col1_2}
\end{equation}
where $B_{m}^{1/2}$ denotes the symmetric square root of $B_{m}$ [exists asymptotically by (\ref{eq:col1_1})].

{\bf Corollary 2.} Suppose that the conditions of Theorem 2 hold with (\ref{eq:thm2_3}) replaced by
\begin{equation}
\liminf\lambda_{\min}(B_{mr})>0\;\;\;{\rm and}\;\;\;\limsup\lambda_{\max}(B_{mr})<\infty,\;\;\;r=1,2,\label{eq:col1_3}
\end{equation}
where $B_{mr}$ denotes the left side of (\ref{eq:thm2_3}). Then, the conclusions of Theorem 2 hold with (\ref{eq:thm1_2}) replaced by (\ref{eq:col1_2}), where $B_{m}={\rm diag}(2B_{m1},B_{m2}/2)$.

The proofs of Corollary 1 and Corollary 2 are deferred to Section 2.2 of the supplement.
\subsection{EBP for area-level causal effects under LMM}
\hspace{4mm}
Let $\hat{E}(\cdot|\cdot)$ denote the conditional expectation, ${\rm E}(\cdot|\cdot)$, with $\psi$ replaced by $\hat{\psi}$. In subsequent derivations, we follow the assumption of Theorem 2 by assuming that $m\rightarrow\infty$ while $n_{i}, 1\leq i\leq m$ are bounded. The following expression, whose derivation is deferred to Section 1.2 of the supplement, can be established:
\begin{eqnarray}
&&\hat{d}_{i}-d_{i}=b_{i0}(W_{i})\alpha_{i0}+b_{i1}(W_{i})\alpha_{i1}
+c_{i1}(W_{i})\sum_{k=1}^{n_{1i}(W_{i})}\epsilon_{ik}(1)\nonumber\\
&&+c_{i0}(W_{i})\sum_{k=n_{1i}(W_{i})+1}^{n_{i}}
\epsilon_{ik}(0)-\frac{1}{N_{i}}\left\{\sum_{k=n_{1i}(W_{i})+1}^{n_{i}}\epsilon_{ik}(1)
-\sum_{k=1}^{n_{1i}(W_{i})}\epsilon_{ik}(0)\right.\nonumber\\
&&\left.+\sum_{j=n_{i}+1}^{N_{i}}\epsilon_{ij}(1)
-\sum_{j=n_{i}+1}^{N_{i}}\epsilon_{ij}(0)\right\}+o_{\rm P}(1)
=\eta_{i}+o_{\rm P}(1),\label{eq:inf_9}
\end{eqnarray}
$\eta_{i}$ defined in an obvious way, where
\begin{eqnarray*}
b_{i0}(W_{i})&=&\{1-n_{1i}(W_{i})/N_{i}\}\{n_{1i}(W_{i})a_{i,1,1}(W_{i})+n_{0i}(W_{i})a_{i,1,0}(W_{i})
-1\}\\
&&-\{1-n_{0i}(W_{i})/N_{i}\}\{n_{1i}(W_{i})a_{i,0,1}(W_{i})+n_{0i}(W_{i})a_{i,0,0}(W_{i})-1\},\\
b_{i1}(W_{i})&=&\{1-n_{1i}(W_{i})/N_{i}\}\{n_{1i}(W_{i})a_{i,1,1}(W_{i})-1\}\\
&&-\{1-n_{0i}(W_{i})/N_{i}\}n_{1i}(W_{i})a_{i,0,1}(W_{i}),\\
c_{ir}(W_{i})&=&\{1-n_{1i}(W_{i})/N_{i}\}a_{i,1,r}(W_{i})-\{1-n_{0i}(W_{i})/N_{i}\}a_{i,0,r}(W_{i}),\;\;\;
r=0,1.
\end{eqnarray*}

{\bf Note.} Although, unlike the partial EBP, the propensity scores are not directly used in the expression of EBP, treatment assignment information is, indeed, involved. This is because the $W_{ij}$s, the indicators of treatment assignments, are involved in the EBP. The $W_{ij}$s may be viewed as the observed propensity scores. What is more, they are free of misspecification of the propensity model. As will be shown (see the next section), the proposed EBP is robust to misspecification of the propensity model.

Also note that, given $W$, $(\alpha_{i0}, \alpha_{i1})$ and all of the sums of the $\epsilon$'s involved in $\eta_{i}$ are independent and normally distributed. Thus, $\eta_{i}|W$ is normal with ${\rm E}(\eta_{i}|W)=0$ and
\begin{eqnarray}
&&{\rm var}(\eta_{i}|W)=\sigma_{0}^{2}b_{i0}^{2}(W_{i})+2\rho\sigma_{0}\sigma_{1}
b_{i0}(W_{i})b_{i1}(W_{i})+\sigma_{1}^{2}b_{i1}^{2}(W_{i})\nonumber\\
&&+\tau^{2}\left\{n_{0i}(W_{i})c_{i0}^{2}(W_{i})+n_{1i}(W_{i})c_{i1}^{2}(W_{i})+
\frac{2N_{i}-n_{i}}{N_{i}^{2}}\right\}\equiv\sigma_{i}^{2}(\phi,W_{i}),\label{eq:inf_10}
\end{eqnarray}
where $\phi=(\sigma_{0}^{2},\sigma_{1}^{2},\rho,\tau^{2})'$. The variance formula (\ref{eq:inf_10}) can be used for inference (see below).

Below we establish two main asymptotic results. The first is asymptotic optimality of the EBP; the second is an asymptotic normality result that can be used for inference.
\subsection{Asymptotic optimality of EBP}
\hspace{4mm}
We first establish a lemma regarding the convergence rate of $\hat{\psi}$.

{\bf Lemma 1.} Suppose that the conditions of Corollary 2 hold. Then, there is a solution, $\hat{\psi}$, to the likelihood solution that satisfies ${\rm P}(|\hat{\psi}-\psi_{*}|>m^{\delta-1/2})\leq c_{h}m^{-\delta h}$ for every $h\geq 2$, where $c_{h}$ is a positive constant that depends on $h$.

The proof of Lemma 1 is given in Section 3.1 of the supplement.

Following Jiang and Torabi (2020), we consider a truncated version of the estimator, $\hat{\psi}$, in Lemma 1. Namely, let $a<b$ be any fixed positive numbers. Define $\hat{\psi}_{\rm t}=\hat{\psi}$ if $\hat{\tau}^{2}>a/m$ and $|\hat{\psi}|\leq bm$; otherwise, let $\hat{\psi}_{\rm t}=\psi_{\rm o}$, a fixed point in $\Psi^{\rm o}$, the interior of $\Psi$. By the consistency of $\hat{\psi}$ (see Corollary 2 and Lemma 1), we have, with probability tending to one, that $\hat{\psi}_{\rm t}=\hat{\psi}$. In fact, as noted by Jiang and Torabi (2020), truncation is only needed in theoretical arguments --- there is no need to truncate the estimator, in practice, because the $\hat{\psi}$ computed from the observe data must satisfy the inequality constraints for some constants $0<a<b$, hence $\hat{\psi}_{\rm t}=\hat{\psi}$ for such choices of $a$ and $b$.

Recall the BP of $d_{i}$ is given by (\ref{eq:bp_8}), with the conditional expectations involved given by (\ref{eq:lmm_12}) and (\ref{eq:lmm_13}). The BP is a function of $Y_{i,{\cal S}}=Y_{i,{\cal S}}(W_{i})$, $W_{i}$ and $\psi$, denoted by
\begin{equation}
\tilde{d}_{i}=g_{i}(Y_{i,{\cal S}},W_{i},\psi)\label{eq:bp_exp1}
\end{equation}
for a known function $g_{i}(\cdot,\cdot,\cdot)$. It is easy to see that $g_{i}$ has continuous partial derivatives of any order with respect to $\psi$. The EBP is then $\hat{d}_{i}=g_{i}(Y_{i,{\cal S}},W_{i},\hat{\psi})$. The following theorem establishes the asymptotic optimality of the EBP in terms of minimizing the mean squared prediction error (MSPE), that is, ${\rm MSPE}(\check{d}_{i})={\rm E}(\check{d}_{i}-d_{i})^{2}$ for any predictor $\check{d}_{i}$ of $d_{i}$.

{\bf Theorem 3.} Suppose that the conditions of Corollary 2 are satisfied. Let $\hat{d}_{i,{\rm t}}$ be $\hat{d}_{i}$ with $\hat{\psi}$ replaced by $\hat{\psi}_{\rm t}$. Then, for any other predictor of $d_{i}$, say, $\check{d}_{i}$, we have
\begin{equation}
\limsup_{m\rightarrow\infty}\{{\rm MSPE}(\hat{d}_{i,{\rm t}})-{\rm MSPE}(\check{d}_{i})\}\leq 0.\label{eq:thm3}
\end{equation}
In other words, $\hat{d}_{i,{\rm t}}$ is asymptotically optimal in terms of the MSPE.

The proof of Theorem 3 deferred to in Section 3.2 of the supplement.
\subsection{Asymptotic normality of EBP}
\hspace{4mm}
For notation simplicity, write $a_{i,r,s}=a_{i,r,s}(W_{i}), r,s=0,1$ [defined via (\ref{eq:lmm_8}), (\ref{eq:lmm_9})] and $n_{ri}=n_{ri}(W_{i}), r=0,1$. Define $a_{ir}=n_{ri}(a_{i,1,r}-a_{i,0,r})$, $d_{ir}=a_{i,1,r}-a_{i,0,r}$, $r=0,1$, and
\begin{eqnarray}
\tau_{i}^{2}(\phi,W_{i})&=&\sigma_{0}^{2}(a_{i0}+a_{i1})^{2}+2\rho\sigma_{0}\sigma_{1}(a_{i0}+a_{i1})(a_{i1}-1)
+\sigma_{1}^{2}(a_{i1}-1)^{2}\nonumber\\
&&+\tau^{2}(n_{0i}d_{i0}^{2}+n_{1i}d_{i1}^{2})\label{eq:thm4_1}
\end{eqnarray}
[$\phi$ defined below (\ref{eq:inf_10}); note that $a_{ir}(W_{i})$, $d_{ir}(W_{i})$ also depend on $\phi$].

{\bf Theorem 4.} Suppose that the conditions of Corollary 2 hold, $m, N_{i}\rightarrow\infty$, $n_{ri}(W_{i})\geq 1, r=0,1$, and $n_{i}$ is bounded. If ${\rm P}[a_{i,1,r}(W_{i})=a_{i,0,r}(W_{i}), r=0,1]\rightarrow 0$, then, we have
\begin{equation}
\frac{\hat{d}_{i}-d_{i}}{\hat{\tau}_{i}(W_{i})}\stackrel{\rm d}{\longrightarrow}N(0,1),\label{eq:thm4_2}
\end{equation}
where $\hat{\tau}_{i}(W_{i})=\tau_{i}(\hat{\phi},W_{i})$ with $\hat{\phi}=(\hat{\sigma}_{0}^{2},\hat{\sigma}_{1}^{2},\hat{\rho},\hat{\tau}^{2})'$.

The proof of Theorem 4 is given in Section 3.3 of the supplement.

The result of Theorem 4 can be used to make inference about the small-area causal effects, such as constructing confidence intervals and hypothesis testing. In case $N_{i}$ is known, one may replace $\hat{\tau}_{i}^{2}(W_{i})$ by $\sigma_{i}^{2}(\hat{\psi},W_{i})$, which is given by (\ref{eq:inf_10}) with $\phi$ replaced by $\hat{\psi}$. The asymptotic distribution does not change with this replacement under the assumptions of Theorem 4; on the other hand, $\sigma_{i}^{2}(\hat{\psi},W_{i})$ may provide more accurate estimation of the (conditional) asymptotic variance in case $N_{i}$ is not very large. We explore this empirically in the next section.
\subsection{Unknown subpopulation size}
\hspace{4mm}
So far, the EBP is developed assuming that the subpopulation size, $N_{i}$, is known (same as in Ranjbar {\it et al.} 2023). In some applications, $N_{i}$ is unknown but can be reasonably assumed to be much larger than $n_{i}$. In fact, it can be shown that, under the same setting of Theorem 4, the result of Theorem 4 remains valid with $\hat{d}_{i}$ replaced by
\begin{eqnarray}
\hat{d}_{i}^{*}&=&\bar{X}_{i}'\hat{\beta}_{1}+\{\hat{a}_{i,1,1}(W_{i})-\hat{a}_{i,0,1}(W_{i})\}\hat{\Delta}_{1i}(Y_{i},W_{i})\nonumber\\
&&+\{\hat{a}_{i,1,0}(W_{i})-\hat{a}_{i,0,0}(W_{i})\}\hat{\Delta}_{0i}(Y_{i},W_{i}),\label{eq:dihst}
\end{eqnarray}
where $\bar{X}_{i}=N_{i}^{-1}\sum_{j=1}^{N_{i}}x_{ij}$, $\hat{a}_{i,r,s}(W_{i})$ is $a_{i,r,s}(W_{i})$ with $\psi$ replaced by $\hat{\psi}$, $r,s=0,1$, $\hat{\Delta}_{1i}(Y_{i},W_{i})=\sum_{k=1}^{n_{1i}}\{Y_{ik}-x_{ik}'(\hat{\beta}_{0}+\hat{\beta}_{1})\}$, and $\hat{\Delta}_{0i}(Y_{i},W_{i})=\sum_{k=n_{1i}+1}^{n_{i}}(Y_{ik}-x_{ik}'\hat{\beta}_{0})$ with $n_{1i}=n_{1i}(W_{i})$ (recall $W_{ik}=1, 1\leq k\leq n_{1i}$ and $W_{ik}=0, n_{1i}+1\leq k\leq n_{i}$).

Note that $\hat{d}_{i}^{*}$ is computable as long as $\bar{X}_{i}$ is known. For example, in the special case where $x_{ij}=1$ (corresponding to an intercept), we have $\bar{X}_{i}=1$. More generally, if the covariates are area-level, that is, if $x_{ij}$ does not depend on $j$, then $\bar{X}_{i}$ is clearly known. This would be the case if the subpopulations, or small areas, are defined according to the values of the covariates. For example, if the covariates correspond to gender, race and age-groups, and different combinations of gender, race and age groups define the small areas, then the covariates are area-level. We call $\hat{d}_{i}^{*}$ asymptotic EBP.

As noted (see the note following Theorem 4), the denominator $\hat{\tau}_{i}(W_{i})$ in (\ref{eq:thm4_2}) does not require knowing $N_{i}$. Thus, $\hat{d}_{i}^{*}$, combined with $\hat{\tau}_{i}(W_{i})$, can be used to construct prediction intervals in case $N_{i}$ is unknown (but much larger than $n_{i}$).
\section{Model/design-based simulation studies: Comparison with naive DR and partial EBP}
\hspace{4mm}
In this section, we carry out a series of model/design-based simulation studies, in the sense that the randomness in the data is due to random effects and errors generated under a certain model (see below) as well as sampling from the finite populations, to investigate finite-sample performance of the proposed EBP method and its comparison with existing alternatives. In view of the DR estimator (DRE; see Section 1, third paragraph), the simulation is designed to examine scenarios where the outcome model or propensity model are either correct or misspecified. Altogether, there are four cases: {\it Case I.} Both outcome model and propensity model are correct; {\it Case II.} Outcome model is misspecified but propensity model is correct; {\it Case III.} Outcome model is correct but propensity model is misspecified; and {\it Case IV.} Both outcome model and propensity model are misspecified.

The objective of interest is the area-specific causal-effects, $d_{i}, 1\leq i\leq m$. A natural comparing method is the DRE of $d_{i}$ based on data from the same area. This is referred to as naive DR estimator, as it does not borrow strength from other areas and sources. Another comparing method is the method proposed by Ranjbar {\it et al.} (2023). As noted in Section 1 (second-to-last paragraph), the method starts with the DRE but allows borrowing of strength from other areas and sources by replacing some unobserved quantities involved in the DRE by their EBP. We refer the latter method as partial EBP.

It should be noted that, in Ranjbar {\it et al.} (2023), the authors developed two estimators of the MSE (mean squared error) for the partial EBP, namely, a plug-in MSE estimator [eq. (16) therein], and a second-order bias-corrected MSE estimator [eq. (19) therein]. However, the bias-corrected MSE estimator is derived assuming that the random intercept and slope are uncorrelated, which is unrealistic under our setting (see Section 3). The bias-corrected MSE estimator is also computationally expensive, especially in large-scale simulation studies. Using Taylor series expansion, we developed a similar bias-corrected MSE estimator to that of Ranjbar {\it et al.} (2023) that fits our setting, which we call Ranjbar-style MSE estimator. In an online supplement to Ranjbar {\it et al.} (2023), the authors also developed a bootstrap MSE estimator, which is computationally even more intensive than their bias-corrected MSE estimator. In the supplement of our paper, we carry out sensitivity simulation studies that compare the accuracy of the SEs (standard errors; i.e., square roots of the MSE estimators) and the inferential performance of the associated prediction intervals based on the plug-in MSE estimator and the Ranjbar-style MSE estimator as well as the bootstrap MSE estimator. The analytical correction has little effect on coverage and interval length. Bootstrap and plug-in average SEs are also close, but the bootstrap intervals are slightly shorter and their coverage is lower by approximately 2.2--3.2 percentage points in the selected settings. Thus, due to the computational intensity of the Ranjbar-style and bootstrap SEs, we use the plug-in SE for partial EBP throughout our main simulation studies.

Also note that the partial EBP method requires knowing the covariate subpopulation, including the subpopulation size, $N_{i}$. Although the EBP method is also developed in this setting, there is an approximate version of the EBP, called asymptotic EBP, that has relaxed the requirements of knowing the covariate subpopulation and $N_{i}$ (see Section 4.5). On the other hand, the naive DR method does not require knowing either the covariate subpopulation or $N_{i}$. In our simulation studies, we consider both the known covariate subpopulation and $N_{i}$ setting, in which case all three methods, the naive DR, partial EBP and EBP, are compared, and a setting where the covariate subpopulation and $N_{i}$ are unknown, in which case the naive DR and asymptotic EBP are compared.

We focus on two aspects of performance measures. The first is accuracy in estimating or, more precisely, predicting the area-specific causal-effects, $d_{i}, 1\leq i\leq m$, as the latter are viewed as unobserved random variables. This is measured by the area-specific MSPE (see Section 1). An overall MSPE is obtained by averaging the MSPE overall the small areas. We also present boxplots of the area-specific MSPEs to provide a more detailed picture. The second aspect, in view of the traditional interest in causal inference, is inferential performance in terms of prediction intervals. Specifically, we evaluate coverage probabilities of the 95\% prediction intervals as well as expected lengths of the prediction intervals. Similarly, the results are presented in terms of the overall coverage probability and expected length, as well as boxplots of the area-specific ones.

Due to the space limit and also similarity of the results under different scenarios, only results under {\it Case I} are presented here in the main manuscript; the results under {\it Cases II}, {\it III} and {\it IV} are deferred to the supplementary material. However, a comprehensive summary of the results under all scenarios is provided (see below).
\subsection{Simulation settings and data generating}
\hspace{4mm}
In all settings, simple random samples of size $n_{i}$ are drawn from the corresponding subpopulation (small area) with population size $N_{i}$. The population sizes $N_{i}$ are chosen so that $N_{i}$ is 1,000 for 70\% of the areas, and 10,000 for 30\% of the areas.

{\it 1. The known $N_{i}$ and covariate subpopulation setting.}
In this case, a main covariate, $x_{ij1}$, is used in all simulation settings. The values of the covariate are generated under $N(0,1)$, and fixed throughout the simulations. Two other covariates are constructed based on $x_{ij1}$, namely, $q_{ij}=x_{ij1}^{2}-1$ and $c_{ij}=x_{ij1}^{3}-3x_{ij1}$. The latter covariates are used under some scenarios (but not all scenarios).

{\it Outcome model.} The potential outcomes, $Y_{ij}(r)$, are generated under the model:
\begin{equation}
Y_{ij}(r)=x_{ij}'(\beta_{0}+r\beta_{1})+\alpha_{i0}+r\alpha_{i1}+\gamma_{\rm quad}q_{ij}+\epsilon_{ij}(r),\;\;r=0,1,\label{eq:sim_outcome}
\end{equation}
$i=1,\dots,m$, $j=1,\dots,N_{i}$, where $x_{ij}=(1,x_{ij1})'$, the fixed effects, $\beta_{s}=(\beta_{s0},\beta_{s1})'$, random effects, $\alpha_{is}$, $s=0,1$ and errors, $\epsilon_{ij}(r)$ satisfy the assumptions in Section 3, with $\sigma_{0}^{2}=1$, $\sigma_{1}^{2}=0.5$, $\rho=0.3$ and $\tau^{2}=1$. The true $\beta$ coefficients are $\beta_{0}=(1,1)'$ and $\beta_{1}=(0.5,0.5)'$.
Here, $\gamma_{\rm quad}$ controls the degree of outcome-model misspecification. We consider $\gamma_{quad}=0, 0.5, 1$. When $\gamma_{\rm quad}=0$, there is no outcome-model misspecification; this is the model that is fitted, always. In other words, the assumed outcome model is (\ref{eq:sim_outcome}) with $\gamma_{\rm quad}=0$. As $\gamma_{\rm quad}$ increases, so does the degree of model misspecification.

{\it Propensity models.} Two propensity models are considered. The first is a noninformative, that is, ${\rm P}(W_{ij}=1|x_{ij})=0.5$. The second is a logistic model with
\begin{equation}
{\rm P}(W_{ij}=1|x_{ij})={\rm logit}^{-1}(\eta_{0}+0.5x_{ij1}+0.25q_{ij}+\delta_{\rm ps}c_{ij}),\label{eq:sim_ps_model}
\end{equation}
where ${\rm logit}^{-1}(x)=e^{x}/(1+e^{x})$. Similarly, here, $\delta_{\rm ps}$ controls the degree of misspecification of the propensity model. The possible values are $0, 0.25, 0.5$ and $1$. $\delta_{\rm ps}=0$ corresponds to a case where the propensity model is correctly specified; as $\delta_{\rm ps}$ increases, so does the degree of model misspecification. Both the noninformative and logistic models are considered under {\it Case I}. Under {\it Cases II--IV}, only the logistic model is considered.

We consider $m=50, 100, 200$, although, in some cases, only $m=100$ is considered. We also consider two settings of $n_{i}$. One is an equal-$n_{i}$ design, in which case either $n_{i}=20$ for all small areas, or $n_{i}=40$ for all small areas. Another one is an unequal-$n_{i}$ design, in which case $n_{i}\in\{10,20,40,80\}$ evenly distributed over the $m$ small areas.

{\it 2. The unknown $N_{i}$ setting.} In this case, we pretend that $N_{i}$ is unknown. The covariate is at area-level, that is, $x_{ij1}=x_{i1}, 1\leq j\leq N_{i}$ so that $\bar{X}_{i}$ is known. The outcome model and propensity model are the same as (\ref{eq:sim_outcome}) and (\ref{eq:sim_ps_model}), respectively, with $x_{ij1}=x_{i1}$, with or without misspecification. The true parameters are the same as above. We consider $m=100$ and the same equal and unequal-$n_{i}$ designs as the above.

All results are based on $N_{\rm sim}=1,000$ simulation runs.
\subsection{Case I: Outcome and propensity models correctly specified}
\hspace{4mm}
In this case, the potential outcomes are generated via (\ref{eq:sim_outcome}) with $\gamma_{\rm quad}=0$, which is the same as the assumed model. In other words, the outcome model is correctly specified. The propensity model is also correctly specified under two scenarios, as follows.
\subsubsection{The known $N_{i}$ and covariate subpopulation setting}
\hspace{4mm}
We first consider the noninformative propensity with ${\rm P}(W_{ij}=1|X)=0.5$ for all $i,j$.

{\it 1. Equal-$n_{i}$ design.} Table \ref{tab:final-correct-equal-n20} presents a summary of performance in terms of simulated MSPE (MSPE), coverage probability (Coverage) and average length (Length) of the 95\% prediction intervals (constructed via the normal approximation), averaged over the $m$ small areas. The bold font numbers indicate the best performer under each $m$ and respective performance measure (for MSPE and Length, the smaller the better; for Coverage, the closer to 0.95 the better). It can be seen that, for MSPE and Length, the best performer is always EBP; for Coverage, the best performer is partial EBP for $m=50$ and $m=100$, and EBP (with the approximate asymptotic variance) for $m=200$. The results for $n_{i}=40$ are presented in Table \ref{tab:final-correct-equal-n40}; EBP remains best in MSPE and Length, while Coverage is closest to 0.95 for partial EBP at $m=100$ and for EBP with the approximate asymptotic variance at $m=50$ and $200$.
\begin{table}[ht]
\centering
\setlength{\belowcaptionskip}{8pt}
\caption{Simulated MSPE, Coverage Probability and Average Length of 95\% Prediction Interval: Noninformative Propensity; Equal-$n_{i}$ Design with $n_i=20$; Numbers in the Parentheses Are Results based on the Approximate Asymptotic Variance for the EBP Method; Bold Font Numbers Indicate Best Performer under the $m$ and Performance Measure.}
\label{tab:final-correct-equal-n20}
\setlength{\tabcolsep}{3.5pt}
\begin{tabular}{r l lll}
\hline
$m$ & Method & MSPE & Coverage & Length\\
\hline
50 & naive DR & 0.2251 & 0.9135 & 1.6991\\
50 & partial EBP & 0.1383 & \textbf{0.9577} & 1.5159\\
50 & EBP & \textbf{0.1318} & 0.9347 (0.9350) & \textbf{1.3534} (1.3550)\\
100 & naive DR & 0.2261 & 0.9121 & 1.6988\\
100 & partial EBP & 0.1378 & \textbf{0.9545} & 1.4908\\
100 & EBP & \textbf{0.1297} & 0.9423 (0.9427) & \textbf{1.3731} (1.3751) \\
200 & naive DR & 0.2243 & 0.9139 & 1.6991\\
200 & partial EBP & 0.1361 & 0.9546 & 1.4779\\
200 & EBP & \textbf{0.1276} & 0.9469 (\textbf{0.9472}) & \textbf{1.3834} (1.3854) \\
\hline
\end{tabular}
\end{table}

\begin{table}[ht]
\centering
\setlength{\belowcaptionskip}{8pt}
\caption{Simulated MSPE, Coverage Probability and Average Length of 95\% Prediction Interval: Noninformative Propensity; Equal-$n_{i}$ Design with $n_i=40$; Numbers in the Parentheses Are Results based on the Approximate Asymptotic Variance for the EBP Method; Bold Font Numbers Indicate Best Performer under the $m$ and Performance Measure.}
\label{tab:final-correct-equal-n40}
\setlength{\tabcolsep}{3.5pt}
\begin{tabular}{r l lll}
\hline
$m$ & Method & MSPE & Coverage & Length \\
\hline
50 & naive DR & 0.1068 & 0.9370 & 1.2405 \\
50 & partial EBP & 0.0833 & 0.9582 & 1.1811 \\
50 & EBP & \textbf{0.0781} & 0.9432 (\textbf{0.9444}) & \textbf{1.0708} (1.0761) \\
100 & naive DR & 0.1072 & 0.9352 & 1.2390 \\
100 & partial EBP & 0.0833 & \textbf{0.9525} & 1.1454 \\
100 & EBP & \textbf{0.0777} & 0.9460 (0.9472) & \textbf{1.0782} (1.0838) \\
200 & naive DR & 0.1079 & 0.9356 & 1.2394 \\
200 & partial EBP & 0.0833 & 0.9497 & 1.1278 \\
200 & EBP & \textbf{0.0773} & 0.9485 (\textbf{0.9498}) & \textbf{1.0808} (1.0864) \\
\hline
\end{tabular}
\end{table}

\begin{figure}[ht]
\centering
\includegraphics[width=0.95\textwidth]{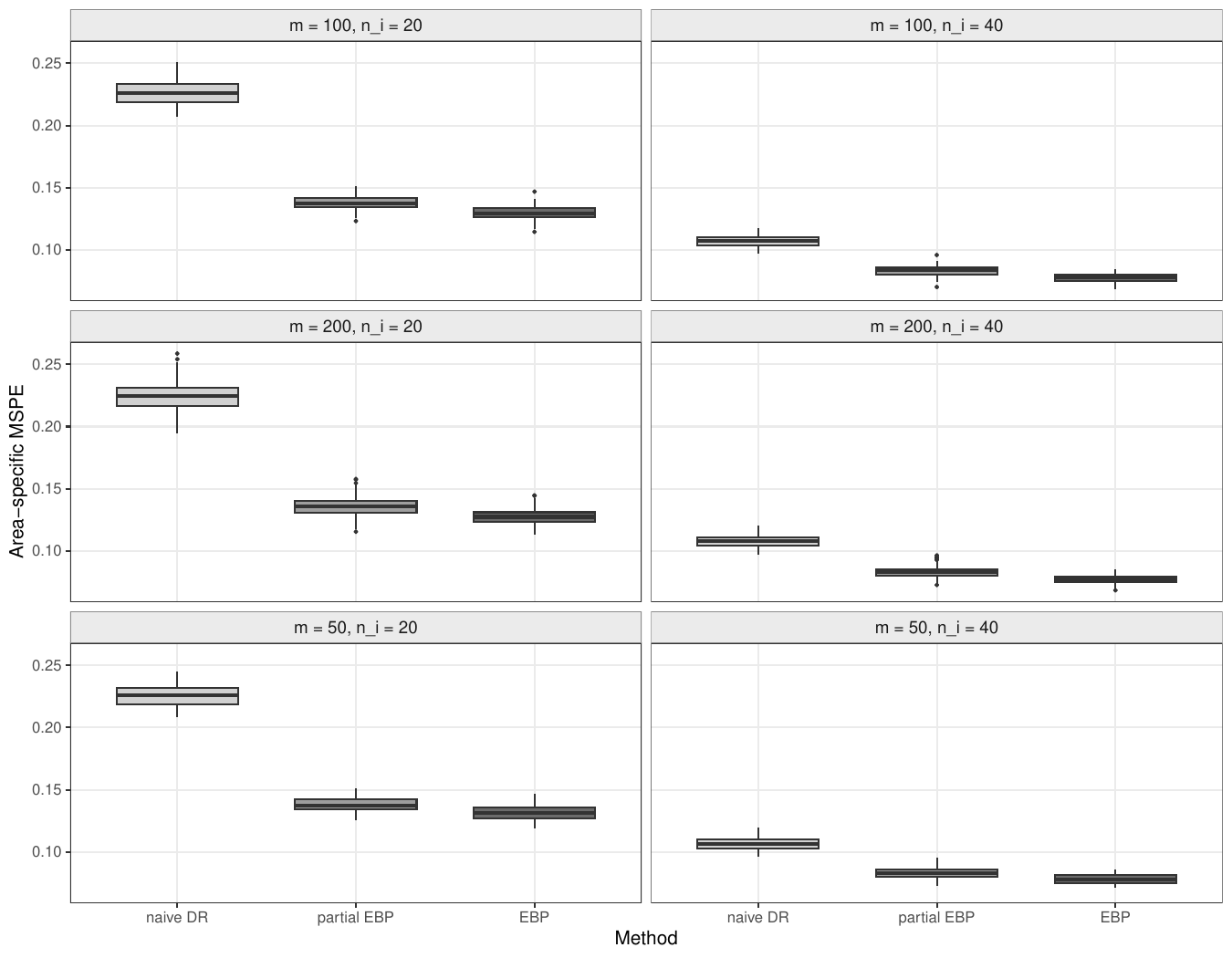}
\caption{\footnotesize Boxplots of Simulated Area-specific MSPE: Noninformative Propensity; Equal-$n_{i}$ Design.}
\label{fig:final-correct-equal-mspe}
\end{figure}

\begin{figure}[ht]
\centering
\includegraphics[width=0.95\textwidth]{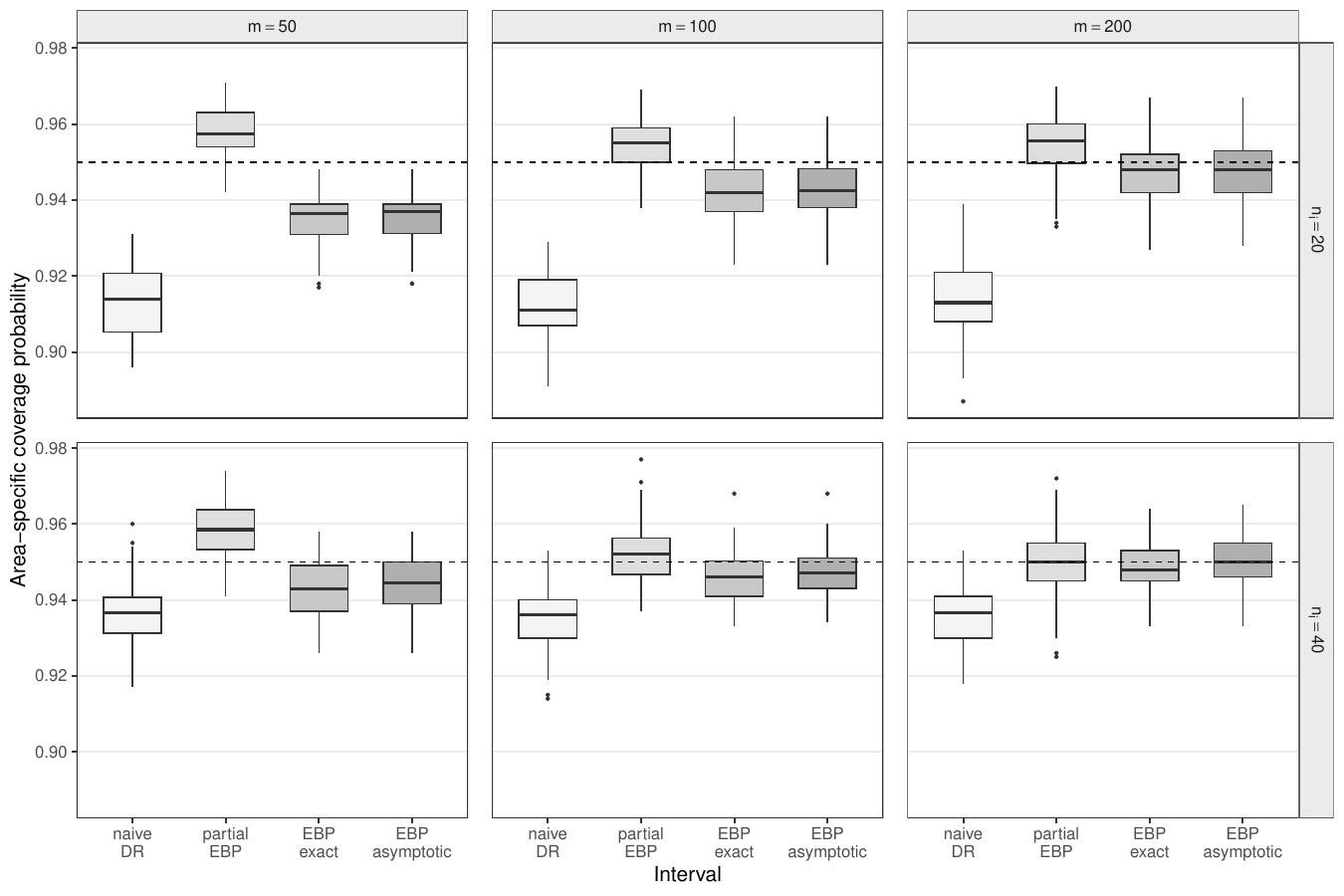}
\caption{\small Boxplots of Simulated Area-specific Coverage Probability of 95\% Prediction Intervals: Noninformative Propensity; Equal-$n_{i}$ Design; EBP exact and EBP asymptotic refer to EBP-based Prediction Interval based on Exact and Approximate Asymptotic Variances, Respectively.}
\label{fig:final-correct-equal-coverage}
\end{figure}

\begin{figure}[ht]
\centering
\includegraphics[width=0.95\textwidth]{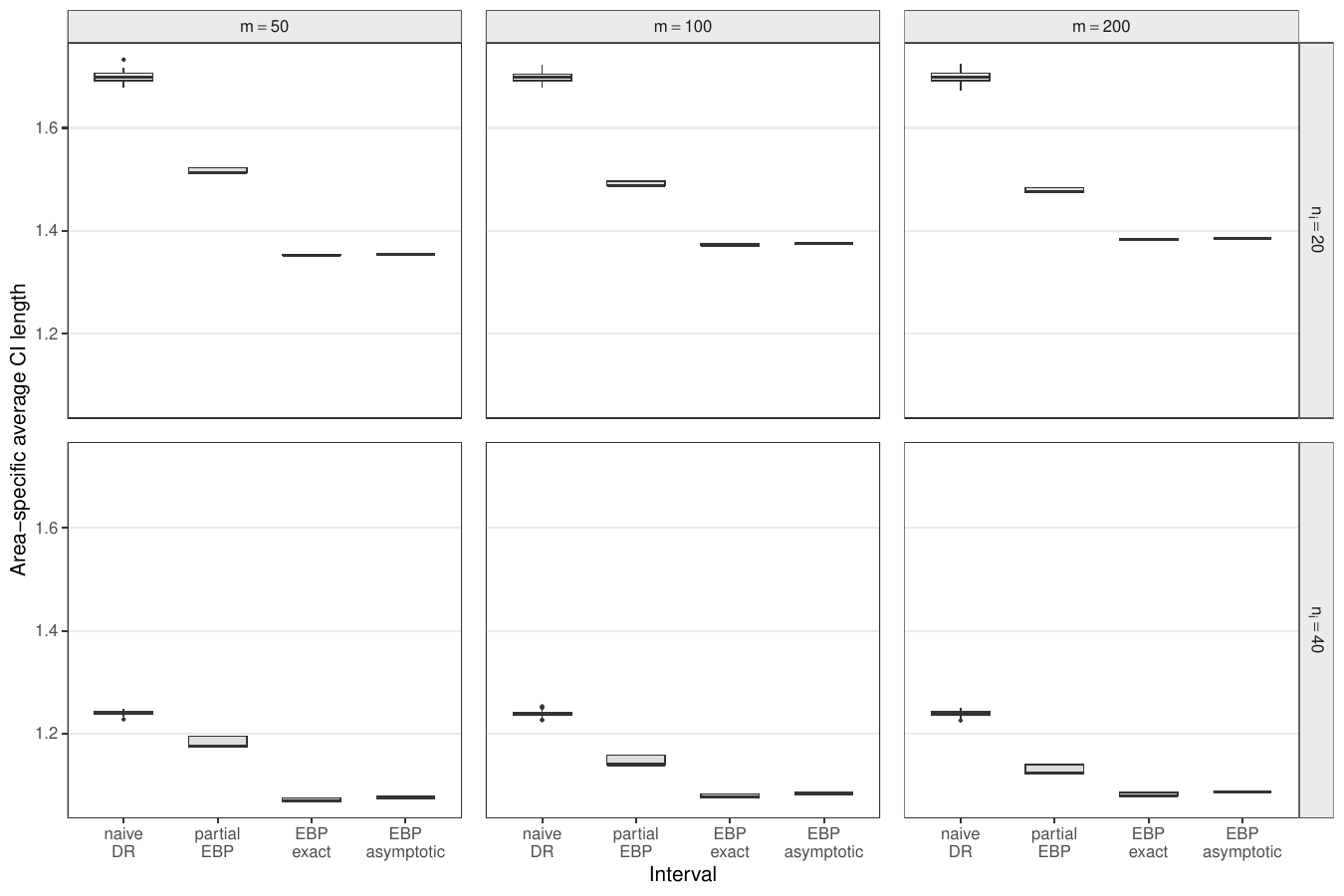}
\caption{\small Boxplots of Simulated Area-specific Average Length of 95\% Prediction Intervals: Noninformative Propensity; Equal-$n_{i}$ Design; EBP exact and EBP asymptotic refer to EBP-based Prediction Interval based on Exact and Approximate Asymptotic Variances, Respectively.}
\label{fig:final-correct-equal-length}
\end{figure}

To have a closer look at the area-specific MSPE, Coverage and Length, we present boxplots in Figures \ref{fig:final-correct-equal-mspe}--\ref{fig:final-correct-equal-length}. The figures show the same pattern as the summary statistics, with EBP being the best performer in MSPE and Length, and partial EBP the best performer in Coverage in 3 out of 6 cases. Another notable feature is that the variation of the area-specific Length of the EBP-based prediction interval [constructed via the approximate asymptotic variance, (\ref{eq:thm4_1})] is much smaller compared to other methods.


\begin{figure}[ht]
\centering
\includegraphics[width=0.85\textwidth]{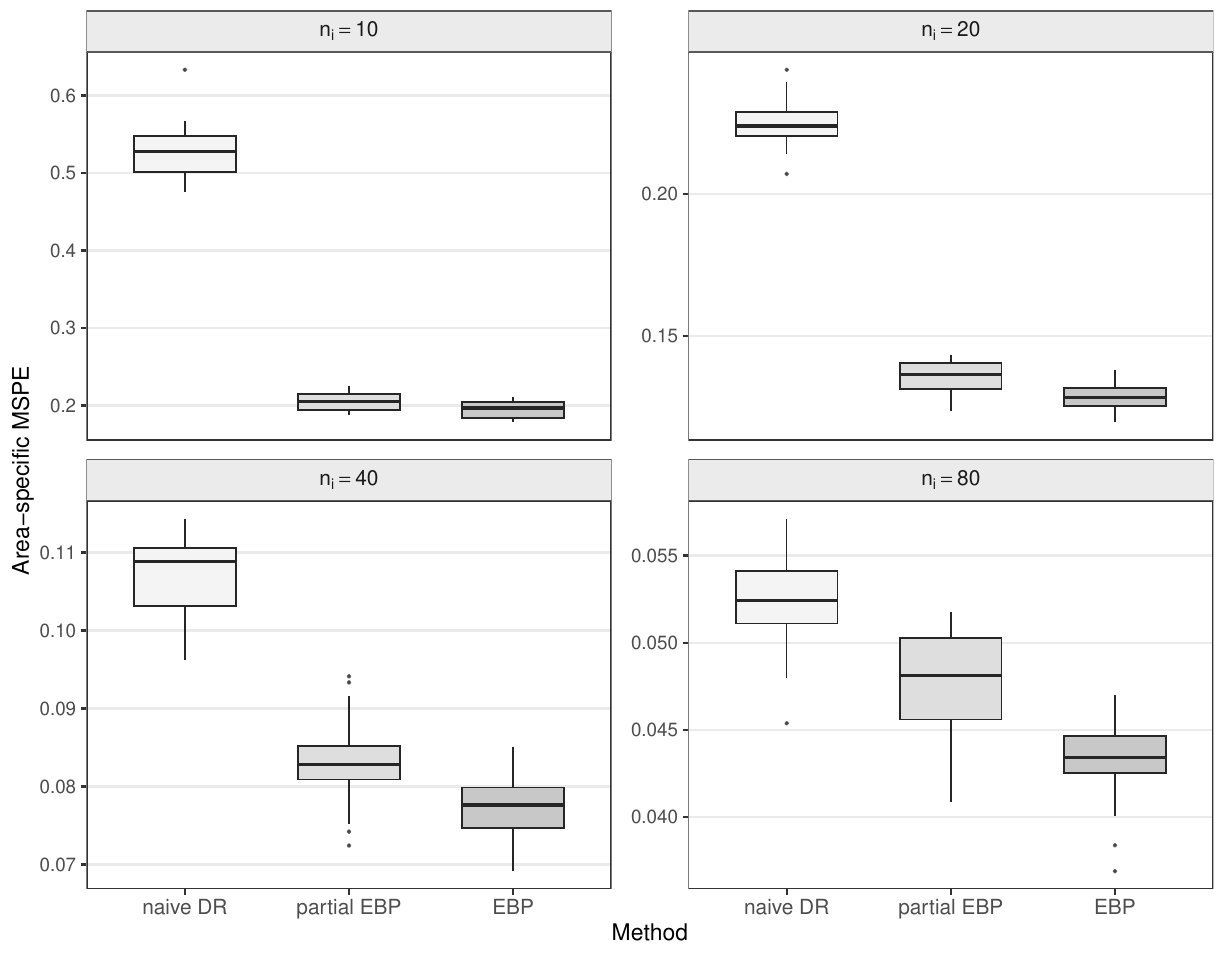}
\caption{Boxplots of Simulated Area-specific MSPE: Noninformative Propensity; $m=100$; Unequal-$n_{i}$ Design; Stratified by $n_{i}$ Values, with 25 Areas for Each $n_{i}$ Value.}
\label{fig:final-correct-variable-mspe}
\end{figure}

\begin{figure}[ht]
\centering
\includegraphics[width=0.85\textwidth]{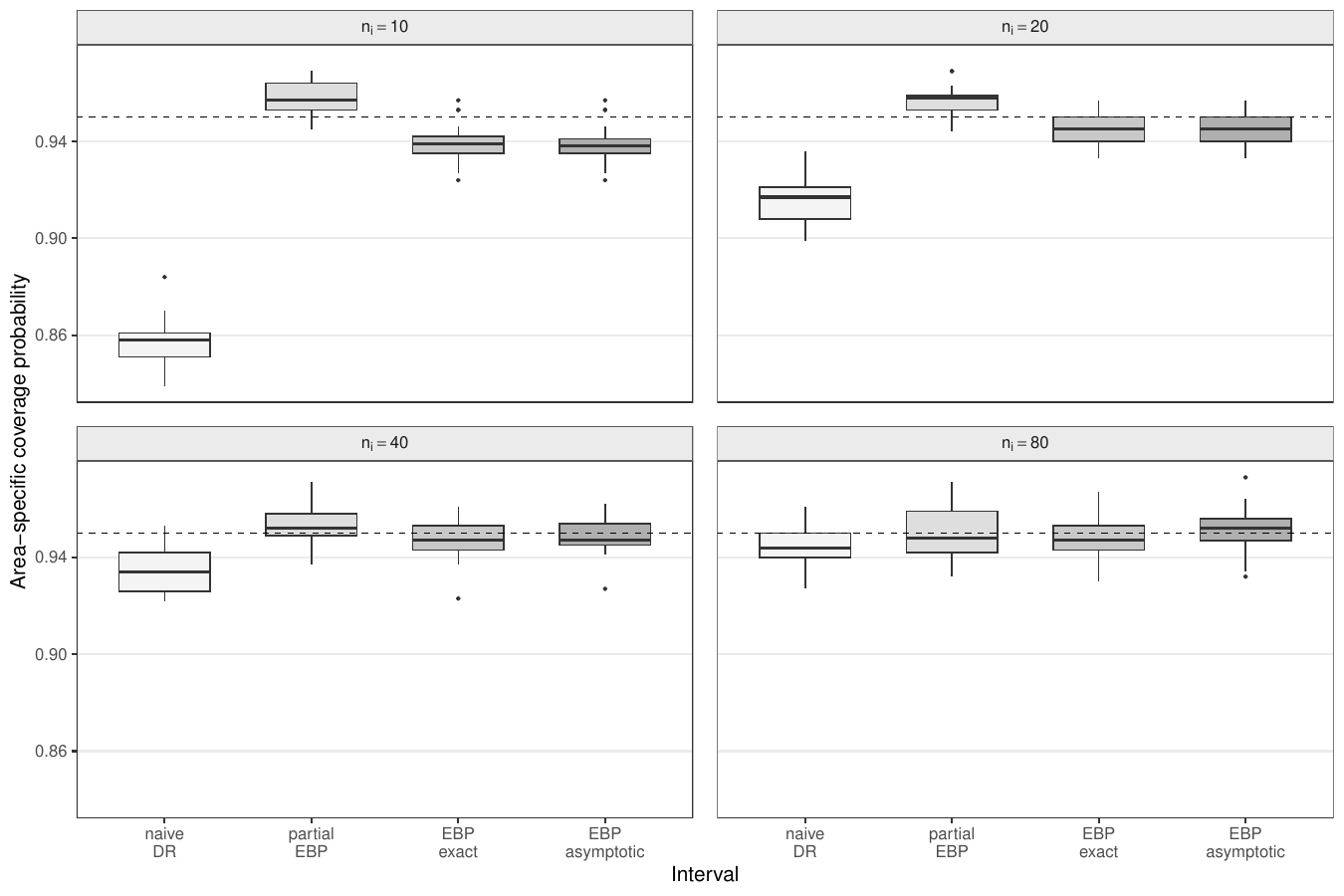}
\caption{Boxplots of Simulated Area-specific Coverage Probability of 95\% Prediction Intervals: Noninformative Propensity; $m=100$; Unequal-$n_{i}$ Design; Stratified by $n_{i}$ Values, with 25 Areas for Each $n_{i}$ Value; EBP exact and EBP asymptotic refer to EBP-based Prediction Interval based on Exact and Approximate Asymptotic Variances, Respectively.}
\label{fig:final-correct-variable-coverage}
\end{figure}

\begin{figure}[ht]
\centering
\includegraphics[width=0.85\textwidth]{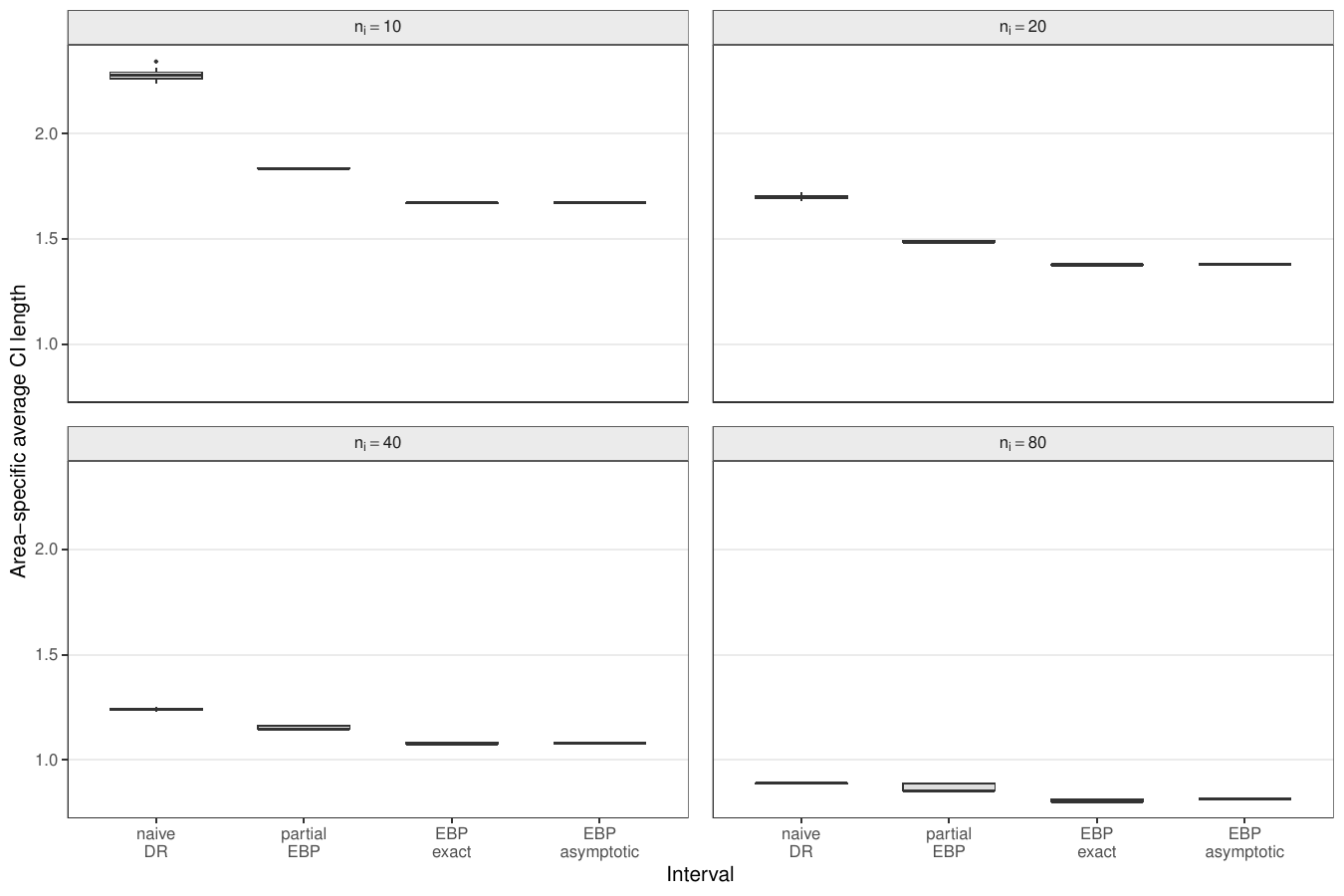}
\caption{Boxplots of Simulated Area-specific Average Length of 95\% Prediction Intervals: Noninformative Propensity; $m=100$; Unequal-$n_{i}$ Design; Stratified by $n_{i}$ Values, with 25 Areas for Each $n_{i}$ Value; EBP exact and EBP asymptotic refer to EBP-based Prediction Interval based on Exact and Approximate Asymptotic Variances, Respectively.}
\label{fig:final-correct-variable-length}
\end{figure}

{\it 2. Unequal-$n_{i}$ design.} Here, we consider $m=100$. Table \ref{tab:final-correct-variable_1} shows the simulated MSPE, Coverage and Length of the prediction interval, with separate summaries for each $n_{i}$ value. The patterns are very much the same as those observed in Table \ref{tab:final-correct-equal-n20} and Table \ref{tab:final-correct-equal-n40}. Namely, EBP has the smallest MSPE at every $n_{i}$ value. The asymptotic EBP prediction interval is slightly the shortest at $n_{i}=10$, whereas the EBP prediction interval is the shortest at $n_{i}=20,40$ and $80$; as for Coverage, partial EBP performs the best in 2 out of the 4 $n_{i}$ values, and EBP asymptotic performs the best in the other 2 $n_{i}$ values. The corresponding area-specific boxplots are presented in Figures \ref{fig:final-correct-variable-mspe}--\ref{fig:final-correct-variable-length}, showing consistent patterns as the summary statistics.

\begin{table}[ht]
\centering
\setlength{\belowcaptionskip}{8pt}
\caption{Simulated MSPE, Coverage Probability and Average Length of 95\% Prediction Interval: Noninformative Propensity; $m=100$; Unequal-$n_{i}$ Design; AAV --- Approximate Asymptotic Variance; Separate Table for Each $n_{i}$; Bold Font Number Indicates Best Performer under the Performance Measure.}
\label{tab:final-correct-variable_1}
\scriptsize
\setlength{\tabcolsep}{3.5pt}
\begin{minipage}[t]{0.48\textwidth}
\centering
\textbf{$n_i=10$}\\[4pt]
\begin{tabular}{lccc}
\hline
Method/interval & MSPE & Coverage & Avg. length \\
\hline
naive DR & 0.5300 & 0.8575 & 2.2775 \\
partial EBP & 0.2049 & \textbf{0.9577} & 1.8351 \\
EBP & \textbf{0.1943} & 0.9388 & 1.6709 \\
EBP (with AAV) & -- & 0.9386 & \textbf{1.6702} \\
\hline
\end{tabular}
\end{minipage}
\hfill
\begin{minipage}[t]{0.48\textwidth}
\centering
\textbf{$n_i=20$}\\[4pt]
\begin{tabular}{lccc}
\hline
Method/interval & MSPE & Coverage & Avg. length \\
\hline
naive DR & 0.2247 & 0.9157 & 1.6988 \\
partial EBP & 0.1358 & 0.9568 & 1.4871 \\
EBP & \textbf{0.1280} & 0.9449 & \textbf{1.3774} \\
EBP (with AAV) & -- & \textbf{0.9450} & 1.3793 \\
\hline
\end{tabular}
\end{minipage}

\vspace{10pt}

\begin{minipage}[t]{0.48\textwidth}
\centering
\textbf{$n_i=40$}\\[4pt]
\begin{tabular}{lccc}
\hline
Method/interval & MSPE & Coverage & Avg. length \\
\hline
naive DR & 0.1072 & 0.9350 & 1.2401 \\
partial EBP & 0.0833 & 0.9529 & 1.1506 \\
EBP & \textbf{0.0772} & 0.9475 & \textbf{1.0766} \\
EBP (with AAV) & -- & \textbf{0.9488} & 1.0822 \\
\hline
\end{tabular}
\end{minipage}
\hfill
\begin{minipage}[t]{0.48\textwidth}
\centering
\textbf{$n_i=80$}\\[4pt]
\begin{tabular}{lccc}
\hline
Method/interval & MSPE & Coverage & Avg. length \\
\hline
naive DR & 0.0524 & 0.9441 & 0.8900 \\
partial EBP & 0.0479 & \textbf{0.9504} & 0.8647 \\
EBP & \textbf{0.0431} & 0.9474 & \textbf{0.8048} \\
EBP (with AAV) & -- & 0.9508 & 0.8147 \\
\hline
\end{tabular}
\end{minipage}
\end{table}


Next, we consider the logistic propensity case. Again, we consider $m=100$. Table \ref{tab:correct-covariate-equal} presents the summary of different performance measures under the equal-$n_{i}$ design. It is seen that EBP is the best performer in MSPE and Length under both $n_{i}$ values; partial EBP performs the best in Coverage under both $n_{i}$ values, although EBP (with the approximate asymptotic variance) comes very close when $n_{i}=40$.
Table \ref{tab:final-correct-variable} presents the summary statistics under the unequal-$n_{i}$ design. Here, we observe similar patterns as in Table \ref{tab:final-correct-variable_1}, that is, EBP has the smallest MSPE at every $n_{i}$ value. The asymptotic EBP interval is slightly the shortest at $n_{i}=10$, whereas the EBP prediction interval is the shortest at $n_{i}=20,40$ and $80$; partial EBP performs the best in Coverage for $n_{i}=10, 20$ and $40$, and EBP performs the best for $n_{i}=80$. The area-specific boxplots, which are omitted, show consistent patterns with the summary statistics.

\begin{table}[ht]
\centering
\setlength{\belowcaptionskip}{8pt}
\caption{Simulated MSPE, Coverage Probability and Average Length of 95\% Prediction Interval: Logistic Propensity; $m=100$; Equal-$n_{i}$ Design; Numbers in the Parentheses Are Results based on the Approximate Asymptotic Variance for the EBP Method; Bold Font Numbers Indicate Best Performer under Respective $n_{i}$ and Performance Measure.}
\label{tab:correct-covariate-equal}
\scriptsize
\setlength{\tabcolsep}{3.5pt}
\begin{tabular}{r l lll}
\hline
$n_i$ & Method & MSPE & Coverage & Length\\
\hline
20 & naive DR & 0.2772 & 0.8917 & 1.7486\\
20 & partial EBP & 0.1482 & \textbf{0.9524} & 1.5224\\
20 & EBP & \textbf{0.1336} & 0.9439 (0.9442) & \textbf{1.3994} (1.4011)\\
40 & naive DR & 0.1224 & 0.9239 & 1.2785\\
40 & partial EBP & 0.0886 & \textbf{0.9486} & 1.1610\\
40 & EBP & \textbf{0.0791} & 0.9470 (0.9482) & \textbf{1.0887} (1.0941)\\
\hline
\end{tabular}
\end{table}

\begin{table}[ht]
\centering
\setlength{\belowcaptionskip}{8pt}
\caption{Simulated MSPE, Coverage Probability and Average Length of 95\% Prediction Interval: Logistic Propensity; $m=100$; Unequal-$n_{i}$ Design; AAV --- Approximate Asymptotic Variance; Separate Table for Each $n_{i}$; Bold Font Number Indicates Best Performer under Respective Performance Measure.}
\label{tab:final-correct-variable}
\scriptsize
\setlength{\tabcolsep}{3.5pt}
\begin{minipage}[t]{0.48\textwidth}
\centering
\textbf{$n_i=10$}\\[4pt]
\begin{tabular}{lccc}
\hline
Method/interval & MSPE & Coverage & Avg. length \\
\hline
naive DR & 1.6338 & 0.8086 & 2.4720 \\
partial EBP & 0.2210 & \textbf{0.9548} & 1.8930 \\
EBP & \textbf{0.2035} & 0.9399 & 1.7128 \\
EBP (with AAV) & -- & 0.9399 & \textbf{1.7118} \\
\hline
\end{tabular}
\end{minipage}
\hfill
\begin{minipage}[t]{0.48\textwidth}
\centering
\textbf{$n_i=20$}\\[4pt]
\begin{tabular}{lccc}
\hline
Method/interval & MSPE & Coverage & Avg. length \\
\hline
naive DR & 0.2774 & 0.8893 & 1.7406 \\
partial EBP & 0.1475 & \textbf{0.9508} & 1.5189 \\
EBP & \textbf{0.1352} & 0.9413 & \textbf{1.4013} \\
EBP (with AAV) & -- & 0.9416 & 1.4030 \\
\hline
\end{tabular}
\end{minipage}

\vspace{10pt}

\begin{minipage}[t]{0.48\textwidth}
\centering
\textbf{$n_i=40$}\\[4pt]
\begin{tabular}{lccc}
\hline
Method/interval & MSPE & Coverage & Avg. length \\
\hline
naive DR & 0.1221 & 0.9235 & 1.2782 \\
partial EBP & 0.0893 & \textbf{0.9496} & 1.1671 \\
EBP & \textbf{0.0792} & 0.9454 & \textbf{1.0875} \\
EBP (with AAV) & -- & 0.9469 & 1.0930 \\
\hline
\end{tabular}
\end{minipage}
\hfill
\begin{minipage}[t]{0.48\textwidth}
\centering
\textbf{$n_i=80$}\\[4pt]
\begin{tabular}{lccc}
\hline
Method/interval & MSPE & Coverage & Avg. length \\
\hline
naive DR & 0.0591 & 0.9396 & 0.9252 \\
partial EBP & 0.0515 & 0.9457 & 0.8745 \\
EBP & \textbf{0.0433} & \textbf{0.9490} & \textbf{0.8095} \\
EBP (with AAV) & -- & 0.9517 & 0.8193 \\
\hline
\end{tabular}
\end{minipage}
\end{table}
\subsubsection{The unknown $N_{i}$ setting}
\hspace{4mm}
As noted (see Section 4.5), partial EBP  does not apply to this case. Thus, here, we compare the predictive and inferential performance of naive DR and asymptotic EBP (EBP asymp), $\hat{d}_{i}^{*}$, given by (\ref{eq:dihst}), with SE given below (\ref{eq:thm4_2}). Here, we present the results under {\it Case I}; results under {\it Case II, III, IV} are deferred to the supplement.

Table \ref{tab:unknown-N-section45-equal} presents the summary statistics for MSPE, Coverage and Length under the equal-$n_{i}$ design, while Table \ref{tab:unknown-N-section45-variable} presents the corresponding results under the unequal-$n_{i}$ design. It is clear that asymptotic EBP performs better in all cases and under all performance measures. Area-specific boxplots, showing patterns consistent with the summary statistics, are deferred to Section 4.1.2 of the supplement.

\begin{table}[ht]
\centering
\setlength{\belowcaptionskip}{8pt}
\caption{\small Simulated MSPE, Coverage Probability and Average Length of 95\% Prediction Interval: Naive DR vs Asymptotic EBP; Unknown $N_{i}$ setting, {\it Case I} Scenario, Equal-\(n_i\) Design.}
\label{tab:unknown-N-section45-equal}
\scriptsize
\setlength{\tabcolsep}{4pt}
\renewcommand{\arraystretch}{1.08}
\par\vspace{4pt}\textit{\(n_i=20\)}\par\vspace{2pt}
\begin{tabular}{lcccccc}
\hline
Method & MSPE & Bias & SD & Mean SE & Coverage & Length \\
\hline
naive DR & 0.1985 & 0.0000 & 0.4455 & 0.4290 & 0.9285 & 1.6816 \\
EBP asymp & \textbf{0.1305} & 0.0000 & 0.3612 & 0.3494 & \textbf{0.9408} & \textbf{1.3698} \\
\hline
\end{tabular}
\par\vspace{6pt}\textit{\(n_i=40\)}\par\vspace{2pt}
\begin{tabular}{lcccccc}
\hline
Method & MSPE & Bias & SD & Mean SE & Coverage & Length \\
\hline
naive DR & 0.0984 & 0.0005 & 0.3137 & 0.3101 & 0.9416 & 1.2156 \\
EBP asymp & \textbf{0.0773} & 0.0005 & 0.2781 & 0.2762 & \textbf{0.9474} & \textbf{1.0828} \\
\hline
\end{tabular}
\end{table}
\begin{table}[ht]
\centering
\setlength{\belowcaptionskip}{8pt}
\caption{\small Simulated MSPE, Coverage Probability and Average Length of 95\% Prediction Interval: Naive DR vs Asymptotic EBP; Unknown $N_{i}$ Setting; {\it Case I} Scenario, Unequal-\(n_i\) Design.}
\label{tab:unknown-N-section45-variable}
\scriptsize
\setlength{\tabcolsep}{3.5pt}
\renewcommand{\arraystretch}{1.08}
\begin{tabular}{clcccccc}
\hline
\(n_i\) & Method & MSPE & Bias & SD & Mean SE & Coverage & Length \\
\hline
10 & naive DR & 0.3944 & -0.0010 & 0.6280 & 0.5767 & 0.8991 & 2.2607 \\
10 & EBP asymp & \textbf{0.1924} & -0.0016 & 0.4387 & 0.4241 & \textbf{0.9406} & \textbf{1.6623} \\
20 & naive DR & 0.1967 & 0.0038 & 0.4436 & 0.4292 & 0.9288 & 1.6824 \\
20 & EBP asymp & \textbf{0.1297} & 0.0018 & 0.3602 & 0.3505 & \textbf{0.9420} & \textbf{1.3741} \\
40 & naive DR & 0.0990 & 0.0026 & 0.3147 & 0.3100 & 0.9404 & 1.2154 \\
40 & EBP asymp & \textbf{0.0788} & 0.0035 & 0.2807 & 0.2753 & \textbf{0.9449} & \textbf{1.0793} \\
80 & naive DR & 0.0480 & 0.0021 & 0.2190 & 0.2214 & 0.9515 & 0.8678 \\
80 & EBP asymp & \textbf{0.0427} & 0.0023 & 0.2066 & 0.2075 & \textbf{0.9512} & \textbf{0.8133} \\
\hline
\end{tabular}
\end{table}
\subsection{A summary of results in all cases/scenarios}
\subsubsection{Known $N_{i}$ and covariate subpopulation}
\hspace{4mm}
Overall, EBP has smaller MSPE in most settings and shorter prediction intervals, while partial EBP often has Coverage closer to the nominal level. These patterns are not uniform across settings. The advantage of EBP in MSPE and Length is more significant under {\it Case I} and {\it Case III} (in which cases the outcome model is correctly specified) than under {\it Case II} and {\cal Case IV} (in which cases the outcome model is misspecified). On the other hand, the advantage of partial EBP in Coverage is more significant under {\it Case II} and {\it Case IV} than under {\it Case I} and {\it Case III} (because, in the latter cases, EBP performs better in Coverage, especially under larger $n_{i}$).

The performance of naive DR and partial EBP, especially the former, worsens as the degree of propensity model misspecification increases; on the other hand, the performance of EBP does not seem to be affected by the misspecification of the propensity model.

Below are more specific summaries under different scenarios.

{\it Case I and Case III.} The patterns under these two scenarios are similar, that is, EBP performs the best in MSPE and Length in all cases, and partial EBP performs the best in Coverage in most cases. The relative performance of EBP and partial EBP in Coverage does not seem to depend much on the degree of propensity model misspecification; rather, it depends more on the sample size of $n_{i}$; namely, for smaller $n_{i}$, partial EBP performs better; for larger $n_{i}$, EBP (with approximate asymptotic variance) performs better.

{\it Case II.} Despite outcome model misspecification, EBP has the smallest MSPE in most settings and retains shorter prediction intervals. Under stronger outcome misspecification, partial EBP has slightly smaller MSPE in some settings. On the other hand, as the degree of outcome model misspecification increases, the Coverage performance of all methods declines, but the decline in Coverage is more substantial for EBP than for naive DR and partial EBP.

{\it Case IV.} Despite that both the outcome model and the propensity model are misspecified, the overall pattern regarding the relative performance of the three comparing methods remain largely the same. EBP has the smallest MSPE in most settings and retains shorter prediction intervals, while partial EBP generally has Coverage closer to the nominal level. Partial EBP has smaller MSPE in some settings with stronger outcome misspecification. On the other hand, as the degrees of outcome model misspecification increases, the advantage in MSPE of EBP over the comparing methods, especially partial EBP, becomes less substantial; likewise, the advantage of partial EBP in Coverage over the comparing methods becomes less significant. The performance of EBP does not seem to be affected by the degree of misspecification of the propensity model; on the other hand, the increase of the degree of propensity model misspecification is seen to worsen the performance of naive DR and partial EBP, especially the former.

Detailed results can be found in Section 5.2.1 and Sections 4.1--4.5 of the supplement.
\subsubsection{Unknown-$N_{i}$}
\hspace{4mm}
In all cases of equal or unequal-$n_{i}$ designs, asymptotic EBP has superior performance over naive DR in MSPE and Coverage. As for Length, asymptotic EBP has superior performance over naive DR in all cases with smaller $n_{i}$ (i.e., $n_{i}\leq 20$), and most cases with larger $n_{i}$ (i.e., $n_{i}\geq 40$). In the overall summary under unequal-$n_{i}$ design, asymptotic EBP outperforms naive DR in all cases and under all performance measures. Detailed results can be found in Section 5.2.2 and Section 4.6 of the supplement.
\section{Design-based and model/design-based simulation studies: Comparison with CSAE-AIPW}
\hspace{4mm}
In this section, we carry out a series of design-based simulation studies, in the sense that, given a population (either provided or generated), the only randomness in the data is due to repeated sampling from the finite populations (small areas). In addition, we also consider design/model-based simulation studies, in which the populations are independently regenerated across simulation and a single set of data are drawn from each generated population. The latter setting is similar to that in the previous section. We compare the predictive performance of the proposed EBP method with the CSAE-AIPW method (Reluga {\it et al.} 2026; see Section 1 for description). 

We consider three scenarios. The first scenario, scenario I, has two sub-scenarios. Scenario I.1 is one in which the finite population was generated, and released, by Reluga {\it et al.} (2026), which we refer to as Reluga population. Here, we have a fixed synthetic population, from which 1,000 samples are repeatedly drawn using the sampling frames described below. Scenario I.2 is essentially repeat scenario I.1 except that 20 new populations, which are different from the Reluga population, were generated independently; for each of the 20 populations, 1,000 samples are repeated drawn as done for the Reluga population. The second scenario, scenario II, is one in which the population was generated by the code published by the above authors. Here, the generated populations are independently simulated in each of the 1,000 simulation runs using the publicly available code, with covariates, treatment assignments, and potential outcomes regenerated per population. One set of sampled data are then drawn from the generated population using the sampling frames described below. The third scenario, scenario III, includes a number of variations under the latter scenario, with the purpose of investigating the robustness of both methods, EBP and CSAE-AIPW. The detailed simulation settings are described in Section 5.1 of the supplementary material.

Within each generated subpopulation (small area), two sampling designs are considered: (i) \textbf{SRS40} denotes simple random sampling without replacement of 40 units per area. (ii)
\textbf{Stratified sampling} follows the original Reluga {\it et al.} design, sampling treated and control units separately with unequal sample sizes across areas. 

Under each scenario, same-sample comparisons, in which both methods used identical samples of outcomes and auxiliary variables, are made to evaluate the performance of the two methods in terms of the average (over the small areas) simulated MSPE based on $K=1,000$ simulation runs. The results are presented in Tables \ref{tab:reluga} -- \ref{tab:mixture}, with boldface number highlighting the better performer under each scenario.

\subsection{A summary of results}
\hspace{4mm}
It is seen that, under scenario I.1, the comparison is in favor of CSAE-AIPW, whereas under all the other scenarios, namely, scenarios I.2, II and III, the latter including various simulation settings (Table \ref{tab:robustness}) and the mixture situations (Table \ref{tab:mixture}), the comparison is in favor of the EBP. In particular, under scenario I.2, EBP performed better under 16 out 20 of the generated populations in case of SRS40, and EBP performed better under all 20 generated populations in case of Stratified sampling. The advantage of EBP is more significant under the stratified sampling scheme.

\begin{table}[H]
\centering
\caption{Average Simulated MSPE. Scenario I and II}
\label{tab:reluga}
\scriptsize
\begin{tabular}{@{}lllll@{}}
\toprule
Scenario & Sampling & EBP & CSAE-AIPW & Further Detail\\
\midrule
I.1 & SRS40 & 0.0806 & \textbf{0.0705} &\\
I.1 & Stratified & 0.1185 & \textbf{0.1054} &\\ \hline
I.2 & SRS40 & \textbf{0.0707} & 0.0769 & 16/20 (80\%) in favor of EBP\\
I.2 & Stratified & \textbf{0.0698} & 0.0846
& 20/20 (100\%) in favor of EBP \\\hline
II & SRS40 & \textbf{0.0628} & 0.0711 &\\
II & Stratified & \textbf{0.0648} & 0.0815 &\\
\bottomrule
\end{tabular}
\end{table}

\begin{table}[H]
\centering
\caption{Average Simulated MSPE. Scenario III: Various Simulation Settings.}
\label{tab:robustness}
\scriptsize
\begin{tabular}{@{}lrr@{}}
\toprule
Simulation setting & SACI EBP & CSAE-AIPW\\
\midrule
Baseline & \textbf{0.076236} & 0.089092 \\
Shared quadratic term ($\gamma=2$) & \textbf{0.289389} & 0.302039 \\
Treated-only quadratic term ($\delta=2$) & \textbf{0.285204} & 0.289511 \\
Heavy-tailed area effects (scaled $t_3$) & \textbf{0.072918} & 0.085316 \\
Heavy-tailed residuals (scaled $t_3$) & \textbf{0.075560} & 0.088483 \\
Heteroskedastic residuals & \textbf{0.118448} & 0.130211 \\
Smaller samples ($n_i=10$) & \textbf{0.198090} & 0.213955 \\
Unequal samples ($n_i\in\{10,20,40,80\}$) & \textbf{0.108792} & 0.122144 \\
Covariate-dependent assignment ($\kappa=1$) & \textbf{0.077895} & 0.096020 \\
Covariate-dependent assignment ($\kappa=3$) & \textbf{0.078631} & 0.167053 \\
Area-dependent assignment ($s=1$) & \textbf{0.094248} & 0.998584 \\
Area-dependent assignment ($s=3$) & \textbf{0.181773} & 2.880434 \\
\bottomrule
\end{tabular}
\end{table}

\begin{table}[H]
\centering
\caption{Average Simulated MSPE. Scenario III: SACI--Reluga Outcome Mixture.}
\label{tab:mixture}
\scriptsize
\begin{tabular}{@{}rrr@{}}
\toprule
$\lambda$ & SACI EBP MSPE & CSAE-AIPW MSPE\\
\midrule
0 & \textbf{0.104045} & 0.189159 \\
0.25 & \textbf{0.073392} & 0.123352 \\
0.5 & \textbf{0.078779} & 0.105774 \\
0.75 & \textbf{0.120103} & 0.151233 \\
1 & \textbf{0.183930} & 0.216858 \\
\bottomrule
\end{tabular}
\end{table}

{\bf Acknowledgement:} Jiming Jiang's research is partially supported by the NSF grant DMS-2210569. The authors report there are no competing interests to declare.


\begin{thebibliography}{99}
\bibitem[Bang and Robins(2005)]{BangRobins2005}
Bang, H. and Robins, J. M. (2005), Doubly robust estimation in missing data and causal inference models, {\it Biometrics} 61, 962--972.
\bibitem[Chambers and Tzavidis(2006)]{ChambersTzavidis2006}
Chambers, R. and Tzavidis, N. (2006), M-quantile models for small area estimation, {\it Biometrika}, 93, 255--268.
\bibitem[Chen and Guestrin(2016)]{ChenGuestrin2016}
Chen, T. and Guestrin, C. (2016), XGBoost: A scalable tree boosting system, in {\it Proceedings of the 22nd ACM SIGKDD International Conference on Knowledge Discovery and Data Mining}, 785--794.
\bibitem[Ito and Sugasawa(2026)]{ItoSugasawa2026}
Ito, T. and Sugasawa, S. (2026), Causal small area estimation with survey-only covariates, arXiv:2605.03285 [math.ST].
\bibitem[Jiang(2001)]{Jiang2001}
Jiang, J. (2001), Mixed-effects models with random cluster sizes,
{\it Stat. Probab. Lett.} 53, 201-206.
\bibitem[Jiang and Nguyen(2021)]{JiangNguyen2021}
Jiang, J. and Nguyen, T. (2021), {\it Linear and Generalized
Linear Mixed Models and Their Applications}, 2nd ed., Springer,
New York.
\bibitem[Jiang et al.(2011)]{JiangNguyenRao2011}
Jiang, J., Nguyen, T. and Rao, J. S. (2011), Best predictive small area estimation, {\it J. Amer. Statist. Assoc.} 106, 732-745.
\bibitem[Jiang and Rao(2025)]{JiangRao2025}
Jiang, J. and Rao, J. S. (2025), {\it Robust Small Area Estimation: Methods, Theory, Applications, and Open Problems}, Chapman \& Hall/CRC.
\bibitem[Jiang and Torabi(2020)]{JiangTorabi2020}
Jiang, J. and Torabi, M. (2020), Sumca: Simple, unified, Monte-Carlo assisted approach to second-order unbiased mean-squared prediction error estimation, {\it J. Roy. Statist. Soc. Ser. B} 82, 467--485.
\bibitem[Jiang et al.(2022)]{JiangWandBhaskaran2022}
Jiang, J., Wand, M. P. and Bhaskaran, A. (2022), Usable and
precise asymptotics for generalized linear mixed model analysis and
design, {\it J. Roy. Statist. Soc. B} 84, 55--82.
\bibitem[Lu and Shiou(2002)]{LuShiou2002}
Lu, T.-T. and Shiou, S.-H. (2002), Inverses of $2\times 2$ block matrices, {\it Comput. Math. Appl.} 43, 119--129.
\bibitem[Neyman(1923)]{Neyman1923}
Neyman, J. (1923), Sur les applications de la th\'{e}orie des probabilit\'{e}s aux experiences agricoles: Essai des principes, {\it Roczniki Nauk Rolniczych} 10, 1--51.
\bibitem[Posbergh et al.(2023)]{PosberghEtAl2023}
Posbergh, C., Roeder, B., Murphy, T., Stewart, W., Kott, R., Hatfield, P., Thomas, V., and Drummond, J. (2023), Ewe and lamb body condition, breeding and production data from the Montana Agricultural Experiment station, Montana State Univ., Bozeman, MT from 1960 - 2012. Ag Data Commons, https://doi.org/10.15482/USDA.ADC/1528462.
\bibitem[Ranjbar et al.(2023)]{RanjbarEtAl2023}
Ranjbar, S., Salvati, N. and Pacini, B. (2023), Estimating heterogeneous causal effects in observational studies using small area predictors, {\it Comput. Stat. Data Anal.} 184, 107742.
\bibitem[Reluga et al.(2026)]{RelugaEtAl2026}
Reluga, K., Kong, D., Ranjbar, S., Salvati, N. and van der Laan, M. (2026), The impact of job stability on monetary poverty in Italy: Causal small area estimation, {\it J. Amer. Statist. Assoc.} 121, 877--890.
\bibitem[Rao and Molina(2015)]{RaoMolina2015}
Rao, J. N. K. and Molina, I. (2015), {\it Small Area Estimation}, 2nd ed., Wiley, New York.
\bibitem[Robins and Rotnitzky(1995)]{RobinsRotnitzky1995}
Robins, J. M. and Rotnitzky, A. (1995), Semiparametric efficiency in multivariate regression models with missing data, {\it J. Amer. Statist. Assoc.} 90, 122--129.
\bibitem[Rubin(1974)]{Rubin1974}
Rubin, D. B. (1974), Estimating causal effects of treatments in randomized and nonrandomized studies, {\it J. Educ. Psychol.} 66, 688.
\end{thebibliography}
\end{document}


\renewcommand{\baselinestretch}{1}
\def\Quote{\begin{quotation}\normalfont\small}
\def\EndQuote{\end{quotation}\rm}
\def\BigHeading{\bfseries\Large}\def\MediumHeading{\bfseries\large}
\def\bct{\begin{center}}
\def\ect{\end{center}}
\font\BigCaps=cmcsc9 scaled \magstep 1
\font\BigSlant=cmsl10    scaled \magstep 1
\def\lbk{\linebreak}
\def\Report{Small Area Functional Estimation}
\def\Author{Ma and Jiang}
\pagestyle{myheadings}
\markboth{\Author}{\Report}
\thispagestyle{empty}
\bct{\BigHeading Supplementary Material}\\\vskip10pt
\BigCaps Jiming Jiang, Yilin Li and Jiangshan Zhang\lbk
\BigSlant University of California, Davis
\ect
\Quote
\vskip-5pt

This supplement provides technical derivations and proofs, additional results of simulation studies, a real-data example, as well as extensions and remarks, for the main manuscript, Small area causal inference, by the same authors. Throughout the supplement, the main manuscript is referred to as SACI.

\EndQuote
\section{Some technical derivations}
\subsection{Derivation of expressions of conditional expectations}
\hspace{4mm}
In the derivations below, the probabilities are conditional on the sampling configuration; in other words, the sampling configuration is considered fixed. Without loss of generality, assume that the first $n_{i}$ units, $j=1,\dots,n_{i}$, are sampled and the rest of the units, $j=n_{i}+1,\dots,N_{i}$ are non-sampled. It follows that  ${\cal S}=\{(i,j):1\leq i\leq m,1\leq j\leq n_{i}\}$, and $W=(W_{ij})_{1\leq i\leq m,1\leq j\leq n_{i}}$. Also let $Y_{\cal S}=(Y_{ij})_{1\leq i\leq m, 1\leq j\leq n_{i}}$ and $Y_{\cal S}^{\rm c}=(Y_{ij}^{c})_{1\leq i\leq m,1\leq j\leq n_{i}}$, where $Y_{ij}=Y_{ij}(W_{ij})$ and $Y_{ij}^{c}=Y_{ij}(1-W_{ij})$. $Y_{\cal
S}$ is the set of responses under either treatment. For any fixed $y_{ij}(r),(i,j)\in{\cal P}\setminus{\cal S}$, $r=0,1$, and $y_{ij}, y_{ij}^{c}, w_{ij},(i.j)\in{\cal S}$,  let $w=(w_{ij})_{1\leq i\leq m,1\leq j\leq n_{i}}$. Then, we have
\begin{eqnarray}
&&{\rm P}[Y_{ij}(r)\leq y_{ij}(r), (i,j)\in{\cal P}\setminus{\cal S}, r=0,1, Y_{ij}^{c}\leq
y_{ij}^{c},Y_{ij}\leq y_{ij}, (i,j)\in{\cal S}, W=w]\nonumber\\
&=&{\rm P}[Y_{ij}(r)\leq y_{ij}(r), (i,j)\in{\cal P}\setminus{\cal S}, r=0,1, Y_{ij}(1
-w_{ij})\leq y_{ij}^{c}, (i,j)\in{\cal S},\nonumber\\
&&Y_{ij}(w_{ij})\leq y_{ij}, (i,j)\in{\cal S}, W=w]\nonumber\\
&=&{\rm P}[Y_{ij}(r)\leq y_{ij}(r), (i,j)\in{\cal P}\setminus{\cal S}, r=0,1,Y_{ij}(1-
w_{ij})\leq y_{ij}^{c}, (i,j)\in{\cal S},\nonumber\\
&&Y_{ij}(w_{ij})\leq y_{ij},(i,j)\in{\cal S}]{\rm P}(W=w).\label{eq:bp_1}
\end{eqnarray}

Denote $Y_{0,1;{\cal P}\setminus{\cal S}}=[Y_{ij}(r)]_{(i,j)\in{\cal P}\setminus{\cal
S},r=0,1}$, $y_{0,1;{\cal P}\setminus{\cal S}}=[y_{ij}(r)]_{(i,j)\in{\cal P}\setminus{\cal S},r=0,1}$, $y=(y_{ij})_{(i,j)\in{\cal S}}$, $y^{c}=(y_{ij}^{c})_{(i,j)\in{\cal S}}$. $Y_{\cal S}(w)=[Y_{ij}(w_{ij})]_{(i,j)\in{\cal S}}$, and $Y_{\cal S}^{c}(w)=[Y_{ij}^{c}(w_{ij})]_{(i,j)
\in{\cal S}}$ with $Y_{ij}^{c}(w_{ij})=Y_{ij}(1-w_{ij})$. Note that $Y_{\cal S}=Y_{\cal S}(W)$ and $Y_{\cal S}^{c}=Y_{\cal
S}^{c}(W)$. For any (Euclidean) random object, $Z$, let $f_{Z}(z)$ denote its pdf with respect to a $\sigma$-finite measure (Lebesgue measure for continuous variables; counting measure for discrete variables) with argument variable $z$. A conditional pdf, like $f_{Z_{1}|Z_{2}}(z_{1}|z_{2})$, is defined analogously. By (\ref{eq:bp_1}), and the arbitrariness of the $y$'s, we have shown that
\begin{equation}
f_{Y_{0,1;{\cal P}\setminus{\cal S}},Y_{\cal S}^{c},Y_{\cal S},W}(y_{0,1;{\cal
P}\setminus{\cal S}},y^{c},y,w)=f_{Y_{0,1;{\cal P}\setminus{\cal S}},Y_{\cal S}^{c}(w),
Y_{\cal S}(w)}(y_{0,1;{\cal P}\setminus{\cal S}},y^{c},y)f_{W}(w).\label{eq:bp_2}
\end{equation}
By integrating out $y_{0,1;{\cal P}\setminus{\cal S}}$ and $y^{c}$ on both sides of (\ref{eq:bp_2}), one gets
\begin{equation}
f_{Y_{\cal S},W}(y,w)=f_{Y_{\cal S}(w)}(y)f_{W}(w).\label{eq:bp_3}
\end{equation}
It follows that the conditional pdf of $[Y_{0,1;{\cal P}\setminus{\cal S}},Y_{\cal S}^{c}]$ given $Y_{\cal S},W$ is
\begin{eqnarray}
f_{Y_{0,1;{\cal P}\setminus{\cal S}},Y_{\cal S}^{c}|Y_{\cal S},W}(y_{0,1;{\cal
P}\setminus{\cal S}},y^{c}|y,w)&=&\frac{f_{Y_{0,1;{\cal P}\setminus{\cal S}},Y_{\cal
S}^{c},Y_{\cal S},W}(y_{0,1;{\cal P}\setminus{\cal S}},y^{c},y,w)}{f_{Y_{\cal S},W}(y,w)}
\nonumber\\
&=&\frac{f_{Y_{0,1;{\cal P}\setminus{\cal S}},Y_{\cal S}^{c}(w),Y_{\cal S}(w)}(y_{0,
1;{\cal P}\setminus{\cal S}},y^{c},y)f_{W}(w)}{f_{Y_{\cal S}(w)}(y)f_{W}(w)}\nonumber\\
&=&\frac{f_{Y_{0,1;{\cal P}\setminus{\cal S}},Y_{\cal S}^{c}(w),Y_{\cal S}(w)}(y_{0,
1;{\cal P}\setminus{\cal S}},y^{c},y)}{f_{Y_{\cal S}(w)}(y)}\nonumber\\
&=&f_{Y_{0,1;{\cal P}\setminus{\cal S}},Y_{\cal S}^{c}(w)|Y_{\cal S}(w)}(y_{0,
1;{\cal P}\setminus{\cal S}},y^{c}|y).\label{eq:bp_4}
\end{eqnarray}

Our characteristics of interest are the area-specific causal effects,
\begin{eqnarray}
d_{i}&=&\frac{1}{N_{i}}\sum_{j=1}^{N_{i}}\{Y_{ij}(1)-Y_{ij}(0)\}\nonumber\\
&=&\frac{1}{N_{i}}\left[\sum_{j=1}^{n_{i}}\{Y_{ij}(1)-Y_{ij}(0)\}+\sum_{j=n_{i}+1}^{N_{i}}
\{Y_{ij}(1)-Y_{ij}(0)\}\right],\label{eq:bp_5}
\end{eqnarray}
$1\leq i\leq m$. It is easy to show that
\begin{eqnarray}
\sum_{j=1}^{n_{i}}\{Y_{ij}(1)-Y_{ij}(0)\}&=&\sum_{1\leq j\leq n_{i},W_{ij}=1}Y_{ij}
-\sum_{1\leq j\leq n_{i},W_{ij}=0}Y_{ij}\nonumber\\
&&+\sum_{1\leq j\leq n_{i},W_{ij}=0}Y_{ij}^{c}-\sum_{1\leq j\leq n_{i},W_{ij}=1}Y_{ij}^{c},
\label{eq:bp_6}
\end{eqnarray}
Combining (\ref{eq:bp_5}) and (\ref{eq:bp_6}), we obtain the following expression:
\begin{eqnarray}
d_{i}&=&\frac{1}{N_{i}}\left(\sum_{1\leq j\leq n_{i},W_{ij}=1}Y_{ij}-\sum_{1\leq j\leq
n_{i},W_{ij}=0}Y_{ij}\right)\nonumber\\
&&+\frac{1}{N_{i}}\left[\sum_{1\leq j\leq n_{i},W_{ij}=0}Y_{ij}^{c}-\sum_{1\leq j\leq
n_{i},W_{ij}=1}Y_{ij}^{c}+\sum_{j=n_{i}+1}^{N_{i}}\{Y_{ij}(1)-Y_{ij}(0)\}\right],
\label{eq:bp_7}
\end{eqnarray}
$1\leq i\leq m$. Therefore, the BP of $d_{i}$ is given by
\begin{eqnarray}
{\rm E}(d_{i}|Y_{\cal S},W)&=&\frac{1}{N_{i}}\left(\sum_{1\leq j\leq n_{i},W_{ij}=1}Y_{ij}-
\sum_{1\leq j\leq n_{i},W_{ij}=0}Y_{ij}\right)\nonumber\\
&&+\frac{1}{N_{i}}\left[\sum_{1\leq j\leq n_{i},W_{ij}=0}{\rm E}(Y_{ij}^{c}|Y_{\cal S},W)
-\sum_{1\leq j\leq n_{i},W_{ij}=1}{\rm E}(Y_{ij}^{c}|Y_{\cal S},W)\right.\nonumber\\
&&\left.+\sum_{j=n_{i}+1}^{N_{i}}{\rm E}\{Y_{ij}(1)-Y_{ij}(0)|Y_{\cal S},W\}\right],
\label{eq:bp_8}
\end{eqnarray}
where, by (\ref{eq:bp_4}), we have the following expressions:
\begin{eqnarray}
{\rm E}(Y_{ij}^{c}|Y_{\cal S},W)&=&\left.\int y_{ij}^{c}f_{Y_{0,1;{\cal P}\setminus{\cal S}},
Y_{\cal S}^{c}|Y_{\cal S},W}(y_{0,1;{\cal P}\setminus{\cal S}},y^{c}|y,w)d y_{0,1;{\cal
P}\setminus{\cal S}}d y^{c}\right|_{y=Y_{\cal S},w=W}\nonumber\\
&=&\left.\int y_{ij}^{c}f_{Y_{0,1;{\cal P}\setminus{\cal S}},Y_{\cal S}^{c}(w)|Y_{\cal S}(w)}(y_{0,
1;{\cal P}\setminus{\cal S}},y^{c}|y)d y_{0,1;{\cal P}\setminus{\cal S}}d y^{c}\right|_{y=
Y_{\cal S}(w),w=W}\nonumber\\
&=&{\rm E}\{Y_{ij}^{c}(w_{ij})|Y_{\cal S}(w)=y\}|_{y=Y_{\cal S}(w),w=W}\nonumber\\
&=&{\rm E}\{Y_{ij}^{c}(w_{ij})|Y_{\cal S}(w)\}|_{w=W},\;\;\;1\leq j\leq n_{i};\label{eq:bp_9}\\
{\rm E}\{Y_{ij}(r)|Y_{\cal S},W)&=&\left.\int y_{ij}(r)f_{Y_{0,1;{\cal P}\setminus{\cal S}},
Y_{\cal S}^{c}|Y_{\cal S},W}(y_{0,1;{\cal P}\setminus{\cal S}},y^{c}|y,w)d y_{0,1;{\cal
P}\setminus{\cal S}}d y^{c}\right|_{y=Y_{\cal S},w=W}\nonumber\\
&=&\left.\int y_{ij}(r)f_{Y_{0,1;{\cal P}\setminus{\cal S}},Y_{\cal S}^{c}(w)|Y_{\cal S}(w)}(y_{0,
1;{\cal P}\setminus{\cal S}},y^{c}|y)d y_{0,1;{\cal P}\setminus{\cal S}}d y^{c}\right|_{y=
Y_{\cal S}(w),w=W}\nonumber\\
&=&{\rm E}\{Y_{ij}(r)|Y_{\cal S}(w)=y\}|_{y=Y_{\cal S}(w),w=W}\nonumber\\
&=&{\rm E}\{Y_{ij}(r)|Y_{\cal S}(w)\}|_{w=W},\;\;\;n_{i}+1\leq j\leq N_{i}, r=0,1.\label{eq:bp_10}
\end{eqnarray}
\subsection{Some derivations in Section 4.2 of SACI}
\hspace{4mm}
By (3) and (4) of SACI, the following expression can be derived: $\hat{d}_{i}-d_{i}=$
\begin{eqnarray}
&&\frac{1}{N_{i}}\left(\sum_{1\leq j\leq n_{i},W_{ij}=0}\{\hat{E}(Y_{ij}^{c}|Y_{\cal S},W)
-Y_{ij}^{c}\}-\sum_{1\leq j\leq n_{i},W_{ij}=1}\{\hat{E}(Y_{ij}^{c}|Y_{\cal S},W)-Y_{ij}^{c}\}\right.
\nonumber\\
&&\left.+\sum_{j=n_{i}+1}^{N_{i}}[\hat{\rm E}\{Y_{ij}(1)|Y_{\cal S},W\}-Y_{ij}(1)]-\sum_{j=n_{i}+1}^{N_{i}}
[\hat{\rm E}\{Y_{ij}(0)|Y_{\cal S},W\}-Y_{ij}(0)]\right).\label{eq:inf_1}
\end{eqnarray}

Note that, under the LMM, (7) of SACI, we have
\begin{equation}
Y_{ij}^{c}=Y_{ij}(1-W_{ij})=x_{ij}'\{\beta_{0}+(1-W_{ij})\beta_{1}\}+\alpha_{i0}+(1-W_{ij})\alpha_{i1}+
\epsilon_{ij}(1-W_{ij}).\label{eq:inf_2}
\end{equation}
On the other hand, by (18) of SACI, we have
\begin{eqnarray}
\hat{E}(Y_{ij}^{c}|Y_{\cal S},W)&=&x_{ij}'\{\hat{\beta}_{0}+(1-W_{ij})\hat{\beta}_{1}\}\nonumber\\
&&+\hat{a}_{i,1-W_{ij},1}(W_{i})\sum_{k=1}^{n_{1i}(W_{i})}\{Y_{ik}-x_{ik}'(\hat{\beta}_{0}+W_{ik}
\hat{\beta}_{1})\}\nonumber\\
&&+\hat{a}_{i,1-W_{ij},0}(W_{i})\sum_{k=n_{1i}(W_{i})+1}^{n_{i}}\{Y_{ik}-x_{ik}'(\hat{\beta}_{0}+W_{ik}
\hat{\beta}_{1})\},\label{eq:inf_3}
\end{eqnarray}
where, by (14) and (15) of SACI, we have
\begin{eqnarray}
\hat{a}_{i,1-W_{ij},1}(W_{i})&=&\frac{\hat{\sigma}_{0}^{2}+(1-W_{ij})\hat{\rho}\hat{\sigma}_{0}
\hat{\sigma}_{1}}{\hat{\tau}^{2}+n_{1i}(W_{i})\hat{\nu}}\left\{1+n_{1i}(W_{i})\hat{\phi}_{i}
+\frac{n_{0i}(W_{i})(\hat{\sigma}_{0}^{2}+\hat{\rho}\hat{\sigma}_{0}\hat{\sigma}_{1})}{\hat{\tau}^{2}
+n_{0i}(W_{i})\hat{\gamma}_{i}}\right\}\nonumber\\
&&+\{\hat{\rho}\hat{\sigma}_{0}\hat{\sigma}_{1}+(1-W_{ij})\hat{\sigma}_{1}^{2}\}\frac{1+n_{1i}(W_{i})
\hat{\phi}_{i}}{\hat{\tau}^{2}+n_{1i}(W_{i})\hat{\nu}},\label{eq:inf_4}\\
\hat{a}_{i,1-W_{ij},0}(W_{i})&=&\frac{\hat{\sigma}_{0}^{2}+(1-W_{ij})\hat{\rho}\hat{\sigma}_{0}
\hat{\sigma}_{1}}{\hat{\tau}^{2}+n_{0i}(W_{i})\hat{\gamma}_{i}}\left\{1+\frac{n_{1i}(W_{i})
(\hat{\sigma}_{0}^{2}+\hat{\rho}\hat{\sigma}_{0}\hat{\sigma}_{1})}{\hat{\tau}^{2}+n_{1i}(W_{i})
\hat{\nu}}\right\}\nonumber\\
&&+\{\hat{\rho}\hat{\sigma}_{0}\hat{\sigma}_{1}+(1-W_{ij})\hat{\sigma}_{1}^{2}\}\frac{n_{1i}(W_{i})(
\hat{\sigma}_{0}^{2}+\hat{\rho}\hat{\sigma}_{0}\hat{\sigma}_{1})}{\{\hat{\tau}^{2}+n_{0i}(W_{i})
\hat{\gamma}_{i}\}\{\hat{\tau}^{2}+n_{1i}(W_{i})\hat{\nu}\}}\label{eq:inf_5}
\end{eqnarray}
with $\hat{\nu}=\hat{\sigma}_{0}^{2}+2\hat{\rho}\hat{\sigma}_{0}\hat{\sigma}_{1}
+\hat{\sigma}_{1}^{2}$,
$$\hat{\gamma}_{i}=\frac{\hat{\sigma}_{0}^{2}\{\hat{\tau}^{2}+n_{1i}(W_{i})(1-\hat{\rho}^{2})
\hat{\sigma}_{1}^{2}\}}{\hat{\tau}^{2}+n_{1i}(W_{i})\hat{\nu}}\;\;{\rm and}\;\;\hat{\phi}_{i}=
\frac{n_{0i}(W_{i})(\hat{\sigma}_{0}^{2}+\hat{\rho}\hat{\sigma}_{0}\hat{\sigma}_{1})^{2}}{\{\hat{\tau}^{2}
+n_{0i}(W_{i})\hat{\gamma}_{i}\}\{\hat{\tau}^{2}+n_{1i}(W_{i})\hat{\nu}\}}.$$
From (\ref{eq:inf_3}), the following expressions can be derived:
\begin{eqnarray}
&&\frac{1}{n_{0i}(W_{i})}\sum_{1\leq j\leq n_{i},W_{ij}=0}\hat{\rm E}(Y_{ij}^{c}|Y_{\cal S},W)=\left\{
\frac{1}{n_{0i}(W_{i})}\sum_{1\leq j\leq n_{i},W_{ij}=0}x_{ij}\right\}'(\hat{\beta}_{0}+\hat{\beta}_{1})
\nonumber\\
&&+\hat{a}_{i,1,1}(W_{i})\sum_{k=1}^{n_{1i}(W_{i})}\{Y_{ik}-x_{ik}'(\hat{\beta}_{0}
+W_{ik}\hat{\beta}_{1})\}\nonumber\\
&&+\hat{a}_{i,1,0}(W_{i})\sum_{k=n_{1i}(W_{i})+1}^{n_{i}}\{Y_{ik}-x_{ik}'(\hat{\beta}_{0}+W_{ik}
\hat{\beta}_{1})\},\label{eq:inf_6}
\end{eqnarray}
\begin{eqnarray}
&&\frac{1}{n_{1i}(W_{i})}\sum_{1\leq j\leq n_{i},W_{ij}=1}\hat{\rm E}(Y_{ij}^{c}|Y_{\cal S},W)=\left\{
\frac{1}{n_{1i}(W_{i})}\sum_{1\leq j\leq n_{i},W_{ij}=1}x_{ij}\right\}'\hat{\beta}_{0}\nonumber\\
&&+\hat{a}_{i,0,1}(W_{i})\sum_{k=1}^{n_{1i}(W_{i})}\{Y_{ik}-x_{ik}'(\hat{\beta}_{0}
+W_{ik}\hat{\beta}_{1})\}\nonumber\\
&&+\hat{a}_{i,0,0}(W_{i})\sum_{k=n_{1i}(W_{i})+1}^{n_{i}}\{Y_{ik}-x_{ik}'(\hat{\beta}_{0}+W_{ik}
\hat{\beta}_{1})\}.\label{eq:inf_7}
\end{eqnarray}

As mentioned in SACI (Section 4.2 of SACI, first paragraph), we consider the scenario such that $m\rightarrow\infty$ and $n_{i}, 1\leq i\leq m$ are bounded. We also assume that the conditions of Corollary 2 of SACI hold. Then, the following expression can be established:
\begin{eqnarray*}
&&\hat{d}_{i}-d_{i}=\frac{1}{N_{i}}\left(\sum_{1\leq j\leq n_{i},W_{ij}=0}\{{\rm E}(Y_{ij}^{c}|Y_{\cal S},W)
-Y_{ij}^{c}\}\right.\\
&&-\sum_{1\leq j\leq n_{i},W_{ij}=1}\{{\rm E}(Y_{ij}^{c}|Y_{\cal S},W)-Y_{ij}^{c}\}+\sum_{j=n_{i}+1}^{N_{i}}[{\rm E}\{Y_{ij}(1)|Y_{\cal S},W\}-Y_{ij}(1)]\\
&&\left.-\sum_{j=n_{i}+1}^{N_{i}}
[{\rm E}\{Y_{ij}(0)|Y_{\cal S},W\}-Y_{ij}(0)]\right)+o_{\rm P}(1).
\end{eqnarray*}
Furthermore, the following expressions can be derived:
\begin{eqnarray*}
&&\frac{1}{n_{0i}(W_{i})}\sum_{1\leq j\leq n_{i},W_{ij}=0}\{{\rm E}(Y_{ij}^{c}|Y_{\cal S},W)-Y_{ij}^{c}\}
=\{n_{1i}(W_{i})a_{i,1,1}(W_{i})-1\}(\alpha_{i0}+\alpha_{i1})\\
&&+n_{0i}(W_{i})a_{i,1,0}(W_{i})\alpha_{i0}+a_{i,1,1}(W_{i})\sum_{k=1}^{n_{1i}(W_{i})}\epsilon_{ik}(1)+a_{i,1,0}(W_{i})\sum_{k=n_{1i}(W_{i})
+1}^{n_{i}}\epsilon_{ik}(0)\\
&&-\frac{1}{n_{0i}(W_{i})}\sum_{1\leq j\leq n_{i},W_{ij}=0}\epsilon_{ij}(1);
\end{eqnarray*}
\begin{eqnarray*}
&&\frac{1}{n_{1i}(W_{i})}\sum_{1\leq j\leq n_{i},W_{ij}=1}\{{\rm E}(Y_{ij}^{c}|Y_{\cal S},W)-Y_{ij}^{c}\}=n_{1i}(W_{i})a_{i,0,1}(W_{i})(\alpha_{i0}+\alpha_{i1})\\
&&+\{n_{0i}(W_{i})a_{i,0,0}(W_{i})-1\}\alpha_{i0}+a_{i,0,1}(W_{i})\sum_{k=1}^{n_{1i}(W_{i})}\epsilon_{ik}(1)+a_{i,0,0}(W_{i})\sum_{k=n_{1i}(W_{i})
+1}^{n_{i}}\epsilon_{ik}(0)\\
&&-\frac{1}{n_{1i}(W_{i})}\sum_{1\leq j\leq n_{i},W_{ij}=1}\epsilon_{ij}(0);
\end{eqnarray*}
\begin{eqnarray*}
&&\frac{1}{N_{i}-n_{i}}\sum_{j=n_{i}+1}^{N_{i}}[{\rm E}\{Y_{ij}(1)|Y_{\cal S},W\}-Y_{ij}(1)]=\{n_{1i}(W_{i})a_{i,1,1}(W_{i})-1\}(\alpha_{i0}+\alpha_{i1})\\
&&+n_{0i}(W_{i})a_{i,1,0}(W_{i})\alpha_{i0}+a_{i,1,1}(W_{i})\sum_{k=1}^{n_{1i}(W_{i})}\epsilon_{ik}(1)+a_{i,1,0}(W_{i})\sum_{k=n_{1i}(W_{i})
+1}^{n_{i}}\epsilon_{ik}(0)\\
&&-\frac{1}{N_{i}-n_{i}}\sum_{j=n_{i}+1}^{N_{i}}\epsilon_{ij}(1);
\end{eqnarray*}
\begin{eqnarray*}
&&\frac{1}{N_{i}-n_{i}}\sum_{j=n_{i}+1}^{N_{i}}[{\rm E}\{Y_{ij}(0)|Y_{\cal S},W\}-Y_{ij}(0)]=n_{1i}(W_{i})a_{i,0,1}(W_{i})(\alpha_{i0}+\alpha_{i1})\\
&&+\{n_{0i}(W_{i})a_{i,0,0}(W_{i})-1\}\alpha_{i0}+a_{i,0,1}(W_{i})\sum_{k=1}^{n_{1i}(W_{i})}\epsilon_{ik}(1)+a_{i,0,0}(W_{i})\sum_{k=n_{1i}(W_{i})
+1}^{n_{i}}\epsilon_{ik}(0)\\
&&-\frac{1}{N_{i}-n_{i}}\sum_{j=n_{i}+1}^{N_{i}}\epsilon_{ij}(0).
\end{eqnarray*}
With these expressions, expression (27) of SACI can be established.
\section{Proofs of Theorem 2, Corollary 1 and Corollary 2}
\subsection{Proof of Theorem 2}
\hspace{4mm}
We verify the conditions of Theorem 1. By (5) of SACI, we have $f_{Y_{\cal S},W}(y,w)=f_{Y_{\cal S}(w)}(y)f_{W}(w)$ [recall $Y_{\cal S}=Y_{\cal S}(W)$]. Furthermore, by (16) of SACI, it is easy to derive that $l=\sum_{i=1}^{m}l_{i}(\psi,\xi_{i},n_{i})$, where $\xi_{ij}=(Y_{ij}(W_{ij}),W_{ij})$, $1\leq j\leq n_{i}$, and
\begin{eqnarray*}
&&l_{i}(\psi,\xi_{i},n_{i})=\log f_{Y_{i,{\cal S}}(w_{i})}(Y_{i,{\cal S}}(w_{i}))|_{w_{i}=W_{i}}=-\frac{1}{2}\{n_{i}\log(2\pi)+\log(|\Sigma_{i}(W_{i})|)\\
&&+(Y_{i,{\cal S}}-X_{i,{\cal S}}\beta_{0}-T_{i,{\cal S}}\beta_{1})'\Sigma_{i}^{-1}(W_{i})(Y_{i,{\cal S}}-X_{i,{\cal S}}\beta_{0}-T_{i,{\cal S}}\beta_{1})\}
\end{eqnarray*}
with $Y_{i,{\cal S}}=Y_{i,{\cal S}}(W_{i})$ and $T_{i,{\cal S}}=T_{i,{\cal S}}(W_{i})$.

Condition (i) is implied by condition (I). Condition (ii) is obvious. As for condition (iii), let $\psi$ be $\psi_{*}$, the true parameter vector. We have
\begin{eqnarray*}
\frac{\partial l_{i}}{\partial\beta_{0}}&=&-2X_{i,{\cal S}}'\Sigma_{i}^{-1}(W_{i})(Y_{i,{\cal S}}-X_{i,{\cal S}}\beta_{0}-T_{i,{\cal S}}\beta_{1}),\\
\frac{\partial l_{i}}{\partial\beta_{1}}&=&-2T_{i,{\cal S}}'\Sigma_{i}^{-1}(W_{i})(Y_{i,{\cal S}}-X_{i,{\cal S}}\beta_{0}-T_{i,{\cal S}}\beta_{1})
\end{eqnarray*}
(see, e.g., Jiang and Nguyen 2021, sec. A.2 for differentiation of matrix expressions). It follows that ${\rm E}(\partial l_{i}/\partial\beta_{r})={\rm E}\{{\rm E}(\partial l_{i}/\partial\beta_{r}|W)\}=0$, $r=0,1$. Next, let $\gamma$ denote one of the variance components, $\sigma_{0}^{2}, \sigma_{1}^{2}, \rho$ and $\tau^{2}$, and $\Delta_{i}=Y_{i,{\cal S}}-X_{i,{\cal S}}\beta_{0}-T_{i,{\cal S}}\beta_{1}$. We have
$$\frac{\partial l_{i}}{\partial\gamma}=-\frac{1}{2}\left\{{\rm tr}\left(\Sigma_{i}^{-1}\frac{\partial\Sigma_{i}}{\partial\gamma}\right)-\Delta_{i}'\Sigma_{i}^{-1}\frac{\partial\Sigma_{i}}{\partial\gamma}\Sigma_{i}^{-1}\Delta_{i}\right\},$$
where $\Sigma_{i}^{-1}=\Sigma_{i}^{-1}(W_{i})$ and $\partial\Sigma_{i}/\partial\gamma=(\partial/\partial\gamma)\Sigma_{i}(W_{i})$. We have
\begin{eqnarray*}
&&{\rm E}\left(\left.\Delta_{i}'\Sigma_{i}^{-1}\frac{\partial\Sigma_{i}}{\partial\gamma}\Sigma_{i}^{-1}\Delta_{i}\right|W\right)={\rm E}\left\{\left.{\rm tr}\left(\Sigma_{i}^{-1}\frac{\partial\Sigma_{i}}{\partial\gamma}\Sigma_{i}^{-1}\Delta_{i}\Delta_{i}'\right)\right|W\right\}\\
&&={\rm tr}\left(\Sigma_{i}^{-1}\frac{\partial\Sigma_{i}}{\partial\gamma}\Sigma_{i}^{-1}{\rm E}(\Delta_{i}\Delta_{i}'|W)\right)={\rm tr}\left(\Sigma_{i}^{-1}\frac{\partial\Sigma_{i}}{\partial\gamma}\right).
\end{eqnarray*}
It follows that ${\rm E}(\partial l_{i}/\partial\gamma|W)=0$, hence ${\rm E}(\partial l_{i}/\partial\gamma)=0$. Thus, condition (iii) holds.

To verify condition (iv), let $\|A\|=\sqrt{\lambda_{\max}(A'A)}$ denote the spectral norm of matrix $A$. First note that, by (19) of SACI and the subsequent expressions of $B_{st}, s,t=1,2$ given in SACI, it can be shown that $\|\Sigma_{i}^{-1}(W_{i})\|$, which depends on the variance components, is uniformly bounded in a neighborhood of the true variance components. Let $X_{ik}=(x_{ijk})_{1\leq j\leq n_{i}}$ denote the $k$th column of $X_{i,{\cal S}}$. For $\psi=\psi_{*}$, it can be shown
$${\rm E}\left\{\left.\left(\frac{\partial l_{i}}{\partial\beta_{0k}}\right)^{2}\right|W\right\}=4X_{ik}'\Sigma_{i}^{-1}X_{ik}\leq 4\|\Sigma_{i}^{-1}\|\cdot|X_{ik}|^{2}\leq cn_{i},$$
$1\leq k\leq p$. Hereafter, $c$ denotes a generic positive constant, whose value may be different at different places. Similarly, it can be shown
$${\rm E}\left\{\left.\left(\frac{\partial l_{i}}{\partial\beta_{1k}}\right)^{2}\right|W\right\}=4T_{ik}'\Sigma_{i}^{-1}T_{ik}\leq 4\|\Sigma_{i}^{-1}\|\cdot|T_{ik}|^{2}\leq cn_{i},$$
$1\leq k\leq p$, where $T_{ik}=(w_{ij}x_{ijk})_{1\leq j\leq n_{i}}$ denotes the $k$th column of $T_{i,{\cal S}}$. It follows, by condition (II), that ${\rm E}(\partial l_{i}/\partial\beta_{rk})^{2}$, $1\leq k\leq p, r=0,1$ are bounded. Similarly, it can be shown that ${\rm E}(\partial l_{i}/\partial\gamma)^{2}$, $\gamma=\sigma_{0}^{2},\sigma_{1}^{2},\rho,\tau^{2}$ are bounded.

Similarly, it can be shown that ${\rm E}(\partial^{2}l_{i}/\partial\beta_{rk}\partial\beta_{sl})^{2}$, ${\rm E}(\partial^{2}l_{i}/\partial\beta_{rk}\partial\gamma)^{2}$, ${\rm E}(\partial^{2}l_{i}/\partial\gamma\partial\nu)^{2}$, $r,s=0,1$, $1\leq k,l\leq p$, $\gamma,\nu\in\{\sigma_{0}^{2},\sigma_{1}^{2},\rho,\tau^{2}\}$ are bounded, using results for the second moments of quadratic forms (e.g., Jiang and Nguyen 2021, sec. B.2).

Finally, let $\psi_{k}$ denote the $k$th component of $\psi$. Using the uniform boundedness of $\|\Sigma_{i}^{-1}(W_{i})\|$ in a neighborhood of $\psi_{*}$, it can be shown that there is $\delta>0$ such that the following quantities are bounded:
$${\rm E}\left(\sup_{|\psi-\psi_{*}|\leq\delta}\left|\frac{\partial^{3}l_{i}}{\partial\psi_{j}\partial\psi_{k}\partial\psi_{l}}\right|\right),\;\;\;1\leq j,k,l\leq 2p+4.$$

For condition (v), again let $\psi=\psi_{*}$. The following expression can be established:
$${\rm E}\left(\frac{\partial l_{i}}{\partial\psi\partial\psi'}\right)=\left(\begin{array}{cc}2B_{i,1}&0\\
0&B_{i,2}/2\end{array}\right),$$
where $B_{i,r}, r=1,2$ are defined above Theorem 2 in SACI. Condition (v) then follows by condition (III) of Theorem 2.

For condition (vi), by the expressions of partial derivatives derived earlier, it is easy to show that ${\rm E}(\partial l_{i}/\partial\psi_{k})^{4}$, $1\leq k\leq 2p+4, 1\leq i\leq m$ are bounded.

Condition (vii) follows from the fact that, in this case, the regularity conditions hold for exchanging the order of differentiation and expectation (e.g., Protter and Morrey 1985).
\subsection{Proofs of Corollary 1 and Corollary 2}
\hspace{4mm}
For Corollary 1, note that (30) of SACI implies that $B_{m}$ is a bounded sequence of matrices. Therefore, for every subsequence of $B_{m}$, by the Bolzano-Weierstrass theorem (e.g., Bartle and Sherbert 2000), there is a further subsequence for which (26) of SACI holds. It follows that, for the further subsequence, we have $\hat{\psi}-\psi_{*}\stackrel{\rm P}{\longrightarrow}0$ and (31) of SACI holds. This implies that $\hat{\psi}-\psi_{*}\stackrel{\rm P}{\longrightarrow}0$ and (31) of SACI must hold for the entire sequence; otherwise, a contradiction can be constructed via the above subsequence argument.

For Corollary 2, the arguments are similar. The boundedness of the sequences $\{B_{m,r}\}$, $r=1,2$ implies that, for every subsequence of $B_{m}={\rm diag}(2B_{m,1},B_{m,2}/2)$, there is a further subsequence such that $B_{m,r}\rightarrow B_{r}, r=1,2$ for the subsequence. It then follows from Theorem 2 that, for the further subsequence, we have $B={\rm diag}(2B_{1},B_{2}/2)>0$; hence, we have $\hat{\psi}-\psi\stackrel{\rm P}{\longrightarrow}0$ and that (31) of SACI holds for $B_{m}={\rm diag}(2B_{m,1},B_{m,2}/2)$, for the further subsequence. Thus, that $\hat{\psi}-\psi\stackrel{\rm P}{\longrightarrow}0$ and (31) of SACI must hold for the entire sequence, or a contradiction will take place due to the subsequence argument.
\section{Proofs of Lemma 1, Theorem 3 and Theorem 4}
\subsection{Proof of Lemma 1}
\hspace{4mm}
We first establish a lemma regarding the convergence rate of $\hat{\psi}$, a solution to the likelihood equation. For any fixed $0<\delta<1/2$, let $b_{m}=m^{\delta}$. For notation simplicity, we drop the subscript $m$, so write $b=b_{m}$ (but keep in mind what it is). Let $S_{b}=\{\psi: |\psi-\psi_{*}|\leq b/\sqrt{m}\}$. It follows that the boundary of $S_{b}$ is $\partial S_{b}=\{\psi:|\psi-\psi_{*}|=b/\sqrt{m}\}$. By Taylor expansion, we have, for $\psi\in\partial S_{b}$,
\begin{eqnarray}
l(\psi)&=&l(\psi_{*})+\frac{\partial l_{*}}{\partial\psi'}(\psi-\psi_{*})+\frac{1}{2}(\psi-\psi_{*})'\frac{\partial^{2}l_{*}}{\partial\psi\partial\psi'}(\psi-\psi_{*})\nonumber\\
&&+\frac{1}{6}\sum_{u,v,w=1}^{2p+4}\frac{\partial^{3}\tilde{l}}{\partial\psi_{u}\partial\psi_{v}\partial\psi_{w}}(\psi_{u}-\psi_{*u})(\psi_{v}-\psi_{*v})(\psi_{w}-\psi_{*w})\nonumber\\
&=&l(\psi_{*})+\left(\frac{1}{\sqrt{m}}\frac{\partial l_{*}}{\partial\psi'}\right)\sqrt{m}(\psi-\psi_{*})\nonumber\\
&&+\frac{1}{2}\sqrt{m}(\psi-\psi_{*})'\left(\frac{1}{m}\frac{\partial^{2}l_{*}}{\partial\psi\partial\psi'}\right)\sqrt{m}(\psi-\psi_{*})
\nonumber\\
&&+\frac{1}{6}\sum_{u,v,w=1}^{2p+4}\left(\frac{1}{m^{3/2}}\frac{\partial^{3}\tilde{l}}{\partial\psi_{u}\partial\psi_{v}\partial\psi_{w}}\right)
\nonumber\\
&&\times\sqrt{m}(\psi_{u}-\psi_{*u})\sqrt{m}(\psi_{v}-\psi_{*v})\sqrt{m}(\psi_{w}-\psi_{*w}),\label{eq:plS1_1}
\end{eqnarray}
where the partial derivatives with $l_{*}$ are evaluated at $\psi_{*}$, and those with $\tilde{l}$ evaluated at a point $\tilde{\psi}$ between $\psi$ and $\psi_{*}$. It can be shown that, for sufficiently large $m$, we have
$$\sup_{\psi\in\partial{\cal S}}\left|\frac{\partial^{3}\tilde{l}_{i}}{\partial\psi_{u}\partial\psi_{v}\partial\psi_{w}}\right|\leq c_{1}(1+|\Delta_{i}|^{2})$$
for some constants $c_{1}>0$, where $\Delta_{i}=Y_{i,{\cal S}}-X_{i,{\cal S}}\beta_{*0}-T_{i,{\cal S}}\beta_{*1}$. Thus, we have
$$\frac{1}{m^{3/2}}\left|\frac{\partial^{3}\tilde{l}}{\partial\psi_{u}\partial\psi_{v}\partial\psi_{w}}\right|\leq \frac{c_{1}}{m^{3/2}}\left(m+\sum_{i=1}^{m}|\Delta_{i}|^{2}\right),$$
$\leq u,v,w\leq 2p+4$, $\psi\in\partial S_{b}$. It follows that, for $\psi\in\partial S_{b}$, the last term on the right side of (\ref{eq:plS1_1}) is bounded in absolute value by
\begin{eqnarray*}
&&\frac{c_{1}}{6m^{3/2}}\left(m+\sum_{i=1}^{m}|\Delta_{i}|^{2}\right)\left(\sum_{u=1}^{2p+4}\sqrt{m}|\psi_{u}-\psi_{*u}|\right)^{3}\\
&\leq&\frac{c_{1}}{6m^{3/2}}(2p+4)^{3/2}\left(m+\sum_{i=1}^{m}|\Delta_{i}|^{2}\right)|\sqrt{m}(\psi-\psi_{*})|^{3}\\
&=&\frac{c_{1}}{6m^{3/2}}(2p+4)^{3/2}b^{3}\left(m+\sum_{i=1}^{m}|\Delta_{i}|^{2}\right).
\end{eqnarray*}
using the Cauchy-Schwarz inequality for the inequality step above. We have $\sum_{i=1}^{m}|\Delta_{i}|^{2}=\sum_{i=1}^{m}{\rm E}(|\Delta_{i}|^{2})+\sum_{i=1}^{m}\Delta_{i3}$ with $\Delta_{i3}=|\Delta_{i}|^{2}-{\rm E}(|\Delta_{i}|^{2})$. It follows that $\sum_{i=1}^{m}|\Delta_{i}|^{2}\leq c_{2}m+\sum_{i=1}^{m}\Delta_{i3}$ for some constant $c_{2}>0$. Thus, on the event
$${\cal A}_{3}=\left\{\left|\sum_{i=1}^{m}\Delta_{i3}\right|\leq m\right\},$$
we have that the last term on the right side of (\ref{eq:plS1_1}) is bounded by
$$\frac{c_{1}}{6m^{1/2}}(c_{2}+2)(2p+4)^{3/2}b^{3}\leq c_{3}b^{3}m^{-1/2},$$
$\forall\psi\in\partial S_{b}$ for some constant $c_{3}>0$

Now consider the second term on the right side of (\ref{eq:plS1_1}). We have
$$\frac{1}{m}\frac{\partial^{2}l_{*}}{\partial\psi\partial\psi'}=\frac{1}{m}\sum_{i=1}^{m}{\rm E}\left(\frac{\partial^{2}l_{i*}}{\partial\psi\partial\psi'}\right)+\frac{1}{m}\sum_{i=1}^{m}\Delta_{i2}=-B_{m}+\frac{1}{m}\sum_{i=1}^{m}\Delta_{i2},$$
where $B_{m}$ is defined above Corollary 1 in SACI, and
$$\Delta_{i2}=\frac{\partial^{2}l_{i*}}{\partial\psi\partial\psi'}-{\rm E}\left(\frac{\partial^{2}l_{i*}}{\partial\psi\partial\psi'}\right).$$
By Corollary 2, we have $\liminf\lambda_{\min}(B_{m})>0$; thus, there is a constant $a>0$ such that $\lambda_{\min}(B_{m})\geq a$ for sufficiently large $m$. It follows that, for sufficiently large $m$, we have $B_{m}\geq\lambda_{\min}(B_{m})I_{2p+4}\geq aI_{2p+4}$; therefore, for sufficiently large $m$, we have, on the event
$${\cal A}_{2}=\left\{\left\|\sum_{i=1}^{m}\Delta_{i2}\right\|\leq\frac{a}{2}m\right\},$$
that $m^{-1}(\partial^{2}l_{*}/\partial\psi\partial\psi')\leq -aI_{2p+4}+\|m^{-1}\sum_{i=1}^{m}\Delta_{i2}\|I_{2p+4}\leq-(a/2)I_{2p+4}$. It follows that, for sufficiently large $m$, the third term on the right side of (\ref{eq:plS1_1}) is bounded by
$$-\frac{a}{4}|\sqrt{m}(\psi-\psi_{*})|^{2}=-\frac{ab^{2}}{4},\;\;\forall\psi\in\partial S_{b}.$$

Finally, on the event
$${\cal A}_{1}=\left\{\left|\frac{\partial l_{*}}{\partial\psi}\right|\leq\frac{ab}{8}\sqrt{m}\right\},$$
the second term on the right side of (\ref{eq:plS1_1}) is bounded by
$$\frac{ab}{8}|\sqrt{m}(\psi-\psi_{*})|=\frac{ab^{2}}{8},\;\;\forall\psi\in\partial S_{b}.$$

In conclusion, on the event ${\cal A}={\cal A}_{1}\cap{\cal A}_{2}\cap{\cal A}_{3}$, we have, for all $\psi\in\partial S_{b}$,
\begin{eqnarray*}
&&l(\psi)\leq l(\psi_{*})+\frac{ab^{2}}{8}-\frac{ab^{2}}{4}+c_{3}b^{3}m^{-1/2}\\
&&=-\frac{ab^{2}}{8}+c_{3}b^{2}m^{\delta-1/2}=-b^{2}\left(\frac{a}{8}-c_{3}m^{\delta-1/2}\right)<0,
\end{eqnarray*}
if $m$ is sufficiently large; therefore, the maximizer of $l(\cdot)$ over $S_{b}$, $\hat{\psi}$, must lie in the interior of $S_{b}$; it follows that $\partial l/\partial\psi|_{\hat{\psi}}=0$ and $|\hat{\psi}-\psi_{*}|\leq b/\sqrt{m}=m^{\delta-1/2}$. Thus, we have
$\{|\hat{\psi}-\psi|>m^{\delta-1/2}\}\subset{\cal A}^{c}$, hence ${\rm P}(|\hat{\psi}-\psi_{*}|>m^{\delta-1/2})\leq{\rm P}({\cal A}^{c})\leq\sum_{s=1}^{3}{\rm P}({\cal A}_{s}^{c})$.

First, we have ${\rm P}({\cal A}_{1}^{c})={\rm P}[|\partial l_{*}/\partial\psi|>(a/8)m^{1/2+\delta}]\leq{\rm P}[\max_{1\leq u\leq 2p+4}|\partial l_{*}/\partial\psi_{u}|>(a/8)m^{1/2+\delta}/\sqrt{2p_+4}]\leq\sum_{1\leq u\leq 2p+4}{\rm P}[|\partial l_{*}/\partial\psi_{u}|>(a/8\sqrt{2p+4})m^{1/2+\delta}]$. Next, by the Marcinkiewicz-Zygmund inequality (e.g., Jiang 2022, p. 161), we have
\begin{eqnarray*}
{\rm P}\left(\left|\frac{\partial l_{*}}{\partial\psi_{u}}\right|>\frac{a}{8\sqrt{2p+4}}m^{1/2+\delta}\right)&\leq&\left(\frac{8\sqrt{2p+4}}{am^{\delta}}\right)^{h}\frac{1}{m^{h/2}}{\rm E}\left(\left|\sum_{i=1}^{m}\frac{\partial l_{i*}}{\partial\psi_{u}}\right|^{h}\right)\\
&\leq&\left(\frac{8\sqrt{2p+4}}{am^{\delta}}\right)^{h}\frac{c_{1h}}{m^{h/2}}{\rm E}\left\{\sum_{i=1}^{m}\left(\frac{\partial l_{i*}}{\partial\psi_{u}}\right)^{2}\right\}^{h/2},
\end{eqnarray*}
where $c_{1h}>0$ is a constant that depends on $h$. Also, by Jensen's inequality, we have
\begin{eqnarray*}
&&\frac{1}{m^{h/2}}{\rm E}\left\{\sum_{i=1}^{m}\left(\frac{\partial l_{i*}}{\partial\psi_{u}}\right)^{2}\right\}^{h/2}={\rm E}\left\{\frac{1}{m}\sum_{i=1}^{m}\left(\frac{\partial l_{i*}}{\partial\psi_{u}}\right)^{2}\right\}^{h/2}\\
&&\leq{\rm E}\left(\frac{1}{m}\sum_{i=1}^{m}\left|\frac{\partial l_{i*}}{\partial\psi_{u}}\right|^{h}\right)=\frac{1}{m}\sum_{i=1}^{m}{\rm E}\left(\left|\frac{\partial l_{i*}}{\partial\psi_{u}}\right|^{h}\right)\leq c_{2h}
\end{eqnarray*}
for some constant $c_{2h}>0$ that depends on $h$. Combining the above result, we have ${\rm P}({\cal A}_{1}^{c})\leq c_{h}m^{-\delta h}$ for some constant $c_{h}>0$ that depends on $h$.

By similar arguments, it can be shown that ${\rm P}({\cal A}_{s}^{c})\leq c_{h}m^{-h/2}$, $s=2,3$. Because $0<\delta<1/2$, the proof is complete.
\subsection{Proof of Theorem 3}
\hspace{4mm}
Because $\tilde{d}_{i}$ is the BP, that is, the conditional expectation, we have
\begin{eqnarray}
&&{\rm MSPE}(\hat{d}_{i,{\rm t}})={\rm E}(\hat{d}_{i,{\rm t}}-d_{i})^{2}={\rm E}(\tilde{d}_{i}-d_{i})^{2}+{\rm E}(\hat{d}_{i,{\rm t}}-\tilde{d}_{i})^{2}\nonumber\\
&&={\rm MSPE}(\tilde{d}_{i})+{\rm E}(\hat{d}_{i,{\rm t}}-\tilde{d}_{i})^{2}.\label{eq:ptm3_1}
\end{eqnarray}

Next, let $0<\delta<1/2$ be a fixed number to be determined latter. Define ${\cal B}=\{|\hat{\psi}-\psi_{*}|\leq m^{\delta-1/2}\}$. Then, we have
\begin{equation}
{\rm E}(\hat{d}_{i,{\rm t}}-\tilde{d}_{i})^{2}={\rm E}\{(\hat{d}_{i,{\rm t}}-\tilde{d}_{i})^{2}1_{\cal B}\}+{\rm E}\{(\hat{d}_{i,{\rm t}}-\tilde{d}_{i})^{2}1_{{\cal B}^{c}}\}=I_{1}+I_{2}.\label{eq:ptm3_2}
\end{equation}

Note that, for sufficiently large $m$, we have $\hat{\psi}_{\rm t}=\hat{\psi}$ on ${\cal B}$. It then follows, by Taylor expansion, that, for sufficiently large $m$, we have, on ${\cal B}$,
$$|\hat{d}_{i,{\rm t}}-\tilde{d}_{i}|\leq\left|\left.\frac{\partial g_{i}}{\partial\psi'}\right|_{\tilde{\psi}}\right||\hat{\psi}-\psi_{*}|\leq cm^{\delta-1/2}$$
for some constant $c>0$, where $\tilde{\psi}$ lies between $\hat{\psi}$ and $\psi_{*}$. It follows that
\begin{equation}
I_{1}\leq c^{2}m^{^{2(\delta-1/2)}}\longrightarrow 0.\label{eq:ptm3_3}
\end{equation}

On the other hand, by the definition of $\hat{\psi}_{\rm t}$, it can be shown that there are constants $c_{1}, c_{2}>0$ such that
$|\hat{d}_{i,{\rm t}}-\tilde{d}_{i}|\leq c_{1}m^{c_{2}}(|\Delta_{i}|^{2}+1)$, where $\Delta_{i}$ is as defined in the proof of Lemma 1. It follows, by Lemma 1, that
$$I_{2}\leq c_{1}^{2}m^{2c_{2}}{\rm E}\{(|\Delta_{i}|^{2}+1)^{2}1_{{\cal B}^{2}}\}\leq c_{1}^{2}m^{2c_{2}}\{{\rm E}(|\Delta_{i}|^{2}+1)^{4}\}^{1/2}{\rm P}({\cal B}^{c})^{1/2}\leq c_{h}m^{2c_{2}-\delta h/2},$$
using the Cauchy-Schwarz inequality for the second step, where $c_{h}$ is a constant. Thus, by choosing $\delta=1/4$ and $h$ sufficiently large, we have $2c_{2}-\delta h/2<0$. It follows that
\begin{equation}
I_{2}\leq c_{h}m^{2c_{2}-\delta h/2}\longrightarrow 0.\label{eq:ptm3_4}
\end{equation}

Combining (\ref{eq:ptm3_2})--(\ref{eq:ptm3_4}), we have shown that
\begin{equation}
\lim_{m\rightarrow\infty}{\rm E}(\hat{d}_{i,{\rm t}}-\tilde{d}_{i})^{2}=0.\label{eq:ptm3_5}
\end{equation}

Now let $\check{d}_{i}$ be any other predictor of $d_{i}$. We have, by (\ref{eq:ptm3_1}) and the definition of the BP,
\begin{eqnarray*}
{\rm MSPE}(\hat{d}_{i,{\rm t}})={\rm MSPE}(\tilde{d}_{i})+{\rm E}(\hat{d}_{i,{\rm t}}-\tilde{d}_{i})^{2}\leq{\rm MSPE}(\check{d}_{i})+{\rm E}(\hat{d}_{i,{\rm t}}-\tilde{d}_{i})^{2},
\end{eqnarray*}
or ${\rm MSPE}(\hat{d}_{i,{\rm t}})-{\rm MSPE}(\check{d}_{i})\leq{\rm E}(\hat{d}_{i,{\rm t}}-\tilde{d}_{i})^{2}$. It follows, by (\ref{eq:ptm3_5}), that
$$\limsup_{m\rightarrow\infty}\{{\rm MSPE}(\hat{d}_{i,{\rm t}})-{\rm MSPE}(\check{d}_{i})\}\leq 0.$$
\subsection{Proof of Theorem 4}
\hspace{4mm}
Recall (40) of SACI, and let $\phi$ be true vector of variance components. Furthermore, as the assumptions imply $n_{i}/N_{i}\rightarrow 0$, it can be shown that $\eta_{i}=\zeta_{i}+o_{\rm P}(1)$, where
\begin{equation}
\zeta_{i}=a_{i0}(W_{i})\alpha_{i0}+a_{i1}(W_{i})\alpha_{i1}+d_{i1}(W_{i})\sum_{k=1}^{n_{1i}(W_{i})}\epsilon_{ik}(1)+d_{i0}(W_{i})\sum_{k=n_{1i}(W_{i})+1}^{n_{i}}\epsilon_{ik}(0).\label{eq:ptm4_1}
\end{equation}
Note that, conditional on $W$, $\zeta_{i}$ is distributed as $N(0,\tau_{i}^{2}(W_{i}))$, with $\tau_{i}^{2}(W_{i})$ evaluated at the true $\phi$. Let ${\cal A}_{i}=\{a_{i,1,r}(W_{i})=a_{i,0,r}(W_{i}), r=0,1\}$. Then, on ${\cal A}_{i}^{c}$, we have $h_{i}=\min_{r=0,1}|d_{ir}|>0$; thus, we have
\begin{eqnarray*}
&&\tau_{i}^{2}(W_{i})\geq\sigma_{0}^{2}|a_{i0}+a_{i1}|^{2}-2\sigma_{0}\sigma_{1}|a_{i0}+a_{i1}|\cdot|a_{i1}-1|+\sigma_{1}^{2}|a_{i1}-1|^{2}+\tau^{2}h_{i}^{2}\\
&&=(\sigma_{0}|a_{i0}+a_{i1}|-\sigma_{1}|a_{i1}-1|)^{2}+\tau^{2}h_{i}^{2}\geq \tau^{2}h_{i}^{2}>0.
\end{eqnarray*}
It follows that, conditional on $W$, $\zeta_{i}/\tau_{i}(W_{i})\sim N(0,1)$; therefore, combined with (40) of SACI, we have $(\hat{d}_{i}-d_{i})/\tau_{i}(W_{i})|W\stackrel{\rm d}{\longrightarrow}N(0,1)$ on ${\cal A}_{i}^{c}$. 

Furthermore, because $n_{i}$ is bounded, and $h_{i}$ depends on $W_{i}$ only through $1\leq n_{ri}\leq n_{i}$, $r=0,1$, there are only a bounded number of distinct values of $h_{i}$. It follows that there is a positive constant, say, $\delta$, such that $h_{i}>0$ implies $h_{i}\geq\delta$. With that, and the consistency of $\hat{\psi}$, it can then be shown that
\begin{equation}
\left.\frac{\hat{d}_{i}-d_{i}}{\hat{\tau}_{i}(W_{i})}\right|W\stackrel{\rm d}{\longrightarrow}N(0,1).\label{eq:ptm4_3}
\end{equation}

For any $x\in R$, by (\ref{eq:ptm4_3}), we have
\begin{equation}
\xi_{i}={\rm E}\left[\left.\frac{\hat{d}_{i}-d_{i}}{\hat{\tau}_{i}(W_{i})}\leq x\right|W\right]\longrightarrow\Phi(x),\label{eq:ptm4_4}
\end{equation}
where $\Phi(\cdot)$ is the cdf of $N(0,1)$. Because $0\leq\xi_{i}\leq 1$, by (\ref{eq:ptm4_4}) and the dominated convergence theorem (e.g., Jiang 2022, Theorem 2.16), we have
\begin{equation}
{\rm P}\left[\frac{\hat{d}_{i}-d_{i}}{\hat{\tau}_{i}(W_{i})}\leq x\right]={\rm E}(\xi_{i})\longrightarrow\Phi(x).\label{eq:ptm4_5}
\end{equation}
By (\ref{eq:ptm4_5}) and the arbitrariness of $x$, the proof is complete.
\section{Additional simulation results for Section 5 of SACI}
\subsection{Additional results under {\it Case I}}
\subsubsection{A sensitivity analysis}
\hspace{4mm}
We also carried out a sensitivity analysis under {\it Case I}, with known $N_{i}$ and covariate subpopulation, to see whether the pattern observed in Table 1 of SACI changes under different values of the parameters governing treatment-effect heterogeneity, residual variation, and random-effect correlation, namely, $\sigma_1^2$, $\tau^2$, and $\rho$. Here, Baseline refers to $(\sigma_1^2=0.5,\tau^2=1,\rho=0.3)$. We then change the values of one of the components of the baseline. The resulting averaged simulated MSPEs, based on $N_{\rm sim}=200$ simulation runs, are presented in Table \ref{tab:sensitivity-mspe} of this supplement. Here, we consider $m=100$; the settings of $N_{i}$ and $n_{i}$ are the same as the above. The pattern is, more or less, the same as that observed in Table 1 of SACI. Here, the percentages in the parentheses indicate the percentage increase in MSPE over the MSPE of EBP.
\begin{table}[h]
\centering
\caption{Sensitivity Analysis: Average Simulated MSPEs under Different Values of Parameters: $m=100$; Outcome Model Correctly Specified; Baseline$=(\sigma_1^2=0.5,\tau^2=1,\rho=0.3)$; \% in the Parentheses Indicates Percentage Increase in MSPE over the MSPE of EBP.}
\label{tab:sensitivity-mspe}
\begin{tabular}{lclll}
\hline
Scenario & $n_i$ & naive DR & partial EBP & EBP \\
\hline
Baseline & 20 & 0.2260 (74.2\%) & 0.1380 (6.4\%) & {\bf 0.1297} \\
Baseline & 40 & 0.1073 (39.4\%) & 0.0830 (7.8\%) & {\bf 0.0770} \\
$\rho=0$ & 20 & 0.2268 (59.7\%) & 0.1459 (2.7\%) & {\bf 0.1420} \\
$\rho=0$ & 40 & 0.1065 (30.2\%) & 0.0865 (5.7\%) & {\bf 0.0818} \\
$\rho=0.6$ & 20 & 0.2253 (109.4\%) & 0.1322 (22.8\%) & {\bf 0.1076} \\
$\rho=0.6$ & 40 & 0.1074 (56.1\%) & 0.0813 (18.2\%) & {\bf 0.0688} \\
$\sigma_1^2=0.1$ & 20 & 0.2245 (248.6\%) & 0.0709 (10.1\%) & {\bf 0.0644} \\
$\sigma_1^2=0.1$ & 40 & 0.1071 (127.4\%) & 0.0535 (13.6\%) & {\bf 0.0471} \\
$\sigma_1^2=1.0$ & 20 & 0.2212 (48.0\%) & 0.1567 (4.8\%) & {\bf 0.1495} \\
$\sigma_1^2=1.0$ & 40 & 0.1072 (25.7\%) & 0.0904 (6.0\%) & {\bf 0.0853} \\
$\tau^2=2.0$ & 20 & 0.4447 (123.7\%) & 0.2079 (4.6\%) & {\bf 0.1988} \\
$\tau^2=2.0$ & 40 & 0.2116 (62.4\%) & 0.1379 (5.8\%) & {\bf 0.1303} \\
\hline
\end{tabular}
\end{table}
\subsubsection{Figures under the unknown $N_{i}$ setting}
\hspace{4mm}
Here, we present area-specific boxplots associated with Section 5.2.2 of SACI.

\begin{figure}[ht]
\centering
\includegraphics[width=0.9\textwidth]{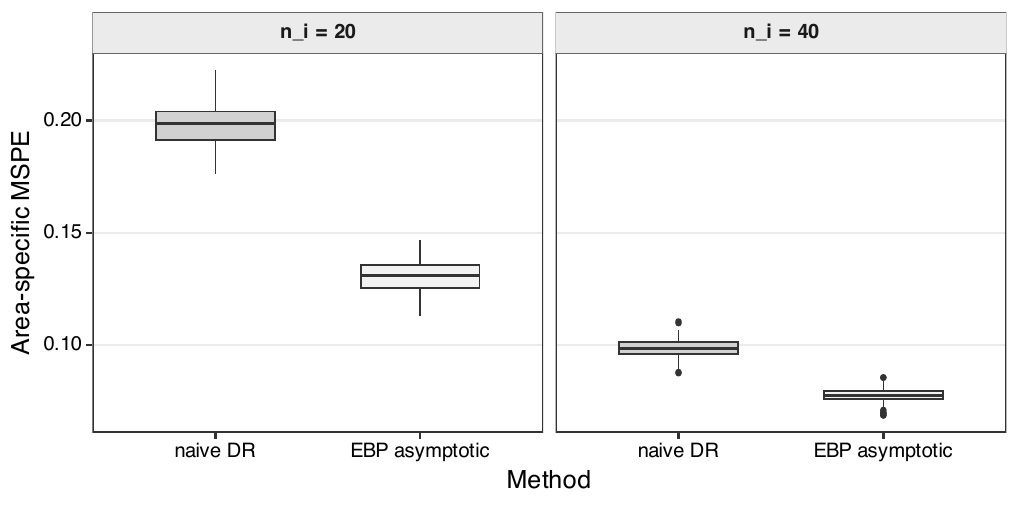}
\caption{Boxplots of Simulated Area-specific MSPE: Unknown $N_{i}$ Setting; Naive DR vs Asymptotic EBP; $m=100$; Equal-\(n_i\) Design.}
\label{fig:unknown-N-equal-mspe-boxplot}
\end{figure}
\begin{figure}[ht]
\centering
\includegraphics[width=0.9\textwidth]{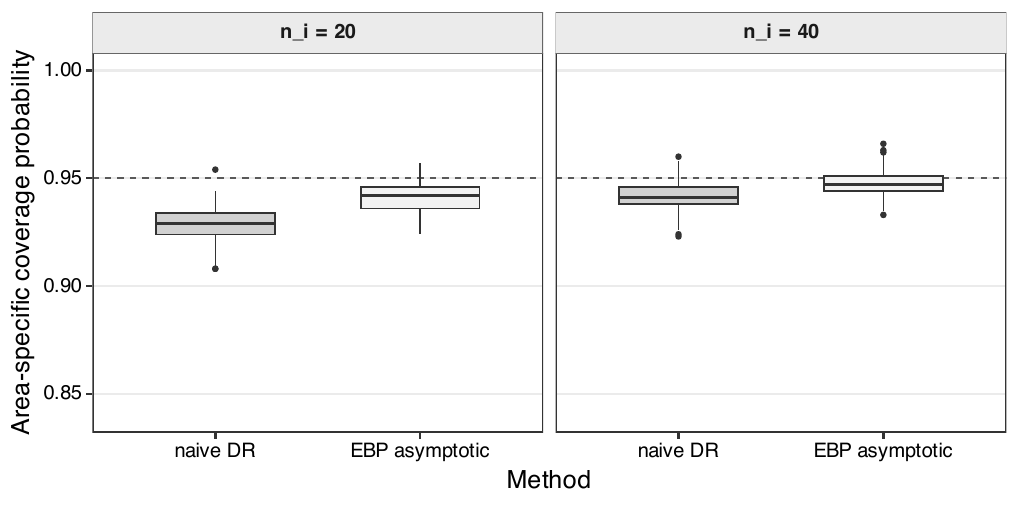}
\caption{Boxplots of Simulated Area-specific Coverage Probability of 95\% Prediction Interval: Unknown $N_{i}$ Setting; Naive DR vs Asymptotic EBP; $m=100$; Equal-\(n_i\) Design.}
\label{fig:unknown-N-equal-coverage-boxplot}
\end{figure}
\begin{figure}[ht]
\centering
\includegraphics[width=0.9\textwidth]{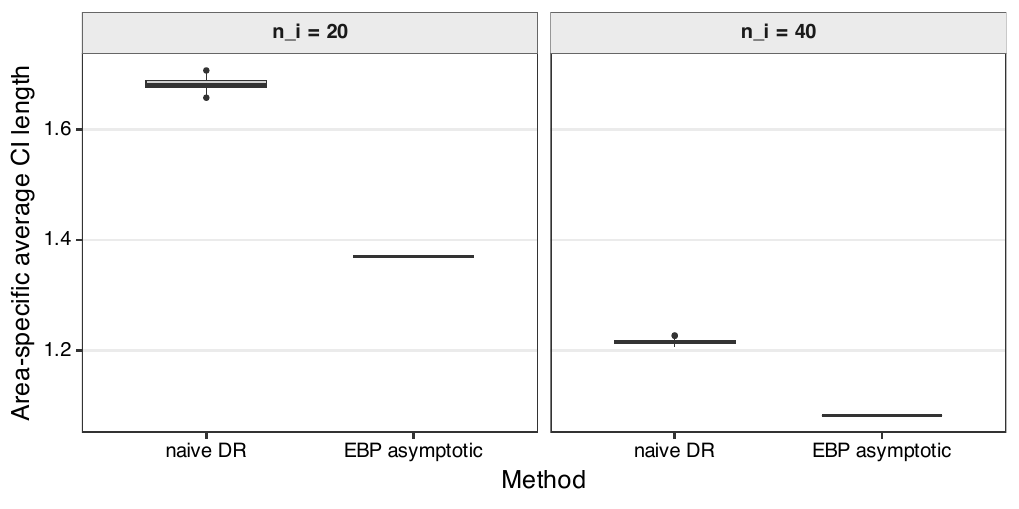}
\caption{Boxplots of Simulated Area-specific Average Length of 95\% Prediction Interval: Unknown $N_{i}$ Setting; Naive DR vs Asymptotic EBP; $m=100$; Equal-\(n_i\) Design.}
\label{fig:unknown-N-equal-length-boxplot}
\end{figure}

\begin{figure}[ht]
\centering
\includegraphics[width=0.9\textwidth]{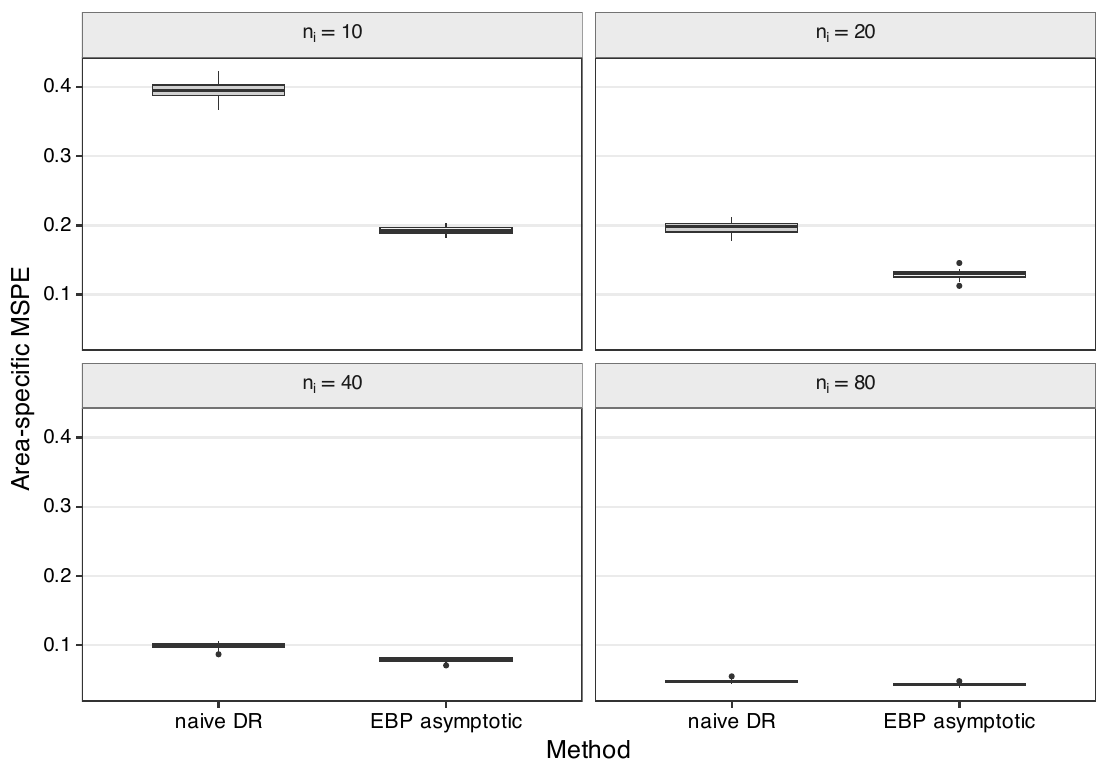}
\caption{Boxplots of Simulated Area-specific MSPE: Unknown $N_{i}$ Setting; Naive DR vs Asymptotic EBP; $m=100$; Unequal-\(n_i\) Design.}
\label{fig:unknown-N-variable-mspe-boxplot}
\end{figure}
\begin{figure}[ht]
\centering
\includegraphics[width=0.9\textwidth]{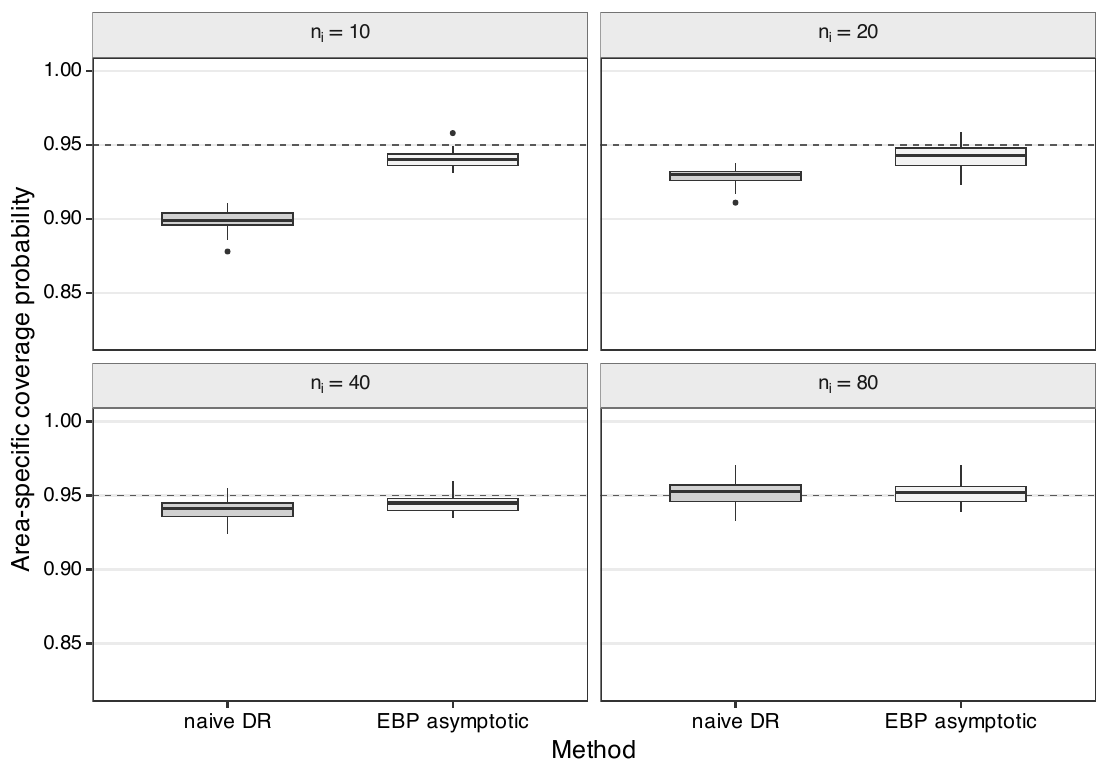}
\caption{\small Boxplots of Simulated Area-specific Coverage Probability of 95\% Prediction Interval: Unknown $N_{i}$ Setting; Naive DR vs Asymptotic EBP; $m=100$; Unequal-\(n_i\) Design.}
\label{fig:unknown-N-variable-coverage-boxplot}
\end{figure}
\begin{figure}[ht]
\centering
\includegraphics[width=0.9\textwidth]{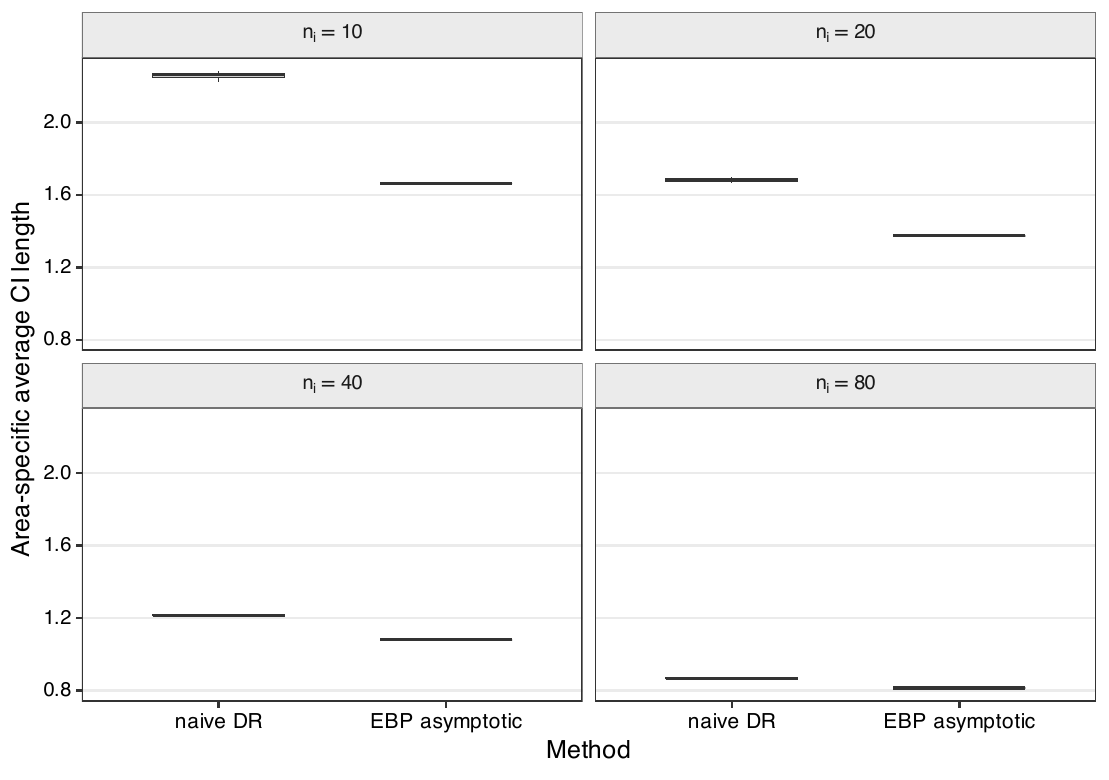}
\caption{Boxplots of Simulated Area-specific Average Length of 95\% Prediction Interval: Unknown $N_{i}$ Setting; Naive DR vs Asymptotic EBP; $m=100$; Unequal-\(n_i\) Design.}
\label{fig:unknown-N-variable-length-boxplot}
\end{figure}
\subsection{{\it Case II}, known $N_{i}$ and covariate subpopulation}
\hspace{4mm}
Here, we consider the scenario where the outcome model is possibly misspecified, with the degree of misspecification controlled by $\gamma_{\rm quad}$; on the other hand, the propensity model, which is the logistic model, is correctly specified. We consider $m=100$.

First consider the equal-$n_{i}$ design. Table \ref{tab:final-outcome-equal-n20} shows the summary statistics of different performance measures for $n_{i}=20$. EBP has the smallest MSPE even when the outcome model is misspecified ($\gamma_{\rm quad}>0$). The finite-population EBP interval is shortest for $\gamma_{\rm quad}=0$ and $0.5$, whereas the asymptotic EBP interval is shortest for $\gamma_{\rm quad}=1$. As for Coverage, partial EBP is the best performer when there is no outcome model misspecification ($\gamma_{\rm quad}=0$) or when there is a moderate outcome model misspecification ($\gamma_{\rm quad}=0.5$); naive DR is the best performer when there is a severe outcome model misspecification ($\gamma_{\rm quad}=1.0$), though partial EBP comes very close in this case.

Table \ref{tab:final-outcome-equal-n40} shows the corresponding summaries for $n_{i}=40$. In this case, EBP continues to be the best performer in Length. For MSPE, EBP is the best performer when there is no outcome model misspecification ($\gamma_{\rm quad}=0$) or when there is moderate outcome model misspecification ($\gamma_{\rm quad}=0.5$); partial EBP is the best performer when there is severe outcome model misspecification ($\gamma_{\rm quad}=1.0$), though EBP comes very close in this case. As for Coverage, partial EBP is the best performer when there is no outcome model misspecification ($\gamma_{\rm quad}=0$), though EBP with the approximate asymptotic variance comes very close in this case; naive DR is the best performer when there is outcome model misspecification ($\gamma_{\rm quad}>0$).

\begin{figure}[ht]
\centering
\includegraphics[width=0.9\textwidth]{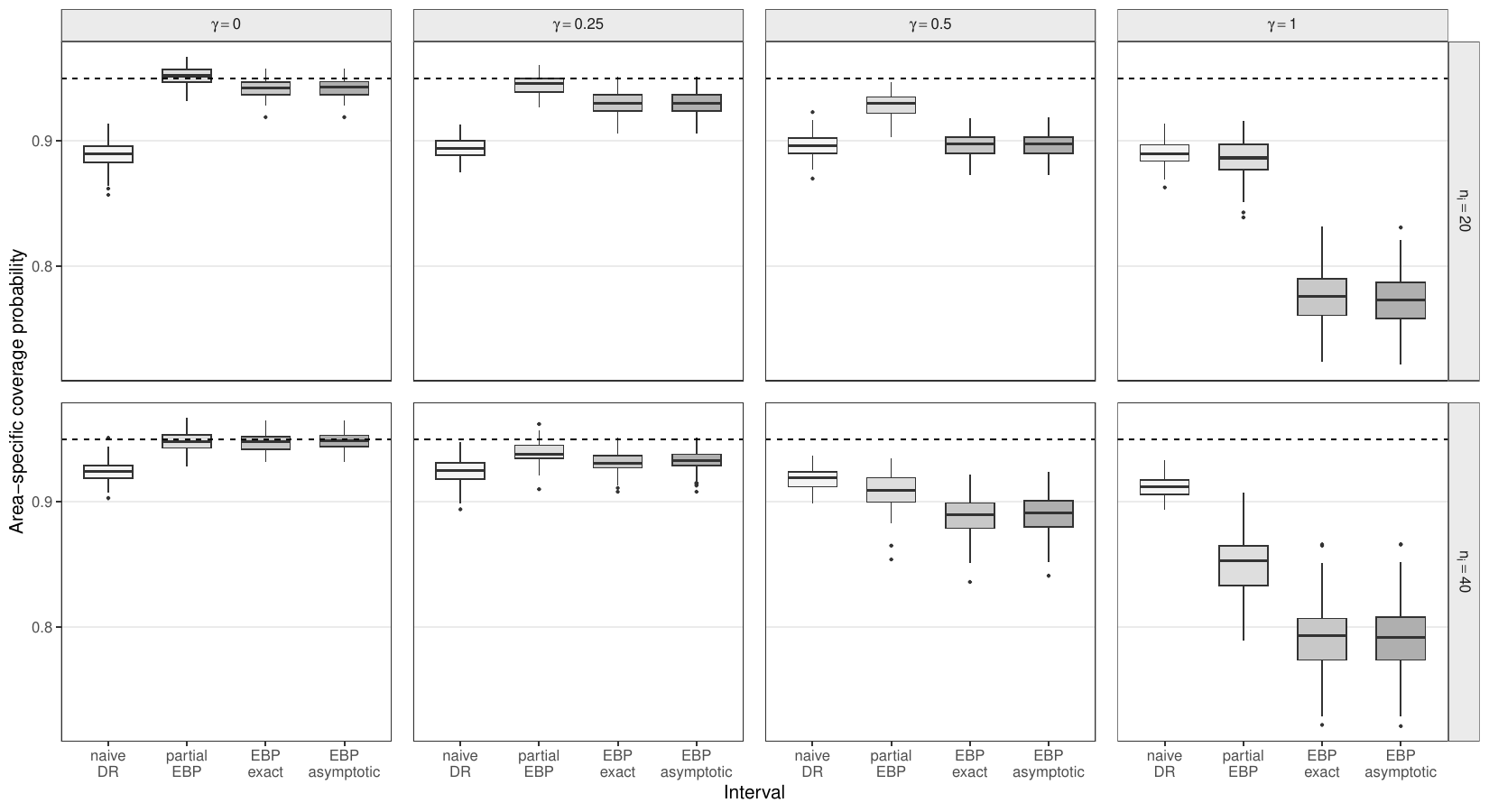}
\caption{\small Boxplots of Simulated Area-specific Coverage Probability of 95\% Prediction Interval with Possible Outcome Model Misspecification and Correctly Specified Logistic Propensity Model: $m=100$, Equal-\(n_i\) Design; EBP exact and EBP asymptotic refer to EBP-based Prediction Interval based on Exact and Approximate Asymptotic Variances, Respectively.}
\label{fig:final-outcome-equal-coverage}
\end{figure}

\begin{figure}[ht]
\centering
\includegraphics[width=0.9\textwidth]{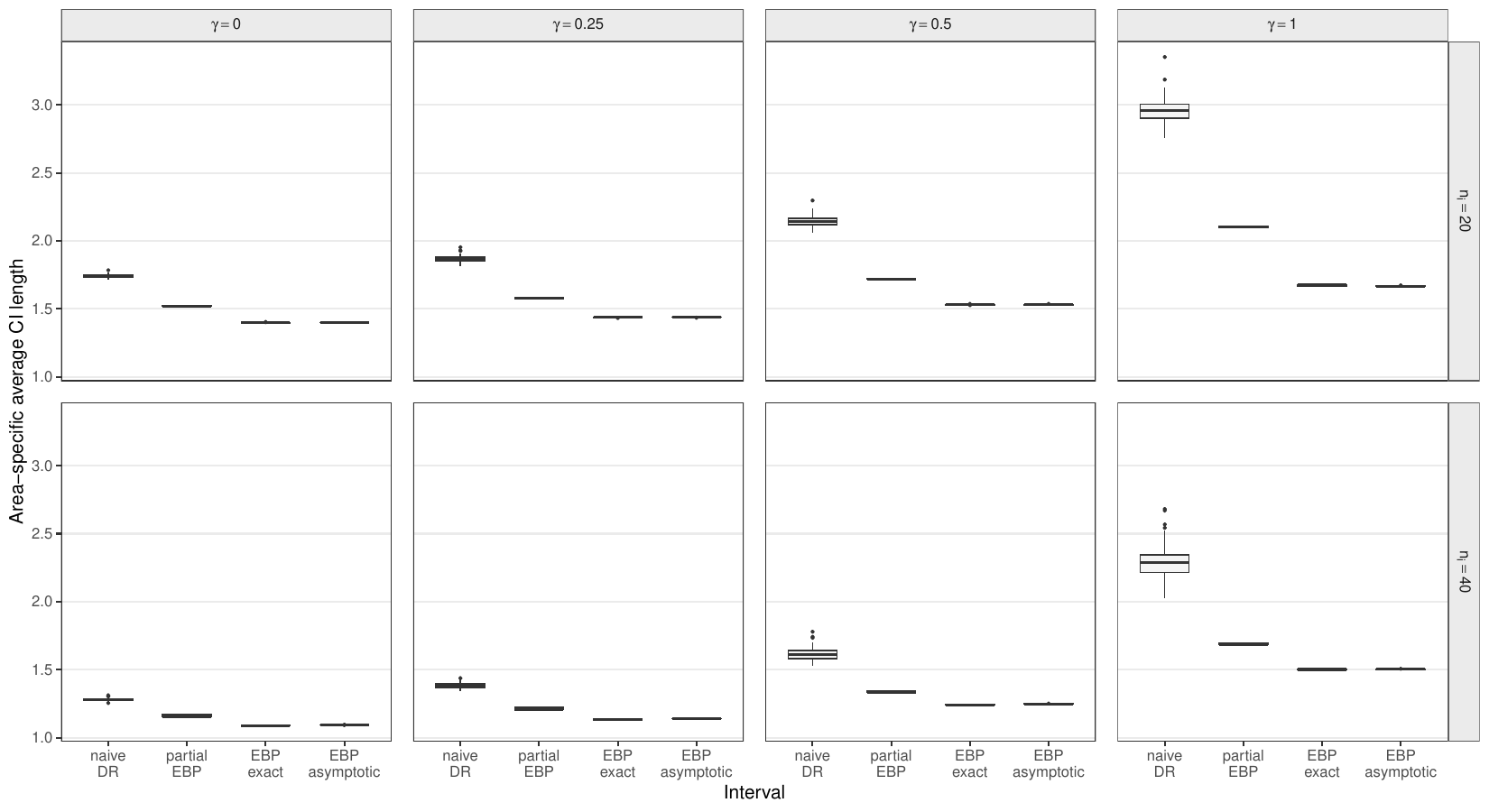}
\caption{Boxplots of Simulated Area-specific Average Length of 95\% Prediction Interval with Possible Outcome Model Misspecification and Correctly Specified Logistic Propensity Model: $m=100$, Equal-\(n_i\) Design; EBP exact and EBP asymptotic refer to EBP-based Prediction Interval based on Exact and Approximate Asymptotic Variances, Respectively.}
\label{fig:final-outcome-equal-length}
\end{figure}

The corresponding boxplots are shown in Figure \ref{fig:final-outcome-equal-mspe}, Figure \ref{fig:final-outcome-equal-mspe-zoom}, Figure \ref{fig:final-outcome-equal-coverage}, and Figure \ref{fig:final-outcome-equal-length}. The figures show patterns consistent with the summary statistics.

\begin{table}[ht]
\centering
\setlength{\belowcaptionskip}{8pt}
\caption{Simulated MSPE, Coverage Probability and Average Length of 95\% Prediction Interval: Possibly Misspecified Outcome Model and Correctly Specified Logistic Propensity Model; $m=100$; Equal-$n_{i}$ Design with $n_{i}=20$; Numbers in the Parentheses Are Results based on the Approximate Asymptotic Variance for the EBP Method; Bold Numbers Indicate Best Performer under Respective $n_{i}$ and Performance Measure.}
\label{tab:final-outcome-equal-n20}
\setlength{\tabcolsep}{3.5pt}
\begin{tabular}{c l lll}
\hline
$\gamma_{\mathrm{quad}}$ & Method & MSPE & Coverage & Length\\
\hline
0 & naive DR & 0.2758 & 0.8893 & 1.7429\\
0 & partial EBP & 0.1486 & \textbf{0.9517} & 1.5227\\
0 & EBP & \textbf{0.1345} & 0.9419 (0.9423) & \textbf{1.3983} (1.4000)\\
0.5 & naive DR & 0.4440 & 0.8963 & 2.1447\\
0.5 & partial EBP & 0.2275 & \textbf{0.9291} & 1.7203\\
0.5 & EBP & \textbf{0.2164} & 0.8974 (0.8974) & \textbf{1.5314} (1.5315)\\
1 & naive DR & 0.9545 & \textbf{0.8899} & 2.9581\\
1 & partial EBP & 0.4431 & 0.8862 & 2.1035\\
1 & EBP & \textbf{0.4385} & 0.7763 (0.7731) & 1.6754 (\textbf{1.6672})\\
\hline
\end{tabular}
\end{table}

\begin{table}[ht]
\centering
\setlength{\belowcaptionskip}{8pt}
\caption{Simulated MSPE, Coverage Probability and Average Length of 95\% Prediction Interval: Possibly Misspecified Outcome Model and Correctly Specified Logistic Propensity Model; $m=100$; Equal-$n_{i}$ Design with $n_{i}=40$; Numbers in the Parentheses Are Results based on the Approximate Asymptotic Variance for the EBP Method; Bold Numbers Indicate Best Performer under Respective $n_{i}$ and Performance Measure.}
\label{tab:final-outcome-equal-n40}
\setlength{\tabcolsep}{3.5pt}
\begin{tabular}{c l lll}
\hline
$\gamma_{\mathrm{quad}}$ & Method & MSPE & Coverage & Length\\
\hline
0 & naive DR & 0.1232 & 0.9239 & 1.2798\\
0 & partial EBP & 0.0891 & \textbf{0.9486} & 1.1605\\
0 & EBP & \textbf{0.0792} & 0.9472 (0.9484) & \textbf{1.0890} (1.0944)\\
0.5 & naive DR & 0.2255 & \textbf{0.9189} & 1.6128\\
0.5 & partial EBP & 0.1580 & 0.9086 & 1.3383\\
0.5 & EBP & \textbf{0.1514} & 0.8900 (0.8916) & \textbf{1.2436} (1.2485)\\
1 & naive DR & 0.5624 & \textbf{0.9119} & 2.2964\\
1 & partial EBP & \textbf{0.3423} & 0.8503 & 1.6886\\
1 & EBP & 0.3429 & 0.7912 (0.7913) & \textbf{1.5019} (1.5026)\\
\hline
\end{tabular}
\end{table}

\begin{figure}[ht]
\centering
\includegraphics[width=0.9\textwidth]{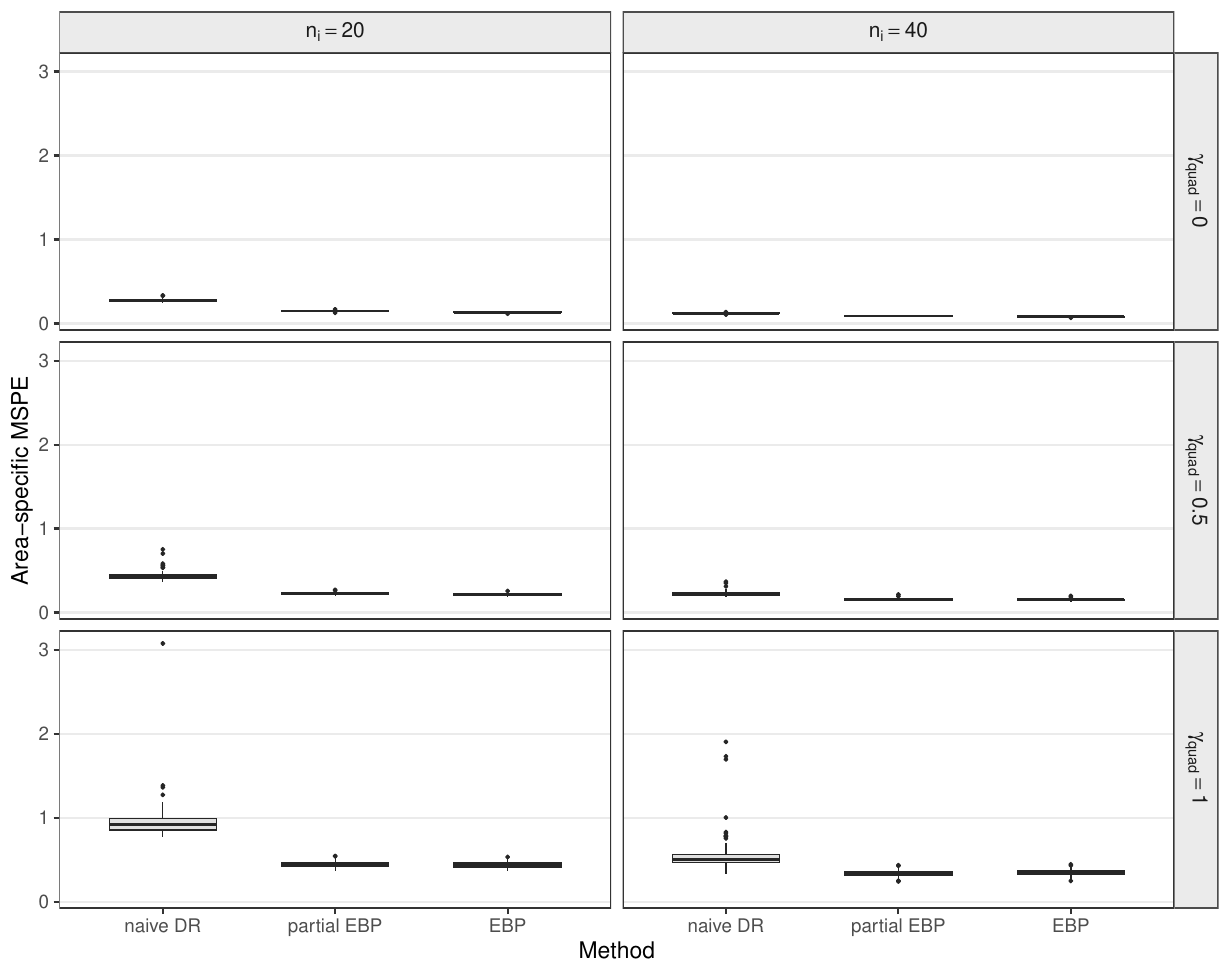}
\begingroup
\renewcommand{\thefigure}{\arabic{figure}(a)}
\caption{Boxplots of Simulated Area-specific MSPE with Possible Outcome Model Misspecification and Correctly Specified Logistic Propensity Model: $m=100$, Equal-\(n_i\) Design; Full Version with Outlying MSPE Values.}
\label{fig:final-outcome-equal-mspe}
\endgroup
\end{figure}
\begin{figure}[ht]
\centering
\begingroup
\addtocounter{figure}{-1}
\renewcommand{\thefigure}{\arabic{figure}(b)}
\includegraphics[width=0.9\textwidth]{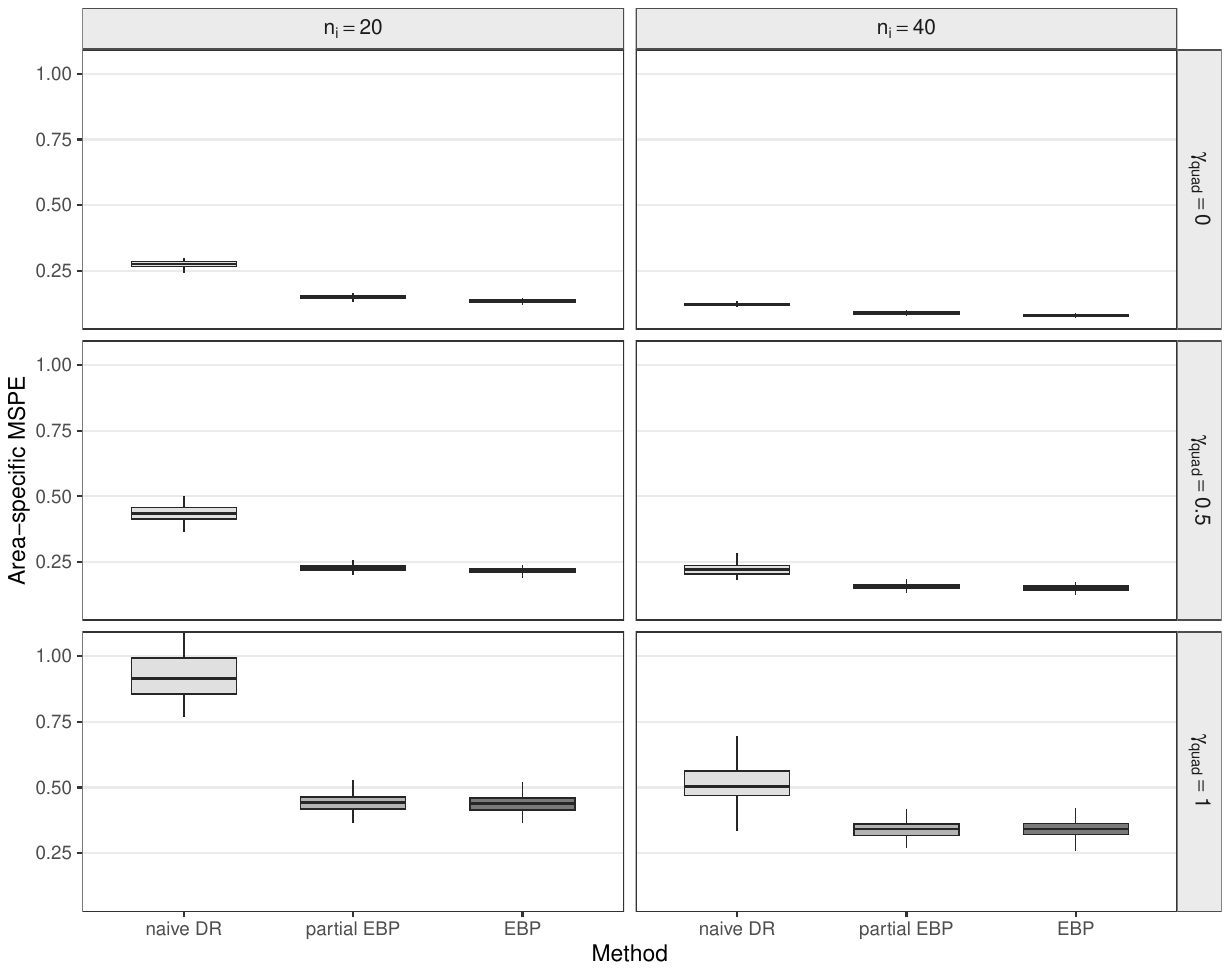}
\caption{Boxplots of Simulated Area-specific MSPE with Possible Outcome Model Misspecification and Correctly Specified Logistic Propensity Model: $m=100$, Equal-\(n_i\) Design; Zoomed Version Displaying the Middle 98\% Range of the Full Version.}
\label{fig:final-outcome-equal-mspe-zoom}
\endgroup
\end{figure}

\begin{figure}[p]
\centering
\includegraphics[width=\textwidth,height=0.88\textheight,keepaspectratio]{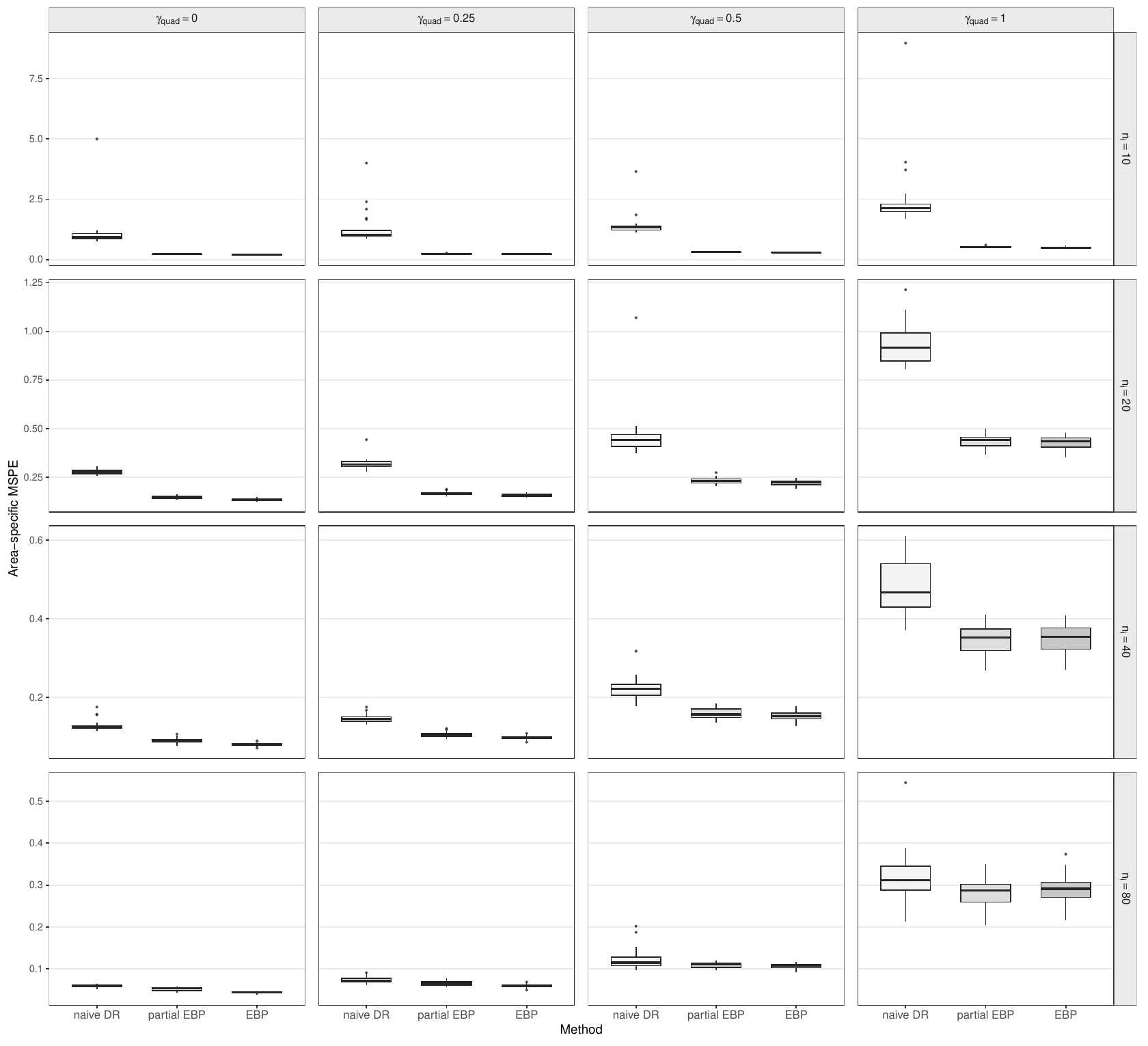}
\caption{Boxplots of Simulated Area-specific MSPE with Possible Outcome Model Misspecification and Correctly Specified Logistic Propensity Model: $m=100$, Unequal-\(n_i\) Design; Rows and Columns Correspond to Different Values of $n_{i}$ and $\gamma_{\rm quad}$, Respectively.}
\label{fig:final-outcome-variable-mspe}
\end{figure}

\begin{figure}[p]
\centering
\includegraphics[width=\textwidth,height=0.88\textheight,keepaspectratio]{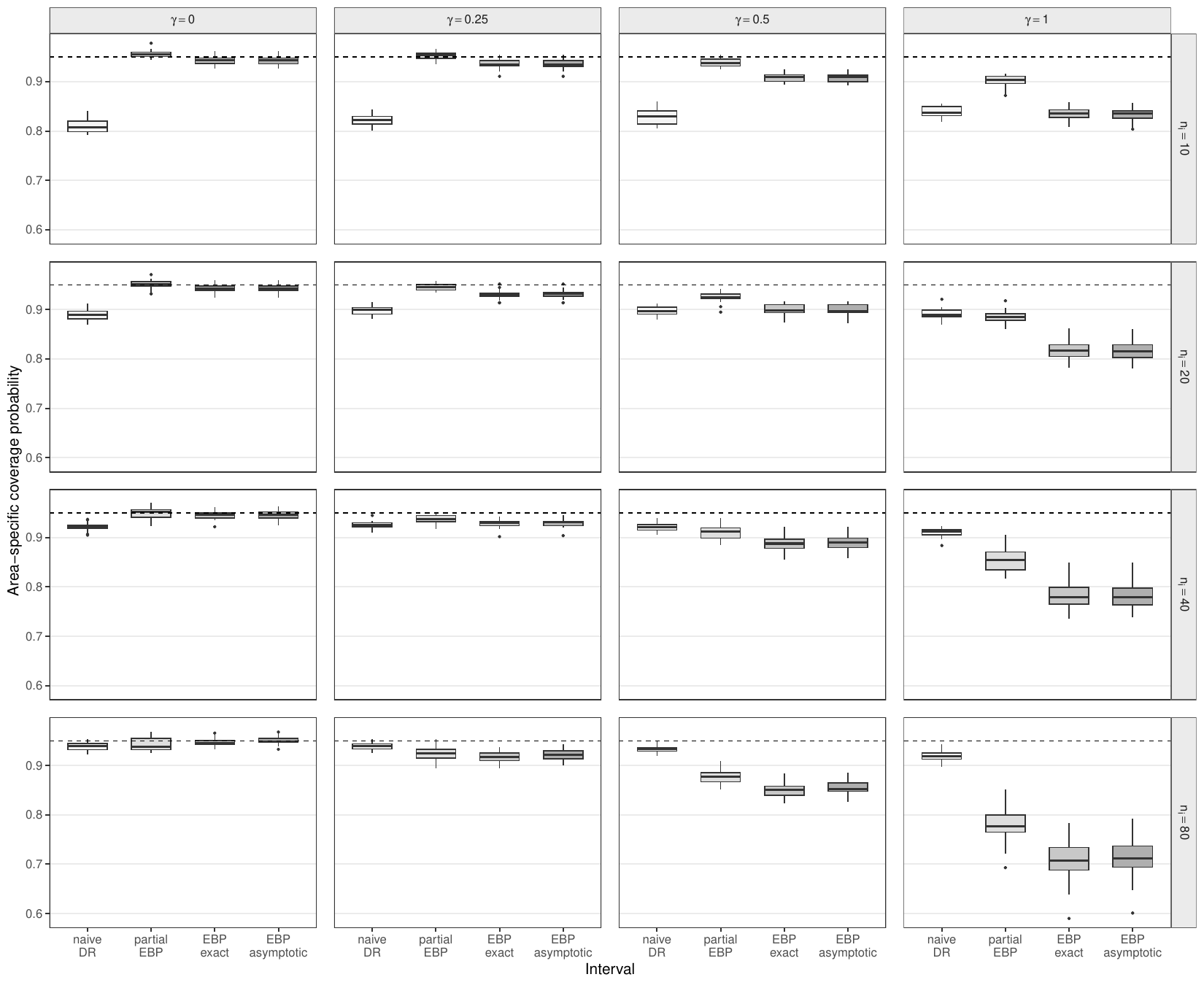}
\caption{\small Boxplots of Simulated Area-specific Coverage Probability of 95\% Prediction Interval with Possible Outcome Model Misspecification and Correctly Specified Logistic Propensity Model: $m=100$, Unequal-\(n_i\) Design; Rows and Columns Correspond to Different Values of $n_{i}$ and $\gamma_{\rm quad}$, Respectively; EBP exact and EBP asymptotic refer to EBP-based Prediction Interval based on Exact and Approximate Asymptotic Variances, Respectively.}
\label{fig:final-outcome-variable-coverage}
\end{figure}

\begin{figure}[p]
\centering
\includegraphics[width=\textwidth,height=0.88\textheight,keepaspectratio]{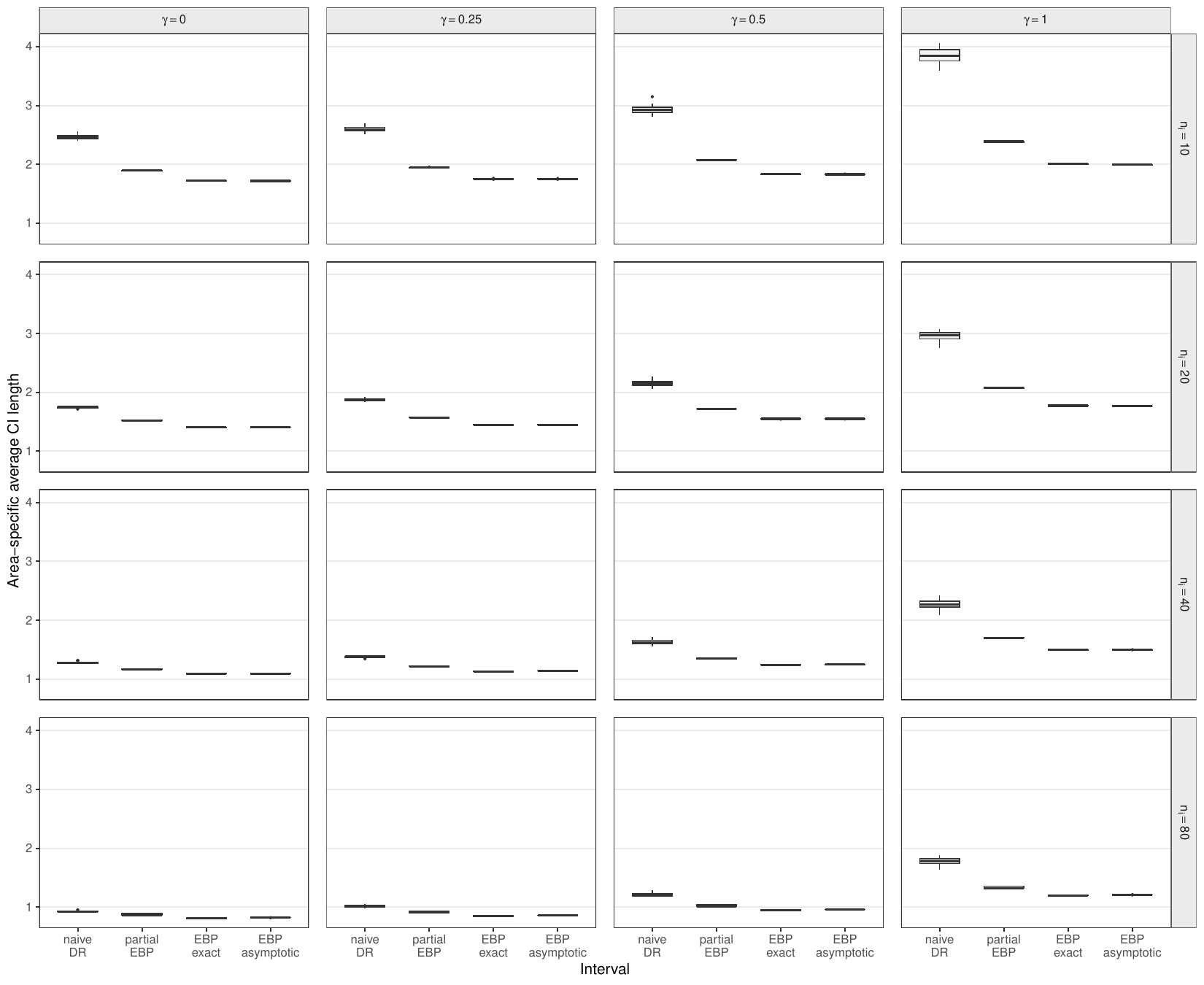}
\caption{Boxplots of Simulated Area-specific Average Length of 95\% Prediction Interval with Possible Outcome Model Misspecification and Correctly Specified Logistic Propensity Model: $m=100$, Unequal-\(n_i\) Design; Rows and Columns Correspond to Different Values of $n_{i}$ and $\gamma_{\rm quad}$, Respectively; EBP exact and EBP asymptotic refer to EBP-based Prediction Interval based on Exact and Approximate Asymptotic Variances, Respectively.}
\label{fig:final-outcome-variable-length}
\end{figure}

Next, we consider the unequal-$n_{i}$ design. Table \ref{tab:final-outcome-variable} presents the summary under different performance measures. In every setting, either the finite-population or the asymptotic EBP interval has the shortest Length; EBP performs the best in MSPE in the vast majority (10 out of 12) of the cases. As for Coverage, partial EBP performs the best in half of the cases (6 out of 12); for the other half, naive DR performs the best in 5 out of the 6 cases and EBP (with approximate asymptotic variance) performs the best in 1 out of the 6 cases. The cases that naive DR performs the best in Coverage are, in general, when $n_{i}$ is not too small ($>10$) and the outcome model is misspecified (keep in mind that the propensity model is correctly specified). The area-specific boxplots, presented in Figures \ref{fig:final-outcome-variable-mspe}--\ref{fig:final-outcome-variable-length}, show consistent patterns with the summaries.

\begin{table}[p]
\centering
\setlength{\belowcaptionskip}{8pt}
\caption{(Part I) Simulated MSPE, Coverage Probability and Average Length of 95\% Prediction Interval: Misspecified Outcome Model and Correctly Specified Logistic Propensity Model; $m=100$; Unequal-$n_{i}$ Design ($n_{i}=10$ and $n_{i}=20$); Separate Table for Each $n_{i}$; Bold Number Indicates the Best Performer under Respective Performance Measure.}
\label{tab:final-outcome-variable}
\setlength{\tabcolsep}{3pt}
\renewcommand{\arraystretch}{1.05}
\par\vspace{4pt}\textit{Panel A: $n_i=10$}\par\vspace{2pt}
\begin{tabular}{c l ccc}
\hline
$\gamma_{\mathrm{quad}}$ & Method/interval & MSPE & Coverage & Avg. length \\
\hline
0 & naive DR & 1.1187 & 0.8092 & 2.4580 \\
0 & partial EBP & 0.2204 & \textbf{0.9563} & 1.8948 \\
0 & EBP exact & \textbf{0.2030} & 0.9425 & 1.7175 \\
0 & EBP asymptotic & -- & 0.9423 & \textbf{1.7166} \\
0.5 & naive DR & 1.4069 & 0.8289 & 2.9329 \\
0.5 & partial EBP & 0.3057 & \textbf{0.9384} & 2.0726 \\
0.5 & EBP exact & \textbf{0.2921} & 0.9081 & 1.8309 \\
0.5 & EBP asymptotic & -- & 0.9076 & \textbf{1.8279} \\
1 & naive DR & 2.5325 & 0.8381 & 3.8469 \\
1 & partial EBP & 0.5184 & \textbf{0.9012} & 2.3837 \\
1 & EBP exact & \textbf{0.5061} & 0.8345 & 2.0047 \\
1 & EBP asymptotic & -- & 0.8326 & \textbf{1.9952} \\
\hline
\end{tabular}
\par\vspace{4pt}\textit{Panel B: $n_i=20$}\par\vspace{2pt}
\begin{tabular}{c l ccc}
\hline
$\gamma_{\mathrm{quad}}$ & Method/interval & MSPE & Coverage & Avg. length \\
\hline
0 & naive DR & 0.2769 & 0.8904 & 1.7432 \\
0 & partial EBP & 0.1454 & \textbf{0.9515} & 1.5199 \\
0 & EBP exact & \textbf{0.1338} & 0.9426 & \textbf{1.4035} \\
0 & EBP asymptotic & -- & 0.9428 & 1.4052 \\
0.5 & naive DR & 0.4620 & 0.8972 & 2.1577 \\
0.5 & partial EBP & 0.2309 & \textbf{0.9254} & 1.7160 \\
0.5 & EBP exact & \textbf{0.2205} & 0.8989 & \textbf{1.5489} \\
0.5 & EBP asymptotic & -- & 0.8989 & 1.5490 \\
1 & naive DR & 0.9367 & \textbf{0.8911} & 2.9506 \\
1 & partial EBP & 0.4358 & 0.8845 & 2.0720 \\
1 & EBP exact & \textbf{0.4286} & 0.8176 & 1.7707 \\
1 & EBP asymptotic & -- & 0.8159 & \textbf{1.7651} \\
\hline
\end{tabular}
\end{table}

\begin{table}[p]
\centering
\ContinuedFloat
\setlength{\belowcaptionskip}{8pt}
\caption{(Part II) Simulated MSPE, Coverage Probability and Average Length of 95\% Prediction Interval: Misspecified Outcome Model and Correctly Specified Logistic Propensity Model; $m=100$; Unequal-$n_{i}$ Design ($n_{i}=40$ and $n_{i}=80$); Separate Table for Each $n_{i}$; Bold Number Indicates the Best Performer under Respective Performance Measure.}
\setlength{\tabcolsep}{3pt}
\renewcommand{\arraystretch}{1.05}
\par\vspace{4pt}\textit{Panel C: $n_i=40$}\par\vspace{2pt}
\begin{tabular}{c l ccc}
\hline
$\gamma_{\mathrm{quad}}$ & Method/interval & MSPE & Coverage & Avg. length \\
\hline
0 & naive DR & 0.1282 & 0.9214 & 1.2805 \\
0 & partial EBP & 0.0901 & \textbf{0.9491} & 1.1675 \\
0 & EBP exact & \textbf{0.0803} & 0.9448 & \textbf{1.0887} \\
0 & EBP asymptotic & -- & 0.9460 & 1.0942 \\
0.5 & naive DR & 0.2240 & \textbf{0.9210} & 1.6264 \\
0.5 & partial EBP & 0.1590 & 0.9115 & 1.3496 \\
0.5 & EBP exact & \textbf{0.1527} & 0.8882 & \textbf{1.2417} \\
0.5 & EBP asymptotic & -- & 0.8897 & 1.2465 \\
1 & naive DR & 0.4858 & \textbf{0.9102} & 2.2718 \\
1 & partial EBP & \textbf{0.3466} & 0.8526 & 1.7037 \\
1 & EBP exact & 0.3495 & 0.7819 & \textbf{1.4962} \\
1 & EBP asymptotic & -- & 0.7819 & 1.4964 \\
\hline
\end{tabular}
\par\vspace{4pt}\textit{Panel D: $n_i=80$}\par\vspace{2pt}
\begin{tabular}{c l ccc}
\hline
$\gamma_{\mathrm{quad}}$ & Method/interval & MSPE & Coverage & Avg. length \\
\hline
0 & naive DR & 0.0591 & 0.9390 & 0.9260 \\
0 & partial EBP & 0.0524 & 0.9433 & 0.8748 \\
0 & EBP exact & \textbf{0.0438} & 0.9475 & \textbf{0.8099} \\
0 & EBP asymptotic & -- & \textbf{0.9505} & 0.8197 \\
0.5 & naive DR & 0.1243 & \textbf{0.9337} & 1.2148 \\
0.5 & partial EBP & 0.1098 & 0.8769 & 1.0232 \\
0.5 & EBP exact & \textbf{0.1071} & 0.8505 & \textbf{0.9474} \\
0.5 & EBP asymptotic & -- & 0.8549 & 0.9578 \\
1 & naive DR & 0.3185 & \textbf{0.9201} & 1.7754 \\
1 & partial EBP & \textbf{0.2799} & 0.7776 & 1.3311 \\
1 & EBP exact & 0.2911 & 0.7040 & \textbf{1.1996} \\
1 & EBP asymptotic & -- & 0.7082 & 1.2084 \\
\hline
\end{tabular}
\end{table}


{\bf Summary.} Despite outcome model misspecification, EBP has the smallest MSPE in most settings and retains shorter prediction intervals. Under stronger outcome misspecification, partial EBP has slightly smaller MSPE in some settings. On the other hand, as the degree of outcome model misspecification increases, the Coverage performance of all methods declines, but the decline in Coverage is more substantial for EBP than for naive DR and partial EBP.
\subsection{{\it Case III}, known $N_{i}$ and covariate subpopulation}
\hspace{4mm}
Here, we consider the scenario where the outcome model is correctly specified; on the other hand, the logistic propensity model is misspecified, with the degree of misspecification controlled by $\delta_{\rm ps}$. Again, we consider $m=100$.

\begin{figure}[ht]
\centering
\includegraphics[width=0.9\textwidth]{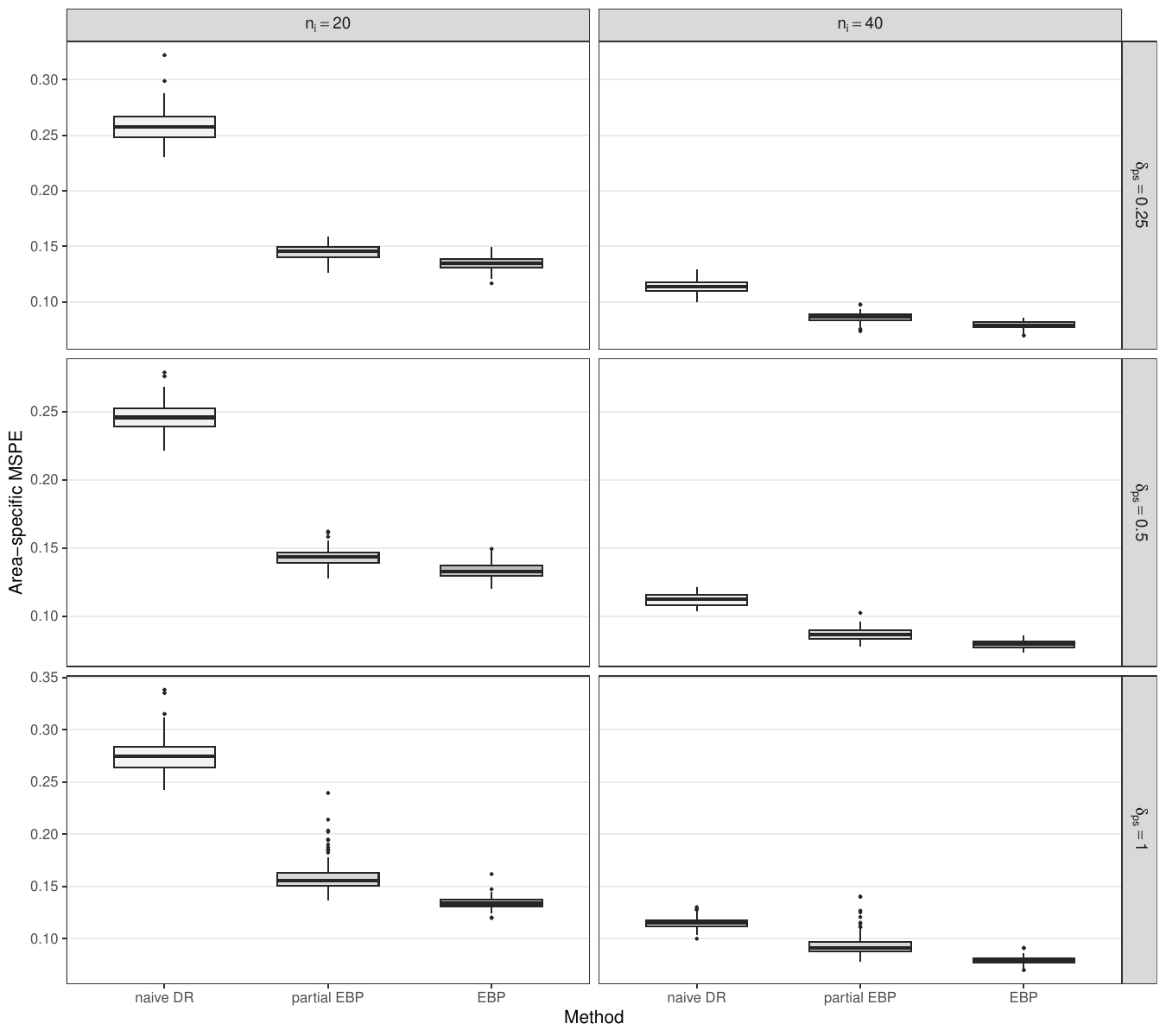}
\begingroup
\renewcommand{\thefigure}{\arabic{figure}}
\caption{Boxplots of Simulated Area-specific MSPE with Correctly Specified Outcome Model and Misspecified Logistic Propensity Model: $m=100$, Equal-\(n_i\) Design.}
\label{fig:propensity-misspec-equal-mspe}
\endgroup
\end{figure}

\begin{figure}[ht]
\centering
\includegraphics[width=0.92\textwidth]{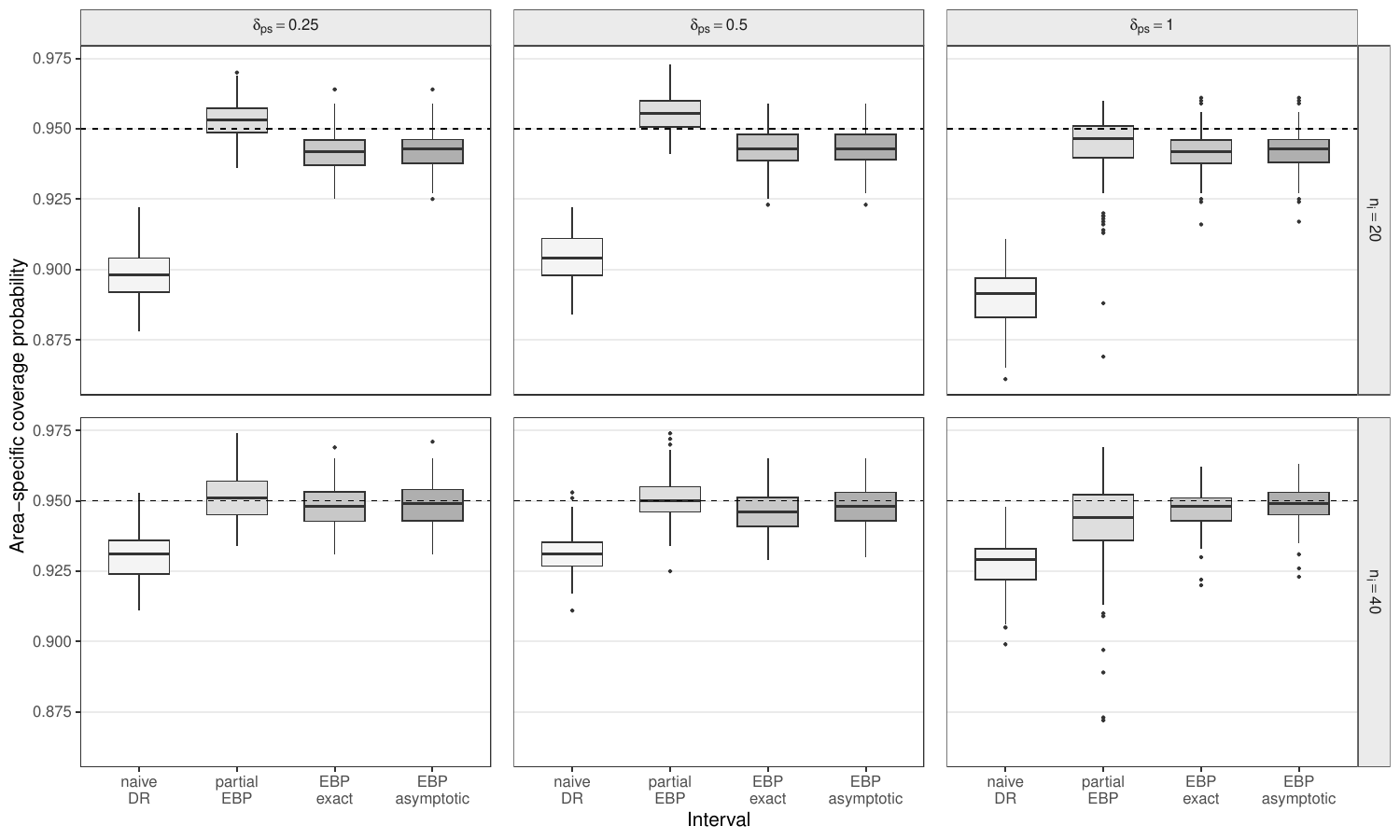}
\caption{Boxplots of Simulated Area-specific Coverage Probability of 95\% Prediction Interval with Correctly Specified Outcome Model and Misspecified Logistic Propensity Model: $m=100$, Equal-\(n_i\) Design; EBP exact and EBP asymptotic refer to EBP-based Prediction Interval based on Exact and Approximate Asymptotic Variances, Respectively.}
\label{fig:propensity-misspec-equal-coverage}
\end{figure}

\begin{figure}[ht]
\centering
\includegraphics[width=0.92\textwidth]{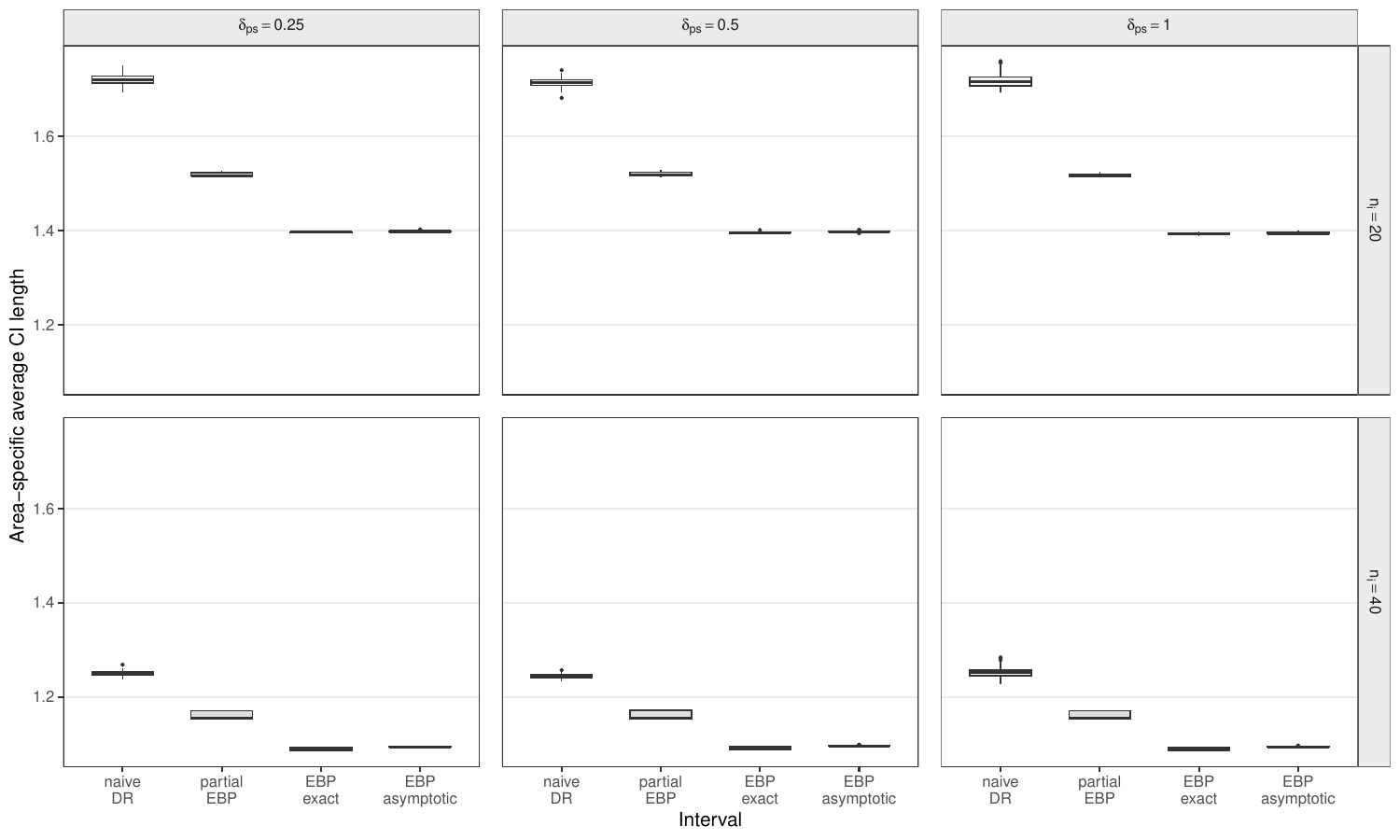}
\caption{Boxplots of Simulated Area-specific Average Length of 95\% Prediction Interval with Correctly Specified Outcome Model and Misspecified Logistic Propensity Model: $m=100$, Equal-\(n_i\) Design; EBP exact and EBP asymptotic refer to EBP-based Prediction Interval based on Exact and Approximate Asymptotic Variances, Respectively.}
\label{fig:propensity-misspec-equal-length}
\end{figure}

First consider the equal-$n_{i}$ design. Tables \ref{tab:propensity-misspec-equal-n20} and \ref{tab:propensity-misspec-equal-n40} show the summary statistics. It is seen that EBP performs the best in MSPE and Length in all cases. As for Coverage, partial EBP performs the best for $n_{i}=20$ and for $n_{i}=40, \delta_{\rm ps}=0.5$, while EBP (approximate asymptotic variance) performs the best for $n_{i}=40$, $\delta_{\rm ps}=0.25$ or $1$. The corresponding boxplots are presented in Figures \ref{fig:propensity-misspec-equal-mspe}--\ref{fig:propensity-misspec-equal-length}. The figures show patterns consistent with the summary statistics, that is, EBP dominates both MSPE and Length; in terms of Coverage, partial EBP and EBP (especially with approximate asymptotic variance) perform similarly, and both perform significantly better than naive DR.

\begin{table}[ht]
\centering
\setlength{\belowcaptionskip}{8pt}
\caption{Simulated MSPE, Coverage Probability and Average Length of 95\% Prediction Interval: Correctly Specified Outcome Model and Misspecified Logistic Propensity Model; $m=100$; Equal-$n_{i}$ Design with $n_{i}=20$; Numbers in the Parentheses Are Results based on the Approximate Asymptotic Variance for the EBP Method; Bold Numbers Indicate Best Performer under Respective $n_{i}$ and Performance Measure.}
\label{tab:propensity-misspec-equal-n20}
\setlength{\tabcolsep}{3.5pt}
\begin{tabular}{c l lll}
\hline
$\delta_{\mathrm{ps}}$ & Method & MSPE & Coverage & Length\\
\hline
0.25 & naive DR & 0.2585 & 0.8980 & 1.7209\\
0.25 & partial EBP & 0.1451 & \textbf{0.9534} & 1.5189\\
0.25 & EBP & \textbf{0.1345} & 0.9421 (0.9424) & \textbf{1.3968} (1.3985)\\
0.5 & naive DR & 0.2466 & 0.9038 & 1.7146\\
0.5 & partial EBP & 0.1433 & \textbf{0.9557} & 1.5202\\
0.5 & EBP & \textbf{0.1330} & 0.9432 (0.9435) & \textbf{1.3966} (1.3983)\\
1 & naive DR & 0.2758 & 0.8902 & 1.7182\\
1 & partial EBP & 0.1597 & \textbf{0.9427} & 1.5174\\
1 & EBP & \textbf{0.1346} & 0.9421 (0.9423) & \textbf{1.3934} (1.3950)\\
\hline
\end{tabular}
\end{table}

\begin{table}[ht]
\centering
\setlength{\belowcaptionskip}{8pt}
\caption{Simulated MSPE, Coverage Probability and Average Length of 95\% Prediction Interval: Correctly Specified Outcome Model and Misspecified Logistic Propensity Model; $m=100$; Equal-$n_{i}$ Design with $n_{i}=40$; Numbers in the Parentheses Are Results based on the Approximate Asymptotic Variance for the EBP Method; Bold Numbers Indicate Best Performer under Respective $n_{i}$ and Performance Measure.}
\label{tab:propensity-misspec-equal-n40}
\setlength{\tabcolsep}{3.5pt}
\begin{tabular}{c l lll}
\hline
$\delta_{\mathrm{ps}}$ & Method & MSPE & Coverage & Length\\
\hline
0.25 & naive DR & 0.1138 & 0.9302 & 1.2500\\
0.25 & partial EBP & 0.0860 & 0.9517 & 1.1593\\
0.25 & EBP & \textbf{0.0788} & 0.9479 (\textbf{0.9491}) & \textbf{1.0890} (1.0944)\\
0.5 & naive DR & 0.1122 & 0.9312 & 1.2438\\
0.5 & partial EBP & 0.0867 & \textbf{0.9509} & 1.1595\\
0.5 & EBP & \textbf{0.0793} & 0.9465 (0.9478) & \textbf{1.0908} (1.0962)\\
1 & naive DR & 0.1149 & 0.9270 & 1.2516\\
1 & partial EBP & 0.0943 & 0.9414 & 1.1595\\
1 & EBP & \textbf{0.0791} & 0.9470 (\textbf{0.9483}) & \textbf{1.0886} (1.0939)\\
\hline
\end{tabular}
\end{table}


\begin{figure}[p]
\centering
\includegraphics[width=\textwidth,height=0.88\textheight,keepaspectratio]{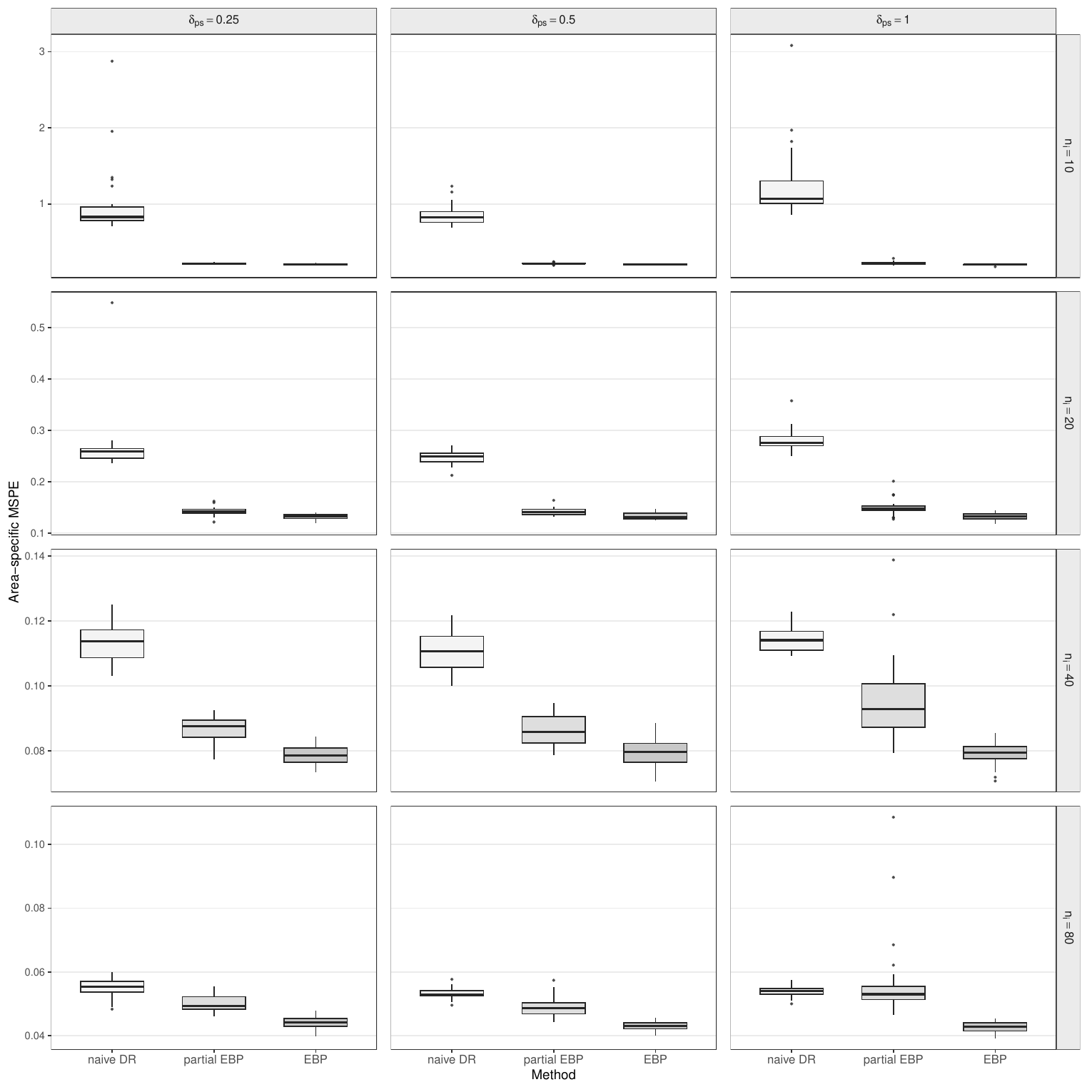}
\begingroup
\renewcommand{\thefigure}{\arabic{figure}}
\caption{Boxplots of Simulated Area-specific MSPE with Correctly Specified Outcome Model and Misspecified Logistic Propensity Model: $m=100$, Unequal-\(n_i\) Design; Rows and Columns Correspond to Different Values of $n_{i}$ and $\delta_{\rm ps}$, Respectively.}
\label{fig:propensity-misspec-variable-mspe}
\endgroup
\end{figure}

\begin{figure}[p]
\centering
\includegraphics[width=\textwidth,height=0.88\textheight,keepaspectratio]{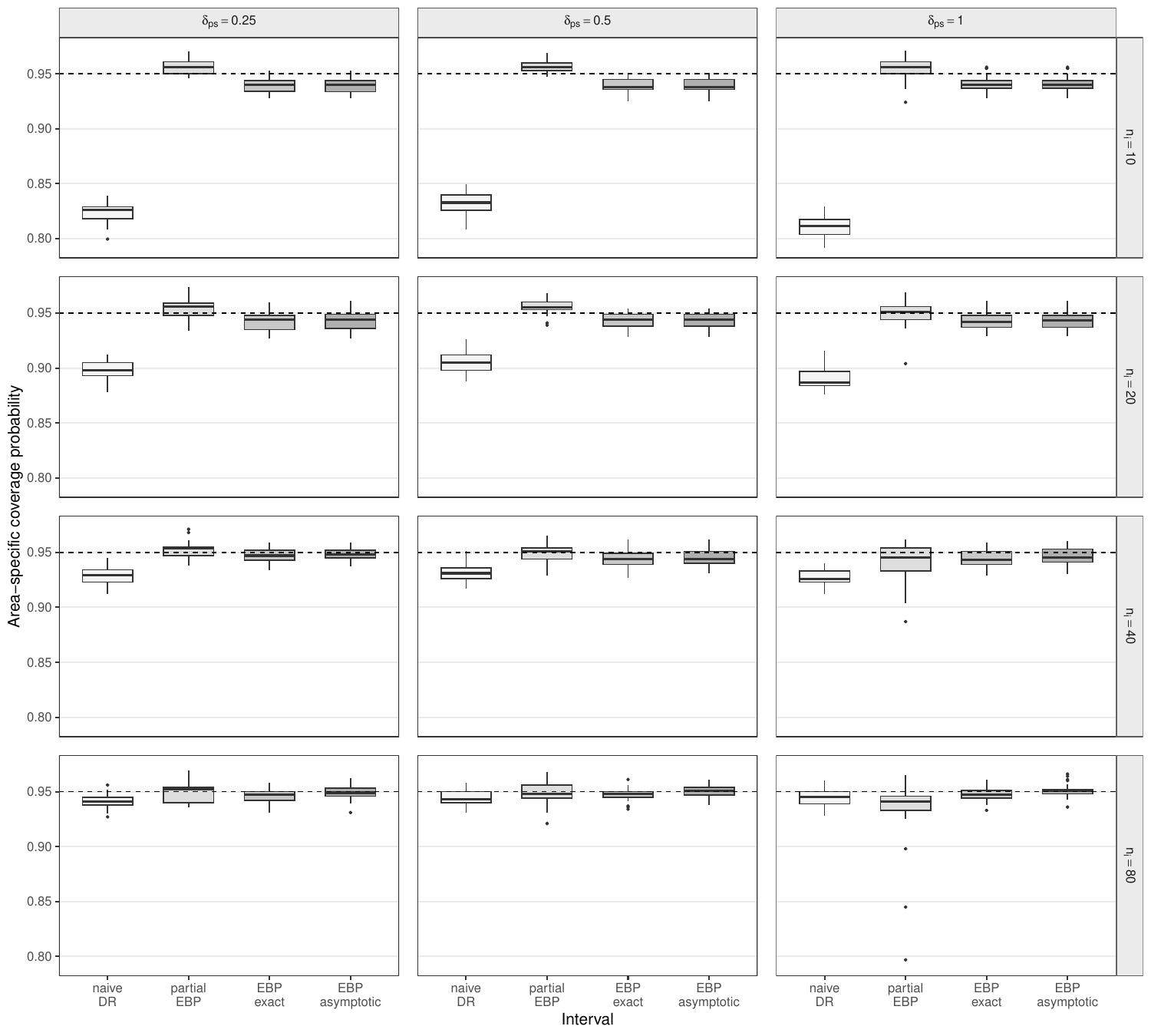}
\caption{Boxplots of Simulated Area-specific Coverage Probability of 95\% Prediction Interval with Correctly Specified Outcome Model and Misspecified Logistic Propensity Model: $m=100$, Unequal-\(n_i\) Design; Rows and Columns Correspond to Different Values of $n_{i}$ and $\delta_{\rm ps}$, Respectively; EBP exact and EBP asymptotic refer to EBP-based Prediction Interval based on Exact and Approximate Asymptotic Variances, Respectively.}
\label{fig:propensity-misspec-variable-coverage}
\end{figure}

\begin{figure}[p]
\centering
\includegraphics[width=\textwidth,height=0.88\textheight,keepaspectratio]{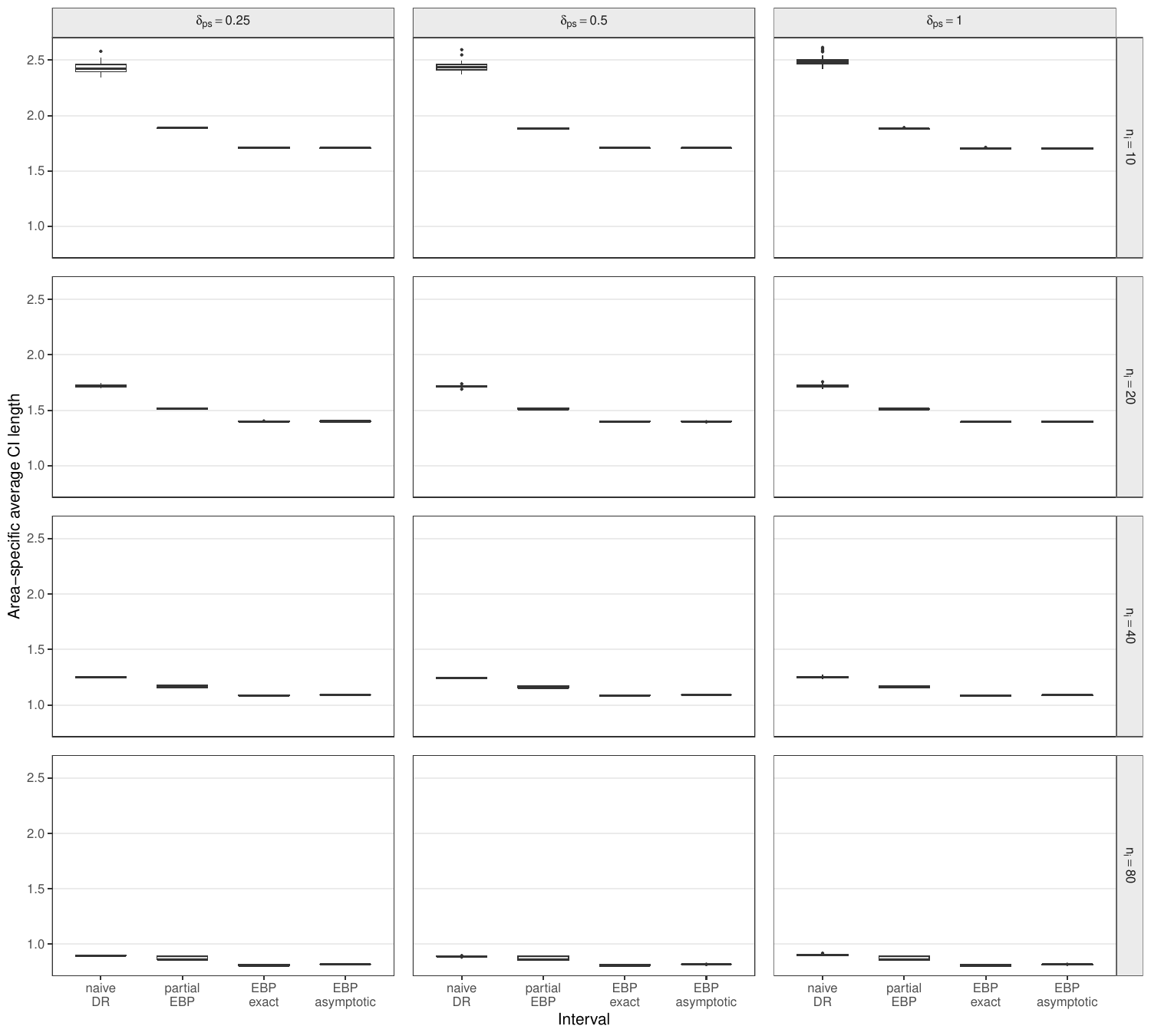}
\caption{Boxplots of Simulated Area-specific Average Length of 95\% Prediction Interval with Correctly Specified Outcome Model and Misspecified Logistic Propensity Model: $m=100$, Unequal-\(n_i\) Design; Rows and Columns Correspond to Different Values of $n_{i}$ and $\delta_{\rm ps}$, Respectively; EBP exact and EBP asymptotic refer to EBP-based Prediction Interval based on Exact and Approximate Asymptotic Variances, Respectively.}
\label{fig:propensity-misspec-variable-length}
\end{figure}

Next, we consider the unequal-$n_{i}$ design. Table \ref{tab:propensity-misspec-variable} presents the summary of different performance measures. EBP has the smallest MSPE in all cases. For each propensity-misspecification level, the asymptotic EBP interval is slightly shortest at $n_{i}=10$, whereas the finite-population EBP interval is shortest at $n_{i}=20,40$ and $80$. For the Coverage, partial EBP performs the best in most (8 out of 12) cases while EBP (with approximate asymptotic variance) performs the best in the rest (4 out of 12) of the cases. A general trend in Coverage is that partial EBP performs better for smaller $n_{i}$ and EBP (with approximate asymptotic variance) performs better for large $n_{i}$. Area-specific boxplots, presented in Figures \ref{fig:propensity-misspec-variable-mspe}--\ref{fig:propensity-misspec-variable-length}, show patterns consistent with the summaries.

\begin{table}[p]
\centering
\setlength{\belowcaptionskip}{8pt}
\caption{(Part I) Simulated MSPE, Coverage Probability and Average Length of 95\% Prediction Interval: Correctly Specified Outcome Model and Misspecified Logistic Propensity Model; $m=100$; Unequal-$n_{i}$ Design ($n_{i}=10$ and $n_{i}=20$); Separate Table for Each $n_{i}$; Bold Number Indicates the Best Performer under Respective Performance Measure.}
\label{tab:propensity-misspec-variable}
\setlength{\tabcolsep}{3pt}
\renewcommand{\arraystretch}{1.05}
\par\vspace{4pt}\textit{Panel A: $n_i=10$}\par\vspace{2pt}
\begin{tabular}{c l ccc}
\hline
$\delta_{\mathrm{ps}}$ & Method/interval & MSPE & Coverage & Avg. length \\
\hline
0.25 & naive DR & 1.0116 & 0.8236 & 2.4320 \\
0.25 & partial EBP & 0.2167 & \textbf{0.9566} & 1.8880 \\
0.25 & EBP exact & \textbf{0.2045} & 0.9396 & 1.7092 \\
0.25 & EBP asymptotic & -- & 0.9394 & \textbf{1.7083} \\
0.5 & naive DR & 0.8522 & 0.8325 & 2.4413 \\
0.5 & partial EBP & 0.2166 & \textbf{0.9568} & 1.8835 \\
0.5 & EBP exact & \textbf{0.2043} & 0.9390 & 1.7085 \\
0.5 & EBP asymptotic & -- & 0.9389 & \textbf{1.7076} \\
1 & naive DR & 1.2527 & 0.8104 & 2.5000 \\
1 & partial EBP & 0.2228 & \textbf{0.9541} & 1.8819 \\
1 & EBP exact & \textbf{0.2008} & 0.9415 & 1.7026 \\
1 & EBP asymptotic & -- & 0.9414 & \textbf{1.7016} \\
\hline
\end{tabular}
\par\vspace{4pt}\textit{Panel B: $n_i=20$}\par\vspace{2pt}
\begin{tabular}{c l ccc}
\hline
$\delta_{\mathrm{ps}}$ & Method/interval & MSPE & Coverage & Avg. length \\
\hline
0.25 & naive DR & 0.2677 & 0.8976 & 1.7199 \\
0.25 & partial EBP & 0.1433 & \textbf{0.9541} & 1.5148 \\
0.25 & EBP exact & \textbf{0.1329} & 0.9432 & \textbf{1.3987} \\
0.25 & EBP asymptotic & -- & 0.9436 & 1.4004 \\
0.5 & naive DR & 0.2473 & 0.9049 & 1.7145 \\
0.5 & partial EBP & 0.1422 & \textbf{0.9554} & 1.5114 \\
0.5 & EBP exact & \textbf{0.1336} & 0.9429 & \textbf{1.3977} \\
0.5 & EBP asymptotic & -- & 0.9432 & 1.3994 \\
1 & naive DR & 0.2804 & 0.8908 & 1.7183 \\
1 & partial EBP & 0.1501 & \textbf{0.9489} & 1.5122 \\
1 & EBP exact & \textbf{0.1329} & 0.9427 & \textbf{1.3954} \\
1 & EBP asymptotic & -- & 0.9431 & 1.3970 \\
\hline
\end{tabular}
\end{table}

\begin{table}[p]
\centering
\ContinuedFloat
\setlength{\belowcaptionskip}{8pt}
\caption{(Part II) Simulated MSPE, Coverage Probability and Average Length of 95\% Prediction Interval: Correctly Specified Outcome Model and Misspecified Logistic Propensity Model; $m=100$; Unequal-$n_{i}$ Design ($n_{i}=40$ and $n_{i}=80$); Separate Table for Each $n_{i}$; Bold Number Indicates the Best Performer under Respective Performance Measure.}
\setlength{\tabcolsep}{3pt}
\renewcommand{\arraystretch}{1.05}
\par\vspace{4pt}\textit{Panel C: $n_i=40$}\par\vspace{2pt}
\begin{tabular}{c l ccc}
\hline
$\delta_{\mathrm{ps}}$ & Method/interval & MSPE & Coverage & Avg. length \\
\hline
0.25 & naive DR & 0.1139 & 0.9286 & 1.2490 \\
0.25 & partial EBP & 0.0865 & 0.9525 & 1.1638 \\
0.25 & EBP exact & \textbf{0.0789} & 0.9473 & \textbf{1.0866} \\
0.25 & EBP asymptotic & -- & \textbf{0.9484} & 1.0920 \\
0.5 & naive DR & 0.1105 & 0.9310 & 1.2430 \\
0.5 & partial EBP & 0.0867 & \textbf{0.9496} & 1.1617 \\
0.5 & EBP exact & \textbf{0.0795} & 0.9443 & \textbf{1.0861} \\
0.5 & EBP asymptotic & -- & 0.9452 & 1.0916 \\
1 & naive DR & 0.1149 & 0.9267 & 1.2511 \\
1 & partial EBP & 0.0953 & 0.9401 & 1.1623 \\
1 & EBP exact & \textbf{0.0791} & 0.9447 & \textbf{1.0845} \\
1 & EBP asymptotic & -- & \textbf{0.9458} & 1.0900 \\
\hline
\end{tabular}
\par\vspace{4pt}\textit{Panel D: $n_i=80$}\par\vspace{2pt}
\begin{tabular}{c l ccc}
\hline
$\delta_{\mathrm{ps}}$ & Method/interval & MSPE & Coverage & Avg. length \\
\hline
0.25 & naive DR & 0.0551 & 0.9416 & 0.8989 \\
0.25 & partial EBP & 0.0500 & \textbf{0.9494} & 0.8710 \\
0.25 & EBP exact & \textbf{0.0438} & 0.9458 & \textbf{0.8089} \\
0.25 & EBP asymptotic & -- & 0.9486 & 0.8187 \\
0.5 & naive DR & 0.0533 & 0.9441 & 0.8919 \\
0.5 & partial EBP & 0.0490 & 0.9495 & 0.8695 \\
0.5 & EBP exact & \textbf{0.0430} & 0.9475 & \textbf{0.8088} \\
0.5 & EBP asymptotic & -- & \textbf{0.9501} & 0.8186 \\
1 & naive DR & 0.0539 & 0.9451 & 0.9026 \\
1 & partial EBP & 0.0573 & 0.9306 & 0.8703 \\
1 & EBP exact & \textbf{0.0429} & 0.9479 & \textbf{0.8081} \\
1 & EBP asymptotic & -- & \textbf{0.9514} & 0.8179 \\
\hline
\end{tabular}
\end{table}


{\bf Summary.} Here, we see similar patterns as {\it Case I}, that is, EBP performs the best in MSPE and Length in all cases, and partial EBP performs the best in Coverage in most cases. The relative performance of EBP and partial EBP in Coverage does not seem to depend much on the degree of propensity model misspecification; rather, it depends more on the sample size of $n_{i}$; namely, for smaller $n_{i}$, partial EBP performs better; for larger $n_{i}$, EBP (with approximate asymptotic variance) performs better.
\subsection{{\it Case IV}, known $N_{i}$ and covariate subpopulation}
\hspace{4mm}
Finally, we consider the scenario where both the outcome model and the propensity model are misspecified. As before, the misspecification of the outcome model is controlled by $\gamma_{\rm quad}$ while the misspecification of the (logistic) propensity model is controlled by $\delta_{\rm ps}$. Once again, we consider $m=100$.

First consider the equal-$n_{i}$ design. Tables \ref{tab:both-misspec-equal-point-n20} and \ref{tab:both-misspec-equal-point-n40} present the average simulated area-specific MSPE for different methods. It can be seen that in the vast majority of the cases, EBP performs the best. In fact, the only cases where partial EBP performs better are, for $n_{i}=20$, the cases where the outcome model is severely misspecified ($\gamma_{\rm quad}=1$) and the propensity model is moderately misspecified ($\delta_{\rm ps}=0.5)$ and, for $n_{i}=40$, the case where the outcome model is severely misspecified ($\gamma_{\rm quad}=1$) and the propensity model is slightly ($\delta_{\rm ps}=0.25$) or moderately ($\delta_{\rm ps}=0.5$) misspecified.

\begin{figure}[h]
\centering
\includegraphics[width=0.92\textwidth,height=0.84\textheight,keepaspectratio]{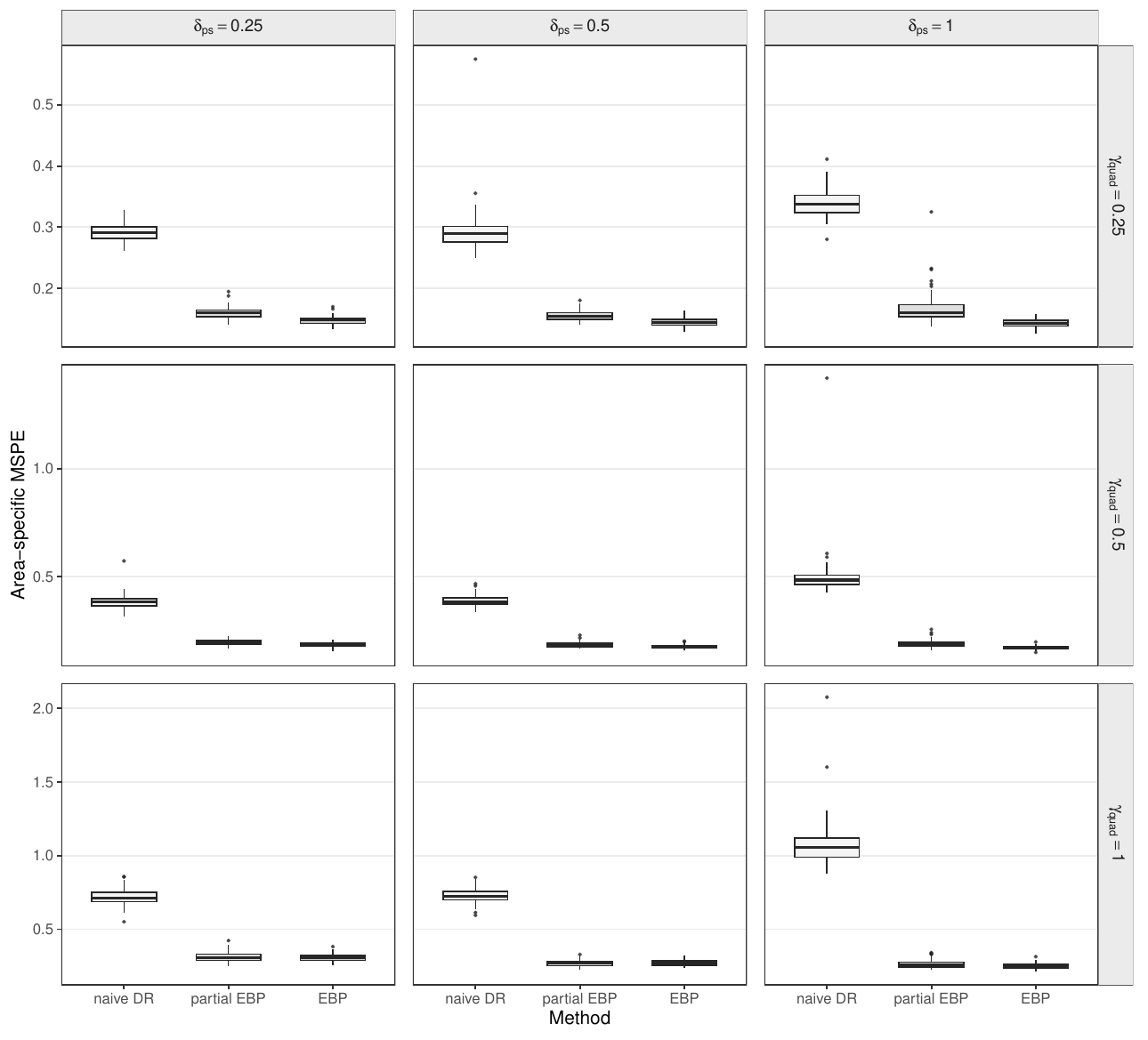}
\caption{Boxplots of Simulated Area-specific MSPE: Misspecified Outcome Model and Logistic Propensity Model: $m=100$, Equal-\(n_i\) Design with $n_{i}=20$; Rows and Columns Correspond to Different Values of \(\gamma_{\mathrm{quad}}\) and \(\delta_{\mathrm{ps}}\), Respectively.}
\label{fig:both-misspec-equal-mspe}
\end{figure}

\begin{figure}[h]
\centering
\includegraphics[width=0.92\textwidth,height=0.84\textheight,keepaspectratio]{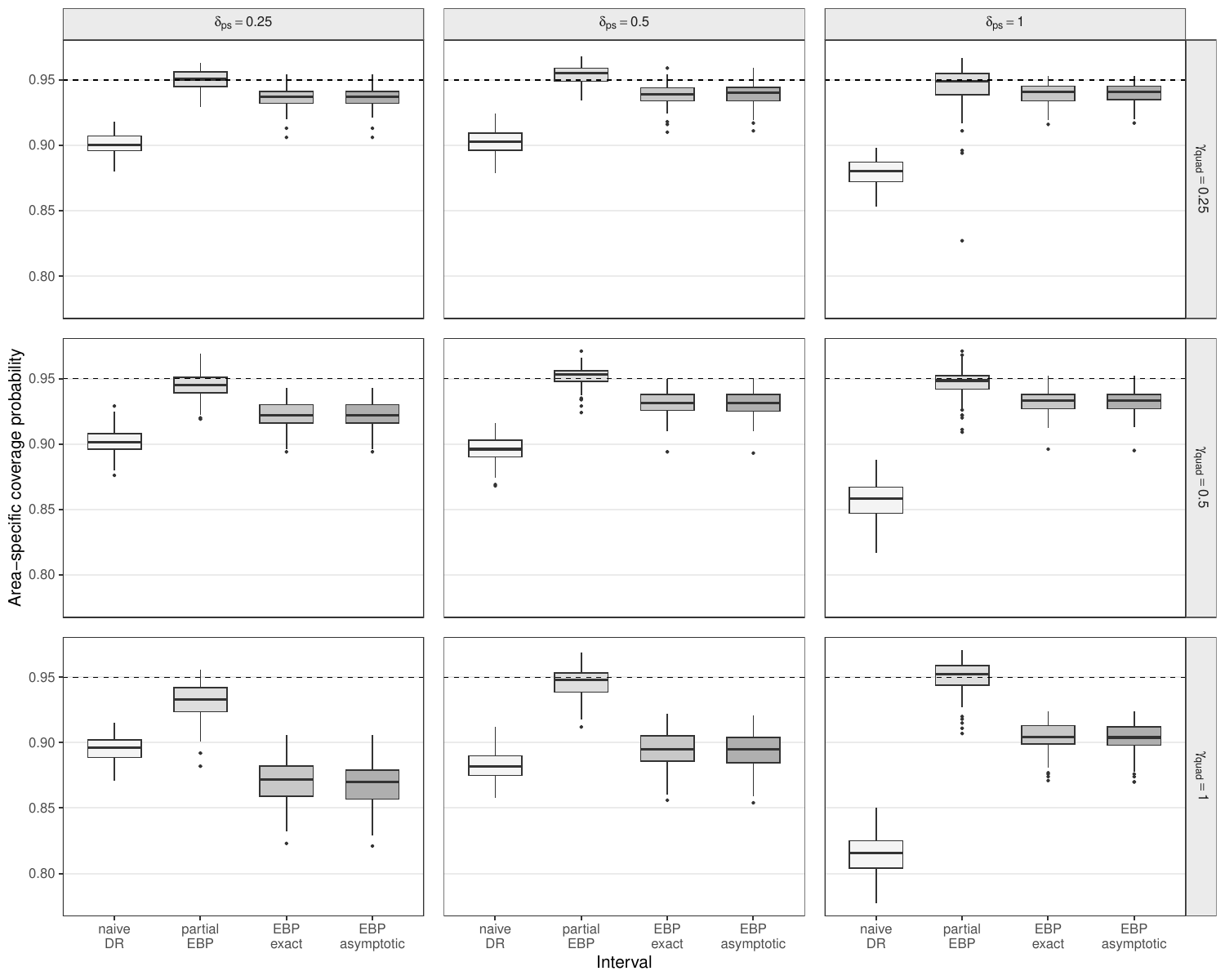}
\caption{Boxplots of Simulated Area-specific Coverage Probability of 95\% Prediction Interval: Misspecified Outcome Model and Logistic Propensity Model: $m=100$, Equal-\(n_i\) Design with $n_{i}=20$; Rows and Columns Correspond to Different Values of \(\gamma_{\mathrm{quad}}\) and \(\delta_{\mathrm{ps}}\), Respectively; EBP exact and EBP asymptotic refer to EBP-based Prediction Interval based on Exact and Approximate Asymptotic Variances, Respectively.}
\label{fig:both-misspec-equal-coverage}
\end{figure}

\begin{figure}[h]
\centering
\includegraphics[width=0.92\textwidth,height=0.84\textheight,keepaspectratio]{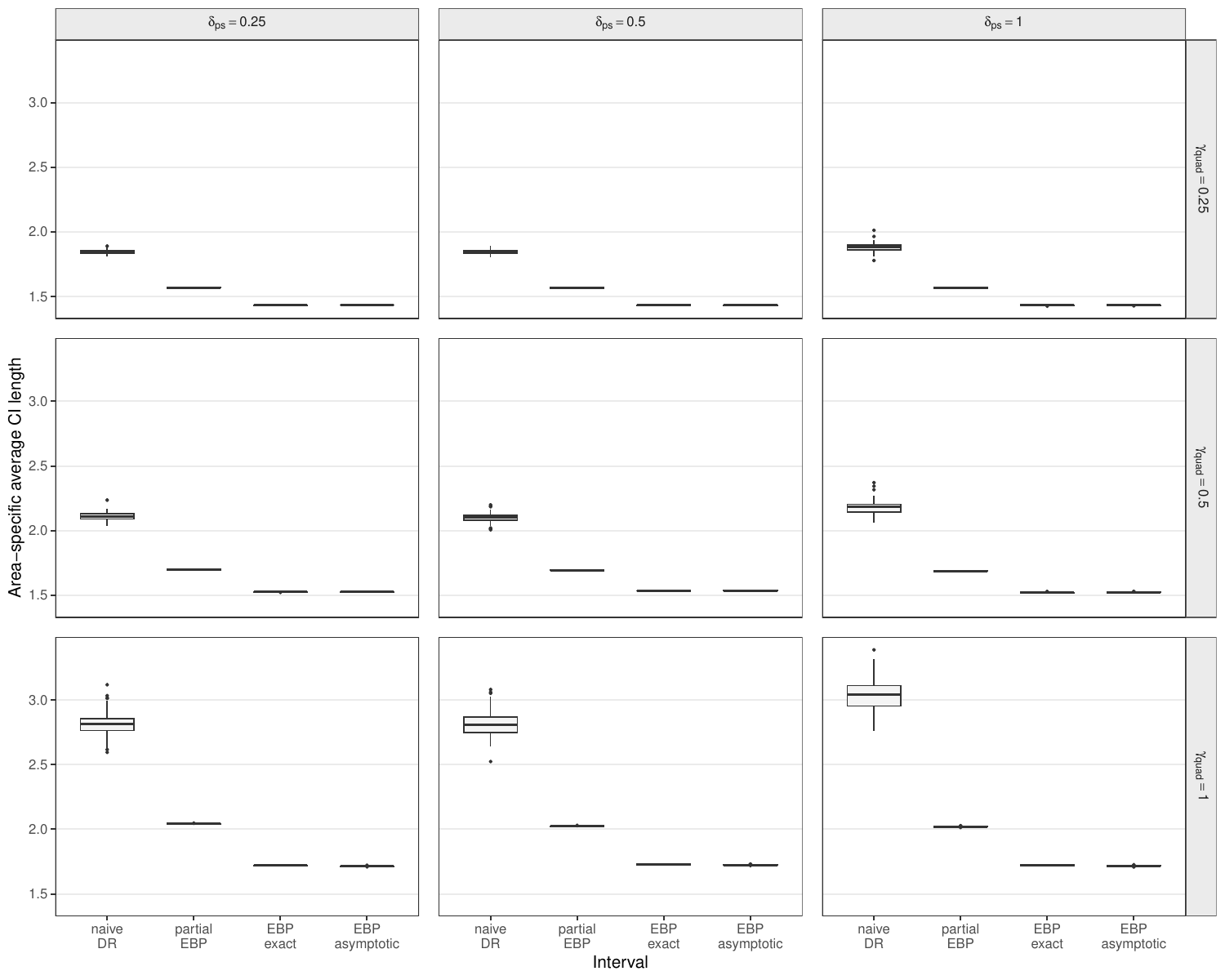}
\caption{Boxplots of Simulated Area-specific Average Length of 95\% Prediction Interval: Misspecified Outcome Model and Logistic Propensity Model: $m=100$, Equal-\(n_i\) Design with $n_{i}=20$; Rows and Columns Correspond to Different Values of \(\gamma_{\mathrm{quad}}\) and \(\delta_{\mathrm{ps}}\), Respectively; EBP exact and EBP asymptotic refer to EBP-based Prediction Interval based on Exact and Approximate Asymptotic Variances, Respectively.}
\label{fig:both-misspec-equal-length}
\end{figure}

\begin{figure}[h]
\centering
\includegraphics[width=0.92\textwidth,height=0.84\textheight,keepaspectratio]{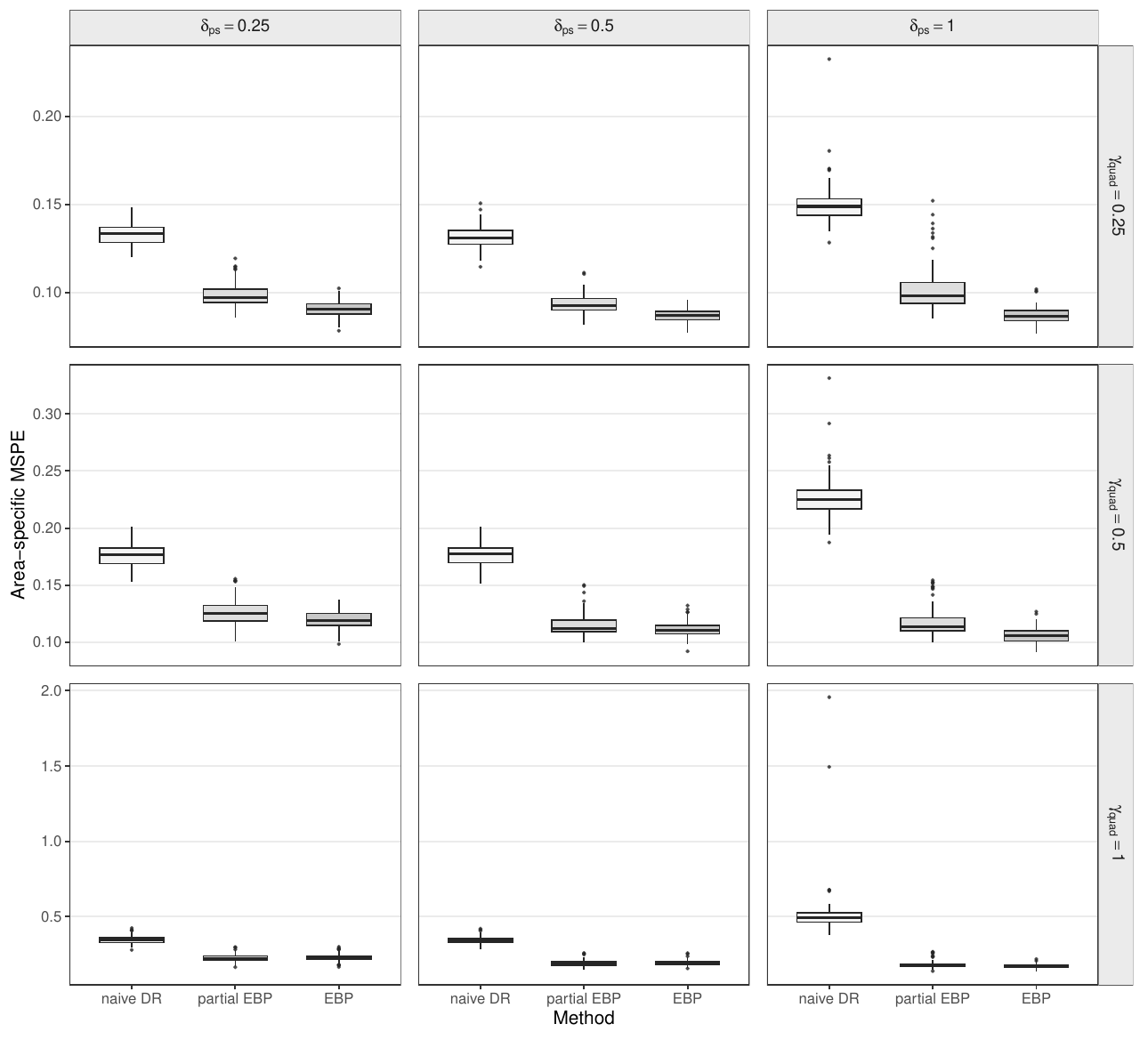}
\caption{Boxplots of Simulated Area-specific MSPE: Misspecified Outcome Model and Logistic Propensity Model: $m=100$, Equal-\(n_i\) Design with $n_{i}=40$; Rows and Columns Correspond to Different Values of \(\gamma_{\mathrm{quad}}\) and \(\delta_{\mathrm{ps}}\), Respectively.}
\label{fig:both-misspec-equal-mspe-n40}
\end{figure}

\begin{figure}[h]
\centering
\includegraphics[width=0.92\textwidth,height=0.84\textheight,keepaspectratio]{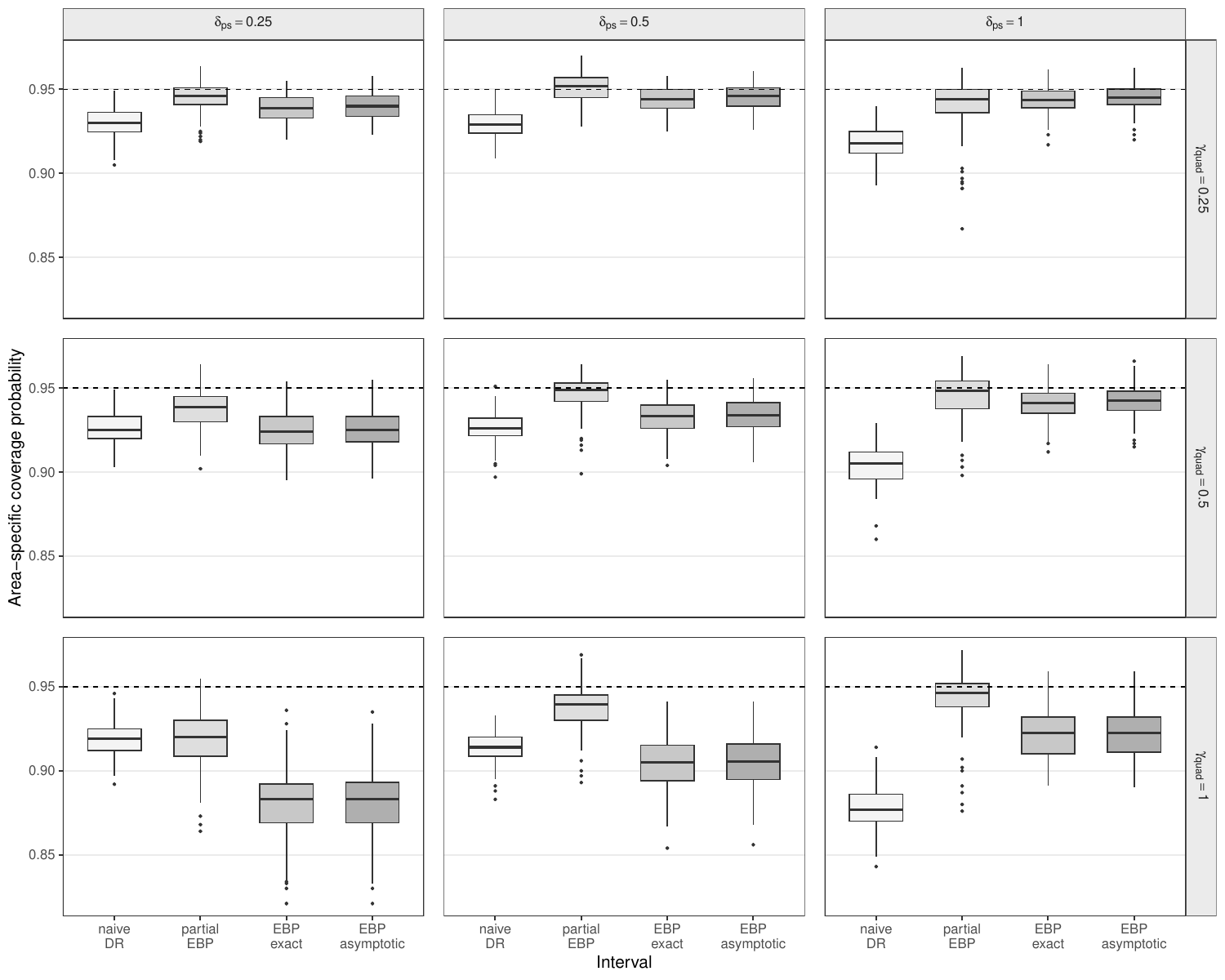}
\caption{Boxplots of Simulated Area-specific Coverage Probability of 95\% Prediction Interval: Misspecified Outcome Model and Logistic Propensity Model: $m=100$, Equal-\(n_i\) Design with $n_{i}=40$; Rows and Columns Correspond to Different Values of \(\gamma_{\mathrm{quad}}\) and \(\delta_{\mathrm{ps}}\), Respectively; EBP exact and EBP asymptotic refer to EBP-based Prediction Interval based on Exact and Approximate Asymptotic Variances, Respectively.}
\label{fig:both-misspec-equal-coverage-n40}
\end{figure}

\begin{figure}[h]
\centering
\includegraphics[width=0.92\textwidth,height=0.84\textheight,keepaspectratio]{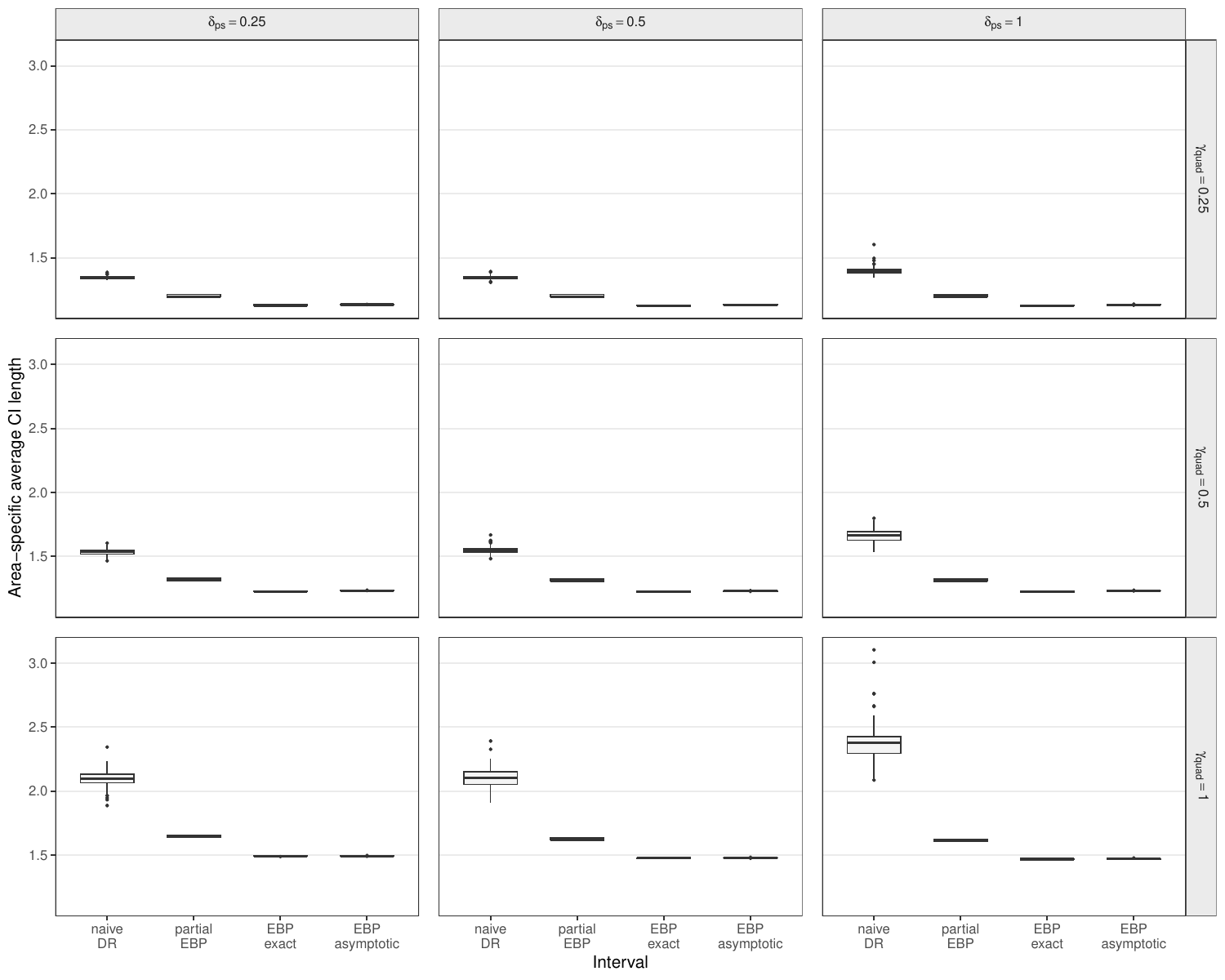}
\caption{Boxplots of Simulated Area-specific Average Length of 95\% Prediction Interval: Misspecified Outcome Model and Logistic Propensity Model: $m=100$, Equal-\(n_i\) Design with $n_{i}=40$; Rows and Columns Correspond to Different Values of \(\gamma_{\mathrm{quad}}\) and \(\delta_{\mathrm{ps}}\), Respectively; EBP exact and EBP asymptotic refer to EBP-based Prediction Interval based on Exact and Approximate Asymptotic Variances, Respectively.}
\label{fig:both-misspec-equal-length-n40}
\end{figure}

Tables \ref{tab:both-misspec-equal-interval-n20} and \ref{tab:both-misspec-equal-interval-n40} present the summary statistics and Coverage and Length. In each setting, either the finite-population or the asymptotic EBP interval has the shortest Length. For Coverage, with $n_{i}=20$, the best performer is always partial EBP; with $n_{i}=40$, partial EBP is the best performer in all but two cases, in which cases the best performers are either naive DR or EBP (with approximate asymptotic variance). Area-specific boxplots, presented in Figures \ref{fig:both-misspec-equal-mspe}--\ref{fig:both-misspec-equal-length-n40}, show patterns consistent with the summary statistics.

\begin{table}[ht]
\centering
\caption{\small Simulated MSPE under Misspecified Outcome Model and Propensity Model; $m=100$; Equal-$n_{i}$ Design with $n_{i}=20$; Bold Numbers Indicate Best Performer under the Setting.}
\label{tab:both-misspec-equal-point-n20}
\begin{tabular}{cc ccc}
\hline
$\gamma_{\mathrm{quad}}$ & $\delta_{\mathrm{ps}}$ & naive DR & partial EBP & EBP \\
\hline
0.2500 & 0.2500 & 0.2916 & 0.1595 & \textbf{0.1475} \\
0.2500 & 0.5000 & 0.2930 & 0.1552 & \textbf{0.1450} \\
0.2500 & 1.0000 & 0.3394 & 0.1665 & \textbf{0.1430} \\
0.5000 & 0.2500 & 0.3828 & 0.1943 & \textbf{0.1838} \\
0.5000 & 0.5000 & 0.3889 & 0.1840 & \textbf{0.1749} \\
0.5000 & 1.0000 & 0.4980 & 0.1898 & \textbf{0.1703} \\
1.0000 & 0.2500 & 0.7184 & 0.3101 & \textbf{0.3068} \\
1.0000 & 0.5000 & 0.7269 & \textbf{0.2702} & 0.2709 \\
1.0000 & 1.0000 & 1.0735 & 0.2615 & \textbf{0.2505} \\
\hline
\end{tabular}
\end{table}

\begin{table}[ht]
\centering
\caption{\small Simulated MSPE under Misspecified Outcome Model and Propensity Model; $m=100$; Equal-$n_{i}$ Design with $n_{i}=40$; Bold Numbers Indicate Best Performer under the Setting.}
\label{tab:both-misspec-equal-point-n40}
\begin{tabular}{cc ccc}
\hline
$\gamma_{\mathrm{quad}}$ & $\delta_{\mathrm{ps}}$ & naive DR & partial EBP & EBP \\
\hline
0.2500 & 0.2500 & 0.1330 & 0.0984 & \textbf{0.0902} \\
0.2500 & 0.5000 & 0.1314 & 0.0933 & \textbf{0.0870} \\
0.2500 & 1.0000 & 0.1501 & 0.1018 & \textbf{0.0866} \\
0.5000 & 0.2500 & 0.1763 & 0.1262 & \textbf{0.1196} \\
0.5000 & 0.5000 & 0.1770 & 0.1151 & \textbf{0.1113} \\
0.5000 & 1.0000 & 0.2263 & 0.1172 & \textbf{0.1060} \\
1.0000 & 0.2500 & 0.3457 & \textbf{0.2255} & 0.2285 \\
1.0000 & 0.5000 & 0.3418 & \textbf{0.1888} & 0.1932 \\
1.0000 & 1.0000 & 0.5201 & 0.1796 & \textbf{0.1717} \\
\hline
\end{tabular}
\end{table}

\begin{table}[ht]
\centering
\setlength{\belowcaptionskip}{8pt}
\caption{Simulated Coverage Probability and Average Length of 95\% Prediction Interval: Misspecified Outcome Model and Propensity Model; $m=100$; Equal-$n_{i}$ Design with $n_{i}=20$; Numbers in the Parentheses Are Results based on the Approximate Asymptotic Variance for the EBP Method; Bold Numbers Indicate Best Performer under Respective $n_{i}$ and Performance Measure.}
\footnotesize
\label{tab:both-misspec-equal-interval-n20}
\setlength{\tabcolsep}{4pt}
\renewcommand{\arraystretch}{1.08}
{\itshape}\par\vspace{3pt}
\begin{tabular*}{0.86\textwidth}{@{\extracolsep{\fill}}cc cccccc@{}}
\hline
$\gamma_{\mathrm{quad}}$ & $\delta_{\mathrm{ps}}$ & \multicolumn{2}{c}{naive DR} & \multicolumn{2}{c}{partial EBP} & \multicolumn{2}{c}{EBP} \\ \cline{3-8}
& & Coverage & Length & Coverage & Length & Coverage & Length\\ \hline
0.25 & 0.25 & 0.9007 & 1.8446 & \textbf{0.9503} & 1.5683 & 0.9364 (0.9366) & \textbf{1.4323} (1.4336)\\
0.25 & 0.5 & 0.9026 & 1.8460 & \textbf{0.9536} & 1.5677 & 0.9385 (0.9387) & \textbf{1.4318} (1.4332)\\
0.25 & 1 & 0.8795 & 1.8810 & \textbf{0.9447} & 1.5671 & 0.9394 (0.9396) & \textbf{1.4319} (1.4332)\\
0.5 & 0.25 & 0.9019 & 2.1082 & \textbf{0.9447} & 1.6980 & 0.9222 (0.9222) & \textbf{1.5255} (1.5257)\\
0.5 & 0.5 & 0.8956 & 2.1006 & \textbf{0.9514} & 1.6946 & 0.9312 (0.9312) & \textbf{1.5333} (1.5339)\\
0.5 & 1 & 0.8575 & 2.1796 & \textbf{0.9462} & 1.6875 & 0.9319 (0.9320) & \textbf{1.5207} (1.5211)\\
1 & 0.25 & 0.8957 & 2.8191 & \textbf{0.9318} & 2.0422 & 0.8697 (0.8679) & 1.7199 (\textbf{1.7144})\\
1 & 0.5 & 0.8825 & 2.8161 & \textbf{0.9460} & 2.0244 & 0.8947 (0.8935) & 1.7289 (\textbf{1.7241})\\
1 & 1 & 0.8149 & 3.0309 & \textbf{0.9501} & 2.0193 & 0.9046 (0.9033) & 1.7221 (\textbf{1.7172})\\
\hline
\end{tabular*}
\end{table}

\begin{table}[ht]
\centering
\setlength{\belowcaptionskip}{8pt}
\caption{Simulated Coverage Probability and Average Length of 95\% Prediction Interval: Misspecified Outcome Model and Propensity Model; $m=100$; Equal-$n_{i}$ Design with $n_{i}=40$; Numbers in the Parentheses Are Results based on the Approximate Asymptotic Variance for the EBP Method; Bold Numbers Indicate Best Performer under Respective $n_{i}$ and Performance Measure.}
\label{tab:both-misspec-equal-interval-n40}
\footnotesize
\setlength{\tabcolsep}{4pt}
\renewcommand{\arraystretch}{1.08}
{\itshape}\par\vspace{3pt}
\begin{tabular*}{0.86\textwidth}{@{\extracolsep{\fill}}cc cccccc@{}}
\hline
$\gamma_{\mathrm{quad}}$ & $\delta_{\mathrm{ps}}$ & \multicolumn{2}{c}{naive DR} & \multicolumn{2}{c}{partial EBP} & \multicolumn{2}{c}{EBP exact} \\ \cline{3-8}
& & Coverage & Length & Coverage & Length & Coverage & Length \\ \hline
0.25 & 0.25 & 0.9299 & 1.3460 & \textbf{0.9449} & 1.2022 & 0.9391 (0.9402) & \textbf{1.1287} (1.1341)\\
0.25 & 0.5 & 0.9296 & 1.3455 & \textbf{0.9505} & 1.2018 & 0.9438 (0.9451) & \textbf{1.1271} (1.1325)\\
0.25 & 1 & 0.9184 & 1.3981 & 0.9395 & 1.2002 & 0.9435 (\textbf{0.9448}) & \textbf{1.1264} (1.1318)\\
0.5 & 0.25 & 0.9253 & 1.5349 & \textbf{0.9369} & 1.3169 & 0.9244 (0.9256) & \textbf{1.2286} (1.2335)\\
0.5 & 0.5 & 0.9261 & 1.5478 & \textbf{0.9462} & 1.3127 & 0.9326 (0.9336) & \textbf{1.2249} (1.2298)\\
0.5 & 1 & 0.9040 & 1.6635 & \textbf{0.9451} & 1.3130 & 0.9405 (0.9415) & \textbf{1.2259} (1.2309)\\
1 & 0.25 & \textbf{0.9186} & 2.0994 & 0.9179 & 1.6474 & 0.8805 (0.8808) & \textbf{1.4924} (1.4942)\\
1 & 0.5 & 0.9139 & 2.1028 & \textbf{0.9377} & 1.6259 & 0.9046 (0.9050) & \textbf{1.4772} (1.4791)\\
1 & 1 & 0.8783 & 2.3833 & \textbf{0.9423} & 1.6157 & 0.9214 (0.9217) & \textbf{1.4691} (1.4711)\\
\hline
\end{tabular*}
\end{table}

\clearpage

\begin{figure}[h]
\centering
\includegraphics[width=0.92\textwidth,height=0.84\textheight,keepaspectratio]{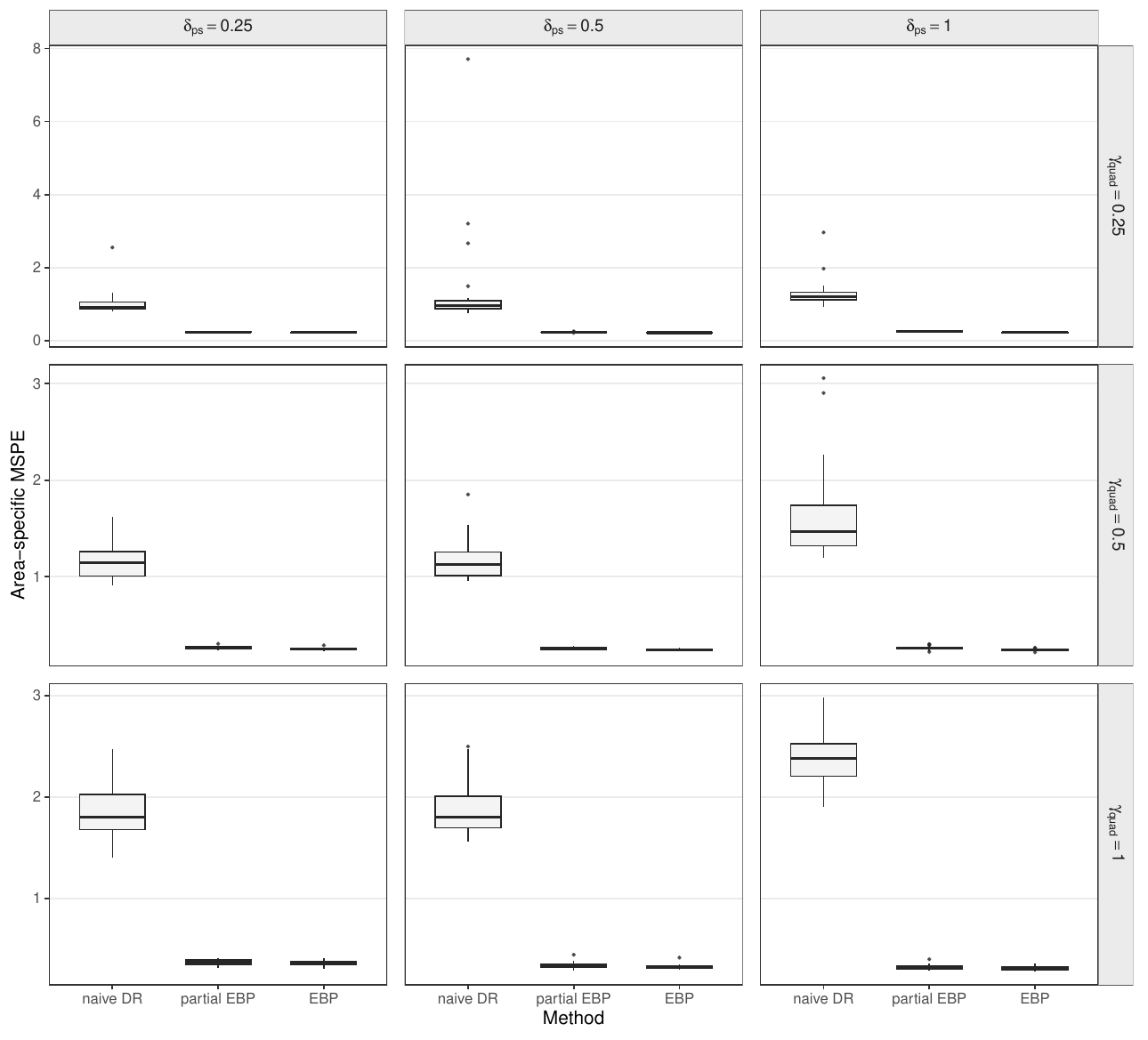}
\caption{Boxplots of Simulated Area-specific MSPE: Misspecified Outcome Model and Logistic Propensity Model: $m=100$, Unequal-\(n_i\) Design (Areas with $n_{i}=10$ Only); Rows and Columns Correspond to Different Values of \(\gamma_{\mathrm{quad}}\) and \(\delta_{\mathrm{ps}}\), Respectively.}
\label{fig:both-misspec-variable-mspe-n10}
\end{figure}

\begin{figure}[h]
\centering
\includegraphics[width=0.92\textwidth,height=0.84\textheight,keepaspectratio]{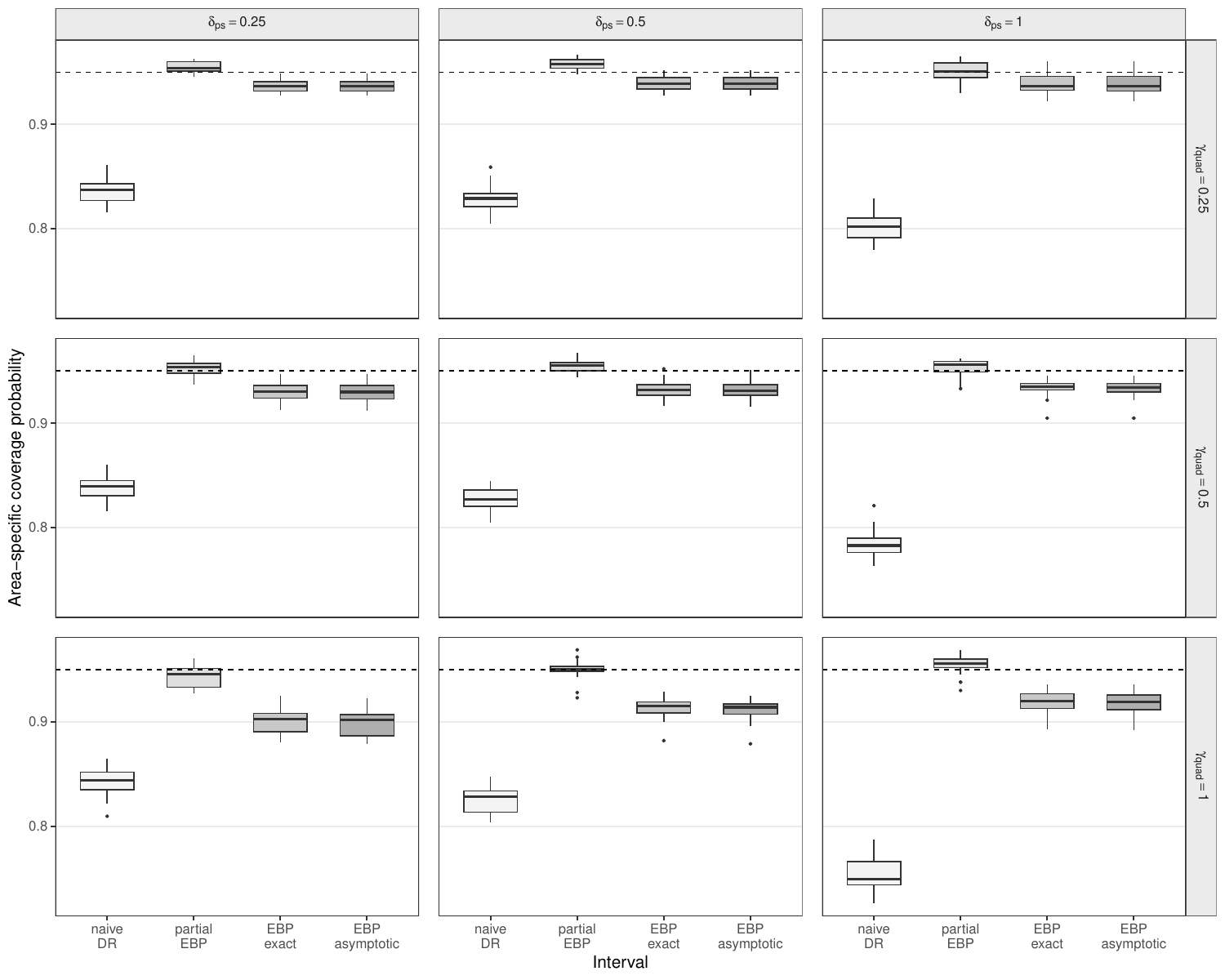}
\caption{Boxplots of Simulated Area-specific Coverage Probability of 95\% Prediction Interval: Misspecified Outcome Model and Logistic Propensity Model: $m=100$, Unequal-\(n_i\) Design (Areas with $n_{i}=10$ Only); Rows and Columns Correspond to Different Values of \(\gamma_{\mathrm{quad}}\) and \(\delta_{\mathrm{ps}}\), Respectively; EBP exact and EBP asymptotic refer to EBP-based Prediction Interval based on Exact and Approximate Asymptotic Variances, Respectively.}
\label{fig:both-misspec-variable-coverage-n10}
\end{figure}

\begin{figure}[h]
\centering
\includegraphics[width=0.92\textwidth,height=0.84\textheight,keepaspectratio]{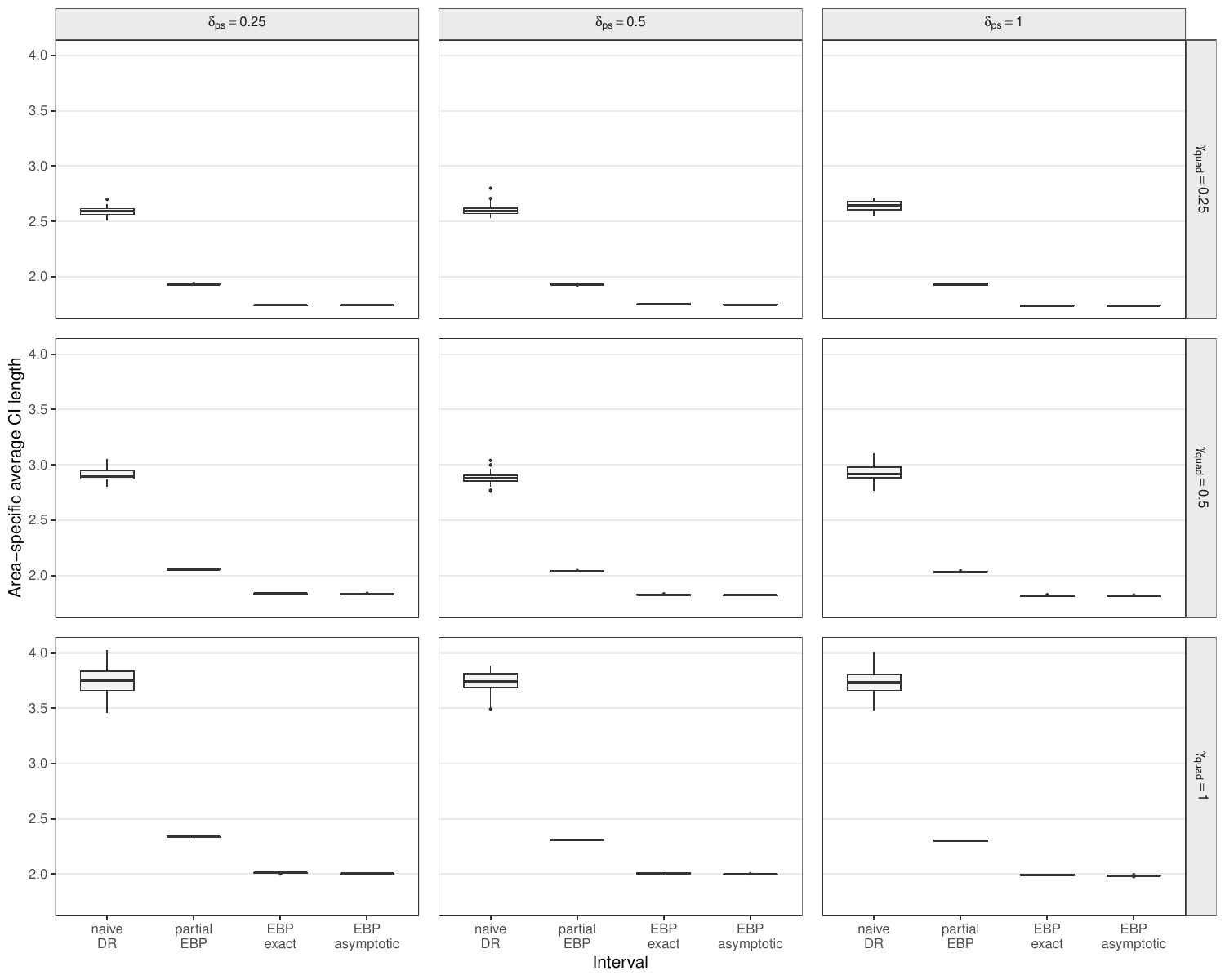}
\caption{Boxplots of Simulated Area-specific Average Length of 95\% Prediction Interval: Misspecified Outcome Model and Logistic Propensity Model: $m=100$, Unequal-\(n_i\) Design (Areas with $n_{i}=10$ Only); Rows and Columns Correspond to Different Values of \(\gamma_{\mathrm{quad}}\) and \(\delta_{\mathrm{ps}}\), Respectively; EBP exact and EBP asymptotic refer to EBP-based Prediction Interval based on Exact and Approximate Asymptotic Variances, Respectively.}
\label{fig:both-misspec-variable-length-n10}
\end{figure}

\begin{figure}[]
\centering
\includegraphics[width=0.92\textwidth,height=0.84\textheight,keepaspectratio]{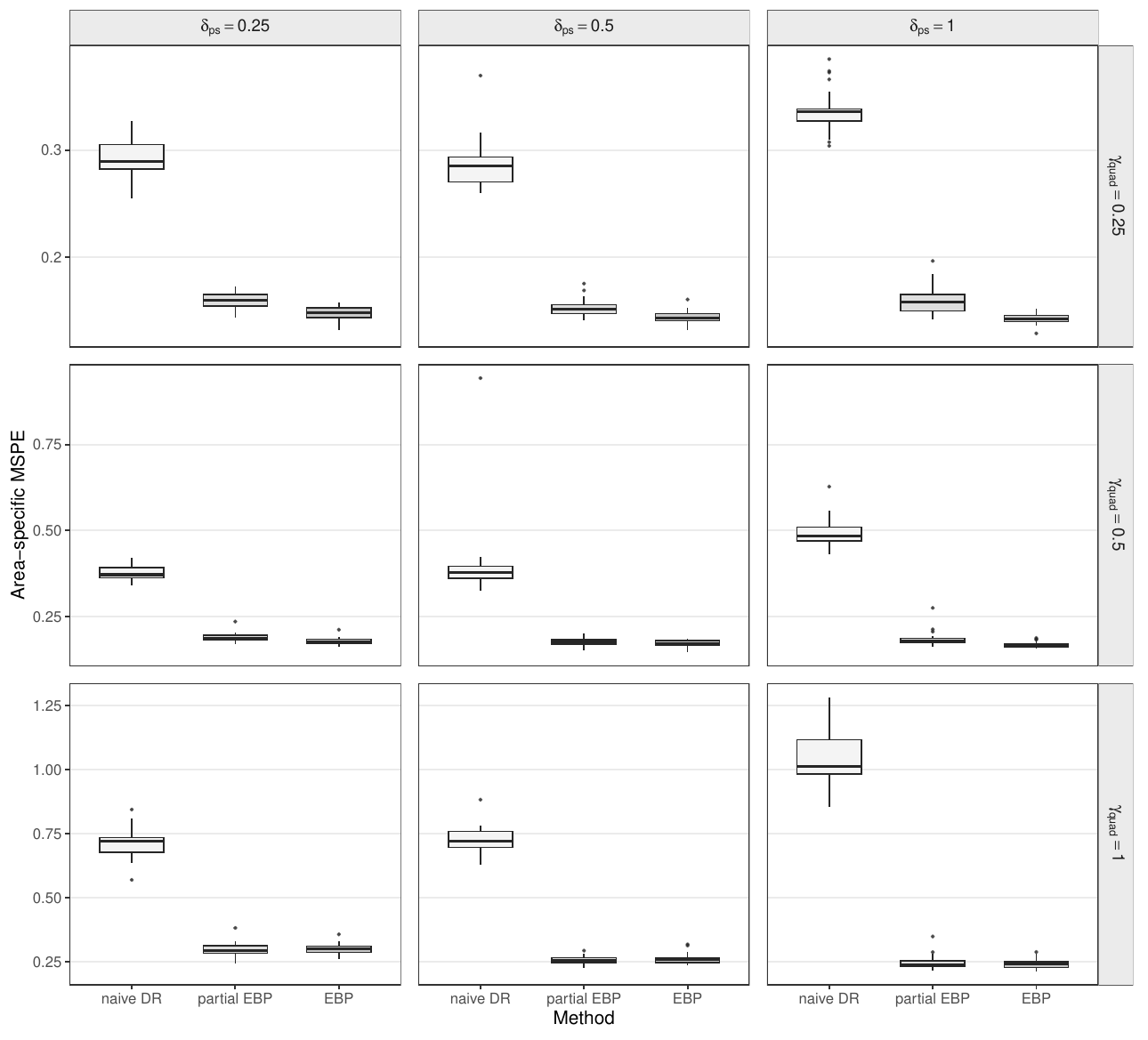}
\caption{Boxplots of Simulated Area-specific MSPE: Misspecified Outcome Model and Logistic Propensity Model: $m=100$, Unequal-\(n_i\) Design (Areas with $n_{i}=20$ Only); Rows and Columns Correspond to Different Values of \(\gamma_{\mathrm{quad}}\) and \(\delta_{\mathrm{ps}}\), Respectively.}
\label{fig:both-misspec-variable-mspe-n20}
\end{figure}

\begin{figure}[h]
\centering
\includegraphics[width=0.92\textwidth,height=0.84\textheight,keepaspectratio]{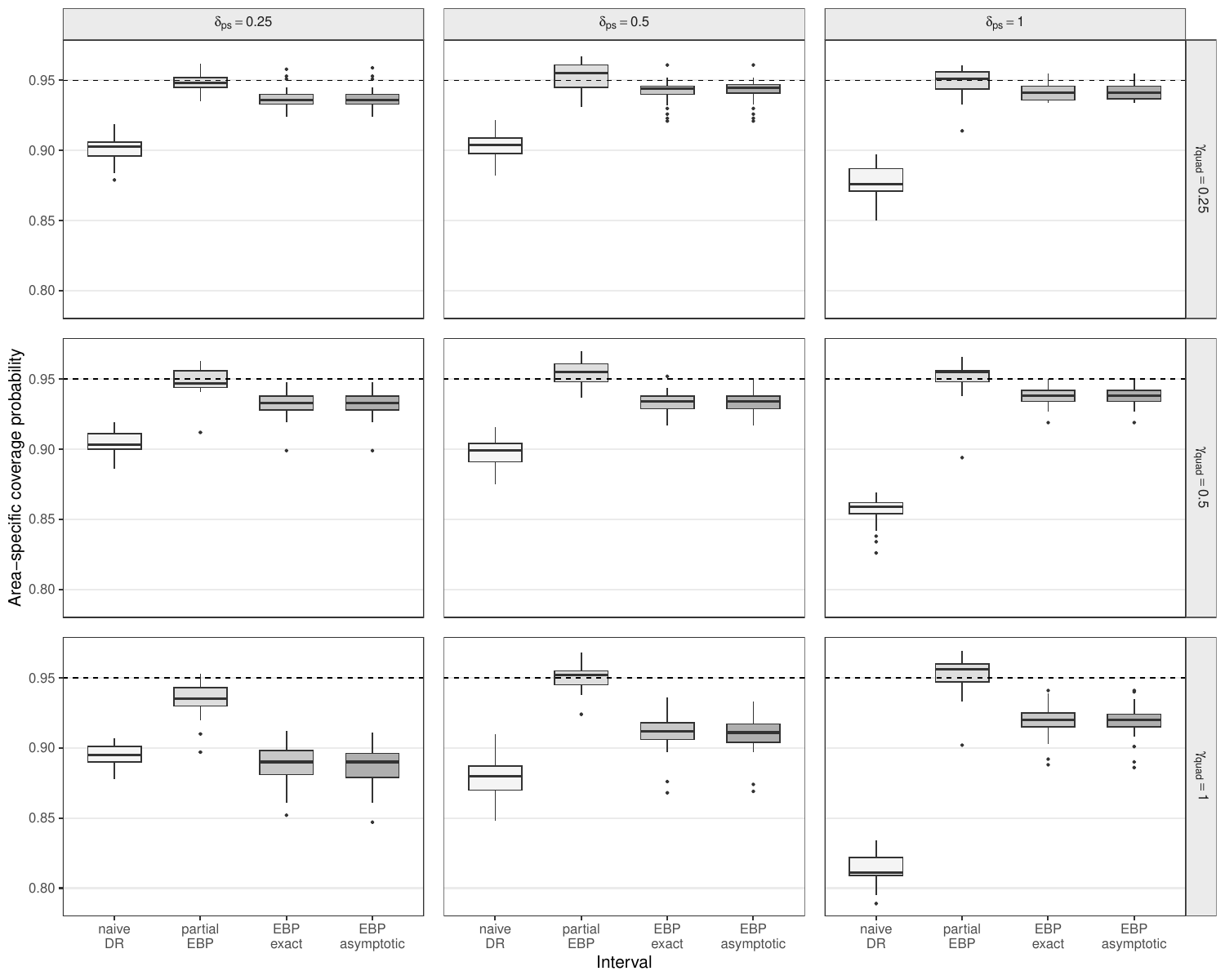}
\caption{Boxplots of Simulated Area-specific Coverage Probability of 95\% Prediction Interval: Misspecified Outcome Model and Logistic Propensity Model: $m=100$, Unequal-\(n_i\) Design (Areas with $n_{i}=20$ Only); Rows and Columns Correspond to Different Values of \(\gamma_{\mathrm{quad}}\) and \(\delta_{\mathrm{ps}}\), Respectively; EBP exact and EBP asymptotic refer to EBP-based Prediction Interval based on Exact and Approximate Asymptotic Variances, Respectively.}
\label{fig:both-misspec-variable-coverage-n20}
\end{figure}

\begin{figure}[h]
\centering
\includegraphics[width=0.92\textwidth,height=0.84\textheight,keepaspectratio]{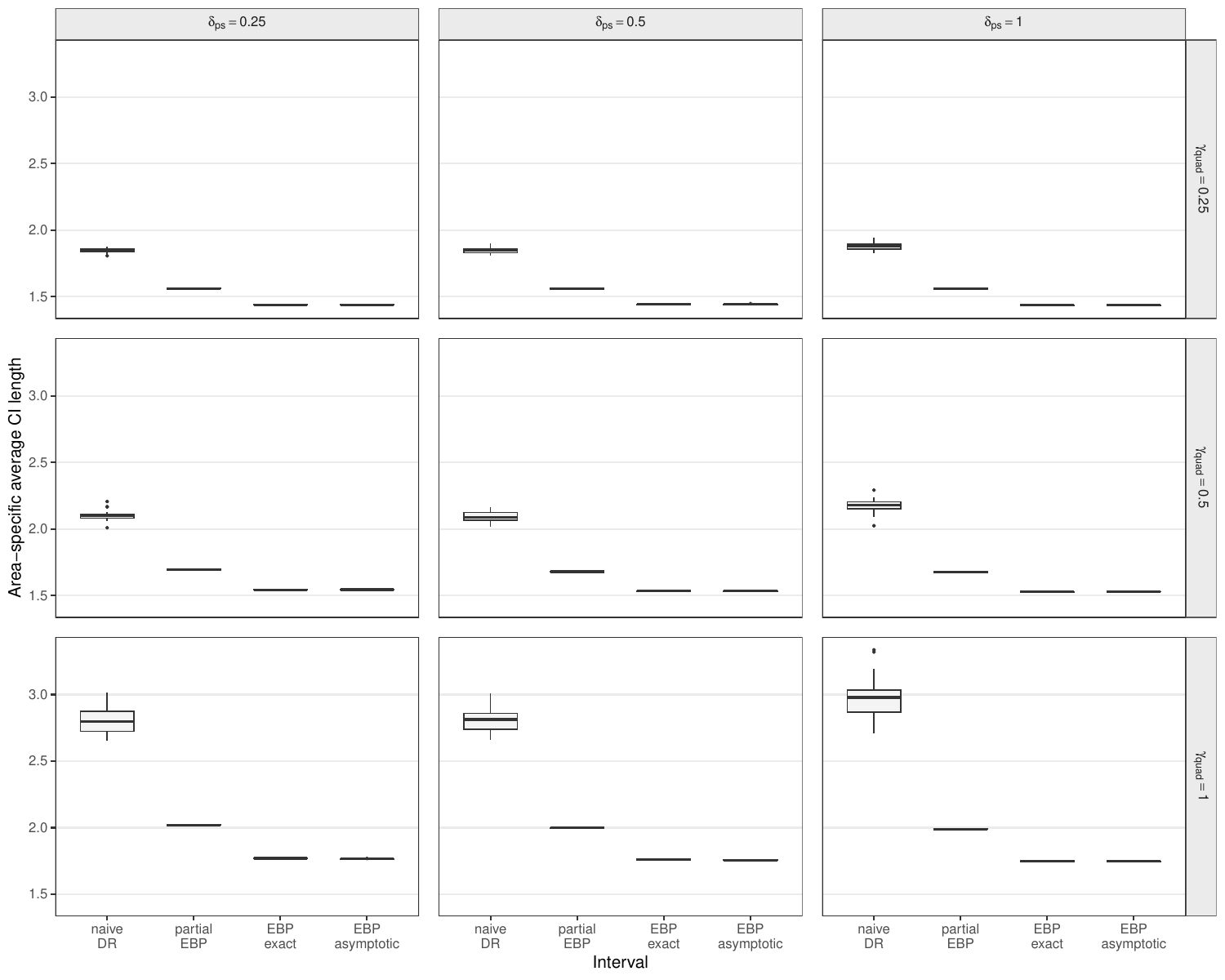}
\caption{Boxplots of Simulated Area-specific Average Length of 95\% Prediction Interval: Misspecified Outcome Model and Logistic Propensity Model: $m=100$, Unequal-\(n_i\) Design (Areas with $n_{i}=20$ Only); Rows and Columns Correspond to Different Values of \(\gamma_{\mathrm{quad}}\) and \(\delta_{\mathrm{ps}}\), Respectively; EBP exact and EBP asymptotic refer to EBP-based Prediction Interval based on Exact and Approximate Asymptotic Variances, Respectively.}
\label{fig:both-misspec-variable-length-n20}
\end{figure}

\begin{figure}[h]
\centering
\includegraphics[width=0.92\textwidth,height=0.84\textheight,keepaspectratio]{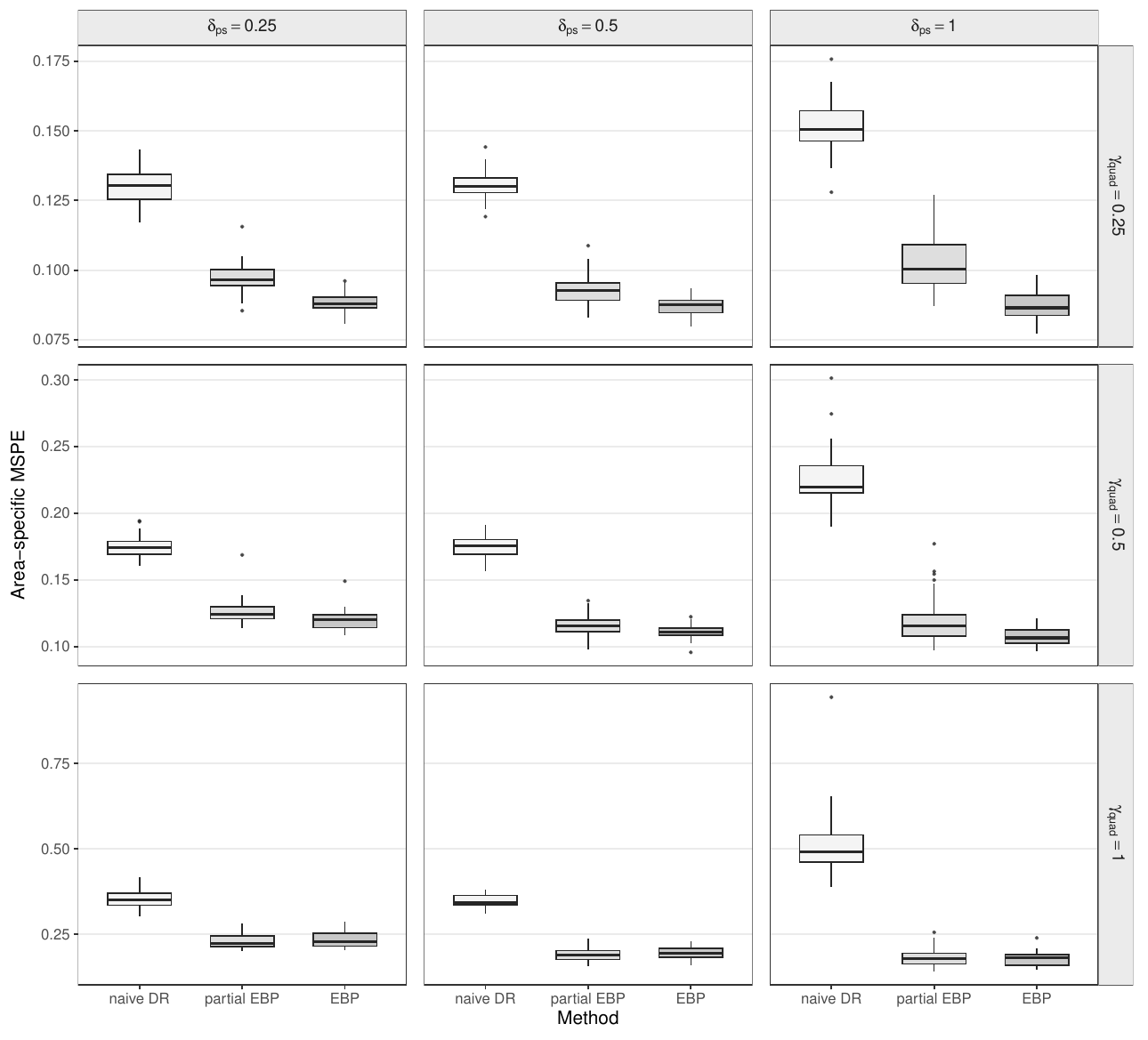}
\caption{Boxplots of Simulated Area-specific MSPE: Misspecified Outcome Model and Logistic Propensity Model: $m=100$, Unequal-\(n_i\) Design (Areas with $n_{i}=40$ Only); Rows and Columns Correspond to Different Values of \(\gamma_{\mathrm{quad}}\) and \(\delta_{\mathrm{ps}}\), Respectively.}
\label{fig:both-misspec-variable-mspe-n40}
\end{figure}

\begin{figure}[h]
\centering
\includegraphics[width=0.92\textwidth,height=0.84\textheight,keepaspectratio]{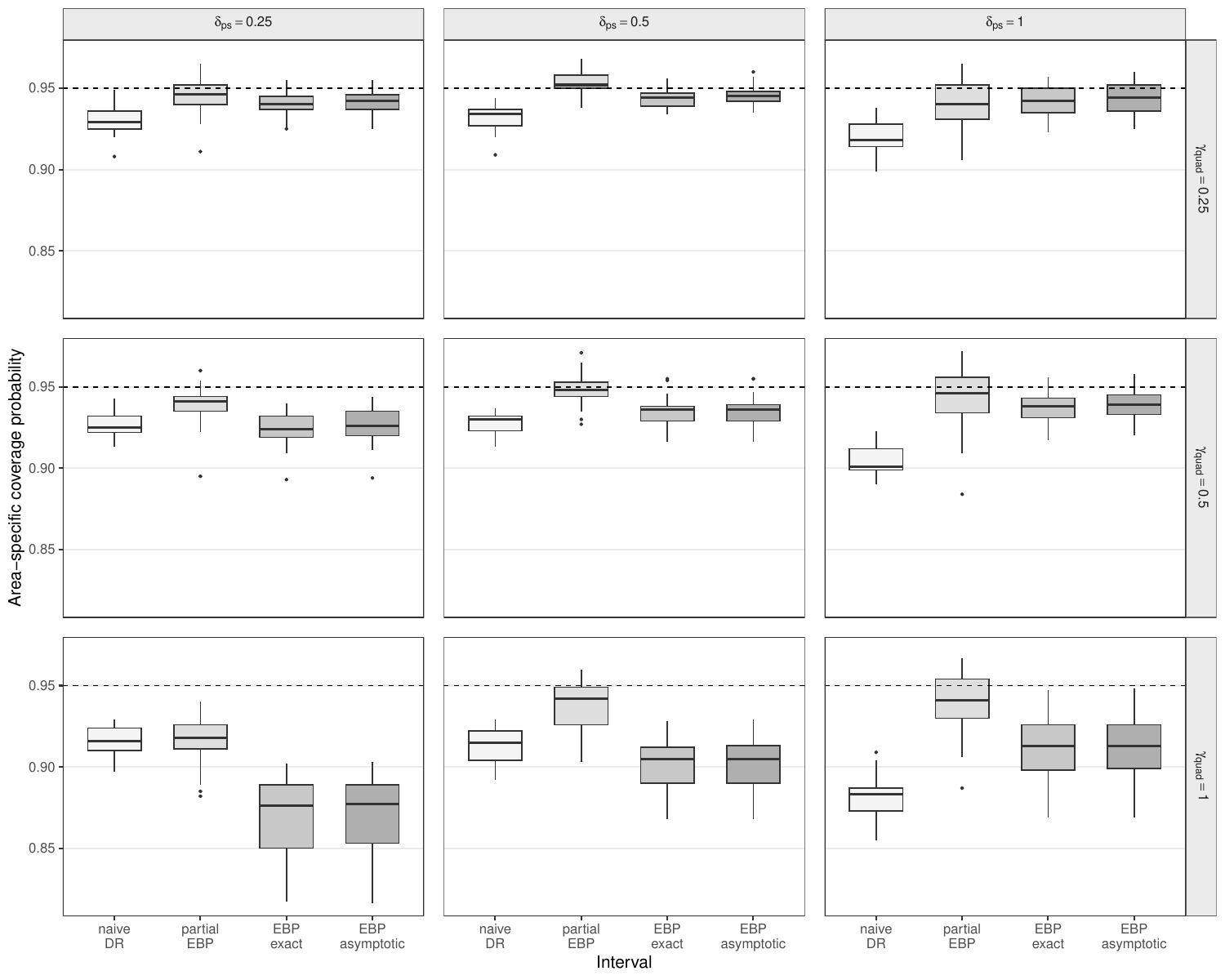}
\caption{Boxplots of Simulated Area-specific Coverage Probability of 95\% Prediction Interval: Misspecified Outcome Model and Logistic Propensity Model: $m=100$, Unequal-\(n_i\) Design (Areas with $n_{i}=40$ Only); Rows and Columns Correspond to Different Values of \(\gamma_{\mathrm{quad}}\) and \(\delta_{\mathrm{ps}}\), Respectively; EBP exact and EBP asymptotic refer to EBP-based Prediction Interval based on Exact and Approximate Asymptotic Variances, Respectively.}
\label{fig:both-misspec-variable-coverage-n40}
\end{figure}

\begin{figure}[h]
\centering
\includegraphics[width=0.92\textwidth,height=0.84\textheight,keepaspectratio]{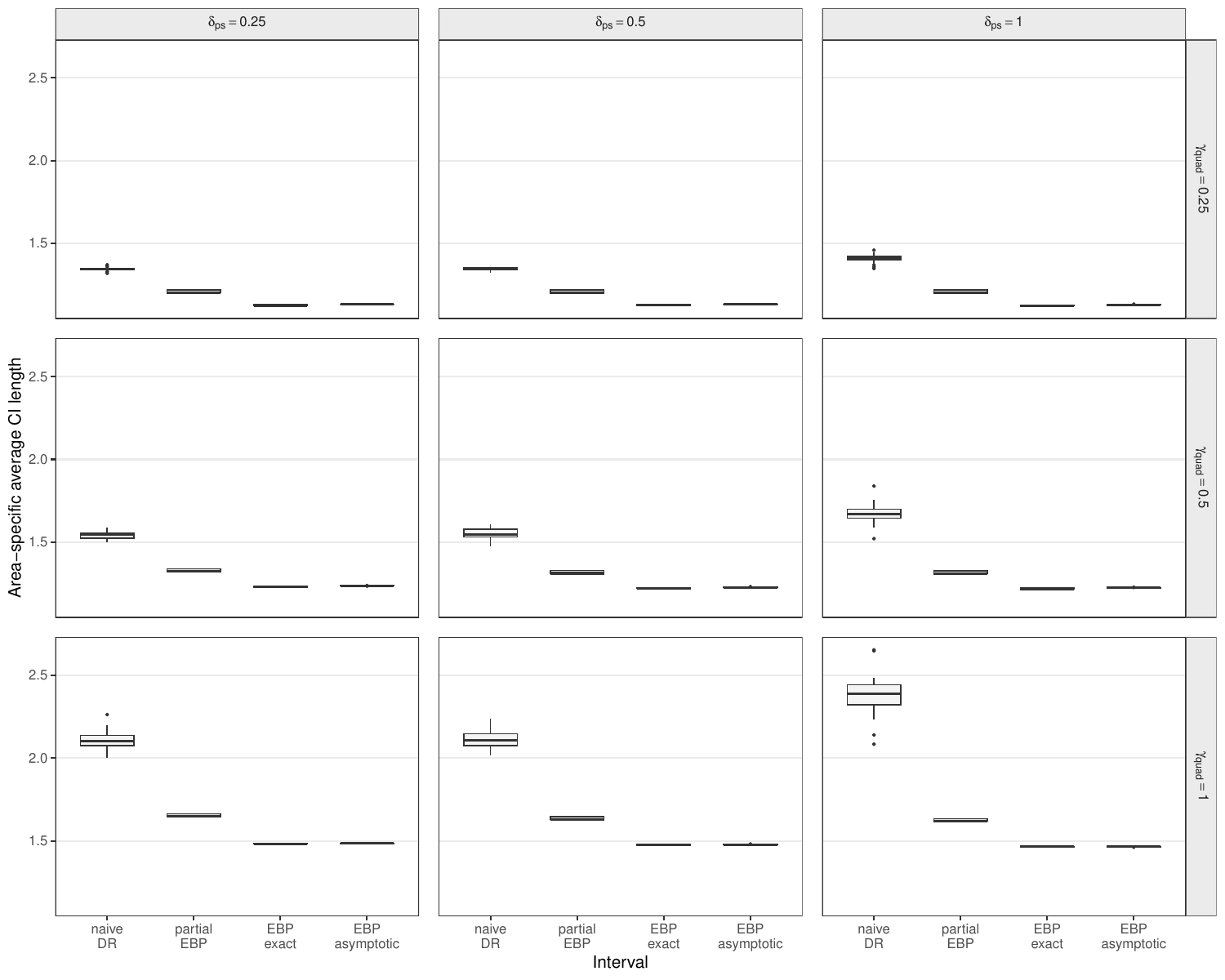}
\caption{Boxplots of Simulated Area-specific Average Length of 95\% Prediction Interval: Misspecified Outcome Model and Logistic Propensity Model: $m=100$, Unequal-\(n_i\) Design (Areas with $n_{i}=40$ Only); Rows and Columns Correspond to Different Values of \(\gamma_{\mathrm{quad}}\) and \(\delta_{\mathrm{ps}}\), Respectively; EBP exact and EBP asymptotic refer to EBP-based Prediction Interval based on Exact and Approximate Asymptotic Variances, Respectively.}
\label{fig:both-misspec-variable-length-n40}
\end{figure}

\begin{figure}[h]
\centering
\includegraphics[width=0.92\textwidth,height=0.84\textheight,keepaspectratio]{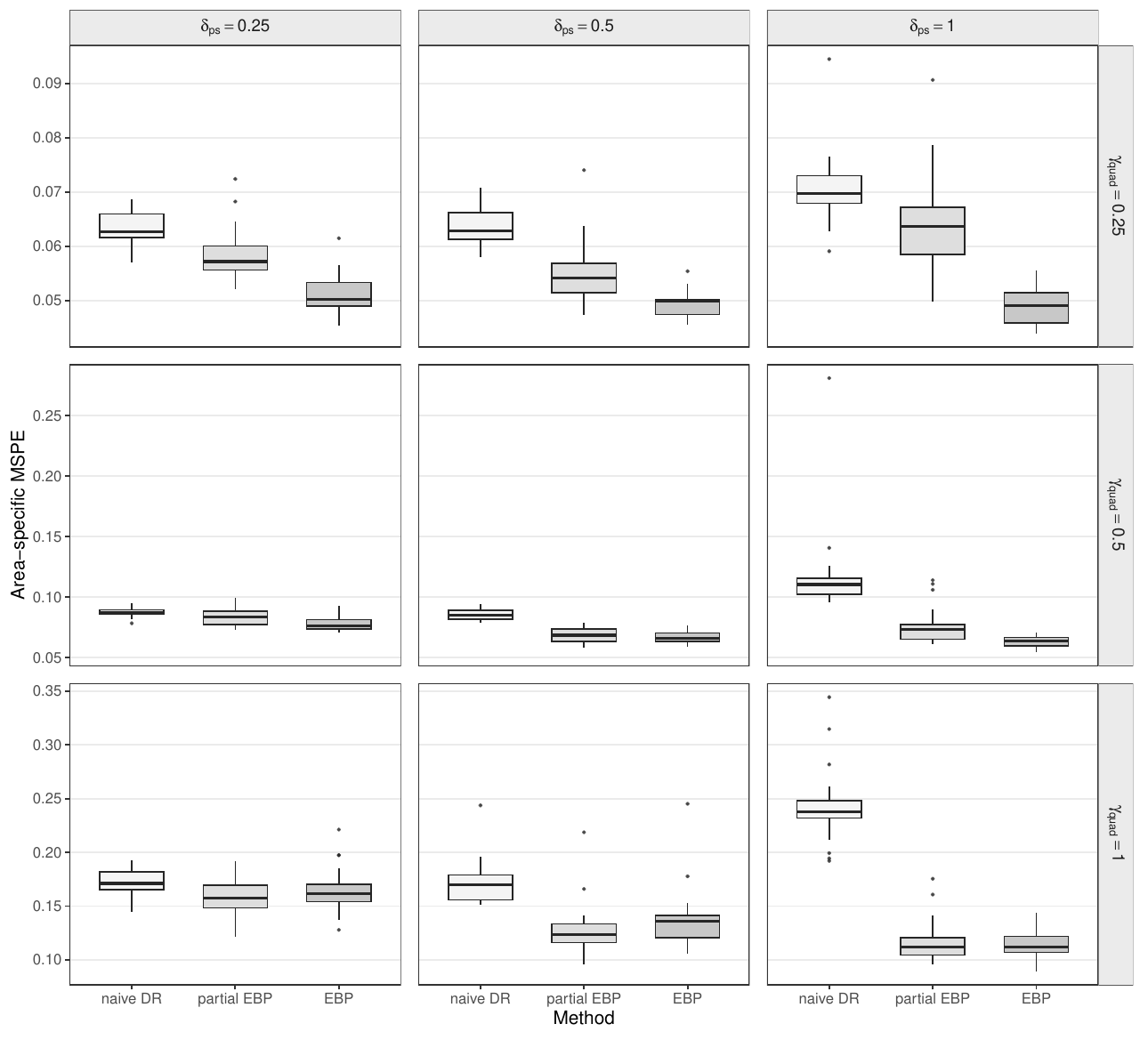}
\caption{Boxplots of Simulated Area-specific MSPE: Misspecified Outcome Model and Logistic Propensity Model: $m=100$, Unequal-\(n_i\) Design (Areas with $n_{i}=80$ Only); Rows and Columns Correspond to Different Values of \(\gamma_{\mathrm{quad}}\) and \(\delta_{\mathrm{ps}}\), Respectively.}
\label{fig:both-misspec-variable-mspe-n80}
\end{figure}

\begin{figure}[h]
\centering
\includegraphics[width=0.92\textwidth,height=0.84\textheight,keepaspectratio]{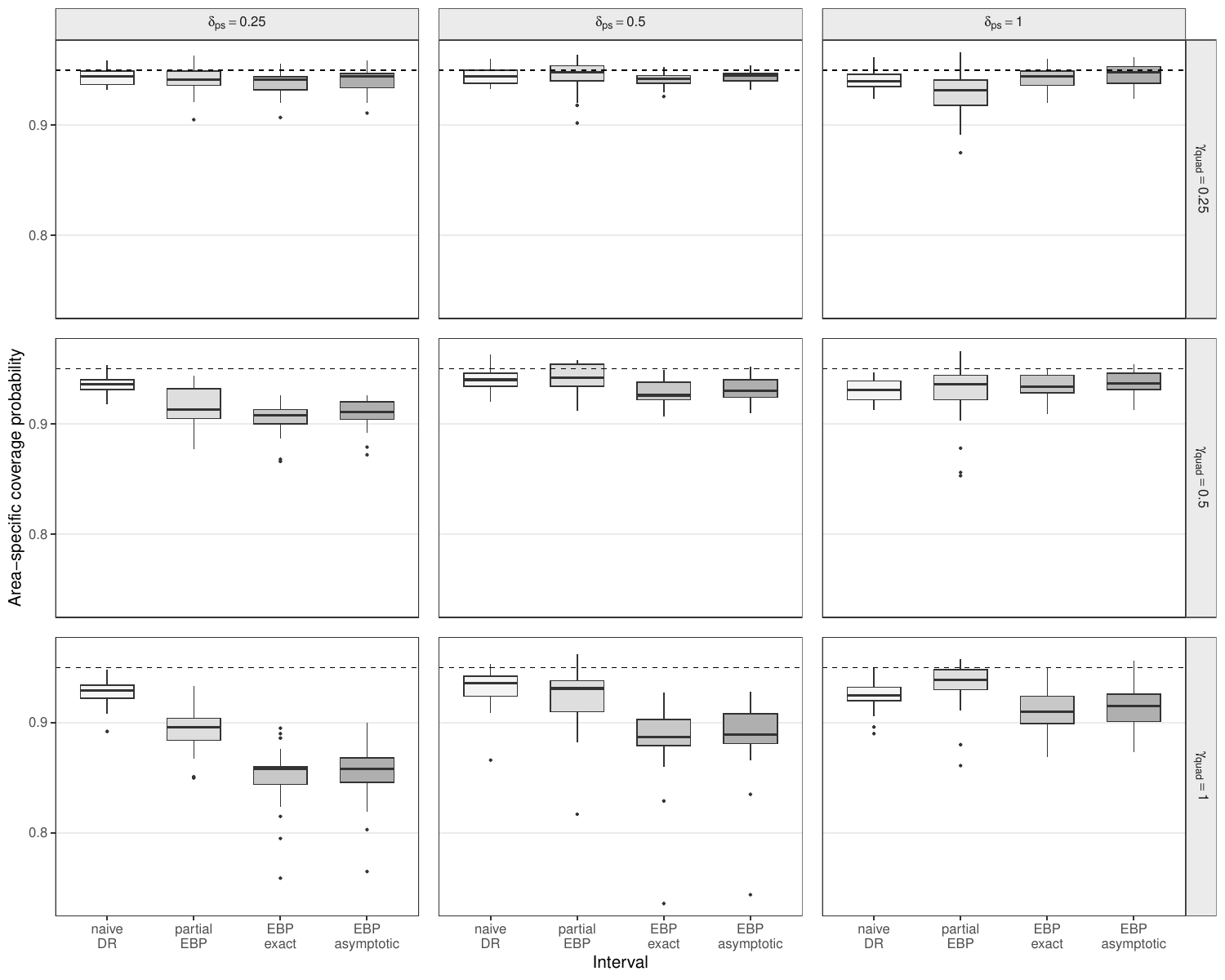}
\caption{Boxplots of Simulated Area-specific Coverage Probability of 95\% Prediction Interval: Misspecified Outcome Model and Logistic Propensity Model: $m=100$, Unequal-\(n_i\) Design (Areas with $n_{i}=80$ Only); Rows and Columns Correspond to Different Values of \(\gamma_{\mathrm{quad}}\) and \(\delta_{\mathrm{ps}}\), Respectively; EBP exact and EBP asymptotic refer to EBP-based Prediction Interval based on Exact and Approximate Asymptotic Variances, Respectively.}
\label{fig:both-misspec-variable-coverage-n80}
\end{figure}

\begin{figure}[h]
\centering
\includegraphics[width=0.92\textwidth,height=0.84\textheight,keepaspectratio]{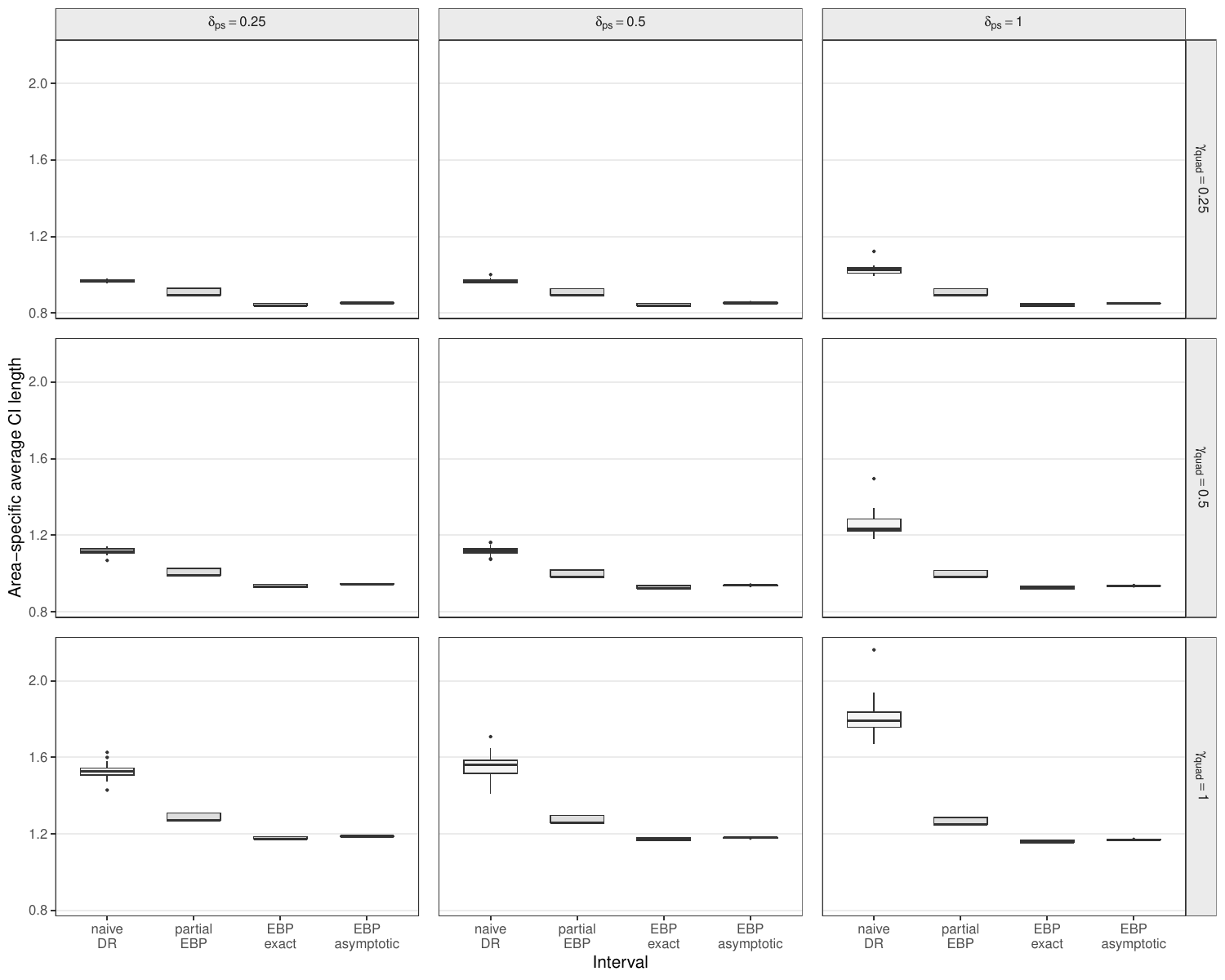}
\caption{Boxplots of Simulated Area-specific Average Length of 95\% Prediction Interval: Misspecified Outcome Model and Logistic Propensity Model: $m=100$, Unequal-\(n_i\) Design (Areas with $n_{i}=80$ Only); Rows and Columns Correspond to Different Values of \(\gamma_{\mathrm{quad}}\) and \(\delta_{\mathrm{ps}}\), Respectively; EBP exact and EBP asymptotic refer to EBP-based Prediction Interval based on Exact and Approximate Asymptotic Variances, Respectively.}
\label{fig:both-misspec-variable-length-n80}
\end{figure}

Now consider the unequal-$n_{i}$ design. Table \ref{tab:both-misspec-variable-point} presents the MSPE summary. It is seen that, for the vast majority of cases (31 out of 36), EBP performs the best. Table \ref{tab:both-misspec-variable-interval} presents the Coverage and Length summaries. In all 36 settings, an EBP-based prediction interval has the shortest average length among the methods compared, with the finite-population variance version shortest in 24 settings and the approximate asymptotic variance version shortest in the remaining 12. In terms of Coverage, partial EBP performs the best in all cases when $n_{i}=10$ or $20$; for $n_{i}=40$ or $80$, partial EBP performs the best for half of the cases, and naive DR or EBP (with approximate asymptotic variance) perform the best for the other half. Area-specific boxplots, presented in Figures \ref{fig:both-misspec-variable-mspe-n10}--\ref{fig:both-misspec-variable-length-n80}, show patterns consistent with the summaries.

\begin{table}[p]
\centering
\setlength{\belowcaptionskip}{8pt}
\caption{(Part I) Simulated MSPE: Misspecified Outcome Model and Logistic Propensity Model; $m=100$; Unequal-$n_{i}$ Design ($n_{i}=10$ and $n_{i}=20$); Separate Table for Each $n_{i}$; Bold Number Indicates the Best Performer under Respective Performance Measure.}
\label{tab:both-misspec-variable-point}
\setlength{\tabcolsep}{3pt}
\renewcommand{\arraystretch}{1.05}
\par\vspace{4pt}\textit{$n_i=10$}\par\vspace{2pt}
\begin{tabular}{cc ccc}
\hline
$\gamma_{\mathrm{quad}}$ & $\delta_{\mathrm{ps}}$ & naive DR & partial EBP & EBP \\
\hline
0.25 & 0.25 & 1.0196 & 0.2294 & \textbf{0.2157} \\
0.25 & 0.5 & 1.3902 & 0.2238 & \textbf{0.2117} \\
0.25 & 1 & 1.2996 & 0.2398 & \textbf{0.2140} \\
0.5 & 0.25 & 1.1611 & 0.2676 & \textbf{0.2519} \\
0.5 & 0.5 & 1.1712 & 0.2568 & \textbf{0.2459} \\
0.5 & 1 & 1.6301 & 0.2608 & \textbf{0.2418} \\
1 & 0.25 & 1.8410 & 0.3729 & \textbf{0.3641} \\
1 & 0.5 & 1.8936 & 0.3391 & \textbf{0.3299} \\
1 & 1 & 2.3766 & 0.3240 & \textbf{0.3153} \\
\hline
\end{tabular}
\par\vspace{4pt}\textit{$n_i=20$}\par\vspace{2pt}
\begin{tabular}{cc ccc}
\hline
$\gamma_{\mathrm{quad}}$ & $\delta_{\mathrm{ps}}$ & naive DR & partial EBP & EBP \\
\hline
0.25 & 0.25 & 0.2904 & 0.1594 & \textbf{0.1478} \\
0.25 & 0.5 & 0.2869 & 0.1525 & \textbf{0.1439} \\
0.25 & 1 & 0.3371 & 0.1594 & \textbf{0.1428} \\
0.5 & 0.25 & 0.3774 & 0.1890 & \textbf{0.1786} \\
0.5 & 0.5 & 0.3989 & 0.1770 & \textbf{0.1715} \\
0.5 & 1 & 0.4940 & 0.1838 & \textbf{0.1672} \\
1 & 0.25 & 0.7138 & 0.2999 & \textbf{0.2996} \\
1 & 0.5 & 0.7273 & \textbf{0.2577} & 0.2617 \\
1 & 1 & 1.0365 & 0.2487 & \textbf{0.2403} \\
\hline
\end{tabular}
\end{table}

\begin{table}[p]
\centering
\ContinuedFloat
\setlength{\belowcaptionskip}{8pt}
\caption{(Part II) Simulated MSPE: Misspecified Outcome Model and Logistic Propensity Model; $m=100$; Unequal-$n_{i}$ Design ($n_{i}=40$ and $n_{i}=80$); Separate Table for Each $n_{i}$; Bold Number Indicates the Best Performer under Respective Performance Measure.}
\setlength{\tabcolsep}{3pt}
\renewcommand{\arraystretch}{1.05}
\par\vspace{4pt}\textit{$n_i=40$}\par\vspace{2pt}
\begin{tabular}{cc ccc}
\hline
$\gamma_{\mathrm{quad}}$ & $\delta_{\mathrm{ps}}$ & naive DR & partial EBP & EBP \\
\hline
0.25 & 0.25 & 0.1299 & 0.0971 & \textbf{0.0887} \\
0.25 & 0.5 & 0.1304 & 0.0928 & \textbf{0.0871} \\
0.25 & 1 & 0.1509 & 0.1028 & \textbf{0.0877} \\
0.5 & 0.25 & 0.1752 & 0.1266 & \textbf{0.1201} \\
0.5 & 0.5 & 0.1749 & 0.1156 & \textbf{0.1109} \\
0.5 & 1 & 0.2271 & 0.1219 & \textbf{0.1073} \\
1 & 0.25 & 0.3504 & \textbf{0.2317} & 0.2346 \\
1 & 0.5 & 0.3466 & \textbf{0.1921} & 0.1975 \\
1 & 1 & 0.5120 & 0.1845 & \textbf{0.1791} \\
\hline
\end{tabular}
\par\vspace{4pt}\textit{$n_i=80$}\par\vspace{2pt}
\begin{tabular}{cc ccc}
\hline
$\gamma_{\mathrm{quad}}$ & $\delta_{\mathrm{ps}}$ & naive DR & partial EBP & EBP \\
\hline
0.25 & 0.25 & 0.0634 & 0.0585 & \textbf{0.0514} \\
0.25 & 0.5 & 0.0635 & 0.0554 & \textbf{0.0494} \\
0.25 & 1 & 0.0708 & 0.0640 & \textbf{0.0490} \\
0.5 & 0.25 & 0.0875 & 0.0838 & \textbf{0.0781} \\
0.5 & 0.5 & 0.0853 & 0.0690 & \textbf{0.0671} \\
0.5 & 1 & 0.1169 & 0.0763 & \textbf{0.0635} \\
1 & 0.25 & 0.1711 & \textbf{0.1575} & 0.1646 \\
1 & 0.5 & 0.1726 & \textbf{0.1285} & 0.1381 \\
1 & 1 & 0.2425 & 0.1167 & \textbf{0.1151} \\
\hline
\end{tabular}
\end{table}

\begin{table}[p]
\centering
\setlength{\belowcaptionskip}{8pt}
\caption{(Part I) Simulated Coverage Probability and Length of 95\% Prediction Interval: Misspecified Outcome Model and Logistic Propensity Model; $m=100$; Unequal-$n_{i}$ Design ($n_{i}=10$ and $n_{i}=20$); Separate Table for Each $n_{i}$; Bold Number Indicates the Best Performer under Respective Performance Measure.}
\label{tab:both-misspec-variable-interval}
\setlength{\tabcolsep}{3pt}
\renewcommand{\arraystretch}{1.05}
\par\vspace{4pt}\textit{$n_i=10$}\par\vspace{2pt}
\begin{tabular}{cccccccccc}
\hline
$\gamma_{\mathrm{quad}}$ & $\delta_{\mathrm{ps}}$ & \multicolumn{2}{c}{naive DR} & \multicolumn{2}{c}{partial EBP} & \multicolumn{2}{c}{EBP exact} & \multicolumn{2}{c}{EBP asymptotic} \\
 &  & Cov. & Len. & Cov. & Len. & Cov. & Len. & Cov. & Len. \\
\hline
0.25 & 0.25 & 0.8354 & 2.5902 & \textbf{0.9548} & 1.9317 & 0.9372 & 1.7428 & 0.9371 & \textbf{1.7414} \\
0.25 & 0.5 & 0.8276 & 2.6078 & \textbf{0.9578} & 1.9318 & 0.9400 & 1.7492 & 0.9399 & \textbf{1.7479} \\
0.25 & 1 & 0.8009 & 2.6436 & \textbf{0.9510} & 1.9303 & 0.9389 & 1.7407 & 0.9388 & \textbf{1.7393} \\
0.5 & 0.25 & 0.8388 & 2.9072 & \textbf{0.9526} & 2.0550 & 0.9300 & 1.8353 & 0.9295 & \textbf{1.8328} \\
0.5 & 0.5 & 0.8271 & 2.8808 & \textbf{0.9542} & 2.0390 & 0.9318 & 1.8244 & 0.9314 & \textbf{1.8219} \\
0.5 & 1 & 0.7835 & 2.9227 & \textbf{0.9529} & 2.0327 & 0.9334 & 1.8173 & 0.9331 & \textbf{1.8147} \\
1 & 0.25 & 0.8426 & 3.7470 & \textbf{0.9429} & 2.3349 & 0.9001 & 2.0129 & 0.8984 & \textbf{2.0047} \\
1 & 0.5 & 0.8258 & 3.7328 & \textbf{0.9494} & 2.3103 & 0.9130 & 2.0050 & 0.9115 & \textbf{1.9970} \\
1 & 1 & 0.7545 & 3.7298 & \textbf{0.9551} & 2.3033 & 0.9187 & 1.9933 & 0.9174 & \textbf{1.9855} \\
\hline
\end{tabular}
\par\vspace{4pt}\textit{$n_i=20$}\par\vspace{2pt}
\begin{tabular}{cccccccccc}
\hline
$\gamma_{\mathrm{quad}}$ & $\delta_{\mathrm{ps}}$ & \multicolumn{2}{c}{naive DR} & \multicolumn{2}{c}{partial EBP} & \multicolumn{2}{c}{EBP exact} & \multicolumn{2}{c}{EBP asymptotic} \\
 &  & Cov. & Len. & Cov. & Len. & Cov. & Len. & Cov. & Len. \\
\hline
0.25 & 0.25 & 0.9012 & 1.8466 & \textbf{0.9487} & 1.5629 & 0.9370 & \textbf{1.4390} & 0.9372 & 1.4404 \\
0.25 & 0.5 & 0.9023 & 1.8483 & \textbf{0.9536} & 1.5620 & 0.9419 & \textbf{1.4424} & 0.9421 & 1.4439 \\
0.25 & 1 & 0.8779 & 1.8800 & \textbf{0.9491} & 1.5605 & 0.9418 & \textbf{1.4359} & 0.9421 & 1.4373 \\
0.5 & 0.25 & 0.9038 & 2.1015 & \textbf{0.9487} & 1.6936 & 0.9311 & \textbf{1.5438} & 0.9310 & 1.5443 \\
0.5 & 0.5 & 0.8974 & 2.0919 & \textbf{0.9539} & 1.6804 & 0.9333 & \textbf{1.5341} & 0.9332 & 1.5346 \\
0.5 & 1 & 0.8560 & 2.1732 & \textbf{0.9507} & 1.6764 & 0.9377 & \textbf{1.5286} & 0.9377 & 1.5291 \\
1 & 0.25 & 0.8952 & 2.8033 & \textbf{0.9337} & 2.0198 & 0.8880 & 1.7688 & 0.8868 & \textbf{1.7646} \\
1 & 0.5 & 0.8802 & 2.8099 & \textbf{0.9501} & 1.9976 & 0.9106 & 1.7608 & 0.9094 & \textbf{1.7567} \\
1 & 1 & 0.8140 & 2.9780 & \textbf{0.9520} & 1.9888 & 0.9191 & 1.7493 & 0.9183 & \textbf{1.7455} \\
\hline
\end{tabular}
\end{table}

\begin{table}[p]
\centering
\ContinuedFloat
\setlength{\belowcaptionskip}{8pt}
\caption{(Part II) Simulated Coverage Probability and Length of 95\% Prediction Interval: Misspecified Outcome Model and Logistic Propensity Model; $m=100$; Unequal-$n_{i}$ Design ($n_{i}=40$ and $n_{i}=80$); Separate Table for Each $n_{i}$; Bold Number Indicates the Best Performer under Respective Performance Measure.}
\setlength{\tabcolsep}{3pt}
\renewcommand{\arraystretch}{1.05}
\par\vspace{4pt}\textit{$n_i=40$}\par\vspace{2pt}
\begin{tabular}{cccccccccc}
\hline
$\gamma_{\mathrm{quad}}$ & $\delta_{\mathrm{ps}}$ & \multicolumn{2}{c}{naive DR} & \multicolumn{2}{c}{partial EBP} & \multicolumn{2}{c}{EBP exact} & \multicolumn{2}{c}{EBP asymptotic} \\
 &  & Cov. & Len. & Cov. & Len. & Cov. & Len. & Cov. & Len. \\
\hline
0.25 & 0.25 & 0.9305 & 1.3426 & \textbf{0.9451} & 1.2072 & 0.9402 & \textbf{1.1261} & 0.9410 & 1.1314 \\
0.25 & 0.5 & 0.9320 & 1.3453 & \textbf{0.9532} & 1.2055 & 0.9438 & \textbf{1.1268} & 0.9452 & 1.1322 \\
0.25 & 1 & 0.9197 & 1.4081 & 0.9410 & 1.2055 & 0.9423 & \textbf{1.1237} & \textbf{0.9439} & 1.1291 \\
0.5 & 0.25 & 0.9265 & 1.5410 & \textbf{0.9386} & 1.3264 & 0.9242 & \textbf{1.2303} & 0.9256 & 1.2353 \\
0.5 & 0.5 & 0.9276 & 1.5487 & \textbf{0.9482} & 1.3159 & 0.9348 & \textbf{1.2221} & 0.9356 & 1.2272 \\
0.5 & 1 & 0.9047 & 1.6728 & \textbf{0.9419} & 1.3139 & 0.9364 & \textbf{1.2190} & 0.9378 & 1.2240 \\
1 & 0.25 & \textbf{0.9166} & 2.1067 & 0.9161 & 1.6518 & 0.8720 & \textbf{1.4824} & 0.8727 & 1.4839 \\
1 & 0.5 & 0.9122 & 2.1124 & \textbf{0.9371} & 1.6345 & 0.9007 & \textbf{1.4752} & 0.9010 & 1.4769 \\
1 & 1 & 0.8809 & 2.3795 & \textbf{0.9398} & 1.6241 & 0.9130 & \textbf{1.4630} & 0.9133 & 1.4649 \\
\hline
\end{tabular}
\par\vspace{4pt}\textit{$n_i=80$}\par\vspace{2pt}
\begin{tabular}{cccccccccc}
\hline
$\gamma_{\mathrm{quad}}$ & $\delta_{\mathrm{ps}}$ & \multicolumn{2}{c}{naive DR} & \multicolumn{2}{c}{partial EBP} & \multicolumn{2}{c}{EBP exact} & \multicolumn{2}{c}{EBP asymptotic} \\
 &  & Cov. & Len. & Cov. & Len. & Cov. & Len. & Cov. & Len. \\
\hline
0.25 & 0.25 & \textbf{0.9436} & 0.9673 & 0.9412 & 0.9055 & 0.9378 & \textbf{0.8426} & 0.9412 & 0.8527 \\
0.25 & 0.5 & \textbf{0.9445} & 0.9696 & 0.9440 & 0.9041 & 0.9411 & \textbf{0.8426} & 0.9439 & 0.8526 \\
0.25 & 1 & 0.9404 & 1.0263 & 0.9274 & 0.9043 & 0.9425 & \textbf{0.8407} & \textbf{0.9450} & 0.8507 \\
0.5 & 0.25 & \textbf{0.9366} & 1.1142 & 0.9158 & 1.0019 & 0.9049 & \textbf{0.9334} & 0.9090 & 0.9439 \\
0.5 & 0.5 & 0.9401 & 1.1173 & \textbf{0.9411} & 0.9940 & 0.9284 & \textbf{0.9274} & 0.9315 & 0.9378 \\
0.5 & 1 & 0.9311 & 1.2572 & 0.9277 & 0.9930 & 0.9340 & \textbf{0.9246} & \textbf{0.9375} & 0.9349 \\
1 & 0.25 & \textbf{0.9270} & 1.5308 & 0.8933 & 1.2839 & 0.8510 & \textbf{1.1776} & 0.8547 & 1.1872 \\
1 & 0.5 & \textbf{0.9312} & 1.5558 & 0.9231 & 1.2704 & 0.8829 & \textbf{1.1702} & 0.8860 & 1.1799 \\
1 & 1 & 0.9233 & 1.8138 & \textbf{0.9334} & 1.2607 & 0.9094 & \textbf{1.1592} & 0.9115 & 1.1690 \\
\hline
\end{tabular}
\end{table}





{\bf Summary.} Despite that both the outcome model and the propensity model are misspecified, the overall pattern regarding the relative performance of the three comparing methods remains largely the same. EBP has the smallest MSPE in most settings and retains shorter prediction intervals, while partial EBP generally has Coverage closer to the nominal level. Partial EBP has smaller MSPE in some settings with stronger outcome misspecification. On the other hand, as the degree of outcome model misspecification increases, the advantage in MSPE of EBP over the comparing methods, especially partial EBP, becomes less significant; likewise, the advantage of partial EBP in Coverage over the comparing methods becomes less obvious. The performance of EBP does not seem to be affected by the degree of misspecification of the propensity model; the increase of the degree of propensity model misspecification is seen to worsen the performance of naive DR and partial EBP, especially the former.
\subsection{Partial EBP MSE sensitivity analyses}
\hspace{4mm}
In this subsection, we compare the plug-in SEs with the Ranjbar-style analytical and parametric-bootstrap SEs as well as the inferential performance of the associated prediction intervals (see the third paragraph of Section 5 of SACI) in simulation studies. Here, we consider a combination of scenarios involving {\it Case I} ($\gamma_{\rm quad}=\delta_{\rm ps}=0$), {\it Case II} ($\gamma_{\rm quad}=0.5, 1.0$, $\delta_{\rm ps}=0$), and {\it Case IV} ($\gamma_{\rm quad}=\delta_{\rm ps}=0.5, 1.0$).
\subsubsection{Plug-in vs Ranjbar-style}
\hspace{4mm}
As mentioned, the Ranjbar-style MSE estimator is obtained by adding a bias-correction term to the plug-in MSE estimator, derived via the Taylor series expansion.

\begin{figure}[ht]
\centering
\includegraphics[width=0.85\textwidth]{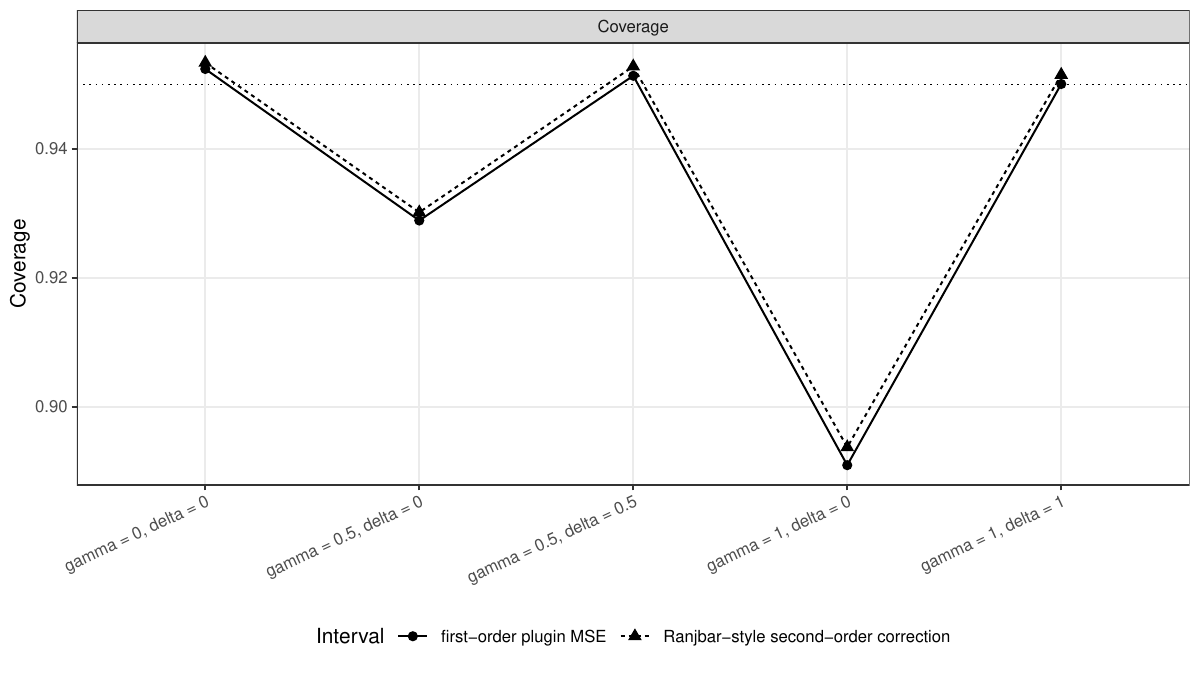}
\caption{\footnotesize Partial EBP Prediction Interval Comparison: Coverage Probability; Plug-in vs Ranjbar-style}
\label{fig:final-partial-sensitivity-coverage}
\end{figure}

\begin{figure}[ht]
\centering
\includegraphics[width=0.85\textwidth]{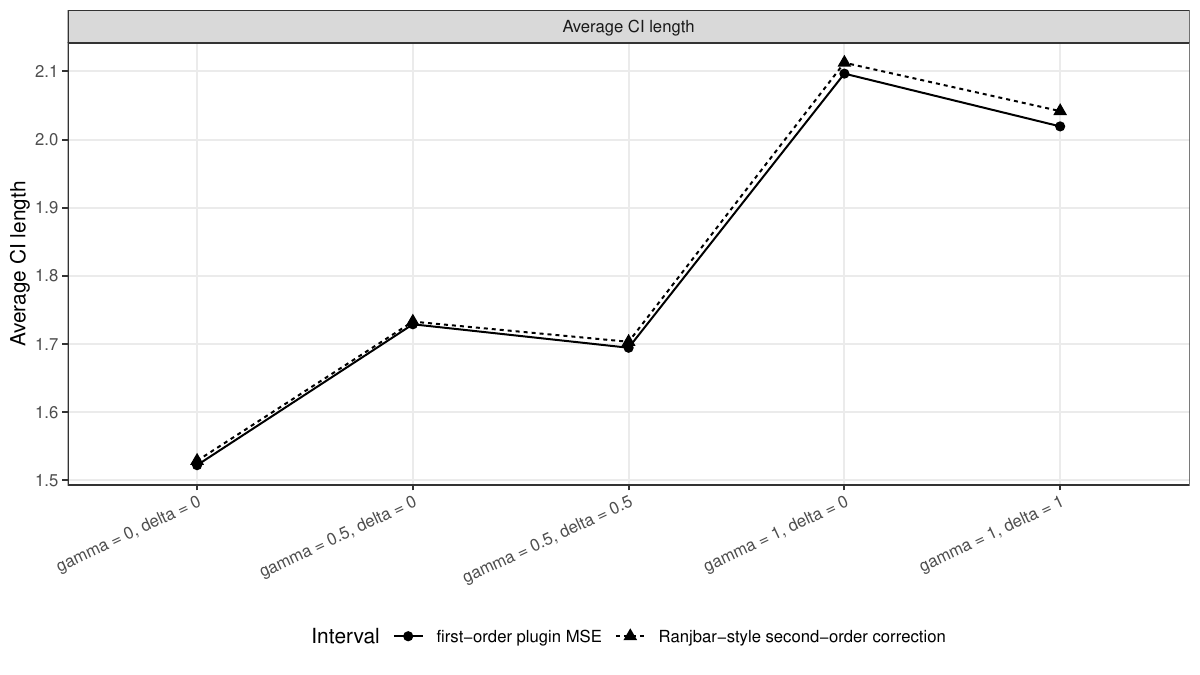}
\caption{\footnotesize Partial EBP Prediction Intervals Comparison: Average Length; Plug-in vs Ranjbar-style}
\label{fig:final-partial-sensitivity-length}
\end{figure}

Table~\ref{tab:final-partial-ebp-sensitivity} presents the simulated average (over the small areas) coverage probability and length of the prediction interval based on the plug-in and Ranjbar-style SEs. It can be seen that the results are very close. Coverage is closer to the nominal 0.95 for plug-in in three settings and for Ranjbar-style in the other two. The plug-in intervals are slightly shorter in all five settings. Some associated plots are presented in Figures \ref{fig:final-partial-sensitivity-coverage} and \ref{fig:final-partial-sensitivity-length}. The symbols and lines are barely distinguishable.

\begin{table}[ht]
\centering
\setlength{\belowcaptionskip}{8pt}
\caption{\small Partial EBP SE and 95\% Prediction Interval: \(m=100\), \(n_i=20\); Confidence Level $=95\%$; Ranjbar---Ranjbar-style, Cov---Coverage, Len---Length; $\Delta$ Cov---Difference in Cov (Higher$-$Lower); $\Delta$ Len---Difference in Len (Higher$-$Lower); Bold Numbers Indicate Coverage Closer to 0.95 and Shorter Length.}
\label{tab:final-partial-ebp-sensitivity}
\setlength{\tabcolsep}{3pt}
\renewcommand{\arraystretch}{1.08}
\begin{tabular}{cc cccccc}
\hline
$\gamma_{\mathrm{quad}}$ & $\delta_{\mathrm{ps}}$ & Plug-in Cov & Plug-in Len & Ranjbar Cov & Ranjbar Len & $\Delta$ Cov & $\Delta$ Len \\
\hline
0.0 & 0.0 & \textbf{0.9524} & \textbf{1.5224} & 0.9534 & 1.5288 & 0.0010 & 0.0064 \\
0.5 & 0.0 & 0.9289 & \textbf{1.7290} & \textbf{0.9302} & 1.7329 & 0.0013 & 0.0039 \\
0.5 & 0.5 & \textbf{0.9514} & \textbf{1.6946} & 0.9528 & 1.7034 & 0.0014 & 0.0088 \\
1.0 & 0.0 & 0.8910 & \textbf{2.0967} & \textbf{0.8938} & 2.1126 & 0.0028 & 0.0160 \\
1.0 & 1.0 & \textbf{0.9501} & \textbf{2.0193} & 0.9515 & 2.0417 & 0.0014 & 0.0224 \\
\hline
\end{tabular}
\end{table}

\subsubsection{Plug-in vs bootstrap}
\hspace{4mm}
As described in Ranjbar {\it et al.} (2023; supplementary material), the bootstrap conditions on the generated finite-population sizes \(N_i\), the covariates \(x_{ij}\), the treatment assignments \(W_{ij}\), and the sampled memberships. Bootstrap outcomes are generated under the fitted outcome LMM. The outcome model is refit in each bootstrap replication,
and the partial EBP estimator is recomputed. The bootstrap MSE estimator is then obtained as
\[
  \widehat{\mathrm{MSE}}_{\mathrm{boot},i}
  =
  \frac{1}{B_{\mathrm{boot}}}
  \sum_{b=1}^{B_{\mathrm{boot}}}
  \{\widehat d_i^{*(b)}-d_i^{*(b)}\}^2.
\]
The propensity scores are treated as fixed because the bootstrap conditions on \(X\) and \(W\).

\begin{figure}[ht]
\centering
\includegraphics[width=0.85\textwidth]{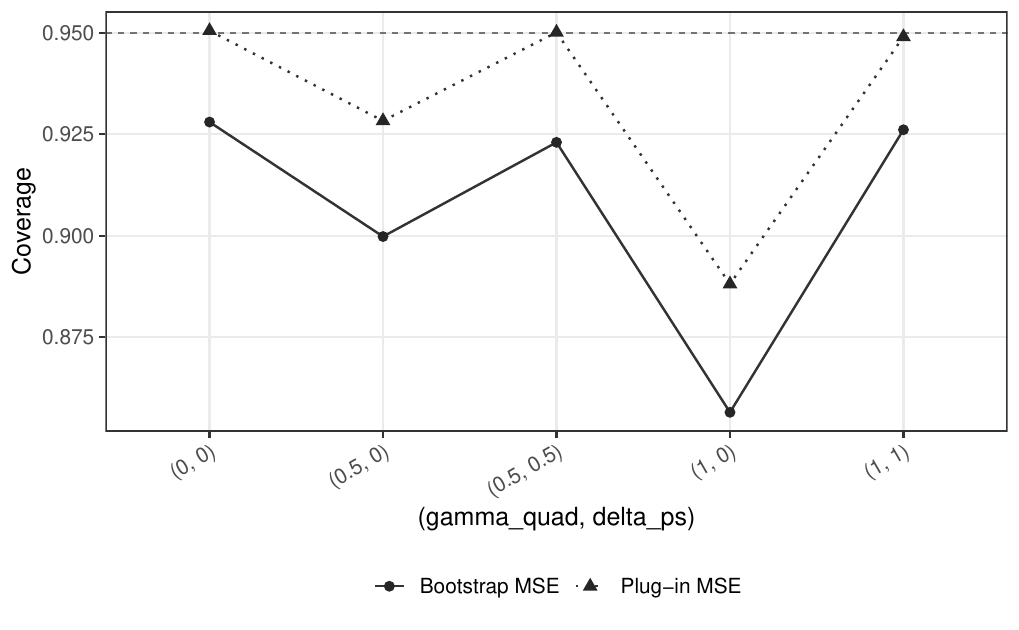}
\caption{\footnotesize Partial EBP Prediction Interval Comparison: Coverage Probability; Plug-in vs Bootstrap}
\label{fig:final-partial-bootstrap-coverage}
\end{figure}

\begin{figure}[ht]
\centering
\includegraphics[width=0.85\textwidth]{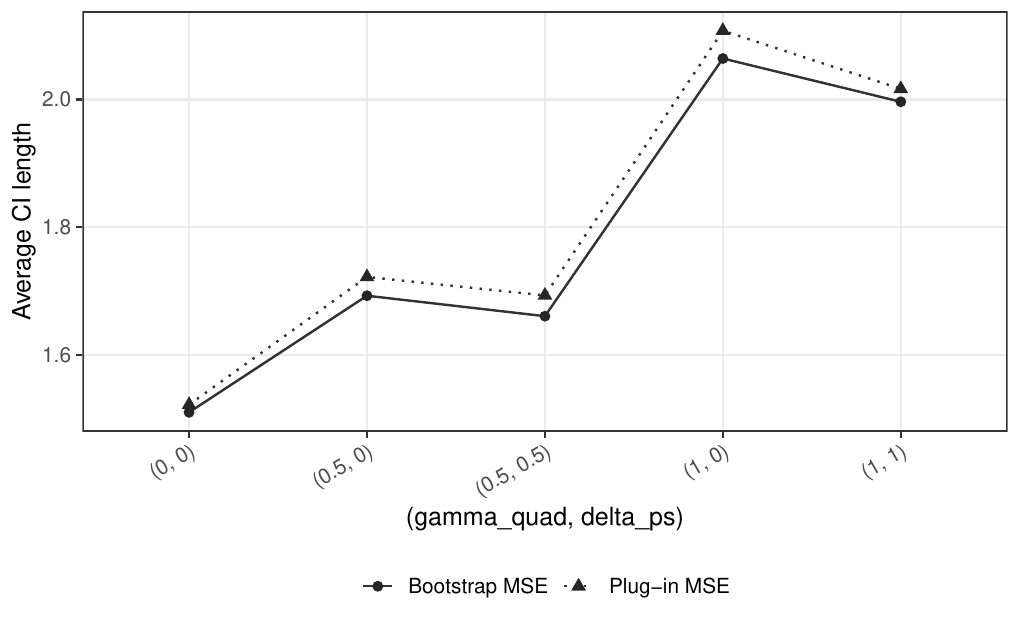}
\caption{\footnotesize Partial EBP Prediction Intervals Comparison: Average Length; Plug-in vs Bootstrap}
\label{fig:final-partial-bootstrap-length}
\end{figure}

Table \ref{tab:final-partial-ebp-bootstrap-sensitivity} presents the simulated average (over the small areas) plug-in and bootstrap SEs as well as the true (simulated) standard deviation (SD), which the SEs intend to estimate. Also presented are the average coverage probability and length of prediction interval based on the plug-in and bootstrap SEs. The average bootstrap SEs are within approximately 3\% of the plug-in SEs and are closer to the empirical SD in these settings. This agreement of averages does not establish the accuracy of individual area-specific MSE estimates. The bootstrap intervals are slightly shorter, but their coverage is lower by approximately 2.2--3.2 percentage points; plug-in coverage is closer to 0.95 in all five settings. With only $B_{\mathrm{boot}}=10$ replications per dataset, variability in the bootstrap MSE approximation may also affect calibration. Shorter intervals alone should therefore not be interpreted as improved inference. Some associated plots are presented in Figures \ref{fig:final-partial-bootstrap-coverage} and \ref{fig:final-partial-bootstrap-length}.

\begin{table}[ht]
\centering
\setlength{\belowcaptionskip}{8pt}
\caption{\small Partial EBP SE and 95\% Prediction Interval: \(m=100\), \(n_i=20\), and \(B_{\mathrm{boot}}=10\); Confidence Level $=95\%$; Boots---Bootstrap, SD---Simulated Standard Deviation; Cov---Coverage, Len---Length; Bold Numbers Indicate Average SE Closer to SD, Coverage Closer to 0.95, and Shorter Length.}
\label{tab:final-partial-ebp-bootstrap-sensitivity}
\small
\setlength{\tabcolsep}{3pt}
\renewcommand{\arraystretch}{1.08}
\begin{tabular}{cc ccccccc}
\hline
\(\gamma_{\mathrm{quad}}\) & \(\delta_{\mathrm{ps}}\) & Plug-in SE & Boots SE & SD & Plug-in Cov & Boots Cov & Plug-in Len & Boots Len \\
\hline
0.0 & 0.0 & 0.3884 & {\bf 0.3853} & 0.3864 & {\bf 0.9505} & 0.9280 & 1.5224 & {\bf 1.5105} \\
0.5 & 0.0 & 0.4394 & {\bf 0.4319} & 0.4296 & {\bf 0.9283} & 0.8998 & 1.7223 & {\bf 1.6929} \\
0.5 & 0.5 & 0.4320 & {\bf 0.4237} & 0.4198 & {\bf 0.9501} & 0.9230 & 1.6934 & {\bf 1.6609} \\
1.0 & 0.0 & 0.5376 & {\bf 0.5265} & 0.5152 & {\bf 0.8881} & 0.8565 & 2.1075 & {\bf 2.0640} \\
1.0 & 1.0 & 0.5144 & {\bf 0.5093} & 0.4943 & {\bf 0.9490} & 0.9261 & 2.0165 & {\bf 1.9963} \\
\hline
\end{tabular}
\end{table}

\subsection{{\it Case II, III, IV}, unknown $N_{i}$ setting}
\hspace{4mm}
To investigate the performance of asymptotic EBP and naive DR under outcome-model and/or propensity-model misspecification, we consider a factorial experiment under the unknown-\(N_i\)
setting. The outcome and propensity perturbations follow the same construction as in the known-\(N_i\) and covariate subpopulation setting. Specifically, with area-level covariate \(x_i\), define \(q_i=x_i^2-1\) and \(r_i=x_i^3-3x_i\). The true outcome model is
\[
  Y_{ij}(a) = x_i'(\beta_0+a\beta_1)+\alpha_{i0}+a\alpha_{i1}
  +\gamma_{\mathrm{quad}}q_i+\epsilon_{ij}(a),
  \qquad a=0,1,
\]
whereas the fitted outcome model omits \(q_i\). The true propensity model is
\[
  \operatorname{logit}\{e_i\}=\eta_0+0.5x_i+0.25q_i
  +\delta_{\mathrm{ps}}r_i,
\]
where \(\eta_0\) is calibrated so that the average true treatment probability is approximately 0.5. The fitted pooled propensity model is \(W\sim x+q\), and therefore omits \(r_i\) when \(\delta_{\mathrm{ps}}>0\). Thus, the factorial grid is
\(\gamma_{\mathrm{quad}},\delta_{\mathrm{ps}}\in\{0,0.25,0.5,1\}\).

\begin{figure}[ht]
\centering
\includegraphics[width=0.86\textwidth]{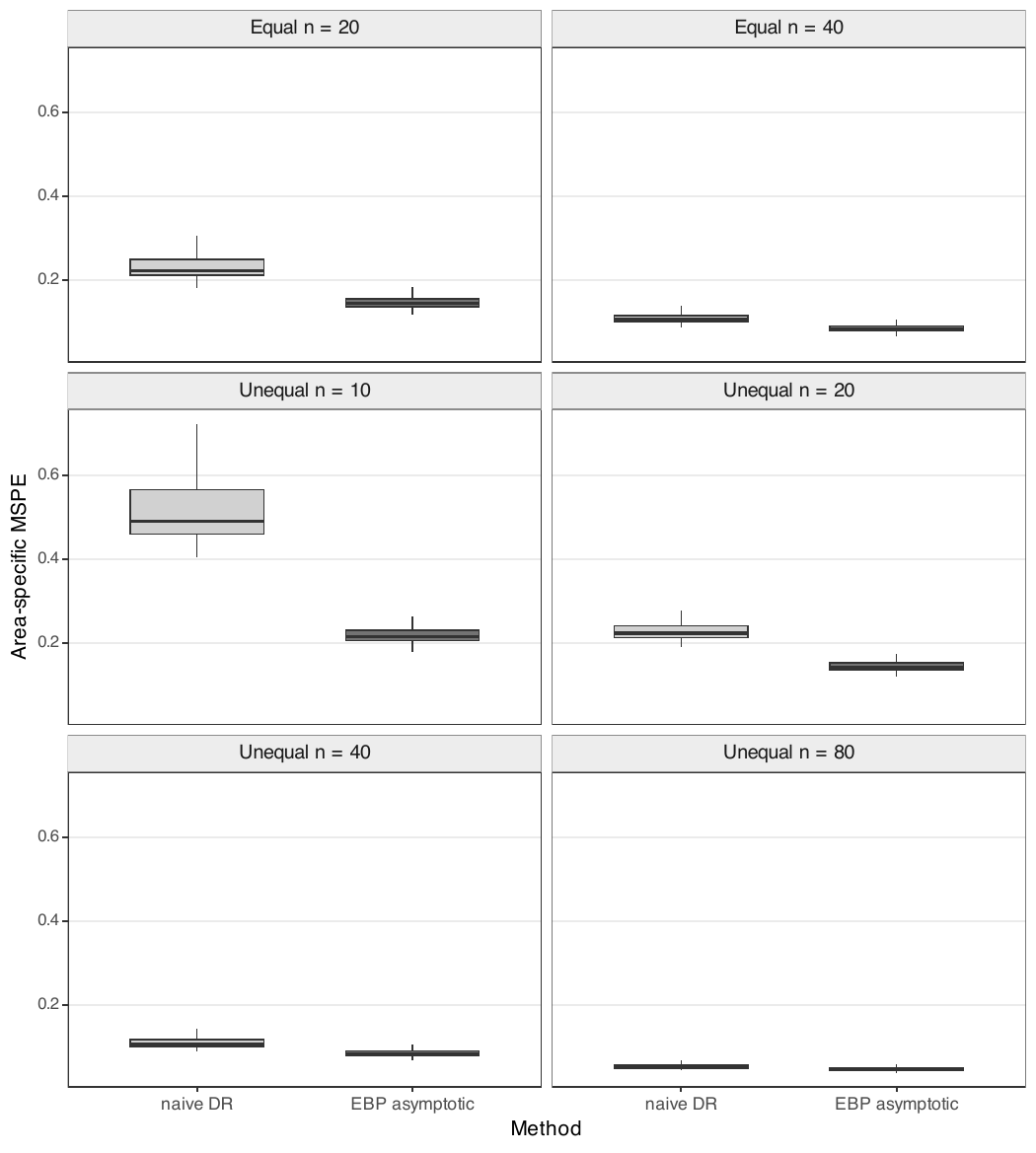}
\caption{Boxplots of Area-specific MSPE: Unknown-\(N_i\) Factorial Simulation Study; Boxplots Based on Full Area-level Summaries across the Factorial Grid, Displayed on Common y-axis Scale within the Figure; Extreme Outlier Removed for Better Visibility.}
\label{fig:unknown-N-factorial-mspe-boxplot}
\end{figure}

\begin{figure}[ht]
\centering
\includegraphics[width=0.86\textwidth]{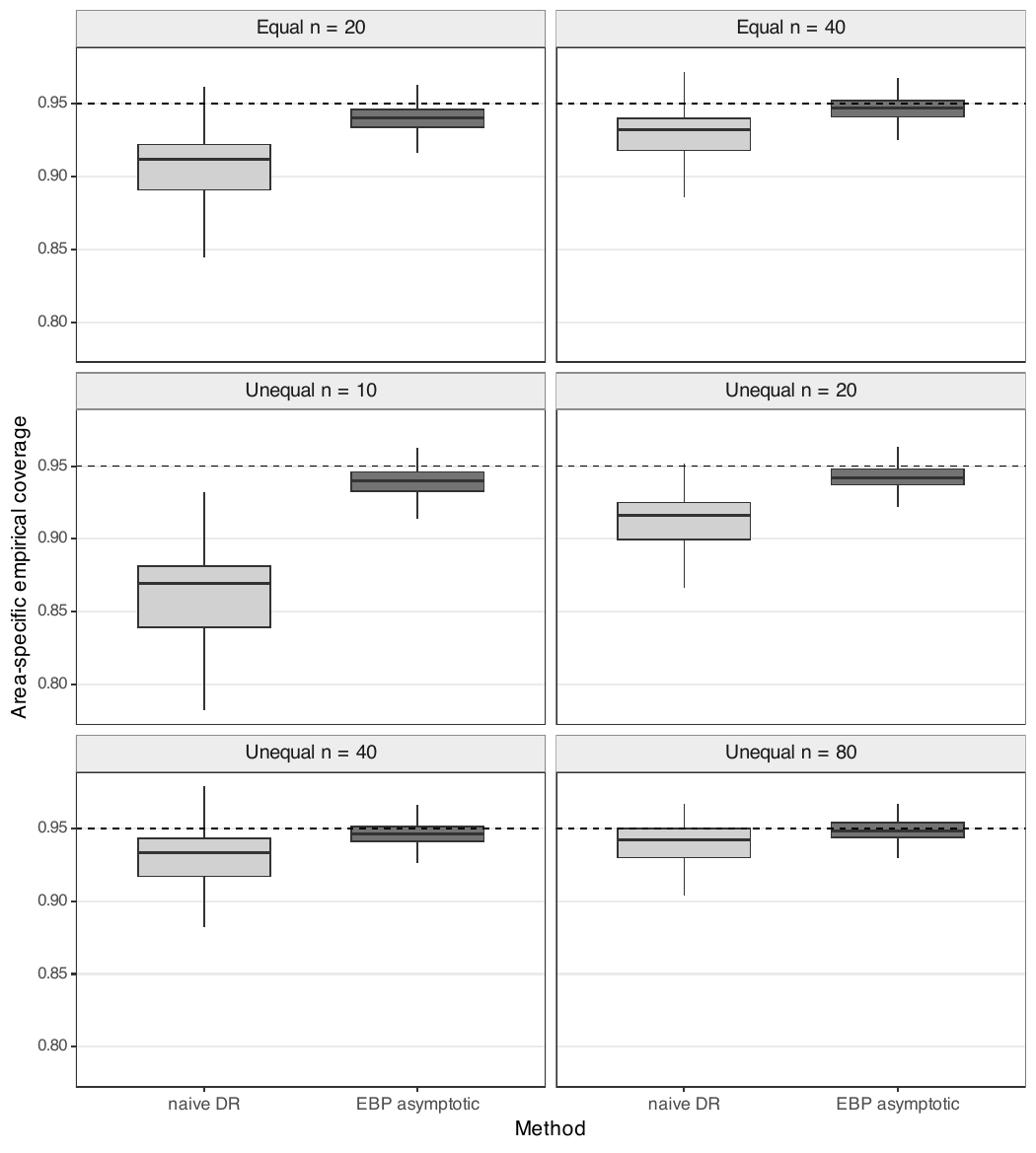}
\caption{Boxplots of Area-specific Coverage Probability of 95\% Prediction Interval: Unknown-\(N_i\) Factorial Simulation Study; Boxplots Based on Full Area-level Summaries across the Factorial Grid, Displayed on Common y-axis Scale within the Figure; Extreme Outlier Removed for Better Visibility.}
\label{fig:unknown-N-factorial-coverage-boxplot}
\end{figure}

\begin{figure}[ht]
\centering
\includegraphics[width=0.86\textwidth]{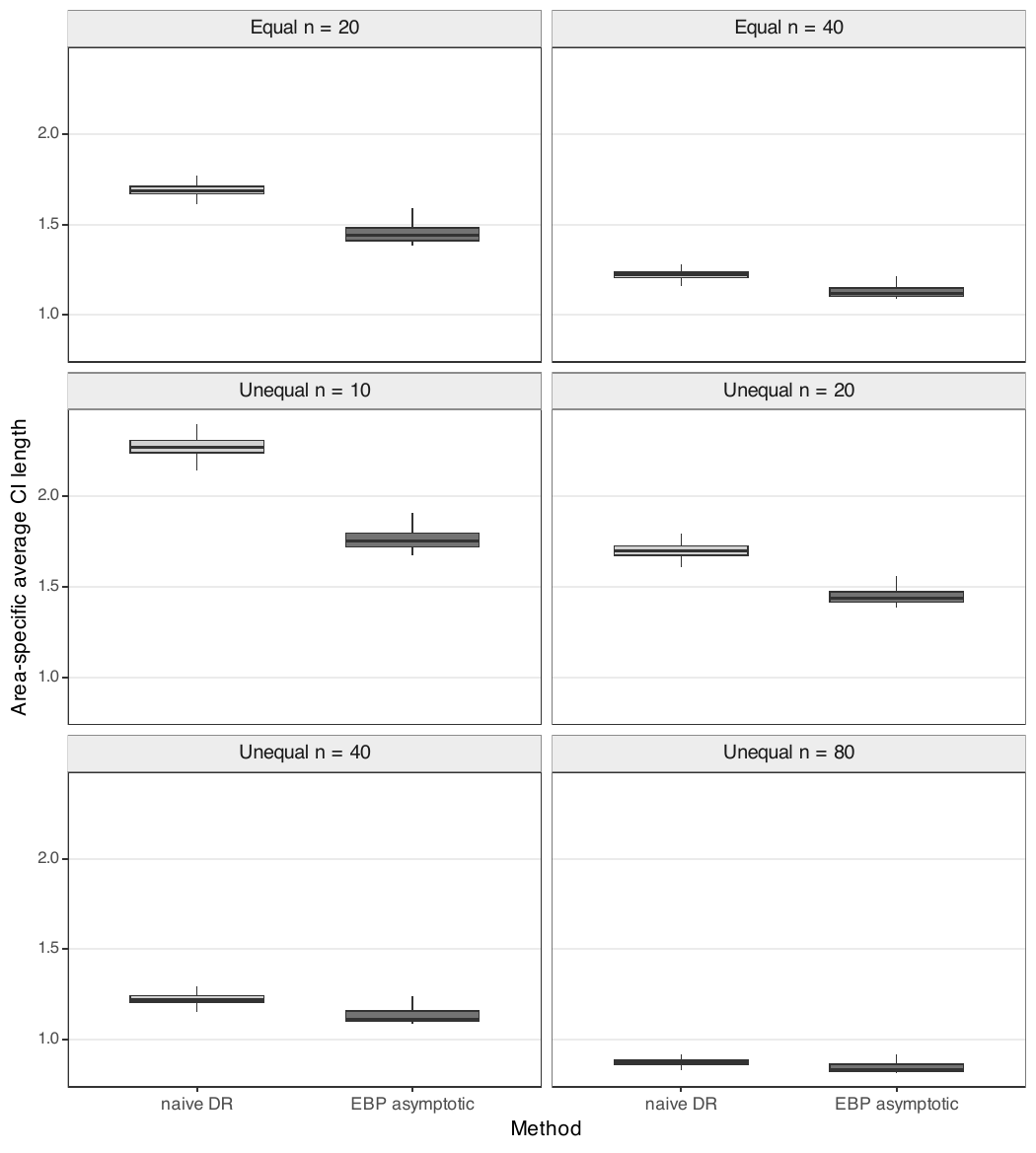}
\caption{Boxplots of Area-specific Average Length of 95\% Prediction Interval: Unknown-\(N_i\) Factorial Simulation Study; Boxplots Based on Full Area-level Summaries across the Factorial Grid, Displayed on Common y-axis Scale within the Figure; Extreme Outlier Removed for Better Visibility.}
\label{fig:unknown-N-factorial-ci-length-boxplot}
\end{figure}

Tables~\ref{tab:unknown-N-factorial-equal-n20}--\ref{tab:unknown-N-factorial-variable-n80} summarize the equal- and unequal-\(n_i\) factorial results. The performance measures, MSPE, Coverage and Length, with the corresponding boldface rules, are the same as in the known-$N_{i}$ and covariate subpopulation setting. For the equal-$n_{i}$ design, each table corresponds to a different sample-size. For the unequal-\(n_i\) factorial design, Table~\ref{tab:unknown-N-factorial-variable-overall} gives an
overall summary averaged over all 100 small areas, pooling the four sample-size groups \(n_i\in\{10,20,40,80\}\); the performance summaries for different sample-size groups are presented in Tables \ref{tab:unknown-N-factorial-variable-n10}--\ref{tab:unknown-N-factorial-variable-n80}. It is seen that, in all cases of equal or unequal-$n_{i}$ designs, asymptotic EBP has superior performance over naive DR in MSPE and Coverage. As for Length, asymptotic EBP has superior performance over naive DR in all cases with smaller $n_{i}$ (i.e., $n_{i}\leq 20$), and most cases with larger $n_{i}$ (i.e., $n_{i}\geq 40$). In the overall summary under unequal-$n_{i}$ design, asymptotic EBP outperforms naive DR in all cases and under all performance measures, MSPE, Coverage and Length. Area-specific boxplots, presented in Figures \ref{fig:unknown-N-factorial-mspe-boxplot}--\ref{fig:unknown-N-factorial-ci-length-boxplot}, show patterns consistent with the summary statistics.

\begin{table}[ht]
\centering
\caption{Unknown-$N_i$ Factorial Simulation Results: Equal-$n_i$ Design with $n_i=20$}
\label{tab:unknown-N-factorial-equal-n20}
\setlength{\tabcolsep}{4.4pt}
\renewcommand{\arraystretch}{1.05}
\begin{tabular}{cccccccc}
\hline
$\gamma$ & $\delta$ & MSPE & MSPE & Coverage & Coverage & Length & Length \\
& & Asy. EBP& Naive DR& Asy. EBP& Naive DR& Asy. EBP & Naive DR \\
\hline
0.00 & 0.00 & \textbf{0.1418} & 0.2385 & \textbf{0.9399} & 0.9035 & \textbf{1.4212} & 1.7175 \\
0.00 & 0.25 & \textbf{0.1410} & 0.2251 & \textbf{0.9388} & 0.9123 & \textbf{1.4147} & 1.6969 \\
0.00 & 0.50 & \textbf{0.1459} & 0.2464 & \textbf{0.9386} & 0.8995 & \textbf{1.4331} & 1.6891 \\
0.00 & 1.00 & \textbf{0.1663} & 0.3388 & \textbf{0.9379} & 0.8391 & \textbf{1.5191} & 1.6514 \\
0.25 & 0.00 & \textbf{0.1475} & 0.2397 & \textbf{0.9373} & 0.9030 & \textbf{1.4278} & 1.7161 \\
0.25 & 0.25 & \textbf{0.1397} & 0.2207 & \textbf{0.9392} & 0.9158 & \textbf{1.4065} & 1.6904 \\
0.25 & 0.50 & \textbf{0.1419} & 0.2291 & \textbf{0.9395} & 0.9103 & \textbf{1.4188} & 1.6834 \\
0.25 & 1.00 & \textbf{0.1668} & 0.3174 & \textbf{0.9356} & 0.8410 & \textbf{1.5197} & 1.6389 \\
0.50 & 0.00 & \textbf{0.1578} & 0.2458 & \textbf{0.9366} & 0.8985 & \textbf{1.4631} & 1.7289 \\
0.50 & 0.25 & \textbf{0.1481} & 0.2265 & \textbf{0.9383} & 0.9118 & \textbf{1.4328} & 1.6971 \\
0.50 & 0.50 & \textbf{0.1464} & 0.2295 & \textbf{0.9389} & 0.9086 & \textbf{1.4311} & 1.7157 \\
0.50 & 1.00 & \textbf{0.1687} & 0.3080 & \textbf{0.9374} & 0.8632 & \textbf{1.5287} & 1.6688 \\
1.00 & 0.00 & \textbf{0.1583} & 0.2359 & \textbf{0.9382} & 0.9053 & \textbf{1.4829} & 1.7179 \\
1.00 & 0.25 & \textbf{0.1585} & 0.2377 & \textbf{0.9392} & 0.9050 & \textbf{1.4756} & 1.7001 \\
1.00 & 0.50 & \textbf{0.1540} & 0.2340 & \textbf{0.9399} & 0.9077 & \textbf{1.4721} & 1.6835 \\
1.00 & 1.00 & \textbf{0.1758} & 0.3018 & \textbf{0.9389} & 0.8652 & \textbf{1.5640} & 1.6663 \\
\hline
\end{tabular}
\end{table}

\begin{table}[ht]
\centering
\caption{Unknown-$N_i$ Factorial Simulation Results: Equal-$n_i$ Design with $n_i=40$}
\label{tab:unknown-N-factorial-equal-n40}

\setlength{\tabcolsep}{4.4pt}
\renewcommand{\arraystretch}{1.05}
\begin{tabular}{cccccccc}
\hline
$\gamma$ & $\delta$ & MSPE & MSPE & Coverage & Coverage & Length & Length \\
& & Asy. EBP & Naive DR & Asy. EBP & Naive DR & Asy. EBP & Naive DR \\
\hline
0.00 & 0.00 & \textbf{0.0859} & 0.1181 & \textbf{0.9460} & 0.9264 & \textbf{1.1250} & 1.2585 \\
0.00 & 0.25 & \textbf{0.0822} & 0.1080 & \textbf{0.9470} & 0.9312 & \textbf{1.1078} & 1.2191 \\
0.00 & 0.50 & \textbf{0.0867} & 0.1208 & \textbf{0.9471} & 0.9182 & \textbf{1.1311} & 1.2181 \\
0.00 & 1.00 & \textbf{0.1054} & 0.1609 & \textbf{0.9448} & 0.8687 & \textbf{1.2370} & 1.2738 \\
0.25 & 0.00 & \textbf{0.0841} & 0.1087 & \textbf{0.9466} & 0.9333 & \textbf{1.1185} & 1.2432 \\
0.25 & 0.25 & \textbf{0.0826} & 0.1074 & \textbf{0.9483} & 0.9317 & \textbf{1.1137} & 1.2207 \\
0.25 & 0.50 & \textbf{0.0882} & 0.1216 & \textbf{0.9438} & 0.9220 & \textbf{1.1281} & 1.2218 \\
0.25 & 1.00 & \textbf{0.1045} & 0.1584 & \textbf{0.9431} & 0.8730 & \textbf{1.2248} & 1.2445 \\
0.50 & 0.00 & \textbf{0.0913} & 0.1174 & \textbf{0.9432} & 0.9254 & \textbf{1.1464} & 1.2637 \\
0.50 & 0.25 & \textbf{0.0924} & 0.1188 & \textbf{0.9424} & 0.9218 & \textbf{1.1310} & 1.2241 \\
0.50 & 0.50 & \textbf{0.0871} & 0.1145 & \textbf{0.9450} & 0.9236 & \textbf{1.1257} & 1.2342 \\
0.50 & 1.00 & \textbf{0.1037} & 0.1507 & \textbf{0.9449} & 0.8806 & 1.2279 & \textbf{1.2079} \\
1.00 & 0.00 & \textbf{0.0890} & 0.1092 & \textbf{0.9463} & 0.9327 & \textbf{1.1450} & 1.2465 \\
1.00 & 0.25 & \textbf{0.0928} & 0.1110 & \textbf{0.9434} & 0.9293 & \textbf{1.1403} & 1.2286 \\
1.00 & 0.50 & \textbf{0.0954} & 0.1141 & \textbf{0.9442} & 0.9267 & \textbf{1.1456} & 1.2300 \\
1.00 & 1.00 & \textbf{0.1066} & 0.1478 & \textbf{0.9447} & 0.8779 & 1.2401 & \textbf{1.1938} \\
\hline
\end{tabular}
\end{table}

\begin{table}[ht]
\centering
\caption{Unknown-\(N_i\) Factorial Simulation Results: Unequal-\(n_i\) Design; Overall Averaging over All 100 Small Areas across Different Sample-size Groups.}
\label{tab:unknown-N-factorial-variable-overall}
\setlength{\tabcolsep}{4.4pt}
\renewcommand{\arraystretch}{1.05}
\begin{tabular}{cccccccc}
\hline
$\gamma$ & $\delta$ & MSPE & MSPE & Coverage & Coverage & Length & Length \\
 &  & Asy. EBP & Naive DR & Asy. EBP & Naive DR & Asy. EBP & Naive DR \\
\hline
0.00 & 0.00 & \textbf{0.1201} & 0.2197 & \textbf{0.9437} & 0.9118 & \textbf{1.2762} & 1.5302 \\
0.00 & 0.25 & \textbf{0.1233} & 0.2305 & \textbf{0.9431} & 0.9031 & \textbf{1.2946} & 1.5503 \\
0.00 & 0.50 & \textbf{0.1261} & 0.2380 & \textbf{0.9424} & 0.9013 & \textbf{1.2976} & 1.5514 \\
0.00 & 1.00 & \textbf{0.1421} & 0.2969 & \textbf{0.9409} & 0.8546 & \textbf{1.3794} & 1.5168 \\
0.25 & 0.00 & \textbf{0.1225} & 0.2225 & \textbf{0.9425} & 0.9097 & \textbf{1.2783} & 1.5247 \\
0.25 & 0.25 & \textbf{0.1214} & 0.2241 & \textbf{0.9428} & 0.9102 & \textbf{1.2677} & 1.5198 \\
0.25 & 0.50 & \textbf{0.1240} & 0.2244 & \textbf{0.9413} & 0.9084 & \textbf{1.2841} & 1.5166 \\
0.25 & 1.00 & \textbf{0.1395} & 0.2884 & \textbf{0.9429} & 0.8556 & \textbf{1.3642} & 1.4943 \\
0.50 & 0.00 & \textbf{0.1292} & 0.2270 & \textbf{0.9409} & 0.9067 & \textbf{1.3030} & 1.5345 \\
0.50 & 0.25 & \textbf{0.1261} & 0.2141 & \textbf{0.9432} & 0.9143 & \textbf{1.2945} & 1.5460 \\
0.50 & 0.50 & \textbf{0.1250} & 0.2258 & \textbf{0.9448} & 0.9088 & \textbf{1.2952} & 1.5028 \\
0.50 & 1.00 & \textbf{0.1386} & 0.2843 & \textbf{0.9424} & 0.8612 & \textbf{1.3577} & 1.4880 \\
1.00 & 0.00 & \textbf{0.1340} & 0.2281 & \textbf{0.9433} & 0.9073 & \textbf{1.3292} & 1.5383 \\
1.00 & 0.25 & \textbf{0.1233} & 0.2087 & \textbf{0.9440} & 0.9187 & \textbf{1.2856} & 1.5102 \\
1.00 & 0.50 & \textbf{0.1319} & 0.2240 & \textbf{0.9414} & 0.9090 & \textbf{1.3056} & 1.5159 \\
1.00 & 1.00 & \textbf{0.1522} & 0.2920 & \textbf{0.9394} & 0.8582 & \textbf{1.4030} & 1.5145 \\
\hline
\end{tabular}
\end{table}

\begin{table}[ht]
\centering
\caption{\small Unknown-$N_i$ Factorial Simulation Results: Unequal-$n_i$ Design, Areas with $n_i=10$}
\label{tab:unknown-N-factorial-variable-n10}
\setlength{\tabcolsep}{4.4pt}
\renewcommand{\arraystretch}{1.05}
\begin{tabular}{cccccccc}
\hline
$\gamma$ & $\delta$ & MSPE & MSPE & Coverage & Coverage & Length & Length \\
 & & Asy. EBP & Naive DR & Asy. EBP & Naive DR & Asy. EBP & Naive DR \\
\hline
0.00 & 0.00 & \textbf{0.2113} & 0.4973 & \textbf{0.9386} & 0.8576 & \textbf{1.7281} & 2.2814 \\
0.00 & 0.25 & \textbf{0.2077} & 0.4727 & \textbf{0.9391} & 0.8788 & \textbf{1.7232} & 2.3166 \\
0.00 & 0.50 & \textbf{0.2118} & 0.4984 & \textbf{0.9375} & 0.8752 & \textbf{1.7234} & 2.3440 \\
0.00 & 1.00 & \textbf{0.2301} & 0.6289 & \textbf{0.9359} & 0.8138 & \textbf{1.7869} & 2.1999 \\
0.25 & 0.00 & \textbf{0.2156} & 0.4959 & \textbf{0.9374} & 0.8582 & \textbf{1.7396} & 2.2822 \\
0.25 & 0.25 & \textbf{0.2094} & 0.4827 & \textbf{0.9391} & 0.8691 & \textbf{1.7084} & 2.2801 \\
0.25 & 0.50 & \textbf{0.2140} & 0.4842 & \textbf{0.9353} & 0.8712 & \textbf{1.7238} & 2.2826 \\
0.25 & 1.00 & \textbf{0.2378} & 0.6526 & \textbf{0.9384} & 0.7922 & \textbf{1.8181} & 2.1417 \\
0.50 & 0.00 & \textbf{0.2339} & 0.5163 & \textbf{0.9303} & 0.8449 & \textbf{1.7671} & 2.2806 \\
0.50 & 0.25 & \textbf{0.2101} & 0.4603 & \textbf{0.9405} & 0.8844 & \textbf{1.7465} & 2.3302 \\
0.50 & 0.50 & \textbf{0.2156} & 0.4938 & \textbf{0.9412} & 0.8704 & \textbf{1.7598} & 2.2723 \\
0.50 & 1.00 & \textbf{0.2362} & 0.6524 & \textbf{0.9378} & 0.8001 & \textbf{1.8225} & 2.2334 \\
1.00 & 0.00 & \textbf{0.2323} & 0.5017 & \textbf{0.9393} & 0.8600 & \textbf{1.8159} & 2.2836 \\
1.00 & 0.25 & \textbf{0.2251} & 0.4689 & \textbf{0.9371} & 0.8750 & \textbf{1.7629} & 2.2655 \\
1.00 & 0.50 & \textbf{0.2364} & 0.5071 & \textbf{0.9334} & 0.8644 & \textbf{1.7884} & 2.3156 \\
1.00 & 1.00 & \textbf{0.2653} & 0.6714 & \textbf{0.9326} & 0.8060 & \textbf{1.8884} & 2.2935 \\
\hline
\end{tabular}
\end{table}

\begin{table}[ht]
\centering
\caption{\small Unknown-$N_i$ Factorial Simulation Results: Unequal-$n_i$ Design, Areas with $n_i=20$}
\label{tab:unknown-N-factorial-variable-n20}
\setlength{\tabcolsep}{4.4pt}
\renewcommand{\arraystretch}{1.05}
\begin{tabular}{cccccccc}
\hline
$\gamma$ & $\delta$ & MSPE & MSPE & Coverage & Coverage & Length & Length \\
 & & Asy. EBP & Naive DR & Asy. EBP & Naive DR & Asy. EBP & Naive DR \\
\hline
0.00 & 0.00 & \textbf{0.1409} & 0.2246 & \textbf{0.9400} & 0.9107 & \textbf{1.4215} & 1.7047 \\
0.00 & 0.25 & \textbf{0.1373} & 0.2219 & \textbf{0.9413} & 0.9198 & \textbf{1.4132} & 1.7316 \\
0.00 & 0.50 & \textbf{0.1393} & 0.2227 & \textbf{0.9417} & 0.9205 & \textbf{1.4129} & 1.7339 \\
0.00 & 1.00 & \textbf{0.1686} & 0.3173 & \textbf{0.9395} & 0.8491 & \textbf{1.5454} & 1.7067 \\
0.25 & 0.00 & \textbf{0.1453} & 0.2404 & \textbf{0.9401} & 0.9026 & \textbf{1.4370} & 1.7141 \\
0.25 & 0.25 & \textbf{0.1387} & 0.2229 & \textbf{0.9422} & 0.9169 & \textbf{1.4126} & 1.7140 \\
0.25 & 0.50 & \textbf{0.1424} & 0.2283 & \textbf{0.9402} & 0.9099 & \textbf{1.4285} & 1.6866 \\
0.25 & 1.00 & \textbf{0.1691} & 0.3067 & \textbf{0.9393} & 0.8620 & \textbf{1.5377} & 1.8061 \\
0.50 & 0.00 & \textbf{0.1457} & 0.2287 & \textbf{0.9430} & 0.9103 & \textbf{1.4592} & 1.7139 \\
0.50 & 0.25 & \textbf{0.1378} & 0.2136 & \textbf{0.9460} & 0.9242 & \textbf{1.4292} & 1.7334 \\
0.50 & 0.50 & \textbf{0.1442} & 0.2372 & \textbf{0.9456} & 0.9083 & \textbf{1.4526} & 1.6802 \\
0.50 & 1.00 & \textbf{0.1769} & 0.3306 & \textbf{0.9348} & 0.8634 & \textbf{1.5441} & 1.7516 \\
1.00 & 0.00 & \textbf{0.1691} & 0.2520 & \textbf{0.9374} & 0.8959 & \textbf{1.5047} & 1.7369 \\
1.00 & 0.25 & \textbf{0.1395} & 0.2138 & \textbf{0.9436} & 0.9184 & \textbf{1.4351} & 1.6910 \\
1.00 & 0.50 & \textbf{0.1443} & 0.2205 & \textbf{0.9451} & 0.9134 & \textbf{1.4576} & 1.6936 \\
1.00 & 1.00 & \textbf{0.1625} & 0.2821 & \textbf{0.9424} & 0.8758 & \textbf{1.5381} & 1.7358 \\
\hline
\end{tabular}
\end{table}

\begin{table}[ht]
\centering
\caption{\small Unknown-$N_i$ Factorial Simulation Results: Unequal-$n_i$ Design, Areas with $n_i=40$}
\label{tab:unknown-N-factorial-variable-n40}
\setlength{\tabcolsep}{4.4pt}
\renewcommand{\arraystretch}{1.05}
\begin{tabular}{cccccccc}
\hline
$\gamma$ & $\delta$ & MSPE & MSPE & Coverage & Coverage & Length & Length \\
 & & Asy. EBP & Naive DR & Asy. EBP & Naive DR & Asy. EBP & Naive DR \\
\hline
0.00 & 0.00 & \textbf{0.0834} & 0.1105 & \textbf{0.9459} & 0.9328 & \textbf{1.1142} & 1.2469 \\
0.00 & 0.25 & \textbf{0.0868} & 0.1254 & \textbf{0.9458} & 0.9236 & \textbf{1.1264} & 1.2557 \\
0.00 & 0.50 & \textbf{0.0865} & 0.1256 & \textbf{0.9455} & 0.9237 & \textbf{1.1209} & 1.2449 \\
0.00 & 1.00 & \textbf{0.1050} & 0.1612 & \textbf{0.9425} & 0.8837 & \textbf{1.2208} & 1.2939 \\
0.25 & 0.00 & \textbf{0.0857} & 0.1094 & \textbf{0.9440} & 0.9328 & \textbf{1.1187} & 1.2439 \\
0.25 & 0.25 & \textbf{0.0926} & 0.1404 & \textbf{0.9402} & 0.9082 & \textbf{1.1236} & 1.2074 \\
0.25 & 0.50 & \textbf{0.0899} & 0.1269 & \textbf{0.9413} & 0.9198 & \textbf{1.1253} & 1.2201 \\
0.25 & 1.00 & \textbf{0.1028} & 0.1561 & \textbf{0.9439} & 0.8682 & 1.2163 & \textbf{1.1916} \\
0.50 & 0.00 & \textbf{0.0918} & 0.1162 & \textbf{0.9420} & 0.9279 & \textbf{1.1453} & 1.2643 \\
0.50 & 0.25 & \textbf{0.0957} & 0.1200 & \textbf{0.9419} & 0.9285 & \textbf{1.1242} & 1.2418 \\
0.50 & 0.50 & \textbf{0.0973} & 0.1291 & \textbf{0.9418} & 0.9154 & \textbf{1.1468} & 1.2147 \\
0.50 & 1.00 & \textbf{0.1046} & 0.1503 & \textbf{0.9454} & 0.8775 & \textbf{1.2369} & 1.2506 \\
1.00 & 0.00 & \textbf{0.0851} & 0.1066 & \textbf{0.9480} & 0.9332 & \textbf{1.1376} & 1.2359 \\
1.00 & 0.25 & \textbf{0.0837} & 0.1035 & \textbf{0.9456} & 0.9348 & \textbf{1.1143} & 1.2184 \\
1.00 & 0.50 & \textbf{0.1047} & 0.1270 & \textbf{0.9376} & 0.9122 & \textbf{1.1533} & 1.2020 \\
1.00 & 1.00 & \textbf{0.1042} & 0.1424 & \textbf{0.9457} & 0.8805 & 1.2388 & \textbf{1.2046} \\
\hline
\end{tabular}
\end{table}

\begin{table}[ht]
\centering
\caption{\small Unknown-$N_i$ Factorial Simulation Results: Unequal-$n_i$ Design, Areas with $n_i=80$}
\label{tab:unknown-N-factorial-variable-n80}
\setlength{\tabcolsep}{4.4pt}
\renewcommand{\arraystretch}{1.05}
\begin{tabular}{cccccccc}
\hline
$\gamma$ & $\delta$ & MSPE & MSPE & Coverage & Coverage & Length & Length \\
 & & Asy. EBP & Naive DR & Asy. EBP & Naive DR & Asy. EBP & Naive DR \\
\hline
0.00 & 0.00 & \textbf{0.0481} & 0.0554 & \textbf{0.9501} & 0.9443 & \textbf{0.8562} & 0.9129 \\
0.00 & 0.25 & \textbf{0.0587} & 0.0968 & \textbf{0.9464} & 0.8896 & 0.8988 & \textbf{0.8687} \\
0.00 & 0.50 & \textbf{0.0651} & 0.0994 & \textbf{0.9450} & 0.8870 & 0.9257 & \textbf{0.8690} \\
0.00 & 1.00 & \textbf{0.0626} & 0.0813 & \textbf{0.9458} & 0.8713 & 0.9496 & \textbf{0.8489} \\
0.25 & 0.00 & \textbf{0.0458} & 0.0517 & \textbf{0.9483} & 0.9437 & \textbf{0.8308} & 0.8796 \\
0.25 & 0.25 & \textbf{0.0452} & 0.0509 & \textbf{0.9496} & 0.9461 & \textbf{0.8267} & 0.8772 \\
0.25 & 0.50 & \textbf{0.0491} & 0.0566 & \textbf{0.9484} & 0.9327 & \textbf{0.8565} & 0.8719 \\
0.25 & 1.00 & \textbf{0.0567} & 0.0667 & \textbf{0.9494} & 0.8958 & 0.9238 & \textbf{0.8983} \\
0.50 & 0.00 & \textbf{0.0483} & 0.0545 & \textbf{0.9480} & 0.9420 & \textbf{0.8530} & 0.8998 \\
0.50 & 0.25 & \textbf{0.0609} & 0.0641 & \textbf{0.9441} & 0.9195 & \textbf{0.8754} & 0.8762 \\
0.50 & 0.50 & \textbf{0.0457} & 0.0517 & \textbf{0.9506} & 0.9400 & \textbf{0.8371} & 0.8693 \\
0.50 & 1.00 & \textbf{0.0552} & 0.0664 & \textbf{0.9502} & 0.8940 & 0.9155 & \textbf{0.8552} \\
1.00 & 0.00 & \textbf{0.0511} & 0.0566 & \textbf{0.9483} & 0.9394 & \textbf{0.8677} & 0.9100 \\
1.00 & 0.25 & \textbf{0.0454} & 0.0503 & \textbf{0.9497} & 0.9463 & \textbf{0.8328} & 0.8705 \\
1.00 & 0.50 & \textbf{0.0459} & 0.0511 & \textbf{0.9492} & 0.9442 & \textbf{0.8398} & 0.8794 \\
1.00 & 1.00 & \textbf{0.0860} & 0.1020 & \textbf{0.9362} & 0.8661 & 0.9833 & \textbf{0.8853} \\
\hline
\end{tabular}
\end{table}

Because treatment assignment is observational in this factorial simulation study, a small number of area-replicates contain sampled observations from only one treatment arm; these non-estimable area contrasts are handled identically for EBP
asymptotic and naive DR. A summary of such cases is provided in Table \ref{tab:unknown-N-factorial-diagnostics}.

\begin{table}[ht]
\centering
\small
\caption{Diagnostics of Unknown-$N_i$ Factorial Simulation Study}
\label{tab:unknown-N-factorial-diagnostics}
\begin{tabular}{lcc}
\hline
Diagnostic & Equal-$n_i$ & Unequal-$n_i$ \\
\hline
Maximum Unavailable Rate & 0.0658 & 0.1608 \\
Total Convergence Warnings & 90 & 296 \\
Failed Fits & 0 & 0 \\
Singular Fits & 0 & 0 \\
99th Percentile Inverse Weight & 167.1365 & 58.8625 \\
Maximum Inverse Weight & 399.2314 & 102.7841 \\
\hline
\end{tabular}

\vspace{2mm}

\begin{minipage}{0.88\textwidth}
\small
{\bf Note:} The unavailable rate is the largest proportion, over the factorial settings, of area-replication pairs in which a sampled area contains only one observed treatment arm, so the area contrast is not estimable for either method. Convergence warnings, failed fits, and singular fits summarize numerical
diagnostics for the fitted mixed models across all factorial settings. The inverse-weight diagnostics summarize the largest realized values of
\(\max\{1/\widehat e_i,1/(1-\widehat e_i)\}\); large values indicate weaker overlap and help explain the instability of naive DR under strong propensity misspecification.
\end{minipage}
\end{table}
\section{Further details for Section 6 of SACI}
\subsection{Data generating processes}
\hspace{4mm}
{\it I. Reluga Data Generating Process (DGP).}
There are $m=41$ areas with $N_i=1000$ units each. The public-code DGP uses
$(X_{1ij},\ldots,X_{10ij})^{\mathsf T}\sim N_{10}(0,\Sigma)$,
$\Sigma_{kk}=1$, $\Sigma_{kl}=0.5$ for $k\ne l$, and an independent area
covariate $X_{11,ij}=X_{s,i}\sim N(0,2.6^2)$, constant within area.
With $c_i^0\sim U(1,2)$ and $c_i^1\sim U(2,3)$, generate
\begin{align}
M_{ijr}&=c_i^r+X_{s,i}+\frac1{10}\sum_{k=1}^{10}
\{X_{kij}\beta_{r1,i,k}+\exp(X_{kij}\beta_{r2,i,k})\},\notag\\
Y_{ij}(r)&=\log|M_{ijr}+\epsilon_{ij}(r)|,\qquad
\epsilon_{ij}(r)\sim N(0,\overline M_r^{\,2}),\quad r=0,1,
\label{eq:reluga-outcome}
\end{align}
where $\overline M_r$ is the mean of $M_{ijr}$ over all population units;
errors are conditionally independent. The area-specific coefficient vectors
and intercepts are drawn as
\begin{align*}
\boldsymbol\beta_{01,i}&\sim N_{10}(0,3^2I),&
\boldsymbol\beta_{02,i}&\sim N_{10}(0,0.3^2I),\\
\boldsymbol\beta_{11,i}&=\boldsymbol\beta_{01,i}+2\boldsymbol v_i,&
\boldsymbol v_i&\sim N_{10}(4\mathbf1,3^2I),\\
\boldsymbol\beta_{12,i}&=\boldsymbol\beta_{02,i}+0.1\boldsymbol t_i,&
\boldsymbol t_i&\sim N_{10}(4\mathbf1,3^2I),
\end{align*}
For treatment assignment, let $\boldsymbol\alpha_i^{\mathrm{PS}}$ have nine
independent $N(2,6^2)$ entries followed by two $N(0,0.25^2)$ entries.
The public-code rule is
$$e_{ij}=\operatorname{logit}^{-1}\!\left(
\left|\frac1{11}\sum_{k=1}^{11}\alpha^{\mathrm{PS}}_{i,\ell(j,k)}X_{kij}\right|\right),\;\;
\ell(j,k)=1+\{(j-1+(k-1)N_i)\bmod11\},$$
$$W_{ij}\sim\operatorname{Bernoulli}(e_{ij}),\;\;
Y_{ij}=W_{ij}Y_{ij}(1)+(1-W_{ij})Y_{ij}(0).$$
The index $\ell$ retains the coefficient indexing in the public implementation;
assignments are independent conditional on covariates and assignment coefficients.
The provided-population comparison fixes the supplied covariates, assignments,
factual outcomes and area truths and draws 1,000 samples. The generated-population
comparison independently redraws the full DGP and draws one sample in each
of 1,000 replications, recalculating truth using (3) of SACI.
The supplied CSV is not assumed to follow exactly this public-code law;
its $X_{11,i}$ values match a standard-normal draw rather than the public-code
standard deviation $2.6$.

SRS40 selects 40 units uniformly without replacement within each area,
independently of existing assignments. Original Reluga stratified sampling
selects units separately within the two treatment groups, using
$n_i^{(1)}=\operatorname{round}(0.02N_i^{(1)})$,
$n_i^{(0)}=\lceil f_i n_i^{(1)}\rceil$,
where $N_i^{(r)}$ is the population arm count and
$f_i\sim U(0.01,0.5)$ for areas 1--25, $U(0.51,1)$ for 26--35 and
$U(0.9,1)$ for 36--41. Zero computed arm counts are replaced by one;
sampling is without replacement within each arm. Both estimators are evaluated using identical samples. Fitted propensity scores are used in the CSAE-AIPW procedure, and inverse-sampling-probability weights are not applied. Results are in raw outcome units; across-row MSPE levels reflect
different populations and sampling allocations.

{II. SACI Data Generating Process.}
The SACI baseline has $m=100$, $N_i=500$, $n_i=40$, $z_{ij}=X_{1ij}\sim N(0,1)$,
and
\begin{align}
Y_{ij}(0)&=1+z_{ij}+\alpha_{i0}+\epsilon_{ij}(0),\notag\\
Y_{ij}(1)&=1+z_{ij}+0.5+0.5z_{ij}+\alpha_{i0}+\alpha_{i1}+\epsilon_{ij}(1),
\label{eq:saci-dgp}
\end{align}
where $\epsilon_{ij}(r)\sim N(0,1)$ and
$G_{00}=1$, $G_{11}=0.5$, $G_{01}=0.3\sqrt{0.5}$.
Covariates are independent across units. Area-effect vectors are independent
$N_2(0,G)$ draws, and residuals are independent across units and potential-outcome
arms. Covariates, area-effect vectors and residuals are mutually independent;
the random intercept and treatment slope have correlation $0.3$ within area.
Each replication generates a new finite population.

Within each area, 40 unit
indices are selected uniformly without replacement; treatment is then assigned
to exactly 20 of these units, chosen uniformly, and the remaining 20 are controls.
Nonsampled assignments are independent Bernoulli$(0.5)$.
This is uniform selection of units followed by forced-balanced treatment
assignment, not treatment-stratified sampling from fixed assignments.

\begingroup
\renewcommand{\labelitemi}{$\bullet$}
\noindent\textbf{1. Outcome modifications.}
Define $q_{ij}=z_{ij}^2-\overline{z^2}$, centered over all population units.
\begin{itemize}
\setlength{\itemsep}{1pt}\setlength{\parsep}{0pt}
\item \textit{Shared quadratic term:}
$Y_{ij}(r)\leftarrow Y_{ij}(r)+\gamma q_{ij}$, $r=0,1$, $\gamma=2$;
the added term cancels from the treatment effect.
\item \textit{Treated-only quadratic term:}
$Y_{ij}(1)\leftarrow Y_{ij}(1)+\delta q_{ij}$, $\delta=2$;
this changes the treatment-effect function. Both quadratic terms are omitted
from the working outcome models.
\item \textit{Heavy-tailed area effects:}
$\boldsymbol{\alpha}_i=\boldsymbol{\xi}_i/\sqrt{Q_i}$,
$\boldsymbol{\xi}_i\sim N_2(0,G)$, $Q_i\sim\chi^2_3$;
covariance remains $G$, with heavier tails and a shared $Q_i$ for both components.
\item \textit{Heavy-tailed residuals:}
$\epsilon_{ij}(r)=\eta_{ij}(r)/\sqrt{Q_{ijr}}$,
$\eta_{ij}(r)\sim N(0,1)$, $Q_{ijr}\sim\chi^2_3$;
residual variance remains one.
\item \textit{Heteroskedastic residuals:}
$\epsilon_{ij}(r)=\sqrt{1+|z_{ij}|}\,\eta_{ij}(r)$,
$\eta_{ij}(r)\sim N(0,1)$, for both arms $r=0,1$.
\end{itemize}
Gaussian draws and mixing variables are independent; residual draws and
mixing variables are independent across units and arms.
Outcome changes retain covariates, assignments and sampled indices within
each replication; truths are recomputed using (3) of SACI.

\smallskip
\noindent\textbf{2. Sampling modifications.}
\begin{itemize}
\setlength{\itemsep}{1pt}\setlength{\parsep}{0pt}
\item \textit{Smaller samples:} $n_i=10$ in every area.
\item \textit{Unequal samples:} $n_i\in\{10,20,40,80\}$, with 25 areas at each size.
\end{itemize}
Retain baseline covariates and potential outcomes. Units are sampled uniformly
without replacement, followed by balanced treatment assignment within each
sampled area; nonsampled assignments are Bernoulli$(0.5)$, as in the baseline.

\smallskip
\noindent\textbf{3. Treatment-assignment modifications.}
\begin{itemize}
\setlength{\itemsep}{1pt}\setlength{\parsep}{0pt}
\item \textit{Covariate-dependent assignment:}
$e_{ij}=\operatorname{logit}^{-1}(\kappa z_{ij})$, $\kappa\in\{1,3\}$;
larger $\kappa$ produces more extreme treatment probabilities.
\item \textit{Area-dependent assignment:}
$e_{ij}=\operatorname{logit}^{-1}(0.5z_{ij}+sZ_i)$,
$Z_i\sim N(0,1)$, $s\in\{1,3\}$;
$Z_i$ is independent of outcome-generating variables and introduces shared
area-level variation of magnitude $s$.
\end{itemize}
Retain baseline potential outcomes and sampled indices, but redraw
$W_{ij}\sim\operatorname{Bernoulli}(e_{ij})$ independently conditional on
the assignment variables, without enforcing balanced sampled arms.
Population XGBoost fits propensity scores using $z_{ij}$ only, omitting
$Z_i$ and area indicators; the area-dependent setting therefore also involves
propensity-model misspecification, not treatment imbalance alone.
All settings use 1,000 paired replications with identical data and samples
for both estimators.
\endgroup

{\it III. SACI--Reluga Outcome Mixture.}
Under a common framework with $m=100$, $N_i=500$ and SRS40, define
\begin{equation}
Y_{ij}^{(\lambda)}(r)=(1-\lambda)S_{ij}(r)+\lambda R_{ij}^{*}(r),
\qquad \lambda\in\{0,0.25,0.5,0.75,1\}.
\label{eq:mixture}
\end{equation}
The component $S_{ij}(r)$ uses the SACI outcome equations in
\eqref{eq:saci-dgp}, with $z_{ij}=X_{1ij}$, Gaussian area effects and
independent Gaussian residuals. The component $R_{ij}(r)$ uses the nonlinear,
log-absolute Reluga public-code outcome law, with its own area coefficients
and residual draws. Before mixing, $R_{ij}^*(r)=a+bR_{ij}(r)$ applies the
same affine transformation to both arms. Independent pilot populations
calibrate $a,b$ to align pooled endpoint means and standard deviations,
so the two components have comparable scales; see Section \ref{app:details} below for more details about the calculations.

Each of 1,000 replications generates a new common covariate population as
in Section~2.1, now with 100 areas of size 500. Treatment follows the
Reluga public-code assignment mechanism adapted to this area size, and
40 units per area are sampled uniformly without replacement; no sampled-arm
balance is imposed. All eleven covariates, assignments, sampled indices
and the two underlying outcome components are shared across $\lambda$.
For each $\lambda$, both potential outcomes are formed by
\eqref{eq:mixture}, factual outcomes are selected by the common $W$,
and the realized area truth is recalculated from (1) of SACI.
Thus $\lambda=0$ uses only SACI outcomes, $\lambda=1$ uses only scaled Reluga
outcomes, and intermediate values mix them. These endpoints share the common
covariates, assignment and sampling framework; they do not reproduce the
complete original benchmarks, which have different population structures
or treatment-assignment schemes.
\subsection{Additional Simulation Details}
\hspace{4mm}
\label{app:details}
Tables 8 and 10 of SACI use XGBoost \texttt{binary:logistic}, learning rate $0.3$, depth 6 and 100 rounds;
Table 9 of SACI uses the package defaults with 100 rounds.
Cross-validation is disabled, and all reported fitted scores lie in $(0,1)$ without clipping. Finite point estimates, including singular mixed-model fits, are retained.

The outcome mixture uses $R^*(r)=a+bR(r)$ with
$a=-0.878754175922069$ and $b=1.76352879671792$.
Thirty independent pilot populations provide pooled arm means and standard
deviations, giving $b=\widehat\sigma_S/\widehat\sigma_R$ and
$a=\widehat\mu_S-b\widehat\mu_R$.
Regenerated Reluga results convert saved fits under $\widetilde Y=a+bY$
to raw units by dividing causal estimates and truths by $b$, and MSPE by
$b^2$; these are not raw-outcome refits.
\section{A real-data example}
\hspace{4mm}
We demonstrate the SACI method using the animal breeding data mentioned in Section 1 (second paragraph) of SACI. Specifically,
we use the lamb body condition, breeding and production data from the Montana Agricultural Experiment Station, Montana State University, Bozeman, MT from 1960–2012 to demonstrate the proposed EBP method. The data is available at
https://doi.org/10.15482/USDA.ADC/1528462

Table \ref{tab:lamb-analysis-sample} provides a summary of the lamb growth data.

\begin{table}[ht]
\centering
\caption{Analysis Samples for Lamb Growth Data}
\label{tab:lamb-analysis-sample}
\begin{tabular}{lr}
\hline
Quantity & Count \\
\hline
Eligible Lambs before Ram-cell Restriction & 25,668 \\
Observed Five-covariate Cells & 221 \\
Retained Ram-containing Cells & 159 \\
Retained Lambs & 25,391 \\
Rams & 10,707 \\
Ewes & 14,684 \\
Cells Containing Both Rams and Ewes & 149 \\
Ram-only Cells & 10 \\
Removed Ewe-only Cells & 62 \\
Lambs Removed by Ram-cell Restriction & 277 \\
\hline
\end{tabular}
\end{table}

We focus on Active lambs (\(\texttt{STATUS}=0\)), whose sex is recorded as Ram (\(\texttt{SEX}=1\)) or Ewe (\(\texttt{SEX}=2\)); in other words, Wether and Unknown sex records are excluded. The
analysis also requires observed birth and weaning measurements and a positive age at weaning. For lamb \(j\) in cell \(i\) (defined below), define $\mathrm{AGE}_{ij}=\mathrm{WEANING\_DAY}_{ij}-\mathrm{BIRTH\_DATE}_{ij}$ and $Y_{ij}=(\mathrm{WEANING\_WT}_{ij}-\mathrm{BIRTH\_WT}_{ij})/\mathrm{AGE}_{ij}$. We refer to \(Y_{ij}\) as average weight improvement at weaning (AWIAW).

Cells are defined using five categorical covariates: \(\texttt{BREED}\),
\(\texttt{LINE}\), grouped \(\texttt{EWE\_AGE}\), grouped \(\texttt{BORN}\),
and grouped \(\texttt{REARING}\). The variable \(\texttt{MOTHERING\_SCORE}\) is excluded because it is unknown for approximately 94\% of otherwise eligible lambs. The value \(\texttt{BREED}=0\) is retained as Rambouillet according to the data dictionary. For the retained categorical covariates, zero, missing, blank, and unsupported values are coded as an additional Unknown category when appropriate. The provisional groupings are as follows. For \(\texttt{EWE\_AGE}\), Young corresponds to age 2, Prime to ages 3--4, Older to ages 5 and older, and Unknown otherwise. For \(\texttt{BORN}\), categories are 1, 2, 3+, and Unknown. For \(\texttt{REARING}\), categories are 1, 2, 3+, and Unknown. For \(\texttt{LINE}\), categories are Low, High, and Unknown. Table 24 of the supplement summarizes the resulting analysis samples. In summary, there are 221 observed five-covariate cells before restricting to Ram-containing cells, that is, we retain only cells containing at least one Ram because the working model includes a Ram-specific cell random effect. This removes 62 Ewe-only cells, corresponding to 277 lambs, and leaves 159 retained cells. Among these, 149 cells contain both Rams and Ewes and 10 are Ram-only cells. In addition, 65 retained cells have fewer than five sampled rams, motivating the use of EBP for borrowing strength.

A potential causal-effect factor is SEX (\(r=0\) for Ram and \(r=1\) for Ewe). The working LMM (see Section 3) can be expressed as $Y_{ij}(r)=x_{i}'(\beta_{0}+r\beta_{1})+\alpha_{i0}+r\alpha_{i1}+\epsilon_{ij}(r)$ with the bivariate, cell-specific, random effects $\alpha_{i}=(\alpha_{i0},\alpha_{i1})'$ distributed as in Section 3. Here $x_i$ contains main-effect indicators for the five
cell-defining categorical variables. The random intercept \(\alpha_{i0}\) is shared by Rams and Ewes in the same cell, while
\(\alpha_{i1}\) is the cell-specific deviation in the Ewe-minus-Ram contrast. The model is fitted by REML. A Phase 1 audit of this corrected model preserves the previously verified data
construction. The fitted model still has 32 fixed-effect columns and rank 32. It converged without optimizer warnings and was nonsingular, with
\(\widehat\sigma_0^2=0.000438\), \(\widehat\sigma_1^2=0.0000528\),
\(\widehat\rho=-0.9918\), and \(\widehat\sigma_\epsilon^2=0.011980\). The estimated correlation coefficient between $\alpha_{i0}$ and $\alpha_{i1}$ is very close to negative $-1$. This suggests that the working model is nearly the same as a reduced model, which replaces $\alpha_{i0}+r\alpha_{i1}$ by $(1-r)\alpha_{i}$ with a univariate $\alpha_{i}\sim N(0,\sigma_{\alpha}^{2})$. A sensitivity analysis was carried out and the results are presented in Section 5.1 of the supplement. The results suggest that the cell-specific asymptotic EBP is relatively insensitive to the near-boundary estimate of $\rho$. The reduced model is also analyzed and the prediction results, and interpretations (see below), are essentially the same, which are deferred to Section 5.2 of the supplement.

\begin{figure}[ht]
\centering
\includegraphics[width=0.88\textwidth]{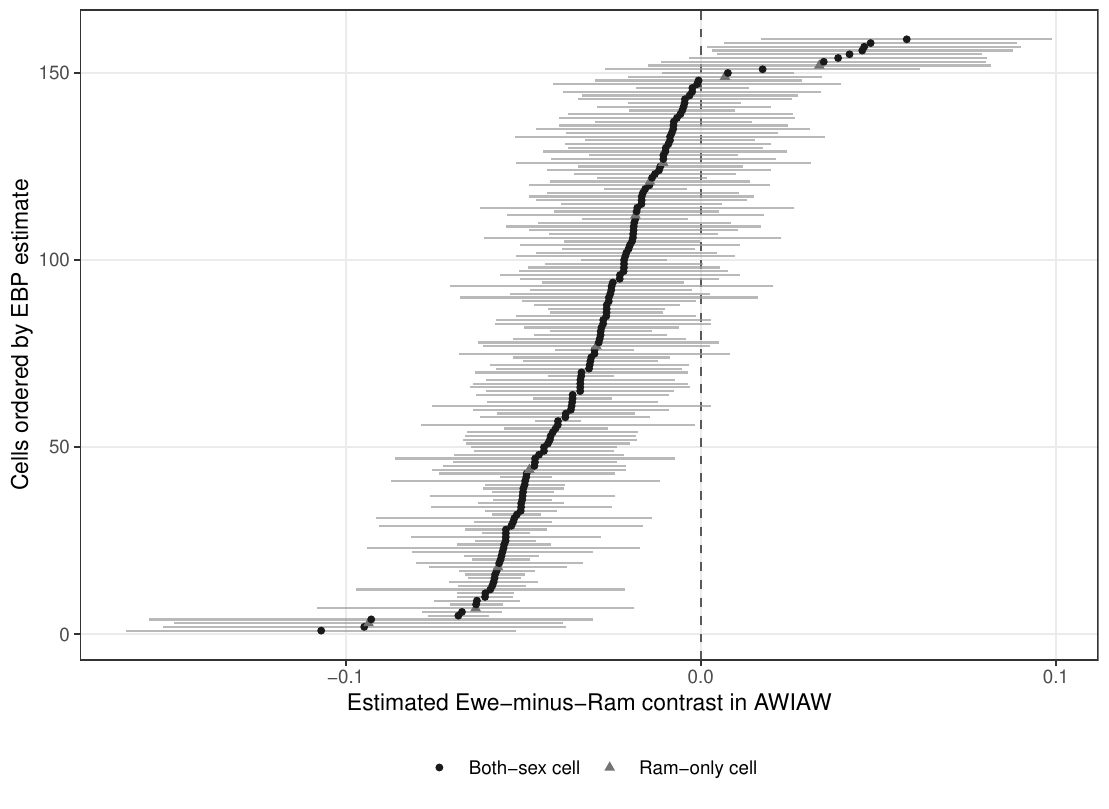}
\caption{Cell-specific 95\% Prediction Intervals for Ewe-minus-Ram Causal Effects based on Asymptotic EBP; Cells Ordered by Values of Asymptotic EBP, including Ram-and-Ewe Cells and Ram-only Cells.}
\label{fig:lamb-section3-forest-ebp}
\end{figure}

\begin{figure}[ht]
\centering
\includegraphics[width=0.88\textwidth]{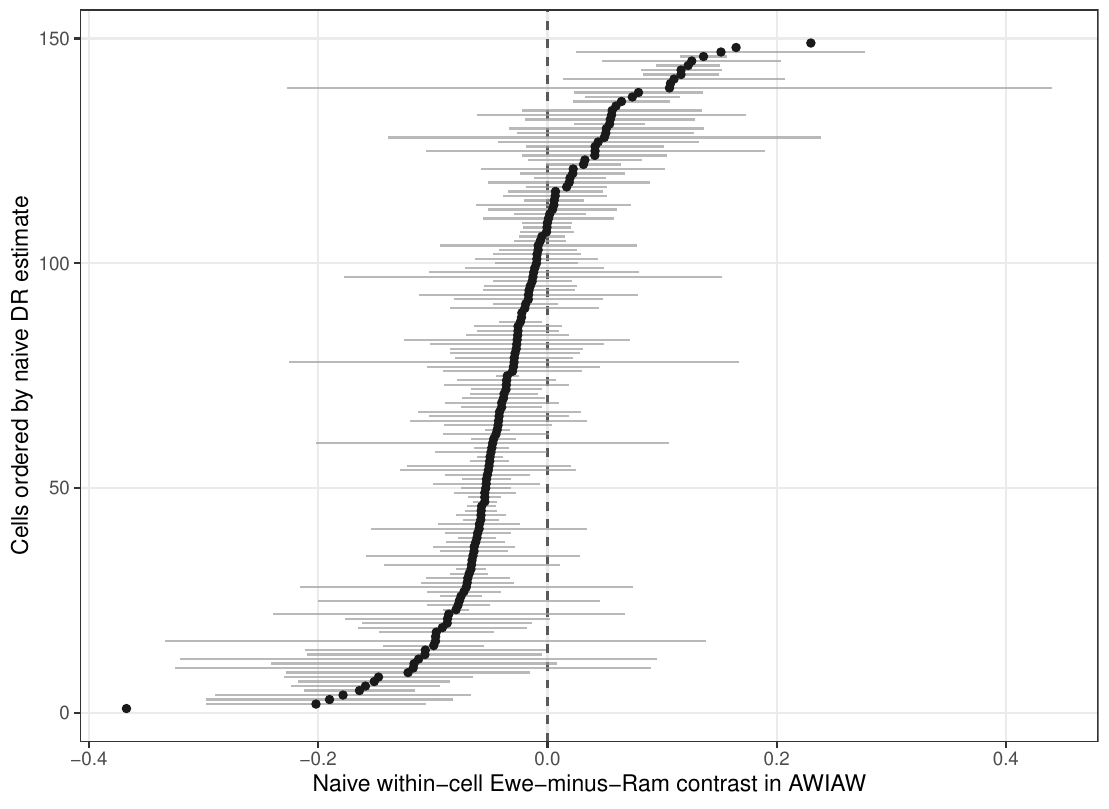}
\caption{Cell-specific 95\% Prediction Intervals for Ewe-minus-Ram Causal Effects based on Naive DR; Cells Ordered by Values of Naive DR Estimates; Panel Includes Ram-and-Ewe Cells Only.}
\label{fig:lamb-section3-forest-naive}
\end{figure}

\begin{figure}[ht]
\centering
\includegraphics[width=0.84\textwidth]{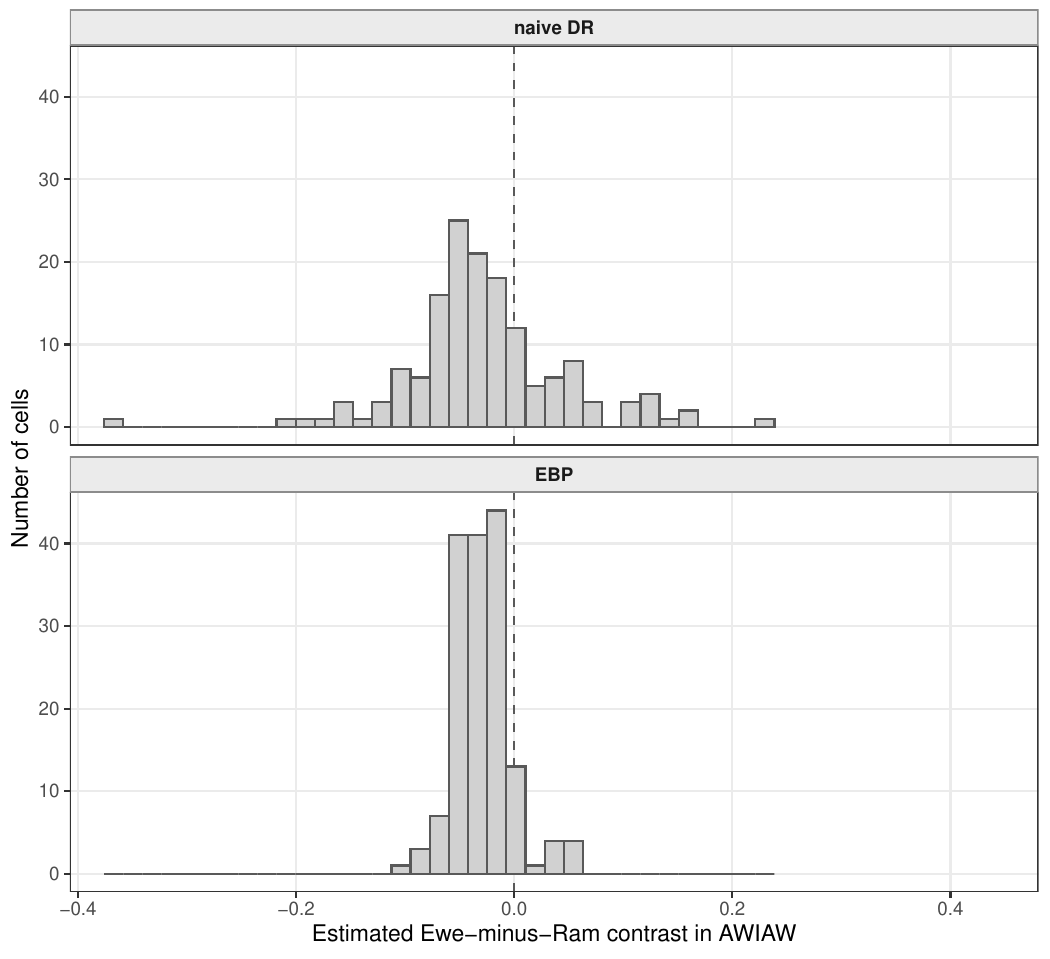}
\caption{Histograms of Asymptotic EBP and Naive DR Estimates of Ewe-minus-Ram Causal Effects: Naive DR Panel Includes Ram-and-Ewe Cells Only; Asymptotic EBP Panel Includes Ram-and-Ewe Cells and Ram-only Cells.}
\label{fig:lamb-section3-distribution}
\end{figure}

\begin{figure}[ht]
\centering
\includegraphics[width=0.84\textwidth]{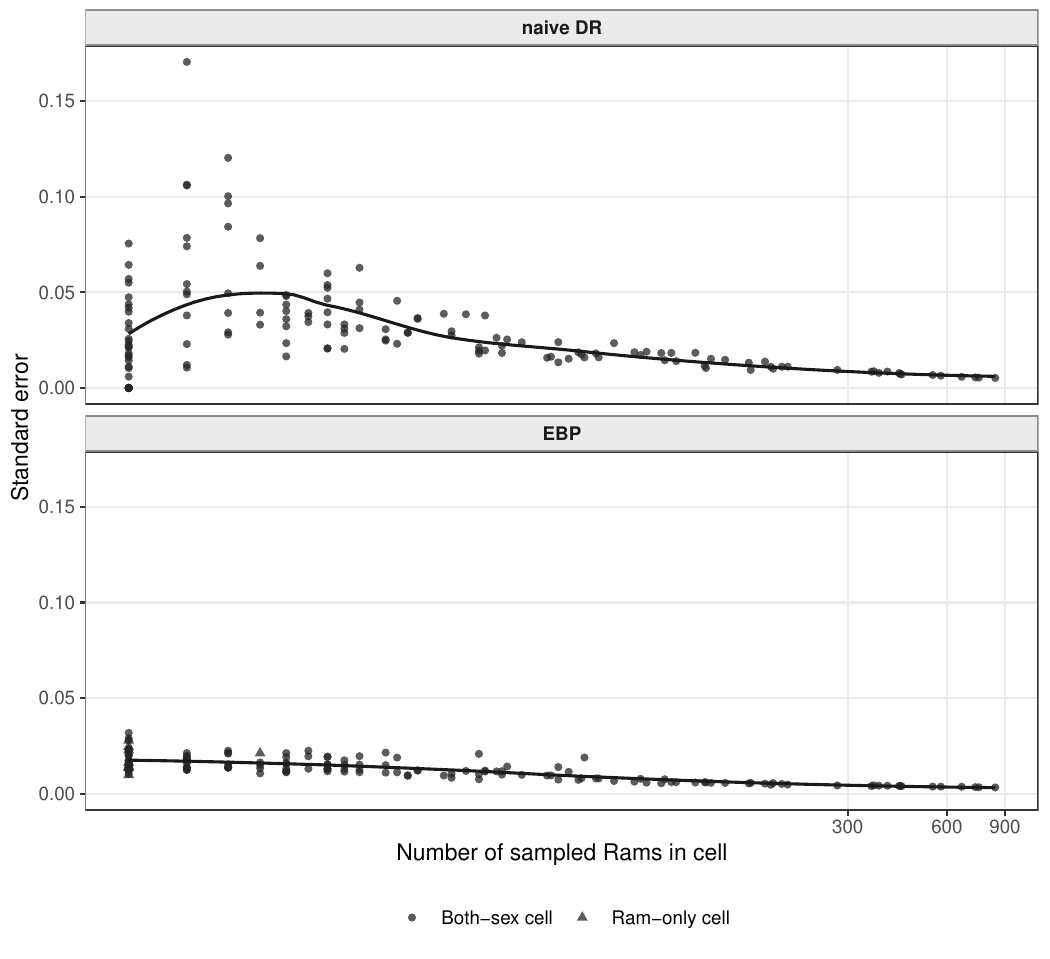}
\caption{Plots of Standard Error vs Number of Sampled Rams: Naive DR and Asymptotic EBP; Asymptotic EBP Panel Includes All Retained Cells; Naive DR Panel Includes Ram-and-Ewe Cells Only.}
\label{fig:lamb-section3-se-ram-count}
\end{figure}

Our primary interest is prediction of the cell-specific causal-effects. Note that $N_{i}$, the subpopulation size, is unknown; thus, the partial EBP method does not apply (see Section 5, fourth paragraph). On the other hand, the asymptotic EBP method (see Section 4.5) and the naive DR method are applicable. In view of the traditional interest in causal inference, here, we focus on prediction intervals. Figures \ref{fig:lamb-section3-forest-ebp} presents the cell-specific 95\% prediction intervals based on asymptotic EBP; the corresponding prediction intervals based on naive DR are presented in Figure \ref{fig:lamb-section3-forest-naive}. It is seen that the asymptotic EBP prediction intervals are, in general, significantly shorter than the naive DR ones (note that the ranges in the x-axis of the two figures are different). This is consistent with our simulation study in Section 5.2, which show that the Length of asymptotic EBP is significantly shorter than that of naive DR. The pattern is further illustrated by Figure \ref{fig:lamb-section3-distribution}, which shows histograms of the naive DR estimates and asymptotic EBP, and Figure \ref{fig:lamb-section3-se-ram-count}, which plots the s.e. of the naive DR estimates and asymptotic EBP against the number of sampled rams in different cells. It is seen that the naive DR estimates have much larger variation than the asymptotic EBP; the s.e.'s of naive DR are not only larger than those of the asymptotic EBP, more significantly, they are more variable than the s.e.'s of the asymptotic EBP.

Summary statistics for the asymptotic EBP and naive DR estimates are presented in Table \ref{tab:lamb-model-ebp-summary}. It is seen that the predicted values are predominantly negative, indicating lower predicted AWIAW for Ewe than for Ram for most retained cells. The asymptotic EBP has significantly smaller mean s.e and mean length of prediction interval (PI), and higher number of ``significant'' PIs (i.e., excluding zero) than the naive DR estimate.
\begin{table}[ht]
\centering
\caption{Summary Statistics for Asymptotic EBP and Naive DR Prediction Intervals}
\label{tab:lamb-model-ebp-summary}
\begin{tabular}{lrr}
\hline
Quantity & Asymp. EBP & naive DR \\
\hline
Available Cells & 159 & 149 \\
Unavailable Cells & 0 & 10 \\
Mean& -0.0291 & -0.0267 \\
Median & -0.0282 & -0.0350 \\
SD & 0.0268 & 0.0748 \\
Minimum & -0.1069 & -0.3673 \\
Maximum & 0.0580 & 0.2298 \\
Mean SE & 0.0132 & 0.0299 \\
Median SE & 0.0130 & 0.0231 \\
Mean length of PI& 0.0516 & 0.1174 \\
PIs Excluding Zero & 95 & 71 \\
\hline
\end{tabular}
\end{table}

Table~\ref{tab:lamb-covariance-sensitivity} summarizes targeted sensitivity diagnostics. The asymptotic EBP is stable under an independent-random-effects specification and when the analysis is restricted to the 149 cells containing both Rams and Ewes. The EBPs based on the full and independent random-effect models have a Pearson
correlation of 0.9810, with mean absolute difference 0.0040 and maximum absolute difference 0.0163. The EBP estimates from the full 159-cell fit and the both-sex-only
refit have a Pearson correlation of 0.9984, with maximum absolute difference 0.0045. Thus, the cell-specific EBP point predictions are relatively insensitive to the near-boundary estimate of the random-effect correlation.

\begin{table}[ht]
\centering
\caption{Covariance-structure Sensitivity Diagnostics for the Lamb Growth Example.}
\label{tab:lamb-covariance-sensitivity}
\begin{tabular}{lr}
\hline
Quantity & Value \\
\hline
\(\widehat\sigma_0^2\) & 0.000438 \\
\(\widehat\sigma_1^2\) & 0.0000528 \\
\(\widehat\rho\) & -0.9918 \\
\(\widehat\sigma_\epsilon^2\) & 0.011980 \\
Full vs. independent EBP Pearson correlation & 0.9810 \\
Full vs. independent mean absolute difference & 0.0040 \\
Full vs. independent maximum absolute difference & 0.0163 \\
Full vs. both-sex-only EBP Pearson correlation & 0.9984 \\
Full vs. both-sex-only maximum absolute difference & 0.0045 \\
Parametric-bootstrap singular refits & 141 / 200 \\
Mean analytical SE & 0.0132 \\
Mean bootstrap SE & 0.0083 \\
\hline
\end{tabular}
\end{table}

We also carried out an analysis under the bivariate cell-level random-effects model. For a lamb in cell $i$, we use the same bivariate cell-level
random-effects structure as in Section 3 of SACI:
\begin{equation}
Y_{ij}(r)
=
\mathbf{x}_i^{\prime}(\mathbf{\beta}_0+r\mathbf{\beta}_1)
+
\alpha_{i0}
+
r\alpha_{i1}
+
\epsilon_{ij}(r),
\qquad r=0,1,
\end{equation}
where $r=0$ denotes Ram and $r=1$ denotes Ewe. The cell-specific
random effects satisfy
\begin{equation}
\begin{pmatrix}
\alpha_{i0}\\
\alpha_{i1}
\end{pmatrix}
\stackrel{\mathrm{ind}}{\sim}
N\left[
\begin{pmatrix}
0\\
0
\end{pmatrix},
\begin{pmatrix}
\sigma_0^2 & \rho\sigma_0\sigma_1\\
\rho\sigma_0\sigma_1 & \sigma_1^2
\end{pmatrix}
\right],
\end{equation}
and
\[
\epsilon_{ij}(r)
\stackrel{\mathrm{ind}}{\sim}
N(0,\sigma_\epsilon^2).
\] Here, $\mathbf{x}_i$ contains the main-effect indicators for the
cell-defining categorical variables. The sex-by-covariate terms allow
the fixed covariate associations to differ between Rams and Ewes.
The random intercept $\alpha_{i0}$ represents the cell-specific
baseline deviation, whereas the random slope $\alpha_{i1}$ represents
the cell-specific random deviation in the Ewe-minus-Ram contrast.
The outcome model is fitted by REML. The fitted model has 32 fixed-effect columns and rank 32; it converged without warnings and was non-singular. The estimated
variance components are
\(\widehat\sigma_0^2=0.000438\),
\(\widehat\sigma_1^2=0.0000528\),
\(\widehat\rho=-0.9918\), and
\(\widehat\sigma_\epsilon^2=0.011980\).

Next, we proceed to obtain the EBP of the cell-specific
Ewe-minus-Ram AWIAW contrasts and the associated uncertainty measures. For cell $i$, the working-model cell means are
$\mu_i(0)=\mathbf{x}_i^{\prime}\mathbf{\beta}_0+\alpha_{i0}$
for Rams and $\mu_i(1)=
\mathbf{x}_i^{\prime}(\mathbf{\beta}_0+\mathbf{\beta}_1)
+\alpha_{i0}+\alpha_{i1}$
for Ewes. Hence, the cell-specific Ewe-minus-Ram AWIAW contrast is
\begin{equation}
d_i
=
\mu_i(1)-\mu_i(0)
=
\mathbf{x}_i^{\prime}\mathbf{\beta}_1+\alpha_{i1}.
\end{equation}
In particular, the random intercept $\alpha_{i0}$ cancels in the
Ewe-minus-Ram contrast. The corresponding empirical best predictor is
\begin{equation}
\widehat d_i
=
\mathbf{x}_i^{\prime}\widehat{\mathbf{\beta}}_1
+
\widehat{\alpha}_{i1}.
\end{equation}
Throughout this real-data application, we interpret $d_i$ as a
cell-specific model-based Ewe-minus-Ram AWIAW contrast rather than as
the causal effect of manipulating biological sex. Negative values
correspond to lower model-predicted AWIAW for Ewes than for Rams within
the corresponding cell.

The observed records are treated as a sample from a large population.
The complete finite-population cell sizes $N_i$ are unavailable, so we
use the asymptotic EBP described in Section 4.5 of SACI. Because $N_i$
is unknown, the partial EBP discussed in Section 5 of SACI does not
apply to this application.

There are 159 retained cells and 25,391 retained lambs, consisting of
10,707 Rams and 14,684 Ewes. Of the 159 cells, 149 contain observations
from both Rams and Ewes, whereas 10 contain Rams only. The asymptotic
EBP is available for all 159 retained cells. In contrast, naive DR is
available only for the 149 cells containing observations from both
sexes, because a direct within-cell Ewe-minus-Ram contrast cannot be
computed for the 10 Ram-only cells.

The mean analytical standard error for the EBP is $0.0132$, compared
with $0.0299$ for naive DR. The corresponding mean lengths of the
pointwise 95\% intervals are $0.0516$ and $0.1174$, respectively.
These intervals are interpreted as pointwise model-based uncertainty
summaries.

The fitted random-effect covariance is close to the boundary, with
$\widehat{\rho}=-0.9918$. We therefore examined the sensitivity of the
cell-specific predictions to alternative covariance specifications.
The EBP estimates based on the full correlated random-effects model
and the independent random-effects model have a Pearson correlation of
$0.9810$, with a mean absolute difference of $0.0040$ and a maximum
absolute difference of $0.0163$. The EBP estimates from the full
159-cell fit and the refit restricted to the 149 cells containing both
Rams and Ewes have a Pearson correlation of $0.9984$, with a maximum
absolute difference of $0.0045$. Thus, although the estimated
random-effect covariance is close to the boundary, the cell-specific
EBP point estimates are relatively stable under these sensitivity
checks.

As an additional sensitivity analysis, we performed a parametric
bootstrap with $B=200$ replicates under the fitted bivariate working
model. There were no failed bootstrap fits and no missing bootstrap
standard errors. However, 141 of the 200 bootstrap refits were
classified as singular. The mean analytical standard error was
$0.0132$, whereas the mean bootstrap standard error was $0.0083$.

These bootstrap results should not be interpreted as validation of the
analytical standard errors. Rather, together with the near-boundary
estimate $\widehat{\rho}=-0.9918$, they indicate that uncertainty
quantification is more sensitive to the fitted random-effect
covariance structure than point estimation. We therefore retain the
analytical intervals as pointwise model-based uncertainty summaries
and use the bootstrap results primarily as a sensitivity diagnostic.

Overall, the real-data application is most informative as an
illustration of cell-specific point prediction and borrowing of
strength. The EBP point estimates are stable under the
covariance-structure sensitivity analyses, whereas interval estimation
is more sensitive because the fitted random-effect covariance is close
to the boundary. 

\begin{figure}[ht]
\centering
\includegraphics[width=0.86\textwidth]{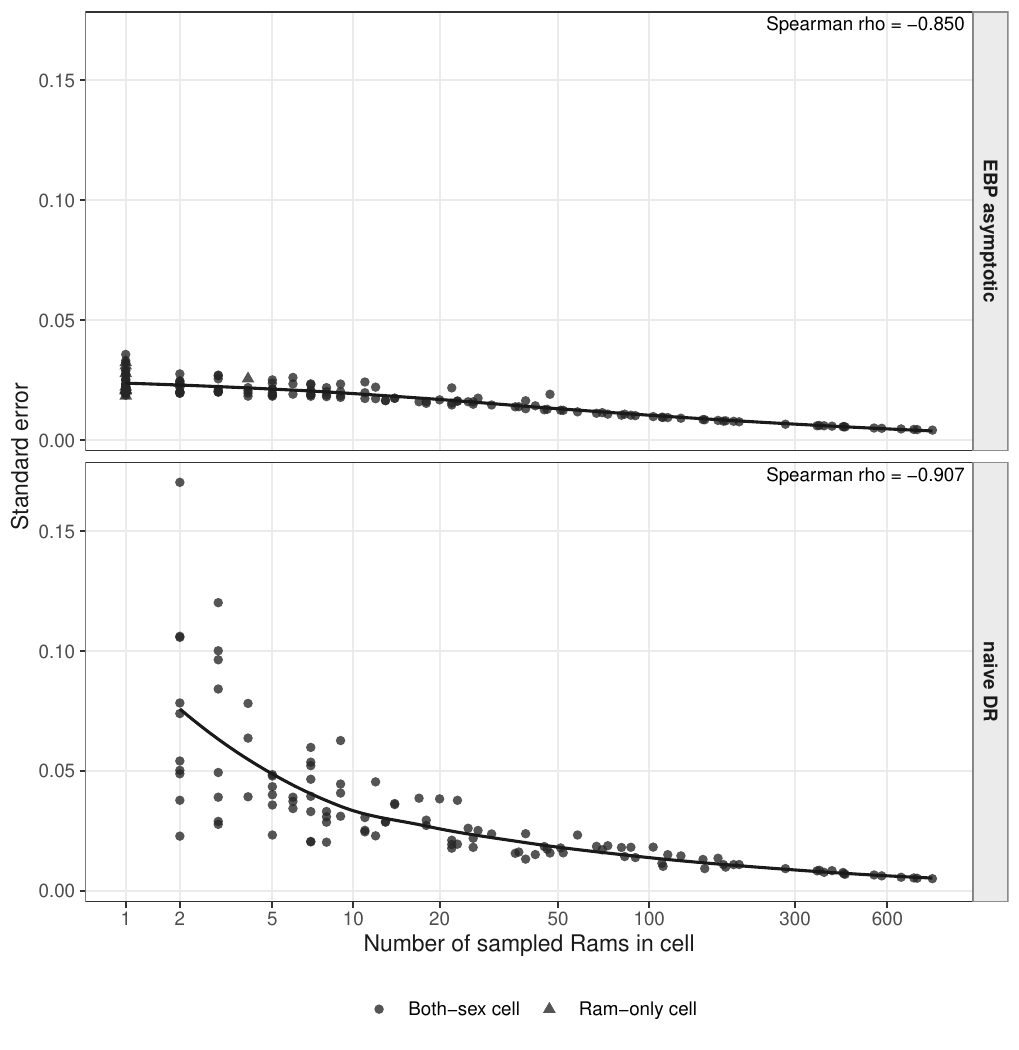}
\caption{Plots of Standard Error vs Number of Sampled Rams: Asymptotic EBP and Naive DR; Asymptotic EBP Panel Includes All Retained Cells; Naive DR Panel Includes Ram-and-Ewe Cells Only.}
\label{fig:lamb-se-ram-count}
\end{figure}


\section{Extensions and concluding remarks}
\hspace{4mm}
Extension of the EBP approach to SACI to more than two categories of treatment is fairly straightforward. Let $Y_{ij}(r), r=0,1,\dots,R$, $1\leq i\leq m, 1\leq j\leq N_{i}$ be the potential outcomes, where $R\geq 1$ is the total number of different treatments, and $r=0$ corresponds to the control. When $R=1$, this is what we have considered. To extend the approach to $R>1$, define $W_{ij}^{c}=\{0,1,\dots,R\}\setminus\{W_{ij}\}$. The observed outcomes are, again, $Y_{ij}=Y_{ij}(W_{ij})$, $(i,j)\in{\cal S}$; the unobserved outcomes are $Y_{ij}(r)$, $r=0,1,\dots,R$, $(i,j)\in{\cal P}\setminus{\cal S}$ and $Y_{ij}(r),r\in W_{ij}^{c}$, $(i,j)\in{\cal S}$. The area-specific causal-effects are
\begin{equation}
d_{ir}=\frac{1}{N_{i}}\sum_{j=1}^{N_{i}}\{Y_{ij}(r)-Y_{ij}(0)\},\;\;\;r=1,\dots,R.\label{eq:ace_ext}
\end{equation}
The LMM introduced in Section 3 can be extended as
\begin{eqnarray}
Y_{ij}(0)&=&x_{ij}'\beta_{0}+\alpha_{i0}+\epsilon_{ij}(0),\nonumber\\
Y_{ij}(r)&=&x_{ij}'(\beta_{0}+\beta_{r})+\alpha_{i0}+\alpha_{ir}+\epsilon_{ij}(r),\;\;1\leq r\leq R.\label{eq:lmm_ext1}
\end{eqnarray}
Model (\ref{eq:lmm_ext1}) can be written in one equation, that is,
\begin{equation}
Y_{ij}(r)=x_{ij}'\beta_{0}+\alpha_{i0}+\sum_{k=1}^{R}1_{(k=r)}(x_{ij}'\beta_{k}+\alpha_{ik})+\epsilon_{ij}(r),\;\;r=0,1,\dots,R.
\label{eq:lmm_ext2}
\end{equation}
This is a LMM involving fixed-effect vectors $\beta_{0},\beta_{1},\dots,\beta_{R}$ and vector-valued random effects $\alpha_{i}=(\alpha_{ir})_{0\leq r\leq R}$. The BP and EBP of the area-specific causal-effects, (\ref{eq:ace_ext}), can be developed similarly as in Sections 2--4. The details are deferred to a future publication.

As another extension, one may allow part of the linear predictors in the LMM, (equation (7)), to be free of the causal-effect. An easy way to make this happen is to set some components of $\beta_{1}$ to be zero. Without loss of generality, let $x_{ij}=(x_{ij,1}',x_{ij,2}')'$ and, correspondingly, $\beta_{r}=(\beta_{r,1},\beta_{r,2})'$, $r=0,1$. If we assume that $\beta_{1,1}=0$, the LMM (7) can be written as
\begin{equation}
Y_{ij}(r)=x_{ij,1}'\beta_{0,1}+x_{ij,2}'(\beta_{0,2}+r\beta_{1,2})+\alpha_{i0}+r\alpha_{i1}+\epsilon_{ij}(r).\label{eq:lmm_ext_2}
\end{equation}
The meaning of (\ref{eq:lmm_ext2}) is clear: $x_{ij,1}'\beta_{0,1}$ is the part of the linear predictor not relevant to the causal-effect, and $x_{ij,1}'(\beta_{0,2}+r\beta_{1,2})$ is the part that is relevant to the causal-effect. The development of the EBP under this extended setting can be simply adopted from the previous development by setting $\beta_{1,1}=0$ (thus, the parameters in $\beta_{1,1}$ do not need to be estimated, because they are known to be zero).

The simulation results show smaller MSPE for EBP in most studied settings and shorter prediction intervals. Partial EBP has smaller MSPE in some settings with stronger outcome misspecification. On the other hand, prediction interval based on the partial EBP method performs better in terms of the coverage probability. It should be noted that more accurate coverage probability does not necessarily mean higher accuracy of the associated SE. For example, in the Section 4.5.2 of the supplement, the average bootstrap SE is closer to the empirical SD of prediction error, but this comparison alone does not establish greater accuracy of the area-specific MSE estimates. Bootstrap coverage is approximately 2.2--3.2 percentage points lower than plug-in coverage. With only 10 bootstrap replications per dataset, uncertainty from the bootstrap approximation is an additional limitation.